\documentclass{article} 
\usepackage{iclr2027_conference,times}

\usepackage{amsmath,amsfonts,bm}

\def\eqref#1{equation~\ref{#1}}

\def\1{\bm{1}}

\DeclareMathAlphabet{\mathsfit}{\encodingdefault}{\sfdefault}{m}{sl}
\SetMathAlphabet{\mathsfit}{bold}{\encodingdefault}{\sfdefault}{bx}{n}

\DeclareMathOperator*{\argmin}{arg\,min}

\usepackage{hyperref}
\usepackage{url}
\usepackage{xcolor}
\usepackage{tabularx}
\usepackage{booktabs}
\usepackage{graphicx}
\usepackage{bm}
\usepackage{amsmath}
\usepackage{amssymb}
\usepackage{subcaption}
\usepackage{wrapfig}

\usepackage{listings}
\usepackage[most]{tcolorbox}   

\definecolor{trSysBg}{HTML}{F4F1EC}\definecolor{trSysLn}{HTML}{8A7E6B}
\definecolor{trUsrBg}{HTML}{EAF1F8}\definecolor{trUsrLn}{HTML}{3E6EA5}
\definecolor{trAstBg}{HTML}{EDF6EE}\definecolor{trAstLn}{HTML}{4C8A54}
\definecolor{trCodeBg}{HTML}{F7F7F4}\definecolor{trCodeFr}{HTML}{D9D9D2}
\definecolor{trKw}{HTML}{9B2D30}\definecolor{trStr}{HTML}{2F7D4F}
\definecolor{trCmt}{HTML}{8A8A8A}\definecolor{trRule}{HTML}{222222}

\newcommand{\mdhead}{\color{trRule}}   
\lstdefinelanguage{yaml}{
  keywords={true,false,null,none,True,False,None},
  keywordstyle=\color{trKw}\bfseries, sensitive=true,
  comment=[l]{\#}, morestring=[b]", morestring=[b]',
}
\lstdefinestyle{codeblock}{
  backgroundcolor=\color{trCodeBg}, basicstyle=\ttfamily\footnotesize,
  keywordstyle=\color{trKw}\bfseries, stringstyle=\color{trStr},
  commentstyle=\color{trCmt}\itshape, showstringspaces=false,
  breaklines=true, breakatwhitespace=false,
  postbreak=\mbox{\textcolor{trCodeFr}{$\hookrightarrow$}\space},
  columns=fullflexible, keepspaces=true, frame=single,
  rulecolor=\color{trCodeFr}, framexleftmargin=3pt,
  xleftmargin=4pt, xrightmargin=2pt, aboveskip=4pt, belowskip=2pt,
  literate={—}{{---}}1 {–}{{--}}1 {’}{{'}}1 {‘}{{'}}1 {“}{{``}}1 {”}{{''}}1
           {×}{{$\times$}}1 {→}{{$\rightarrow$}}1 {≈}{{$\approx$}}1
           {…}{{\ldots}}1 {°}{{$^\circ$}}1
}
\newenvironment{tightitem}
  {\begin{list}{\textbullet}{\setlength{\leftmargin}{1.4em}%
     \setlength{\itemsep}{0pt}\setlength{\parsep}{1pt}\setlength{\topsep}{2pt}}}
  {\end{list}}
\tcbset{trbase/.style={breakable, enhanced, sharp corners=downhill,
    boxrule=0.6pt, left=6pt, right=6pt, top=4pt, bottom=5pt,
    fonttitle=\bfseries\footnotesize\color{white}, coltitle=white,
    attach boxed title to top left={xshift=6pt, yshift=-2pt},
    boxed title style={sharp corners, boxrule=0pt, arc=1pt,
      left=4pt, right=4pt, top=1pt, bottom=1pt},
    before skip=6pt, after skip=6pt, fontupper=\small}}
\newtcolorbox{sysbox}[1]{trbase, colback=trSysBg, colframe=trSysLn, colbacktitle=trSysLn, title={#1}}
\newtcolorbox{userbox}[1]{trbase, colback=trUsrBg, colframe=trUsrLn, colbacktitle=trUsrLn, title={#1}}
\newtcolorbox{asstbox}[1]{trbase, colback=trAstBg, colframe=trAstLn, colbacktitle=trAstLn, title={#1}}

\definecolor{dscolor}{HTML}{1A5276}

\newcommand{\dsname}[1]{{\ttfamily\bfseries\color{dscolor}#1}}

\newcommand{\HighSpeedFlow}{\dsname{HighSpeedFlow}}      
\newcommand{\RealData}{\dsname{RealData}}                
\newcommand{\MixedParam}{\dsname{MixedParam}}            
\newcommand{\Heterogeneous}{\dsname{Heterogeneous}}      
\newcommand{\Partial}{\dsname{Partial}}                  
\newcommand{\MixedDomain}{\dsname{MixedDomain}}          
\newcommand{\MultiPhysics}{\dsname{MultiPhysics}}        
\newcommand{\ThreeD}{\dsname{3D}}                        
\newcommand{\IrregularTime}{\dsname{IrregularTime}}      
\newcommand{\LagDynamic}{\dsname{LagInletOutlet}}            
\newcommand{\LagFSI}{\dsname{LagFSI}}                    
\newcommand{\Extreme}{\dsname{Extreme}}                  
\newcommand{\MixedRes}{\dsname{MixedRes}}                
\newcommand{\ExtLagOne}{\dsname{ExtLag1}}                
\newcommand{\ExtLagTwo}{\dsname{ExtLag2}}                
\newcommand{\MixedDim}{\dsname{MixedDim}}                
\newcommand{\MixedBC}{\dsname{MixedBC}}                  
\newcommand{\LongHorizon}{\dsname{LongHorizon}}          
\newcommand{\Stochastic}{\dsname{Stochastic}}            
\newcommand{\MultiPhase}{\dsname{MultiPhase}}            
\newcommand{\MixedPDE}{\dsname{MixedPDE}}                
\newcommand{\LargeLag}{\dsname{LargeLag}}                
\newcommand{\SimToReal}{\dsname{Sim2Real}}              
\newcommand{\HighDimensional}{\dsname{HighDimensional}}  
\newcommand{\ComplexGeometry}{\dsname{ComplexGeometry}}  

\newcommand{\pub}{\textcolor{green!55!black}{Public}}
\newcommand{\ours}{\textcolor{blue!70!black}{Ours}}

\usepackage{colortbl}
\definecolor{taxfill}{HTML}{DCE6F0}
\newcommand{\taxmark}{\cellcolor{taxfill}$\checkmark$}

\title{AutoPDEBench: Benchmarking LLM Auto-Research for Neural PDE Solver Design}

\author{Ruoyan Li, Wei Wang, Yizhou Sun  \\
Department of Computer Science\\
University of California, Los Angeles\\
}

\iclrfinalcopy 
\begin{document}

\maketitle

\begin{abstract}
    Partial differential equations (PDEs) are essential for modeling complex physical systems, and neural solvers have recently emerged as powerful data-driven tools for numerically solving them. However, existing neural solvers struggle with domain-specific challenges, such as varying parameters and high-speed flows, necessitating specialized architectures. Manually designing these specialized solver architectures is a highly iterative, time-consuming process requiring deep expertise, creating a significant bottleneck in scientific discovery. We propose leveraging autonomous AI research agents to automate the synthesis of specialized solvers. To support this, we introduce AutoPDEBench, a benchmark dedicated to LLM-driven automated research for PDE solver design. The benchmark includes 25 challenging datasets 
    featuring both novel and actively studied physical scenarios. We evaluate a suite of general-purpose models (transformer, ROM, and graph-based) alongside a multi-agent instantiation of the iterative automated research pipeline, which serves as an agentic baseline. Empirical results show that the iterative automated research system significantly outperforms the general-purpose neural solver baselines. Our findings demonstrate the viability of using AI agents to automatically design neural solvers for complex physical systems. AutoPDEBench provides a foundational testbed to accelerate agent-driven scientific discovery in physics and engineering. Code and datasets are fully open-sourced at \href{https://github.com/RuoyanLi2002/AutoPDEBench-Benchmarking-LLM-Auto-Research-for-Neural-PDE-Solver-Design}{Github}.
\end{abstract}

\section{Introduction}
Partial differential equations (PDEs) are the foundational language of physics. They provide the essential mathematical framework to describe how complex systems evolve over space and time. Recent advancements have seen neural solvers demonstrate remarkable performance in learning PDE solutions purely from data. While these models achieve strong results across many standard datasets, others present domain-specific challenges that necessitate specialized architectures. For example, when evaluating datasets with mixed PDE parameters, standard neural solvers cannot reliably map initial conditions to target trajectories; the lack of explicit parameter inputs renders the predictions ill-posed. Encountering this exact scenario, \citet{takamoto2023learningneuralpdesolvers} recognized that solvers must explicitly encode the underlying PDE parameters. To address this, they introduced a parameter-guided channel attention mechanism that conditions the neural solver on these varying physical coefficients. High-speed flows present another domain-specific challenge. Because these systems involve extreme gradients and discontinuities, such as shock waves, existing neural solvers typically suffer from numerical instability or non-physical smoothing. To overcome this, \citet{helwig2026twophasedeeplearningframework} introduced NeuralCFL, a framework that explicitly predicts an adaptive timestep size to stabilize the simulation during rapid state changes. 

Traditionally, human researchers invest a substantial amount of time designing these domain-specific solvers, a process that demands deep specialized expertise and significant computational resources. Because this manual engineering is highly iterative and uniquely tailored to each physical system, it severely limits the pace and scalability of scientific discovery. This critical bottleneck motivates the development of autonomous AI research agents capable of automatically synthesizing specialized solver pipelines. By dynamically adapting to complex physical dynamics, these agents promise to accelerate the modeling process and reduce the reliance on human intervention.

To this end, we introduce AutoPDEBench, a benchmark dedicated to LLM-driven automated research for designing specialized solvers to tackle domain-specific challenges. AutoPDEBench comprises 25 challenging datasets, encompassing both actively studied and entirely novel physical scenarios. These datasets present distinct, idiosyncratic challenges that general-purpose neural solvers either struggle with or fail to handle entirely, thereby necessitating specialized architectures. To establish robust baselines, we first evaluate a suite of versatile general-purpose models, including transformer-based, reduced-order modeling (ROM), and graph-based methods. We then introduce a multi-agent instantiation of the iterative automated research pipeline to serve as an agentic baseline. Our empirical results demonstrate that this auto-research system achieves superior performance compared to the general solver baselines, underscoring the immense potential of utilizing AI agents to automatically design neural solvers for complex physical systems. However, outperforming general-purpose baselines does not mean the task is fully solved. In classical numerical analysis, the search for schemes with minimal local truncation error is an ongoing pursuit, as per-step errors inevitably accumulate over long time horizons and are often amplified by nonlinear dynamics. Neural solvers face this exact vulnerability, with single-step prediction errors compounding severely during autoregressive rollouts. We therefore position AutoPDEBench as an open-ended testbed designed to continually challenge agents to discover increasingly accurate solvers.

Our contributions are as follows:
\begin{itemize}
    \item \textbf{Problem Identification:} We highlight that the domain of neural PDE solvers encompasses a diverse array of physical systems characterized by idiosyncratic challenges. Traditionally, addressing these unique constraints has demanded extensive manual effort to design specialized architectures.
    \item \textbf{Proposed Benchmark:} We introduce AutoPDEBench, the first comprehensive benchmark featuring 25 challenging datasets with unique physical constraints. This collection spans both actively researched problems and entirely novel domains introduced to the community for the first time.
    \item \textbf{Empirical Analysis:} We conduct extensive evaluations comparing versatile, general-purpose neural PDE solvers against a multi-agent instantiation of the iterative automated research pipeline. Our results demonstrate the profound potential of utilizing LLM agents to autonomously design specialized solver pipelines for complex physical systems.
\end{itemize}

\begin{figure}[t]
    \centering
    \includegraphics[width=0.95\textwidth]{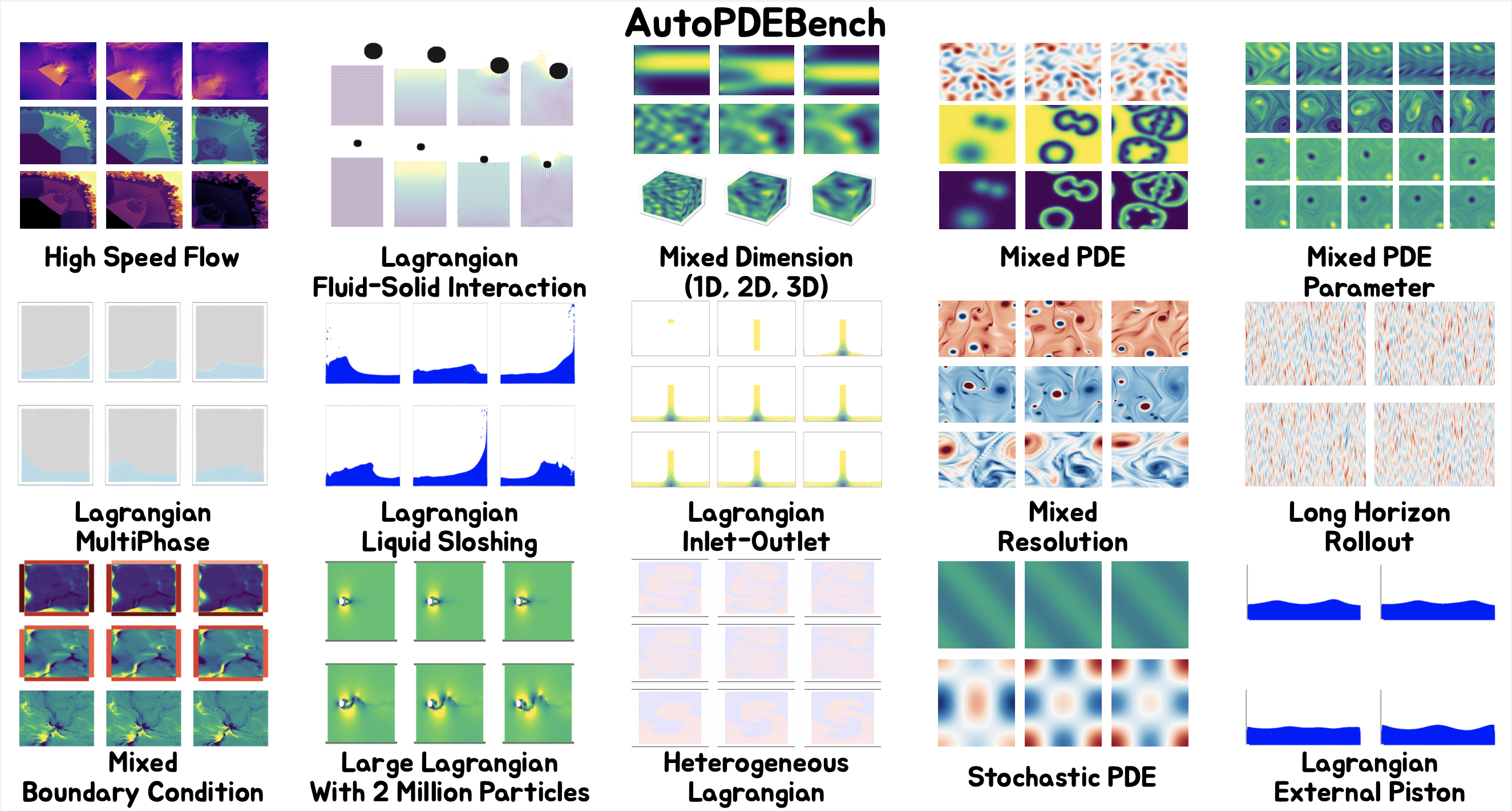}
    \caption{Overview of the example datasets proposed by AutoPDEBench. AutoPDEBench consists of 25 challenging datasets, each presenting specific domain challenges and necessitating specialized solvers. It features both entirely novel and currently understudied datasets. We intentionally reduced the resolution of all figures to maintain a reasonable file size. High-quality versions are available in our GitHub repository. 
    }
    \label{fig:dataset}
\end{figure}

\section{Related Work}
We identify three key areas closely related to our work.

\textbf{Neural PDE Solvers. } The evolution of neural PDE solvers originated with early architectures designed to learn solution operators on regular Euclidean grids~\citep{lu2021learning, li2021fourier, li2020multipolegraphneuraloperator, kovachki2023neuraloperator, rahman2023uno, tripura2023wavelet, tran2023factorized}. To overcome the geometric limitations of uniform grid-based representations, subsequent research rapidly expanded model capabilities to accommodate unstructured meshes~\citep{pfaff2021learning, boussif2022magnet, li2023geometryinformed, han2022learning, lino2021simulatingcontinuummechanicsmultiscale} and Lagrangian particle-based simulations~\citep{sanchezgonzalez2020learningsimulatecomplexphysics, toshev2024neuralsphimprovedneural, toshev2023learning, prantl2022guaranteed, rochman-sharabi2025a}. As these methodologies matured, the community introduced comprehensive benchmark suites~\citep{wang2026fdbenchmodularfairbenchmark, toshev2024lagrangebench, gupta2022towards, ohana2024well, koehler2024apebench, takamoto2024pdebenchextensivebenchmarkscientific} to establish standardized evaluation strategies and curate large-scale datasets across a wide variety of physical systems. Building upon these established baselines, recent solvers have sought to push performance boundaries across multiple complex axes. These advancements include architectures capable of generalizing across varying PDE parameters~\citep{takamoto2023learningneuralpdesolvers, brandstetter2022message} and diverse boundary conditions~\citep{horie2022physicsembedded, wang2024beno, saad2023guiding}, handling highly complex geometries~\citep{wu2024Transolver, luotransolver++, wen2025goat, mousavi2025rigno}, adapting seamlessly across 1D, 2D, and 3D spatial dimensions~\citep{chen2025omniarch}, and unifying Eulerian field and Lagrangian particle representations~\citep{alkin2024upt, alkin2025abupt}. Most recently, this trajectory has culminated in the exploration of physics foundation models~\citep{hao2024dpot, mccabe2024multiple, ye2024pdeformer, herde2024poseidon, shen2024ups, zhouunisolver, wang2025mixtureofexpertsoperatortransformerlargescale}, which are pre-trained on massive, heterogeneous datasets to achieve strong zero-shot and few-shot generalization capabilities across unseen dynamic regimes.

Beyond forward temporal simulation, neural network methodologies have been widely adopted across adjacent domains in scientific machine learning. These broader applications include solving inverse problems~\citep{li2025flow, zheng2025inversebench, molinaro2023neuralinverseoperatorssolving, li2025selfguideddiffusionmodelaccelerating, wang2024LNO} and developing hybrid frameworks that integrate neural networks directly into classical numerical schemes to enhance their performance and mitigate truncation errors~\citep{list2022learned, sun2023neural, greenfeld2019learning, sappl2019deep}.


\textbf{LLM for PDE Solvers. } Researchers have extensively utilized LLMs to autonomously generate traditional numerical PDE solvers~\citep{li2025codepde, wu2025automatedcodedevelopmentpde, hang2026pdeagentbenchmultimetricmultilibrarybenchmark, dong2026autopdereliableagenticpde}. These approaches typically translate high-level mathematical formulations and boundary conditions directly into executable simulation scripts. This automation significantly reduces the manual programming effort and tedious debugging traditionally required for complex scientific computing. Recent studies have extended this paradigm to neural architectures. For example, \citet{wuwu2025pinnsagent, song2026can} leverage LLM agents to design neural solvers; however, their evaluations remain confined to standard benchmark datasets. Our work diverges significantly by deploying LLM agents to tackle complex datasets characterized by unique physical constraints that require domain-specific solvers. Because our benchmark introduces entirely novel, previously unstudied physical systems, the LLM cannot simply retrieve memorized templates from its pre-training data. Instead, the agent must autonomously identify the underlying physical bottlenecks and synthesize novel architectural solutions to overcome them.


\textbf{LLM Auto Research. } The integration of Large Language Models (LLMs) into the scientific workflow has evolved along a spectrum of increasing autonomy. Initially, research focused on utilizing LLMs as assistive tools to augment human effort in specific, localized tasks, such as conducting literature reviews~\citep{agarwal2025litllmtoolkitscientificliterature, gao2025sciencehierarchographyhierarchicalorganization, skarlinski2024languageagentsachievesuperhuman, Go2025LiRAAM}, generating preliminary ideas~\citep{yang2025moosechem, qiu2025aiideabench2025, ghafarollahi2024sciagents}, and assisting with experimental planning and iteration~\citep{ferreira2026llmsbeatclassicalhyperparameter, guo2026autollmresearchtrainingresearchagents}. As their capabilities expanded, LLMs transitioned into more autonomous roles, functioning as analytical agents capable of complex information processing with minimal human intervention. In this capacity, they have been deployed for automated machine learning task modeling~\citep{Huang2023MLAgentBenchEL, Li2024MLRCopilotAM, chan2025mlebench} and the identification of governing equations from observational data~\citep{chen2026physgym, merler-etal-2024-context}. Most recently, the field has advanced toward the LLM scientists paradigm. These systems operate as general-purpose, active agents capable of orchestrating the entire scientific discovery process end-to-end~\citep{mitchener2025kosmosaiscientistautonomous, novikov2025alphaevolvecodingagentscientific, Brixi2026, ghareeb2025robinmultiagentautomatingscientific, tang2025airesearcherautonomousscientificinnovation, schmidgall2025agentlaboratoryusingllm}. By independently navigating from novel ideation and autonomous experimental execution to synthesizing genuine findings into full research paper drafts, these pipelines represent a significant step toward fully autonomous natural science discovery.

\begin{figure}[t]
    \centering
    \includegraphics[width=0.95\textwidth]{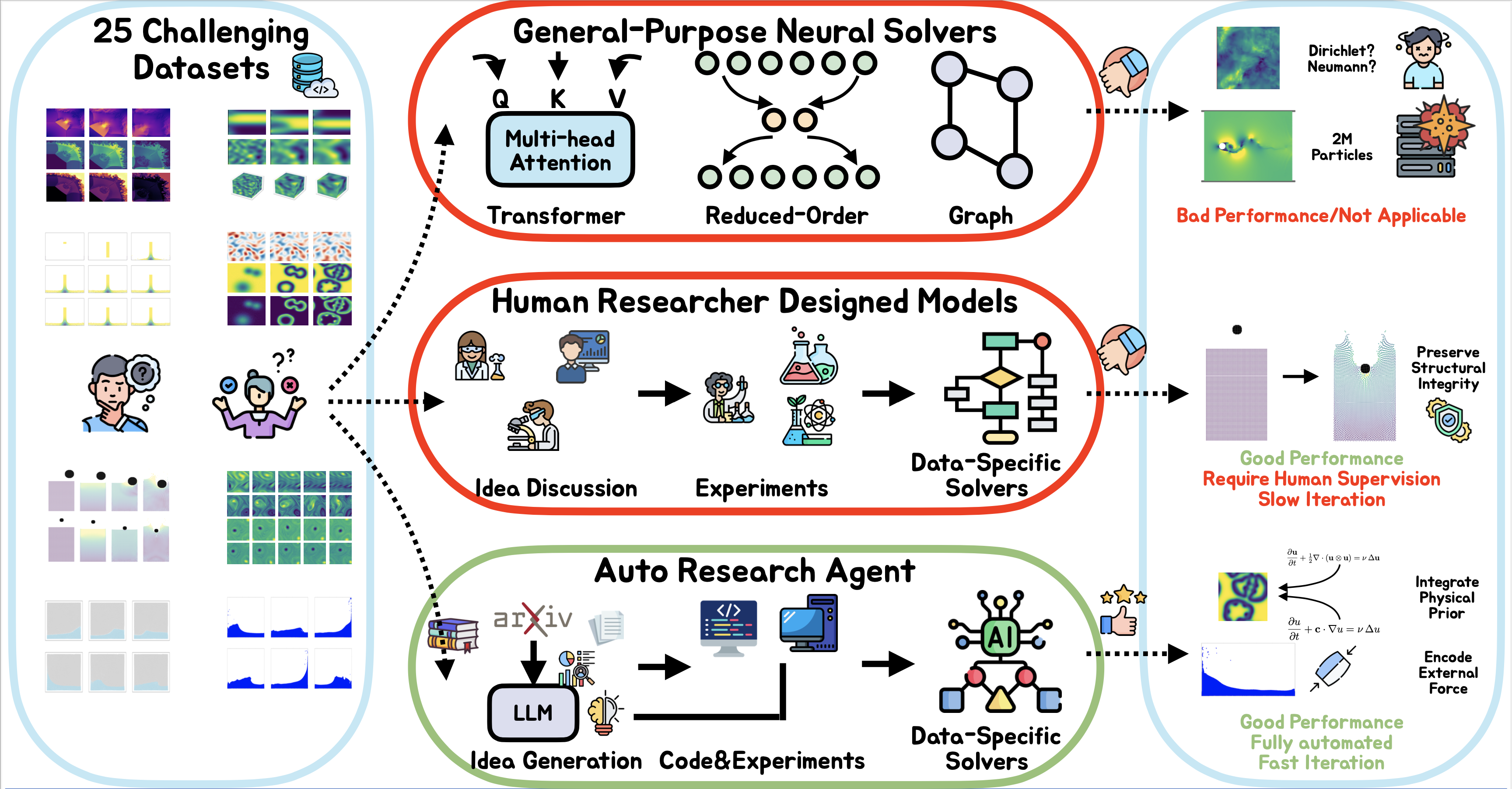}
    
    \caption{Overall framework of AutoPDEBench. We propose 25 challenging datasets, each presenting unique domain-specific challenges that necessitate specialized solvers. While general-purpose neural solvers can often be applied, they inherently lack the mechanisms to handle these specific constraints, leading to suboptimal performance. Traditionally, researchers manually design specialized solvers for these tasks, which requires extensive domain expertise and is highly time-consuming. We argue that automated research agents offer a compelling alternative, leveraging LLMs to autonomously design neural solvers capable of addressing these dataset-specific challenges.}
    \label{fig:framework}
\end{figure}

\section{AutoPDEBench}
\subsection{Problem Formulation}
We present the overall framework in Figure~\ref{fig:framework}. We consider the simulation of a physical system where our goal is to learn a neural simulator $\Phi_\theta$ that predicts a future output state $y \in \mathcal{Y}$ from an input state $x \in \mathcal{X}$. The model also accepts optional supplementary information $c \in \mathcal{C}$, which can include a variable timestep $\Delta t$, boundary conditions, PDE parameters, or the governing PDE equations. Thus, the simulator is defined as $\Phi_\theta : \mathcal{X} \times \mathcal{C} \to \mathcal{Y}$, yielding the prediction rule:
\begin{equation}
y = \Phi_\theta(x, c).
\end{equation}
Each dataset is split into disjoint $\mathcal{D}_{\text{train}}$, $\mathcal{D}_{\text{val}}$, and $\mathcal{D}_{\text{test}}$, and $\mathcal{L}(\Phi_\theta; \mathcal{D})$ denotes evaluation metrics evaluated on split $\mathcal{D}$. We evaluate models under a strict hardware constraint. Given a fixed GPU with memory budget $B$, our objective is to achieve the best possible performance. We place no restrictions on training or inference time, FLOPs, or parameter count.


\textbf{General-purpose solver.} Suppose $h^\star$ is the optimal hyperparameter configuration evaluated on the validation dataset that respects the memory budget $B$:
\begin{equation}
    h^\star = \argmin_{h \,:\, \mathrm{Mem}(\Phi, h) \le B} \mathcal{L}\big(\Phi_{\theta_h};\, \mathcal{D}_{\text{val}}\big), \qquad \theta_{h^\star} = \mathrm{Train}(\Phi, h^\star, \mathcal{D}_{\text{train}}).
\end{equation}
We then use this configuration to report the final test performance $\mathcal{L}(\Phi_{\theta_{h^\star}}; \mathcal{D}_{\text{test}})$.

\textbf{Auto-research agents.} We abstract the research process as an agentic system $\mathcal{A}$ that, at each round, maps a context $\mathcal{I}$ to a candidate solution. The context $\mathcal{I}$ aggregates the information available to the agent, such as the dataset description and the modeling objective. The agentic system is a mapping $\mathcal{A} : \mathcal{I} \to \mathcal{M} \times \mathcal{F} \times \mathcal{H}$ producing a model file $m \in \mathcal{M}$ (architecture $\Phi_m$), a loss file $\ell \in \mathcal{F}$ (training objective $\mathcal{L}^{\text{train}}$, which governs optimization but does not alter the fixed evaluation metrics), and a config file $h \in \mathcal{H}$ (hyperparameters). Over rounds $i = 1, \dots, N$, the context $\mathcal{I}_i$ is updated with the latest history, and the agent produces
\begin{equation}
    (m_i, \ell_i, h_i) = \mathcal{A}(\mathcal{I}_i), \qquad \theta_i = \mathrm{Train}\big(\Phi_{m_i}, \mathcal{L}^{\text{train}}, h_i, \mathcal{D}_{\text{train}}\big),
\end{equation}
subject to the memory limit $\mathrm{Mem}(\Phi_{m_i}, h_i) \le B$. Each candidate is trained on $\mathcal{D}_{\text{train}}$ under its generated objective $\mathcal{L}^{\text{train}}$ and hyperparameters $h_i$, and the agent observes only the validation outcome $r_i = \mathcal{L}(\Phi_{m_i, \theta_i}; \mathcal{D}_{\text{val}})$, never any quantity on $\mathcal{D}_{\text{test}}$. After $N$ rounds, it commits its best validated configuration $i^\star = \argmin_{i} r_i$, and the reported performance is its evaluation on the held-out test split, $\mathcal{L}(\Phi_{m_{i^\star}, \theta_{i^\star}}; \mathcal{D}_{\text{test}})$.

\subsection{Datasets}
We propose 25 challenging datasets, each presenting unique constraints that necessitate domain-specific solvers. This collection includes both entirely novel and currently understudied physical systems. Example visualizations and overarching descriptions are provided in Figure~\ref{fig:dataset} and Table~\ref{tab:datasets}, respectively. For comprehensive details regarding data generation, governing physics, problem setups, and additional visualizations, we refer readers to the dedicated appendices for each dataset.

\begin{table}[t]
\centering
\scriptsize
\setlength{\tabcolsep}{4pt}
\renewcommand{\arraystretch}{1.15}
\begin{tabularx}{\textwidth}{@{} l X c @{}}
\toprule
\textbf{Dataset} & \textbf{Description} & \textbf{Source} \\
\midrule
\HighSpeedFlow\,\textsuperscript{\S\ref{appendix:highspeedflow}}     & Supersonic flows that exceed the speed of sound and create sudden changes such as shock waves. & \pub  \\
\RealData\,\textsuperscript{\S\ref{appendix:realdata}}               & Real-world data consists of noisy measurements collected using time-resolved PIV. & \pub \\
\MixedParam\,\textsuperscript{\S\ref{appendix:mixedparam}}           & Datasets with mixed PDE parameters, where the governing coefficients vary across trajectories. & \ours  \\
\Heterogeneous\,\textsuperscript{\S\ref{appendix:heterogeneous}}     & Lagrangian particle datasets with particles possessing distinct physical attributes such as density. & \ours \\
\Partial\,\textsuperscript{\S\ref{appendix:partial}}                 & Datasets with partially observed initial conditions that require predicting the full trajectory. & \ours  \\
\MixedDomain\,\textsuperscript{\S\ref{appendix:mixeddomain}}         & Datasets featuring grid, mesh, and particle representations of the same underlying physical system. & \ours \\
\MultiPhysics\,\textsuperscript{\S\ref{appendix:multiphysics}}       & Multiphysics simulations that capture the interactions between multiple physical processes. & \pub  \\
\ThreeD\,\textsuperscript{\S\ref{appendix:threed}}                   & Spatiotemporal datasets featuring three-dimensional volumetric observations. & \pub \\
\IrregularTime\,\textsuperscript{\S\ref{appendix:irregulartime}}     & Datasets featuring observations recorded at non-uniform time intervals. & \ours  \\
\LagDynamic\,\textsuperscript{\S\ref{appendix:lagdynamic}}           & Lagrangian inlet-outlet data lacking 1-to-1 particle correspondence. & \ours \\
\LagFSI\,\textsuperscript{\S\ref{appendix:lagfsi}}                   & Lagrangian fluid-solid interaction datasets where both phases are represented by Lagrangian particles. & \ours  \\
\Extreme\,\textsuperscript{\S\ref{appendix:extreme}}                 & Datasets with extreme scale capturing 3D supernova blastwaves across seven orders of magnitude. & \pub \\
\MixedRes\,\textsuperscript{\S\ref{appendix:mixedres}}               & Generalization datasets that capture PDE dynamics on diverse domains and resolutions. & \ours  \\
\ExtLagOne\,\textsuperscript{\S\ref{appendix:extlag1}}               & Lagrangian particle datasets with liquid sloshing as external force. & \ours \\
\ExtLagTwo\,\textsuperscript{\S\ref{appendix:extlag2}}               & Datasets capturing Lagrangian fluid dynamics driven by the external force of a moving piston. & \ours  \\
\MixedDim\,\textsuperscript{\S\ref{appendix:mixeddim}}               & Datasets capturing identical PDE dynamics in 1D, 2D, and 3D. & \ours \\
\MixedBC\,\textsuperscript{\S\ref{appendix:mixedbc}}                 & Datasets capturing identical PDE under Dirichlet, Neumann, and periodic boundary conditions. & \ours  \\
\LongHorizon\,\textsuperscript{\S\ref{appendix:longhorizon}}         & Datasets featuring extremely long-horizon, 5,000-step rollouts. & \ours \\
\Stochastic\,\textsuperscript{\S\ref{appendix:stochastic}}           & Stochastic PDE dataset & \ours  \\
\MultiPhase\,\textsuperscript{\S\ref{appendix:multiphase}}           & Lagrangian particle datasets modeling multiphase liquid-gas interactions. & \ours \\
\MixedPDE\,\textsuperscript{\S\ref{appendix:mixedpde}}               & Datasets comprising a diverse range of PDEs, & \ours  \\
\LargeLag\,\textsuperscript{\S\ref{appendix:largelag}}               & Large-scale, high-resolution Lagrangian dataset featuring millions of particles. & \ours \\
\SimToReal\,\textsuperscript{\S\ref{appendix:sim2real}}              & Datasets designed for sim-to-real transfer. & \pub  \\
\HighDimensional\,\textsuperscript{\S\ref{appendix:highdimensional}} & 5D gyrokinetic dataset capturing high-dimensional PDE dynamics. & \ours \\
\ComplexGeometry\,\textsuperscript{\S\ref{appendix:complexgeometry}} & Datasets featuring flow around complex geometries. & \pub  \\
\bottomrule
\end{tabularx}
\caption{Overview of datasets in AutoPDEBench. Detailed descriptions and visualizations for each dataset are provided in their individual Appendix section.}
\label{tab:datasets}
\end{table}

\subsection{Evaluation Protocol}
Our evaluation protocol assesses final model performance on the test dataset. Given our GPU resource constraints, each model was tuned to maximize performance within these limits. Our primary evaluation metric is RMSE. We also use dataset-specific metrics to evaluate additional aspects such as physical consistency and practical utility. The additional evaluation results for each dataset are provided in their individual appendix sections, while detailed descriptions of these supplementary metrics can be found in Appendix~\ref{appendix:metrics}.

\subsection{General-Purpose Neural Solver Baselines}
To establish our baselines, we consider transformer-based, reduced-order modeling, and graph-based neural solvers. We selected these architectural paradigms for their broad generality and structural flexibility. Unlike convolution-based methods, which are fundamentally constrained by the requirement for uniform spatial grids, these selected models feature adaptable input-output mappings. This flexibility allows them to universally process highly complex, unstructured, and dimensionally diverse data representations. Given their capacity to handle such generalized inputs, we select GNOT~\citep{Hao2023GNOTAG}, Transolver~\citep{wu2024Transolver}, Transolver++~\citep{luotransolver++}, LNO~\citep{wang2024LNO}, and AMG~\citep{li2024harnessing} as our representative baselines.

Foundation models, pretrained on massive amounts of data, would theoretically serve as ideal baselines. However, current foundation models~\citep{hao2024dpot, mccabe2024multiple, ye2024pdeformer, herde2024poseidon, shen2024ups, zhouunisolver, wang2025mixtureofexpertsoperatortransformerlargescale} are predominantly restricted to grid-based domains. They lack exposure to the Eulerian mesh and Lagrangian particle representations that are central to our work. Furthermore, their rigid input-output architectures preclude them from generalizing across our highly unstructured data, which features diverse spatial dimensions and varied discretization schemes.

\subsection{Multi-Agent Instantiation of Auto Research Pipeline}
AutoPDEBench is agnostic to the internal design of $\mathcal{A}$. Any system that maps $\mathcal{I}_i$ to $(m_i, \ell_i, h_i)$ under the protocol above can be evaluated, including more elaborate multi-agent frameworks with dynamic delegation or parallel search. We provide a reference instantiation of $\mathcal{A}$ as a role-specialized multi-agent system of three LLM-driven components: a planner, a coder, and a debugger. The planner consumes the dataset description $\tau_{\text{data}}$, the modeling objective $\tau_{\text{obj}}$, the hardware constraints $\tau_{\text{const}}$, the coding interface specification $\tau_{\text{interface}}$ (editable files and program structure), and the trial history $\tau_{\text{history}}$, and produces a detailed step-by-step plan
\begin{equation}
    \tau_{\text{plan}} = \mathcal{A}_{\text{planner}}\big(\tau_{\text{data}}, \tau_{\text{obj}}, \tau_{\text{const}}, \tau_{\text{interface}}, \tau_{\text{history}}\big),
\end{equation}
a natural-language specification of the model architecture, loss functions, and hyperparameter choices to attempt this round, together with its rationale. The coder then translates this plan into executable files, conditioning on the hardware constraints, the coding interface specification, and the plan:
\begin{equation}
    (m_i, \ell_i, h_i) = \mathcal{A}_{\text{coder}}\big(\tau_{\text{const}}, \tau_{\text{interface}}, \tau_{\text{plan}}\big).
\end{equation}
We then attempt $\theta_i = \mathrm{Train}(\Phi_{m_i}, \mathcal{L}^{\text{train}}, h_i, \mathcal{D}_{\text{train}})$. If training fails, raising an execution error or violating the memory budget $\mathrm{Mem}(\Phi_{m_i}, h_i) \le B$, we invoke the debugger, which revises the files given the current code, the hardware constraints, the coding interface specification, the plan, and the observed error $e$:
\begin{equation}
    (m_i, \ell_i, h_i) \leftarrow \mathcal{A}_{\text{debugger}}\big(m_i, \ell_i, h_i, \tau_{\text{const}}, \tau_{\text{interface}}, \tau_{\text{plan}}, e\big).
\end{equation}
This step repeats until training completes successfully or a fixed retry budget is exhausted, in which case the round is recorded as a failure.

\begin{table}[t]
\centering
\small
\renewcommand{\arraystretch}{1.2}
\resizebox{\textwidth}{!}{%
\begin{tabular}{@{} l | r@{}l r@{}l r@{}l r@{}l r@{}l | r@{}l @{}}
\toprule
\textbf{Dataset} & \multicolumn{2}{c}{\textbf{GNOT}} & \multicolumn{2}{c}{\textbf{Transolver}} & \multicolumn{2}{c}{\textbf{Transolver++}} & \multicolumn{2}{c}{\textbf{LNO}} & \multicolumn{2}{c}{\textbf{AMG}} & \multicolumn{2}{c}{\textbf{Agent}} \\
\midrule
\HighSpeedFlow     & $1.826 $&$\times 10^{1}$ & $3.232 $&$\times 10^{1}$ & $3.271 $&$\times 10^{1}$ & $4.482 $&$\times 10^{1}$ & $2.819 $&$\times 10^{1}$ & $\bm{5.066 }$&$\bm{\times 10^{0}}$ \\
\RealData          & $5.638 $&$\times 10^{-3}$ & $5.626 $&$\times 10^{-3}$ & $5.618 $&$\times 10^{-3}$ & $6.946 $&$\times 10^{-3}$ & $4.889 $&$\times 10^{-3}$ & $\bm{4.458 }$&$\bm{\times 10^{-3}}$ \\
\MixedParam        & $3.417 $&$\times 10^{0}$ & $4.359 $&$\times 10^{0}$ & $4.329 $&$\times 10^{0}$ & $4.346 $&$\times 10^{0}$ & $3.128 $&$\times 10^{0}$ & $\bm{1.345 }$&$\bm{\times 10^{0}}$ \\
\Heterogeneous     & $4.207 $&$\times 10^{-3}$ & $4.155 $&$\times 10^{-3}$ & $3.963 $&$\times 10^{-3}$ & $3.951 $&$\times 10^{-3}$ & $3.900 $&$\times 10^{-3}$ & $\bm{3.311 }$&$\bm{\times 10^{-3}}$ \\
\Partial           & $1.978 $&$\times 10^{0}$ & $3.636 $&$\times 10^{0}$ & $3.636 $&$\times 10^{0}$ & $3.636 $&$\times 10^{0}$ & $2.111 $&$\times 10^{0}$ & $\bm{8.633 }$&$\bm{\times 10^{-2}}$ \\
\MixedDomain       & $1.784 $&$\times 10^{-3}$ & $\bm{1.739 }$&$\bm{\times 10^{-3}}$ & $1.775 $&$\times 10^{-3}$ & $1.785 $&$\times 10^{-3}$ & $6.491 $&$\times 10^{-2}$ & $1.785 $&$\times 10^{-3}$ \\
\MultiPhysics      & $6.902 $&$\times 10^{-2}$ & $1.298 $&$\times 10^{-1}$ & $1.345 $&$\times 10^{-1}$ & $4.971 $&$\times 10^{-1}$ & $7.303 $&$\times 10^{-2}$ & $\bm{5.141} $&$\bm{\times 10^{-2}}$ \\
\ThreeD            & $2.230 $&$\times 10^{-1}$ & $2.279 $&$\times 10^{-1}$ & $2.257 $&$\times 10^{-1}$ & $2.238 $&$\times 10^{-1}$ & \multicolumn{2}{c|}{$-$} & $\bm{7.182 }$&$\bm{\times 10^{-2}}$ \\
\IrregularTime     & $2.577 $&$\times 10^{0}$ & $2.623 $&$\times 10^{0}$ & $2.621 $&$\times 10^{0}$ & $2.622 $&$\times 10^{0}$ & $2.415 $&$\times 10^{0}$ & $\bm{6.511 }$&$\bm{\times 10^{-1}}$ \\
\LagDynamic        & \multicolumn{2}{c}{$-$} & \multicolumn{2}{c}{$-$} & \multicolumn{2}{c}{$-$} & $6.105 $&$\times 10^{-1}$ & \multicolumn{2}{c|}{$-$} & $\bm{5.530 }$&$\bm{\times 10^{-1}}$ \\
\LagFSI            & $7.780 $&$\times 10^{-5}$ & $\bm{7.574 }$&$\bm{\times 10^{-5}}$ & $8.583 $&$\times 10^{-5}$ & $8.455 $&$\times 10^{-5}$ & $8.513 $&$\times 10^{-5}$ & $2.215 $&$\times 10^{-4}$ \\
\Extreme           & \multicolumn{2}{c}{$-$} & $2.289 $&$\times 10^{5}$ & $2.190 $&$\times 10^{5}$ & $4.813 $&$\times 10^{5}$ & \multicolumn{2}{c|}{$-$} & $\bm{1.340 }$&$\bm{\times 10^{5}}$ \\
\MixedRes          & $3.629 $&$\times 10^{0}$ & $3.616 $&$\times 10^{0}$ & $3.564 $&$\times 10^{0}$ & $3.481 $&$\times 10^{0}$ & \multicolumn{2}{c|}{$-$} & $\bm{3.455 }$&$\bm{\times 10^{0}}$ \\
\ExtLagOne         & $2.176 $&$\times 10^{-2}$ & $3.526 $&$\times 10^{-2}$ & $1.729 $&$\times 10^{-2}$ & $1.407 $&$\times 10^{-2}$ & $1.892 $&$\times 10^{-2}$ & $\bm{1.954 }$&$\bm{\times 10^{-3}}$ \\
\ExtLagTwo         & $1.365 $&$\times 10^{-3}$ & $1.456 $&$\times 10^{-3}$ & $1.551 $&$\times 10^{-3}$ & $1.640 $&$\times 10^{-3}$ & $1.566 $&$\times 10^{-3}$ & $\bm{5.483 }$&$\bm{\times 10^{-4}}$ \\
\MixedDim          & $3.615 $&$\times 10^{-2}$ & $3.380 $&$\times 10^{-2}$ & $3.357 $&$\times 10^{-2}$ & $5.047 $&$\times 10^{-2}$ & \multicolumn{2}{c|}{$-$} & $\bm{9.417 }$&$\bm{\times 10^{-3}}$ \\
\MixedBC           & $1.383 $&$\times 10^{-1}$ & $8.500 $&$\times 10^{-2}$ & $8.498 $&$\times 10^{-2}$ & $8.511 $&$\times 10^{-2}$ & $2.064 $&$\times 10^{-1}$ & $\bm{4.435 }$&$\bm{\times 10^{-2}}$ \\
\LongHorizon       & $5.674 $&$\times 10^{0}$ & $4.787 $&$\times 10^{0}$ & $4.756 $&$\times 10^{0}$ & $\bm{2.807 }$&$\bm{\times 10^{0}}$ & $7.088 $&$\times 10^{0}$ & $1.781 $&$\times 10^{1}$ \\
\Stochastic        & $3.989 $&$\times 10^{-3}$ & $3.688 $&$\times 10^{-3}$ & $3.171 $&$\times 10^{-3}$ & $5.447 $&$\times 10^{-3}$ & $3.442 $&$\times 10^{-3}$ & $\bm{2.324} $&$\bm{\times 10^{-3}}$ \\
\MultiPhase        & $8.231 $&$\times 10^{-4}$ & $9.163 $&$\times 10^{-4}$ & $8.810 $&$\times 10^{-4}$ & $8.203 $&$\times 10^{-4}$ & $8.130 $&$\times 10^{-4}$ & $\bm{7.581 }$&$\bm{\times 10^{-4}}$ \\
\MixedPDE          & $8.792 $&$\times 10^{-2}$ & $7.565 $&$\times 10^{-2}$ & $7.535 $&$\times 10^{-2}$ & $7.684 $&$\times 10^{-2}$ & $8.845 $&$\times 10^{-2}$ & $\bm{2.701 }$&$\bm{\times 10^{-2}}$ \\
\LargeLag          & $1.300 $&$\times 10^{-2}$ & $2.753 $&$\times 10^{-2}$ & $2.027 $&$\times 10^{-2}$ & $1.599 $&$\times 10^{-1}$ & \multicolumn{2}{c|}{$-$} & $\bm{5.666 }$&$\bm{\times 10^{-3}}$ \\
\SimToReal         & $1.987 $&$\times 10^{-3}$ & $\bm{1.783 }$&$\bm{\times 10^{-3}}$ & $1.809 $&$\times 10^{-3}$ & $8.306 $&$\times 10^{-3}$ & $2.705 $&$\times 10^{-3}$ & $6.734 $&$\times 10^{-3}$ \\
\HighDimensional   & \multicolumn{2}{c}{$-$} & \multicolumn{2}{c}{$-$} & \multicolumn{2}{c}{$-$} & \multicolumn{2}{c}{$-$} & \multicolumn{2}{c|}{$-$} & $3.412 $&$\times 10^{-4}$ \\
\ComplexGeometry   & $7.638 $&$\times 10^{-2}$ & $8.295 $&$\times 10^{-2}$ & $9.702 $&$\times 10^{-2}$ & $1.032 $&$\times 10^{-1}$ & $1.448 $&$\times 10^{-1}$ & $1.366 $&$\times 10^{-1}$ \\
\bottomrule
\end{tabular}
}
\caption{Performance comparison of general-purpose neural solvers and agent-designed models. We report RMSE as the primary evaluation metric; additional metrics for each dataset are provided in the Appendix. Entries marked with a dash ("-") indicate that the model is either incompatible with the dataset or exceeds computational constraints (e.g., prohibitive training time). We use gpt-5.5 as the backbone.
}
\label{tab:results}
\end{table}

\section{Experiments}
\label{sec:exp}


We present the main results in Table~\ref{tab:results}, where we use RMSE as the primary evaluation metric and refer the readers to the Appendix of each dataset for additional metrics. Overall, we find that even this baseline multi-agent system achieves significantly better performance than general-purpose neural solvers. This underscores the value of tailoring architectures to address dataset-specific challenges and validates the efficacy of utilizing autonomous research agents for this task.

However, the agent struggles with highly complex datasets that remain open challenges in the literature. For instance, the \LagFSI\  dataset models a sphere dropping into a body of water, with both phases discretized using Lagrangian particles. This scenario necessitates innovative solutions for two-way fluid-solid coupling, alongside methods to strictly enforce the rigid integrity of the sphere, which are non-trivial challenges that existing Lagrangian neural solvers have yet to successfully resolve. Consequently, the agent's generated design performs significantly worse than the general-purpose baselines on this dataset. Another failure case is \LongHorizon, which requires accurate autoregressive rollouts over 5,000 steps. While this horizon is substantially longer than those typically explored in existing research, it accurately reflects the strict demands of real-world engineering applications. Here again, the agent-designed model fails to surpass general-purpose baselines. We hypothesize that these shortcomings stem from the agent's exclusive reliance on its pre-trained knowledge to formulate solutions for entirely novel problems. Future work should therefore investigate mechanisms to dynamically integrate external domain expertise and latest scientific literature into the model design.

\subsection{LLM Backbone Sensitivity Analysis}
\label{sec:llm}
The primary results of our study are based on gpt-5.5 as the backbone for the agentic systems, as it was the frontier model at the time of this research. However, it is necessary to benchmark alternative LLM architectures. To achieve this, we replace the default gpt-5.5 backbone with a diverse suite of state-of-the-art LLMs, ensuring that the agentic framework, prompt templates, and search spaces remain strictly identical. For this experiment and all subsequent analyses, we evaluate on \HighSpeedFlow, \LagDynamic, and \MixedParam, which serve as representative benchmarks for Eulerian data, Lagrangian particle data, and generalization tasks.

The comparative results are presented in Figure~\ref{fig:llm}, with supplementary metrics detailed in Appendix~\ref{appendix:llm}. Several flagship models—namely gpt-5.5, gpt-5.6-sol, fable-5, opus-4.8, and grok-4.5—consistently demonstrate strong performance across all datasets. In contrast, models such as deepseek-v4 and qwen-3.7-max exhibit high variance, excelling on certain tasks while underperforming on others. Meanwhile, older architectures like gpt-4o struggle significantly across these evaluations. We hypothesize that this inconsistent performance stems from limited exposure to these specific physical datasets during pre-training.

\begin{wrapfigure}[25]{r}{0.5\textwidth}   
    \centering
    \begin{subfigure}[b]{\linewidth}
        \centering
        \includegraphics[width=\linewidth]{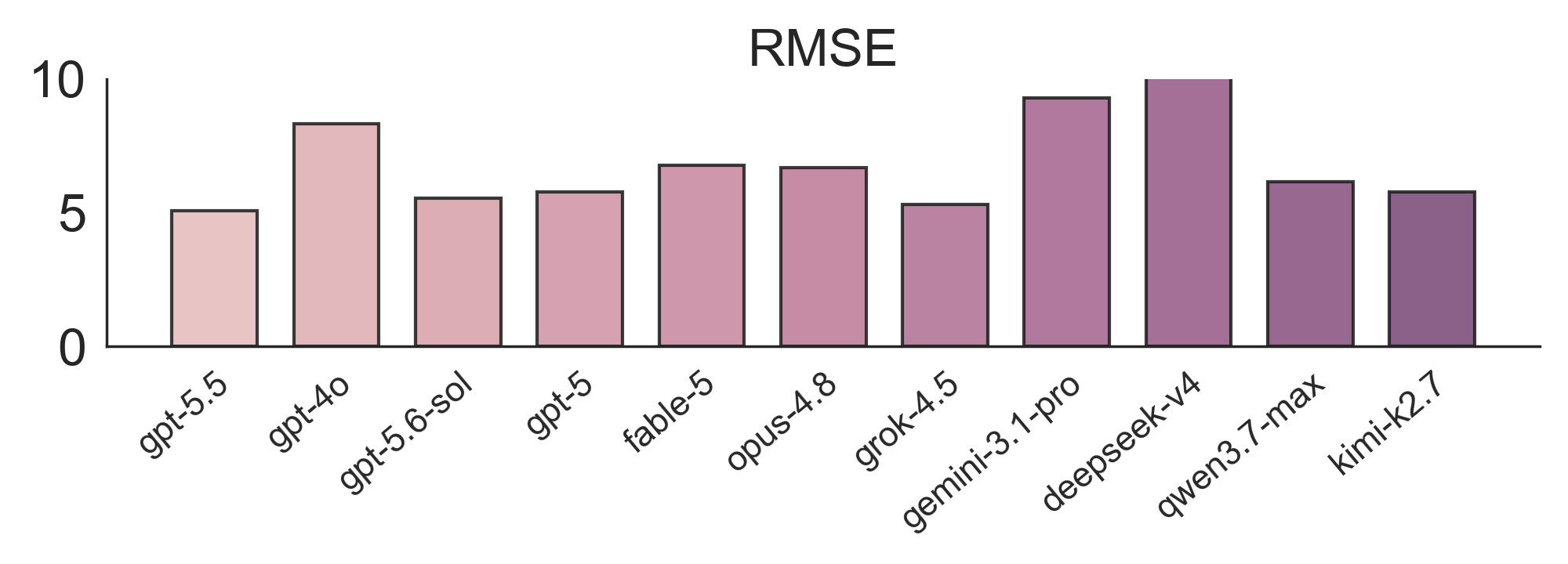}
        \caption{\HighSpeedFlow}
    \end{subfigure}

    \begin{subfigure}[b]{\linewidth}
        \centering
        \includegraphics[width=\linewidth]{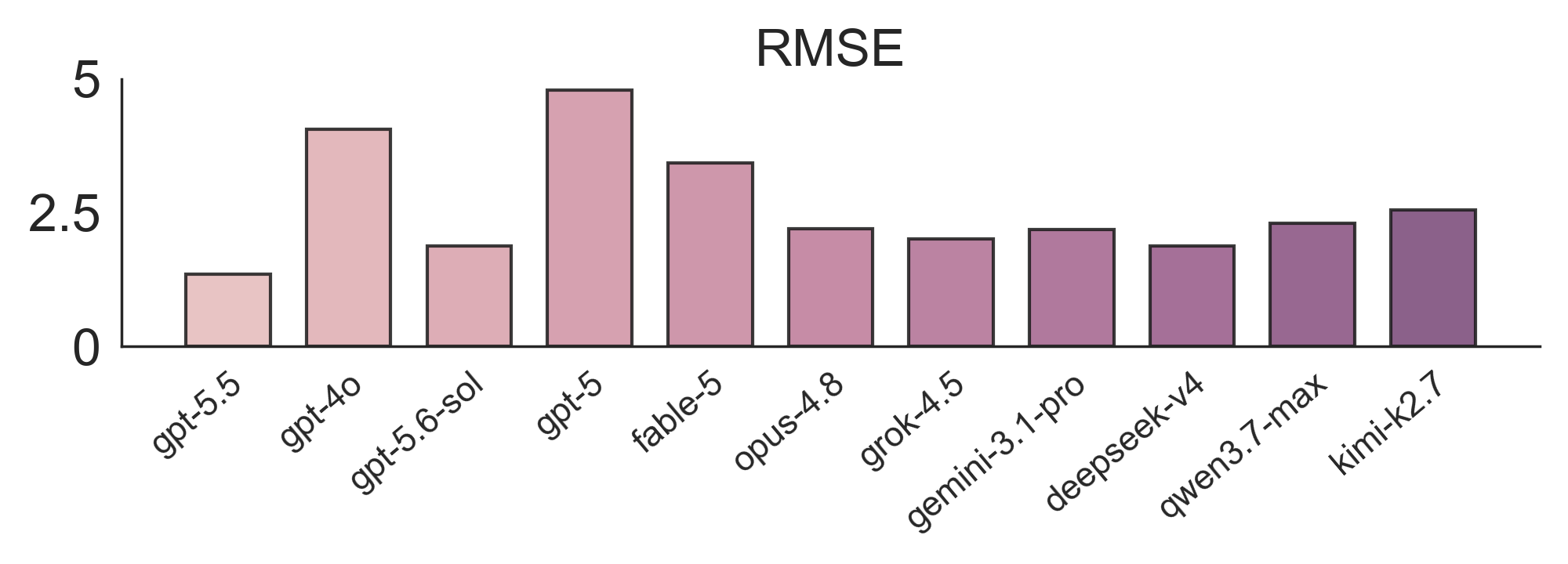}
        \caption{\MixedParam}
    \end{subfigure}

    \begin{subfigure}[b]{\linewidth}
        \centering
        \includegraphics[width=\linewidth]{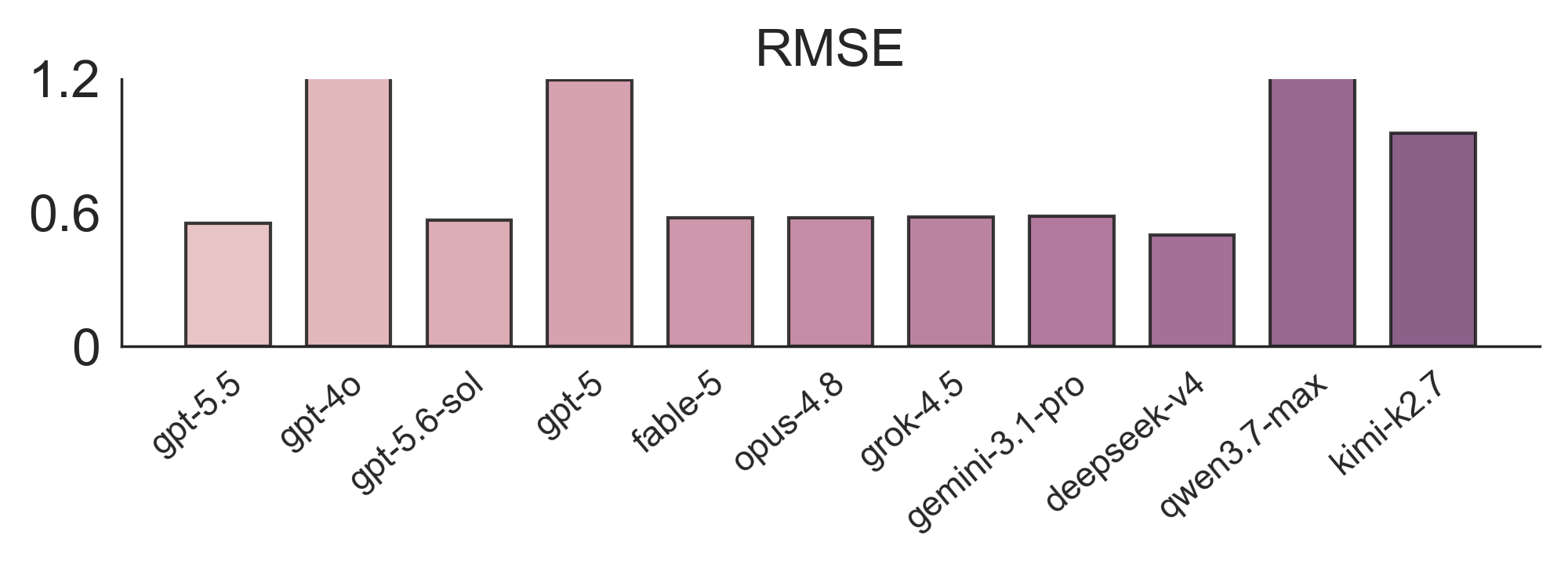}
        \caption{\LagDynamic}
    \end{subfigure}

    \caption{Performance comparison of various SOTA LLMs serving as the agent's backbone.}
    \label{fig:llm}
\end{wrapfigure}

Notably, model performance on these physical tasks does not strictly align with general LLM leaderboards. For instance, while gpt-5.6-sol typically outperforms gpt-5.5 on broad intelligence benchmarks, it underperforms or merely matches gpt-5.5 across all three evaluated cases. This discrepancy indicates that general capabilities do not automatically transfer to these specialized domains, highlighting promising open research directions for targeted agentic training and the development of more robust agentic systems.

\subsection{Comparison with Domain-Specific Baselines}
\label{sec:domain_baselines}

In our primary evaluation, general-purpose neural solvers are benchmarked across the proposed datasets. While these general models represent current state-of-the-art and are capable of broad physical generalization, researchers have also engineered highly specialized, domain-specific architectures tailored to unique dataset characteristics. This experiment aims to contextualize our automated agent’s performance against these specialized baselines.

We selected a representative subset of experimental scenarios that feature well-documented, domain-specific baselines in the existing literature. An exhaustive comparison across all tasks is intractable due to the significant engineering overhead required to reproduce proprietary codebases, combined with the fact that several of our custom, challenge-driven datasets lack existing domain-specific counterparts. Instead, this evaluation serves as a targeted stress test to demonstrate whether the automated agent can match or exceed the performance of manually designed, domain-specific solvers.

\begin{figure}[t]
    \centering
    \begin{subfigure}[b]{0.32\textwidth}
        \centering
        \includegraphics[width=\textwidth]{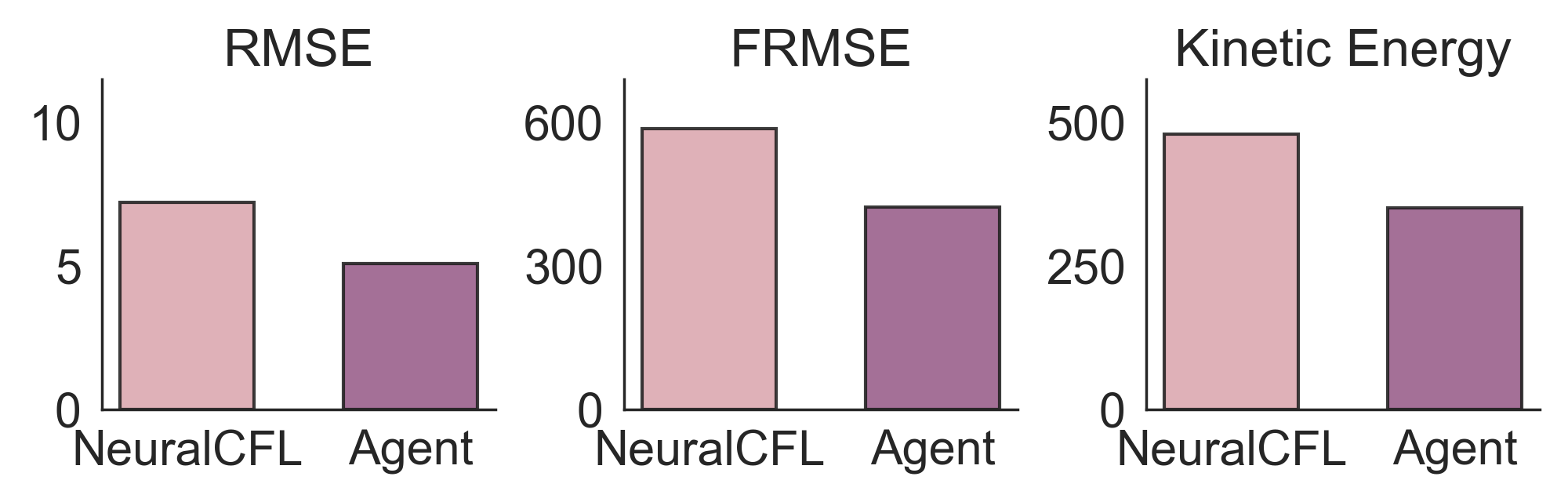} 
        \caption{\HighSpeedFlow}
    \end{subfigure}
    \hfill
    \begin{subfigure}[b]{0.32\textwidth}
        \centering
        \includegraphics[width=\textwidth]{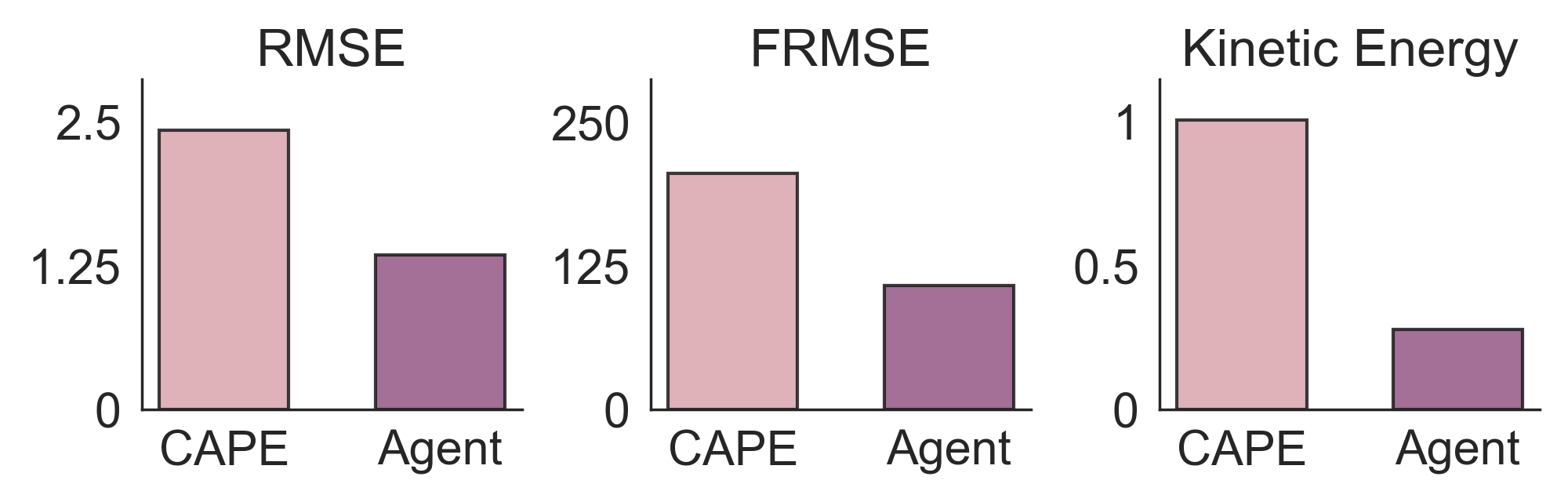} 
        \caption{\MixedParam}
    \end{subfigure}
    \hfill
    \begin{subfigure}[b]{0.32\textwidth}
        \centering
        \includegraphics[width=\textwidth]{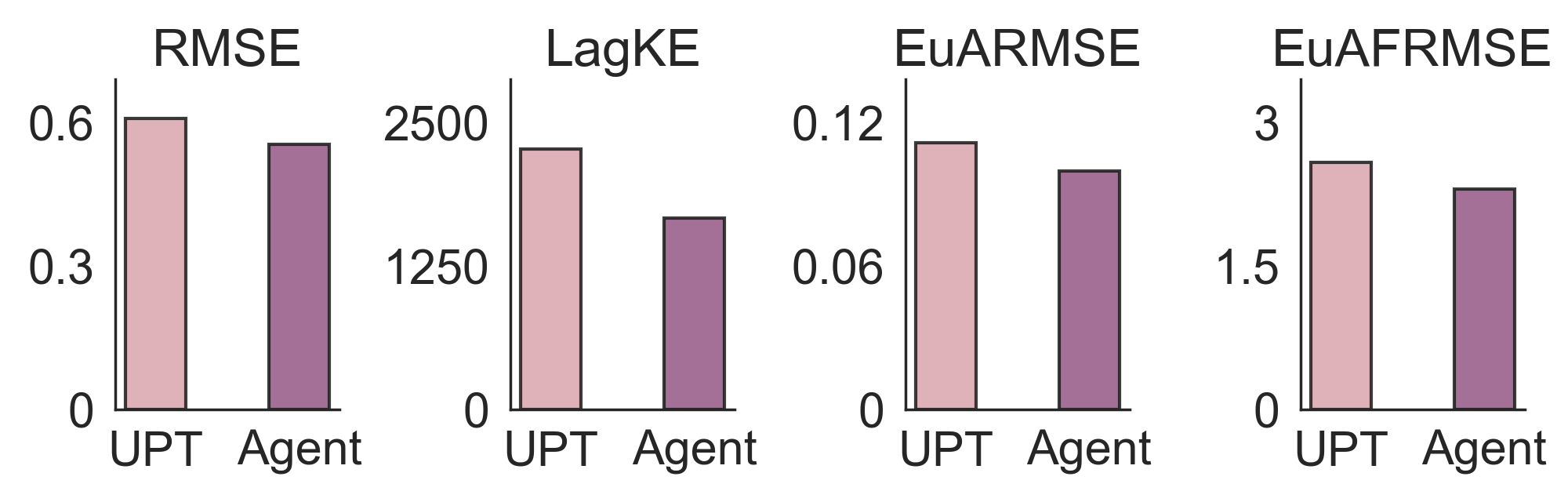} 
        \caption{\LagDynamic}
    \end{subfigure}
    \caption{Performance comparison against domain-specific baselines. The automated agent's designs are evaluated against human engineered architectures across three representative datasets.
    }
    \label{fig:domainbaselines}
\end{figure}

We present the comparative results in Figure~\ref{fig:domainbaselines}. For the \HighSpeedFlow\ domain, we benchmark against NeuralCFL~\citep{helwig2026twophasedeeplearningframework}, which introduces a two-phase deep learning architecture to dynamically adapt time steps. For \MixedParam, we compare our method with CAPE~\citep{takamoto2023learningneuralpdesolvers}, a model that leverages parameter-guided channel attention to improve generalization in PDE parameters. Finally, for \LagDynamic, we evaluate against UPT~\citep{alkin2024upt}, a scalable transformer-based framework that demonstrates superior performance in Lagrangian field learning. The results demonstrate that the agent's design consistently outperforms these human-designed, dataset-specific baselines across all evaluated domains.

\subsection{Transferability Across PDEs}
\label{sec:transfer_pde}

The primary objective of our automated agent framework is to generate neural architectures that capture foundational mathematical and physical principles. Preliminary analyses indicate that these agent-designed architectures are transferable across PDEs. To empirically validate this observation, we evaluate the out-of-distribution transferability of the generated solvers. Specifically, architectures originally optimized by the agent for a target PDE are trained directly on entirely different PDEs that share similar dataset-specific challenges.

\begin{figure}[t]
    \centering
    \begin{subfigure}[b]{0.32\textwidth}
        \centering
        \includegraphics[width=\textwidth]{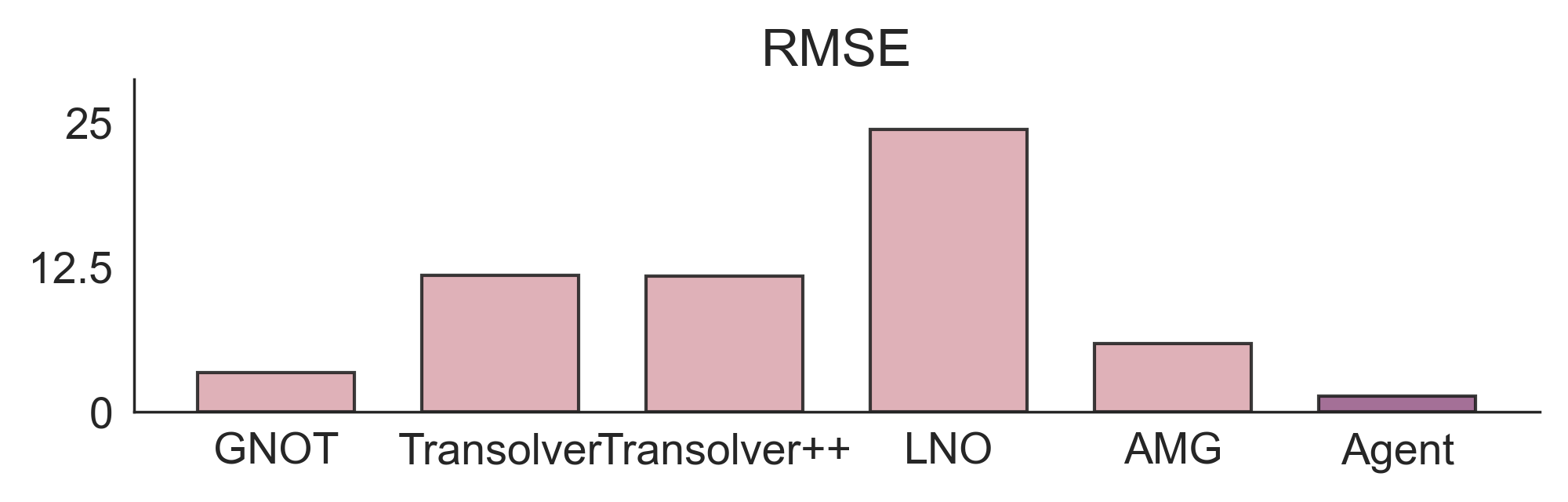} 
        \caption{\HighSpeedFlow}
    \end{subfigure}
    \hfill
    \begin{subfigure}[b]{0.32\textwidth}
        \centering
        \includegraphics[width=\textwidth]{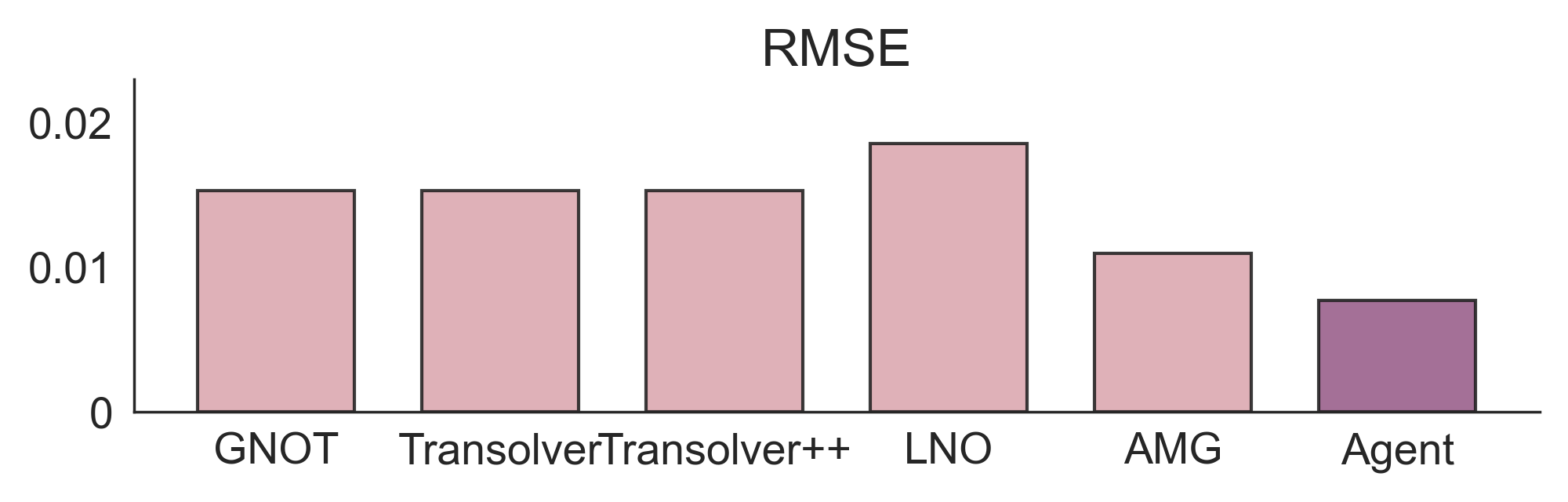} 
        \caption{\MixedParam}
    \end{subfigure}
    \hfill
    \begin{subfigure}[b]{0.32\textwidth}
        \centering
        \includegraphics[width=\textwidth]{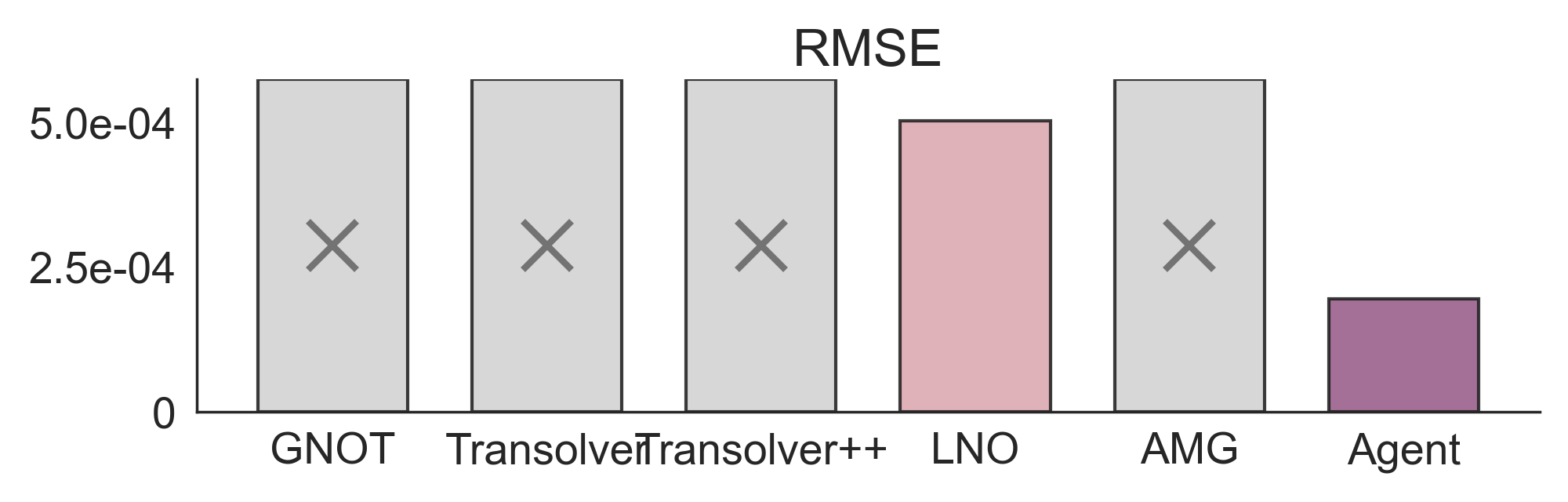} 
        \caption{\LagDynamic}
    \end{subfigure}
    \caption{Transferability of agent-designed architectures across PDEs. Models originally optimized by the agent for specific target PDEs are evaluated on entirely different physical systems sharing similar data characteristics.
    }
    \label{fig:transfer}
\end{figure}

The results are presented in Figure~\ref{fig:transfer}. For \HighSpeedFlow, we test on the coal dust explosion dataset from \citet{helwig2026twophasedeeplearningframework}, which models the multiphase fluid dynamics of a normal shock wave sweeping over a layer of coal dust, triggering interfacial instabilities that evolve into turbulent vortices and drive gas-particle mixing. For \MixedParam, we evaluate on the 2D Burgers' equation with kinematic viscosity varying between 5 and 50. Finally, for \LagDynamic, we test on a 2D wave dataset where regular waves are generated at the left inlet using predefined velocities, and the right boundary features a slanted wall. The results demonstrate that the agent's designs successfully transfer to these fundamentally distinct physical scenarios, validating that they effectively resolve dataset-specific issues without being structurally overfitted to their original equations. Additional metric evaluations are provided in Appendix~\ref{appendix:transfer}.

\subsection{Agent Long-Horizon Scalability}
\label{sec:long_horizon}
Our base experimental pipeline limits the agent's optimization process to 5 iterations, which empirical testing identifies as a pragmatic threshold for achieving stable performance under reasonable resource constraints. This supplementary experiment extends the horizon to 30 iterations to evaluate long-term performance gains.

\begin{figure}[h]
    \centering
    \begin{subfigure}[b]{0.32\textwidth}
        \centering
        \includegraphics[width=\textwidth]{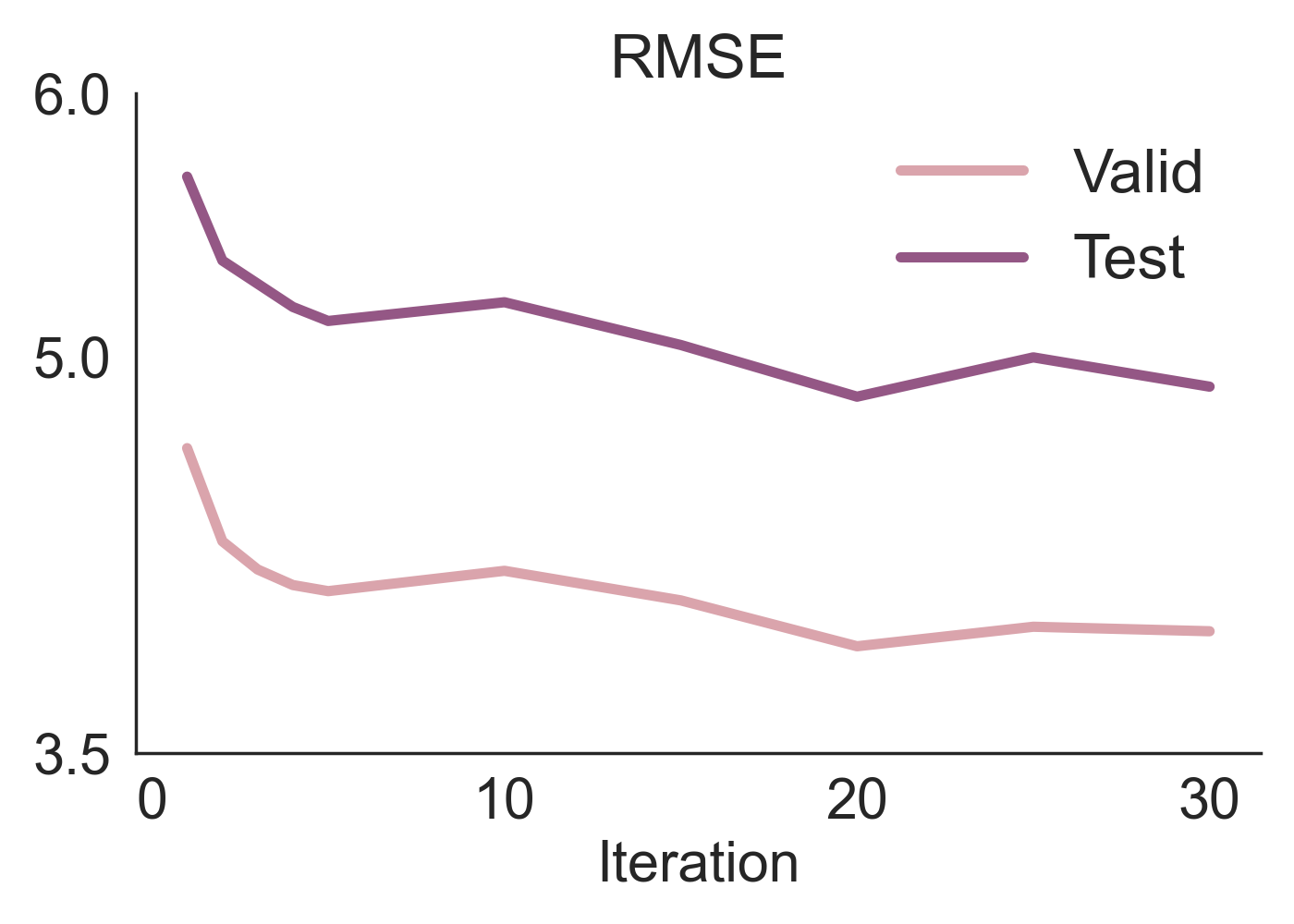} 
        \caption{\HighSpeedFlow}
    \end{subfigure}
    \hfill
    \begin{subfigure}[b]{0.32\textwidth}
        \centering
        \includegraphics[width=\textwidth]{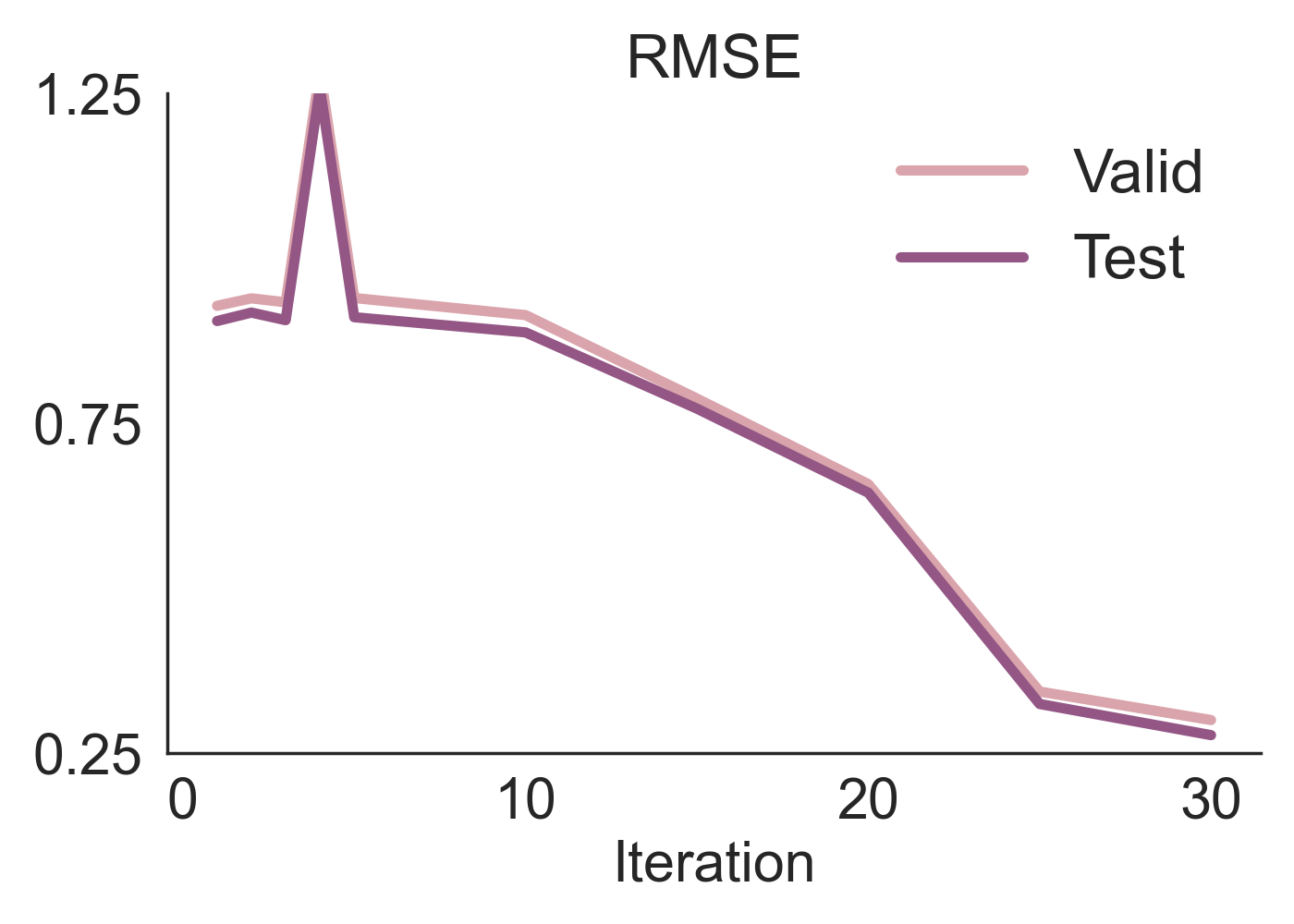} 
        \caption{\MixedParam}
    \end{subfigure}
    \hfill
    \begin{subfigure}[b]{0.32\textwidth}
        \centering
        \includegraphics[width=\textwidth]{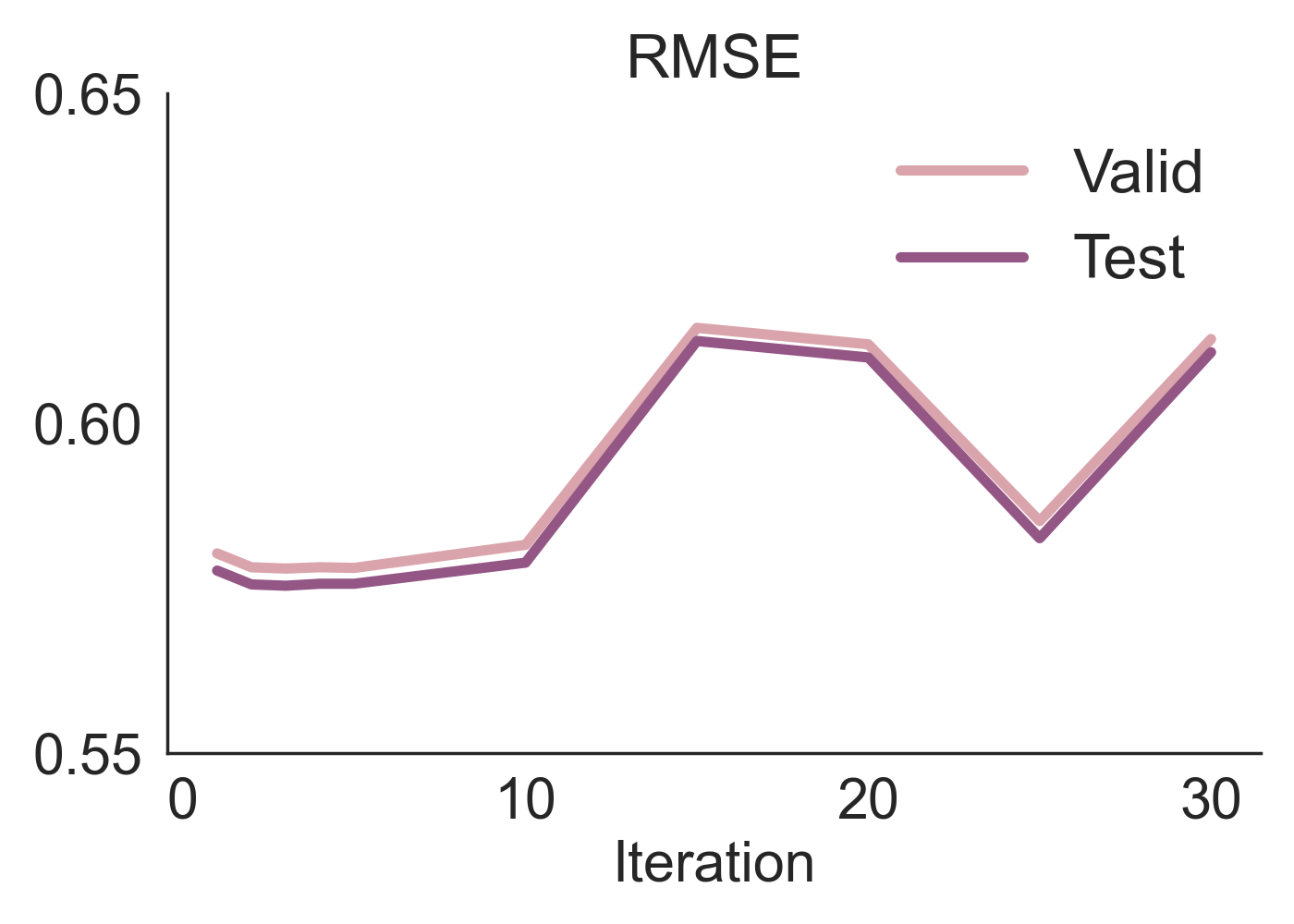} 
        \caption{\LagDynamic}
    \end{subfigure}
    \caption{Model performance across long-horizon optimization. The computational budget is extended to 30 iterations to evaluate long-term behavioral dynamics. 
    }
    \label{fig:iteration}
\end{figure}

Results (Figure~\ref{fig:iteration}, Appendix~\ref{appendix:longhorizon}) show the agent successfully improves design for \HighSpeedFlow\ and \MixedParam, but struggles on \LagDynamic. Notably, performance on \MixedParam\ improves significantly after iteration 10 as the agent discovers superior architectures. Conversely, on \HighSpeedFlow\ and \LagDynamic, it relies on conservative hyperparameter tuning and minor tweaks. These findings suggest future agent designs must systematically identify model limitations.



\section{Conclusion}
We introduce a comprehensive benchmark for evaluating LLM-driven automated research in neural PDE solvers. Our benchmark provides 25 challenging datasets featuring complex physical constraints. Our empirical evaluations demonstrate that an automated research agent can successfully design specialized solver pipelines that consistently outperform versatile baselines. We refer readers to Appendix~\ref{appendix:broad} for broader impact and Appendix~\ref{appendix:limitation} for limitation and future work.

\newpage

\bibliography{iclr2027_conference}

\begin{thebibliography}{92}
\providecommand{\natexlab}[1]{#1}
\providecommand{\url}[1]{\texttt{#1}}
\expandafter\ifx\csname urlstyle\endcsname\relax
  \providecommand{\doi}[1]{doi: #1}\else
  \providecommand{\doi}{doi: \begingroup \urlstyle{rm}\Url}\fi

\bibitem[Agarwal et~al.(2025)Agarwal, Sahu, Puri, Laradji, Dvijotham, Stanley, Charlin, and Pal]{agarwal2025litllmtoolkitscientificliterature}
Shubham Agarwal, Gaurav Sahu, Abhay Puri, Issam~H. Laradji, Krishnamurthy~DJ Dvijotham, Jason Stanley, Laurent Charlin, and Christopher Pal.
\newblock Litllm: A toolkit for scientific literature review, 2025.
\newblock URL \url{https://arxiv.org/abs/2402.01788}.

\bibitem[Alkin et~al.(2024)Alkin, Fürst, Schmid, Gruber, Holzleitner, and Brandstetter]{alkin2024upt}
Benedikt Alkin, Andreas Fürst, Simon Schmid, Lukas Gruber, Markus Holzleitner, and Johannes Brandstetter.
\newblock Universal physics transformers.
\newblock \emph{arXiv preprint arXiv:2402.12365}, 2024.

\bibitem[Alkin et~al.(2025)Alkin, Bleeker, Kurle, Kronlachner, Sonnleitner, Dorfer, and Brandstetter]{alkin2025abupt}
Benedikt Alkin, Maurits Bleeker, Richard Kurle, Tobias Kronlachner, Reinhard Sonnleitner, Matthias Dorfer, and Johannes Brandstetter.
\newblock {AB}-{UPT}: Scaling neural {CFD} surrogates for high- fidelity automotive aerodynamics simulations via anchored- branched universal physics transformers.
\newblock \emph{Transactions on Machine Learning Research}, 2025.
\newblock ISSN 2835-8856.
\newblock URL \url{https://openreview.net/forum?id=nwQ8nitlTZ}.

\bibitem[Boussif et~al.(2022)Boussif, Bengio, Benabbou, and Assouline]{boussif2022magnet}
Oussama Boussif, Yoshua Bengio, Loubna Benabbou, and Dan Assouline.
\newblock {MA}gnet: Mesh agnostic neural {PDE} solver.
\newblock In Alice~H. Oh, Alekh Agarwal, Danielle Belgrave, and Kyunghyun Cho (eds.), \emph{Advances in Neural Information Processing Systems}, 2022.
\newblock URL \url{https://openreview.net/forum?id=bx2roi8hca8}.

\bibitem[Brandstetter et~al.(2022)Brandstetter, Worrall, and Welling]{brandstetter2022message}
Johannes Brandstetter, Daniel~E. Worrall, and Max Welling.
\newblock Message passing neural {PDE} solvers.
\newblock In \emph{International Conference on Learning Representations}, 2022.
\newblock URL \url{https://openreview.net/forum?id=vSix3HPYKSU}.

\bibitem[Brixi et~al.(2026)Brixi, Durrant, Ku, Naghipourfar, Poli, Sun, Brockman, Chang, Fanton, Gonzalez, King, Li, Merchant, Nguyen, Ricci-Tam, Romero, Schmok, Taghibakhshi, Vorontsov, Yang, Deng, Gorton, Nguyen, Wang, Pearce, Simon, Adams, Amador, Ashley, Baccus, Dai, Dillmann, Ermon, Guo, Herschl, Ilango, Janik, Lu, Mehta, Mofrad, Ng, Pannu, Ré, St.~John, Sullivan, Tey, Viggiano, Zhu, Zynda, Balsam, Collison, Costa, Hernandez-Boussard, Ho, Liu, McGrath, Powell, Pinglay, Burke, Goodarzi, Hsu, and Hie]{Brixi2026}
Garyk Brixi, Matthew~G. Durrant, Jerome Ku, Mohsen Naghipourfar, Michael Poli, Gwanggyu Sun, Greg Brockman, Daniel Chang, Alison Fanton, Gabriel~A. Gonzalez, Samuel~H. King, David~B. Li, Aditi~T. Merchant, Eric Nguyen, Chiara Ricci-Tam, David~W. Romero, Jonathan~C. Schmok, Ali Taghibakhshi, Anton Vorontsov, Brandon Yang, Myra Deng, Liv Gorton, Nam Nguyen, Nicholas~K. Wang, Michael~T. Pearce, Elana Simon, Etowah Adams, Zachary~J. Amador, Euan~A. Ashley, Stephen~A. Baccus, Haoyu Dai, Steven Dillmann, Stefano Ermon, Daniel Guo, Michael~H. Herschl, Rajesh Ilango, Ken Janik, Amy~X. Lu, Reshma Mehta, Mohammad R.~K. Mofrad, Madelena~Y. Ng, Jaspreet Pannu, Christopher Ré, John St.~John, Jeremy Sullivan, Joseph Tey, Ben Viggiano, Kevin Zhu, Greg Zynda, Daniel Balsam, Patrick Collison, Anthony~B. Costa, Tina Hernandez-Boussard, Eric Ho, Ming-Yu Liu, Thomas McGrath, Kimberly Powell, Sudarshan Pinglay, Dave~P. Burke, Hani Goodarzi, Patrick~D. Hsu, and Brian~L. Hie.
\newblock Genome modelling and design across all domains of life with evo 2.
\newblock \emph{Nature}, 2026.
\newblock \doi{10.1038/s41586-026-10176-5}.
\newblock URL \url{https://doi.org/10.1038/s41586-026-10176-5}.

\bibitem[Chan et~al.(2025)Chan, Chowdhury, Jaffe, Aung, Sherburn, Mays, Starace, Liu, Maksin, Patwardhan, Madry, and Weng]{chan2025mlebench}
Jun~Shern Chan, Neil Chowdhury, Oliver Jaffe, James Aung, Dane Sherburn, Evan Mays, Giulio Starace, Kevin Liu, Leon Maksin, Tejal Patwardhan, Aleksander Madry, and Lilian Weng.
\newblock {MLE}-bench: Evaluating machine learning agents on machine learning engineering.
\newblock In \emph{The Thirteenth International Conference on Learning Representations}, 2025.
\newblock URL \url{https://openreview.net/forum?id=6s5uXNWGIh}.

\bibitem[Chang et~al.(2015)Chang, Funkhouser, Guibas, Hanrahan, Huang, Li, Savarese, Savva, Song, Su, Xiao, Yi, and Yu]{chang2015shapenet}
Angel~X. Chang, Thomas Funkhouser, Leonidas Guibas, Pat Hanrahan, Qixing Huang, Zimo Li, Silvio Savarese, Manolis Savva, Shuran Song, Hao Su, Jianxiong Xiao, Li~Yi, and Fisher Yu.
\newblock Shapenet: An information-rich 3d model repository, 2015.
\newblock URL \url{https://arxiv.org/abs/1512.03012}.

\bibitem[Chen et~al.(2025)Chen, Zhou, Li, Wang, Gao, Shi, Zhang, and Li]{chen2025omniarch}
Tianyu Chen, Haoyi Zhou, Ying Li, Hao Wang, Chonghan Gao, Rongye Shi, Shanghang Zhang, and Jianxin Li.
\newblock Omniarch: Building foundation model for scientific computing.
\newblock In \emph{Forty-second International Conference on Machine Learning}, 2025.
\newblock URL \url{https://openreview.net/forum?id=UlprLwWYKP}.

\bibitem[Chen et~al.(2026)Chen, Pi{e}kos, Ostaszewski, Laakom, and Schmidhuber]{chen2026physgym}
Yimeng Chen, Piotr Pi{e}kos, Mateusz Ostaszewski, Firas Laakom, and J{\"u}rgen Schmidhuber.
\newblock Physgym: Benchmarking {LLM}s in interactive physics discovery with controlled priors.
\newblock In \emph{The Thirty-ninth Annual Conference on Neural Information Processing Systems Datasets and Benchmarks Track}, 2026.
\newblock URL \url{https://openreview.net/forum?id=w8uII2qAmd}.

\bibitem[Domínguez et~al.(2022)Domínguez, Fourtakas, Altomare, Canelas, Tafuni, García-Feal, Martínez-Estévez, Mokos, Vacondio, Crespo, Rogers, Stansby, and Gómez-Gesteira]{DOMINGUEZ2022867}
J.M. Domínguez, G.~Fourtakas, C.~Altomare, R.B. Canelas, A.~Tafuni, O.~García-Feal, I.~Martínez-Estévez, A.~Mokos, R.~Vacondio, A.J.C. Crespo, B.D. Rogers, P.K. Stansby, and M.~Gómez-Gesteira.
\newblock Dualsphysics: from fluid dynamics to multiphysics problems.
\newblock \emph{Computational Particle Mechanics}, 9\penalty0 (5):\penalty0 867--895, 2022.
\newblock ISSN 2196-4386.
\newblock \doi{https://doi.org/10.1007/s40571-021-00404-2}.
\newblock URL \url{https://www.sciencedirect.com/science/article/pii/S219643862500631X}.

\bibitem[Dong et~al.(2026)Dong, Zhang, Wang, Hao, Zhuang, Liu, Wang, Liu, and Jin]{dong2026autopdereliableagenticpde}
Huanshuo Dong, Keyao Zhang, Hong Wang, Zhezheng Hao, Zhiwei Zhuang, Ziyan Liu, Jiacong Wang, Gengyuan Liu, and Xin Jin.
\newblock Autopde: Reliable agentic pde solving via explicitly represented solver strategies, 2026.
\newblock URL \url{https://arxiv.org/abs/2606.10752}.

\bibitem[Ferreira et~al.(2026)Ferreira, Wobbe, Krishnakumar, Hutter, and Zela]{ferreira2026llmsbeatclassicalhyperparameter}
Fabio Ferreira, Lucca Wobbe, Arjun Krishnakumar, Frank Hutter, and Arber Zela.
\newblock Can llms beat classical hyperparameter optimization algorithms? a study on autoresearch, 2026.
\newblock URL \url{https://arxiv.org/abs/2603.24647}.

\bibitem[Gao et~al.(2025)Gao, Shah, Wang, Huang, and Khashabi]{gao2025sciencehierarchographyhierarchicalorganization}
Muhan Gao, Jash Shah, Weiqi Wang, Kuan-Hao Huang, and Daniel Khashabi.
\newblock Science hierarchography: Hierarchical organization of science literature, 2025.
\newblock URL \url{https://arxiv.org/abs/2504.13834}.

\bibitem[Ghafarollahi \& Buehler(2024)Ghafarollahi and Buehler]{ghafarollahi2024sciagents}
Alireza Ghafarollahi and Markus~J Buehler.
\newblock Sciagents: Automating scientific discovery through bioinspired multi-agent intelligent graph reasoning.
\newblock \emph{Advanced Materials}, pp.\  2413523, 2024.

\bibitem[Ghareeb et~al.(2025)Ghareeb, Chang, Mitchener, Yiu, Szostkiewicz, Laurent, Razzak, White, Hinks, and Rodriques]{ghareeb2025robinmultiagentautomatingscientific}
Ali~Essam Ghareeb, Benjamin Chang, Ludovico Mitchener, Angela Yiu, Caralyn~J. Szostkiewicz, Jon~M. Laurent, Muhammed~T. Razzak, Andrew~D. White, Michaela~M. Hinks, and Samuel~G. Rodriques.
\newblock Robin: A multi-agent system for automating scientific discovery, 2025.
\newblock URL \url{https://arxiv.org/abs/2505.13400}.

\bibitem[Gingold \& Monaghan(1977)Gingold and Monaghan]{10.1093/mnras/181.3.375}
R.~A. Gingold and J.~J. Monaghan.
\newblock Smoothed particle hydrodynamics: theory and application to non-spherical stars.
\newblock \emph{Monthly Notices of the Royal Astronomical Society}, 181\penalty0 (3):\penalty0 375--389, 12 1977.
\newblock ISSN 0035-8711.
\newblock \doi{10.1093/mnras/181.3.375}.
\newblock URL \url{https://doi.org/10.1093/mnras/181.3.375}.

\bibitem[Go et~al.(2025)Go, Ly, Sogaard, Tabatabaei, de~Rijke, and Chen]{Go2025LiRAAM}
Gregory Go, Khang Ly, Anders Sogaard, Amin~Reza Tabatabaei, Maarten de~Rijke, and Xinyi Chen.
\newblock Lira: A multi-agent framework for reliable and readable literature review generation.
\newblock In \emph{AAAI Conference on Artificial Intelligence}, 2025.
\newblock URL \url{https://api.semanticscholar.org/CorpusID:281886200}.

\bibitem[Greenfeld et~al.(2019)Greenfeld, Galun, Basri, Yavneh, and Kimmel]{greenfeld2019learning}
Daniel Greenfeld, Meirav Galun, Ronen Basri, Irad Yavneh, and Ron Kimmel.
\newblock Learning to optimize multigrid {PDE} solvers.
\newblock In \emph{International Conference on Machine Learning}, pp.\  2415--2423. PMLR, 2019.

\bibitem[Guo et~al.(2026)Guo, Chawla, Wiest, and Zhang]{guo2026autollmresearchtrainingresearchagents}
Taicheng Guo, Nitesh~V. Chawla, Olaf Wiest, and Xiangliang Zhang.
\newblock Autollmresearch: Training research agents for automating llm experiment configuration - learning from cheap, optimizing expensive, 2026.
\newblock URL \url{https://arxiv.org/abs/2605.11518}.

\bibitem[Gupta \& Brandstetter(2022)Gupta and Brandstetter]{gupta2022towards}
Jayesh~K Gupta and Johannes Brandstetter.
\newblock Towards multi-spatiotemporal-scale generalized pde modeling.
\newblock \emph{arXiv preprint arXiv:2209.15616}, 2022.

\bibitem[Han et~al.(2022)Han, Huang, Ma, Li, Tenenbaum, and Gan]{han2022learning}
Jiaqi Han, Wenbing Huang, Hengbo Ma, Jiachen Li, Joshua~B Tenenbaum, and Chuang Gan.
\newblock Learning physical dynamics with subequivariant graph neural networks.
\newblock \emph{arXiv preprint arXiv:2210.06876}, 2022.

\bibitem[Hang et~al.(2026)Hang, Yashengjiang, Li, Dong, Wei, Hao, Ma, Bai, Kai, Yue, Si, Jiang, Yao, Hu, Zhang, Liu, Shen, Ren, Liu, Xu, Li, Yao, Dong, and Wang]{hang2026pdeagentbenchmultimetricmultilibrarybenchmark}
Zhen Hang, Yushan Yashengjiang, Junhui Li, Huanshuo Dong, Yang Wei, Zhezheng Hao, Jiangtao Ma, Songlin Bai, Haozhong Kai, Xihang Yue, Gangzong Si, Dongming Jiang, Chao Yao, Zhanhua Hu, Jiangqing Zhang, Pengwei Liu, Yaomin Shen, Xingyu Ren, Lei Liu, Zikang Xu, Han Li, Qingsong Yao, Hande Dong, and Hong Wang.
\newblock Pdeagent-bench: A multi-metric, multi-library benchmark for pde solver generation, 2026.
\newblock URL \url{https://arxiv.org/abs/2605.09636}.

\bibitem[Hao et~al.(2023)Hao, Ying, Wang, Su, Dong, Liu, Cheng, Zhu, and Song]{Hao2023GNOTAG}
Zhongkai Hao, Chengyang Ying, Zhengyi Wang, Hang Su, Yinpeng Dong, Songming Liu, Ze~Cheng, Jun Zhu, and Jian Song.
\newblock Gnot: A general neural operator transformer for operator learning.
\newblock \emph{ArXiv}, abs/2302.14376, 2023.
\newblock URL \url{https://api.semanticscholar.org/CorpusID:257232579}.

\bibitem[Hao et~al.(2024)Hao, Su, Liu, Berner, Ying, Su, Anandkumar, Song, and Zhu]{hao2024dpot}
Zhongkai Hao, Chang Su, Songming Liu, Julius Berner, Chengyang Ying, Hang Su, Anima Anandkumar, Jian Song, and Jun Zhu.
\newblock Dpot: Auto-regressive denoising operator transformer for large-scale pde pre-training.
\newblock \emph{arXiv preprint arXiv:2403.03542}, 2024.

\bibitem[Helwig et~al.(2026)Helwig, Adavi, Zhang, Lin, Chim, Vizzini, Yu, Hasnain, Biswas, Holloway, Singh, Anand, Guhathakurta, and Ji]{helwig2026twophasedeeplearningframework}
Jacob Helwig, Sai~Sreeharsha Adavi, Xuan Zhang, Yuchao Lin, Felix~S. Chim, Luke~Takeshi Vizzini, Haiyang Yu, Muhammad Hasnain, Saykat~Kumar Biswas, John~J. Holloway, Narendra Singh, N.~K. Anand, Swagnik Guhathakurta, and Shuiwang Ji.
\newblock A two-phase deep learning framework for adaptive time-stepping in high-speed flow modeling, 2026.
\newblock URL \url{https://arxiv.org/abs/2506.07969}.

\bibitem[Herde et~al.(2024)Herde, Raonic, Rohner, K{\"a}ppeli, Molinaro, de~Bezenac, and Mishra]{herde2024poseidon}
Maximilian Herde, Bogdan Raonic, Tobias Rohner, Roger K{\"a}ppeli, Roberto Molinaro, Emmanuel de~Bezenac, and Siddhartha Mishra.
\newblock Poseidon: Efficient foundation models for {PDE}s.
\newblock In \emph{The Thirty-eighth Annual Conference on Neural Information Processing Systems}, 2024.
\newblock URL \url{https://openreview.net/forum?id=JC1VKK3UXk}.

\bibitem[Horie \& MITSUME(2022)Horie and MITSUME]{horie2022physicsembedded}
Masanobu Horie and NAOTO MITSUME.
\newblock Physics-embedded neural networks: Graph neural {PDE} solvers with mixed boundary conditions.
\newblock In \emph{NeurIPS 2022 AI for Science: Progress and Promises}, 2022.
\newblock URL \url{https://openreview.net/forum?id=gyIRLZPGiG}.

\bibitem[Hu et~al.(2026)Hu, Feng, Liu, Yan, Deng, Gao, Zheng, Zheng, Yu, Wang, Li, Ma, Zhou, Lu, Fan, and Wu]{hu2026realpdebench}
Peiyan Hu, Haodong Feng, Hongyuan Liu, Tongtong Yan, Wenhao Deng, Tianrun Gao, Rong Zheng, Haoren Zheng, Chenglei Yu, Chuanrui Wang, Kaiwen Li, Zhi-Ming Ma, Dezhi Zhou, Xingcai Lu, Dixia Fan, and Tailin Wu.
\newblock Realpdebench: A benchmark for complex physical systems with real-world data.
\newblock In \emph{The Fourteenth International Conference on Learning Representations}, 2026.
\newblock URL \url{https://openreview.net/forum?id=y3oHMcoItR}.
\newblock Oral Presentation.

\bibitem[Huang et~al.(2023)Huang, Vora, Liang, and Leskovec]{Huang2023MLAgentBenchEL}
Qian Huang, Jian Vora, Percy Liang, and Jure Leskovec.
\newblock Mlagentbench: Evaluating language agents on machine learning experimentation.
\newblock In \emph{International Conference on Machine Learning}, 2023.
\newblock URL \url{https://api.semanticscholar.org/CorpusID:263671541}.

\bibitem[Jimenez et~al.(2024)Jimenez, Yang, Wettig, Yao, Pei, Press, and Narasimhan]{jimenez2024swebench}
Carlos~E Jimenez, John Yang, Alexander Wettig, Shunyu Yao, Kexin Pei, Ofir Press, and Karthik~R Narasimhan.
\newblock {SWE}-bench: Can language models resolve real-world github issues?
\newblock In \emph{The Twelfth International Conference on Learning Representations}, 2024.
\newblock URL \url{https://openreview.net/forum?id=VTF8yNQM66}.

\bibitem[Koehler et~al.(2024)Koehler, Niedermayr, Westermann, and Thuerey]{koehler2024apebench}
Felix Koehler, Simon Niedermayr, R{\"u}diger Westermann, and Nils Thuerey.
\newblock Apebench: A benchmark for autoregressive neural emulators of pdes.
\newblock \emph{Advances in Neural Information Processing Systems}, 37:\penalty0 120252--120310, 2024.

\bibitem[Kovachki et~al.(2023)Kovachki, Li, Liu, Azizzadenesheli, Bhattacharya, Stuart, and Anandkumar]{kovachki2023neuraloperator}
Nikola Kovachki, Zongyi Li, Burigede Liu, Kamyar Azizzadenesheli, Kaushik Bhattacharya, Andrew Stuart, and Anima Anandkumar.
\newblock Neural operator: Learning maps between function spaces with applications to pdes.
\newblock \emph{JMLR}, 24\penalty0 (1), 2023.

\bibitem[Li et~al.(2024{\natexlab{a}})Li, Patel, Wang, Wang, and Du]{Li2024MLRCopilotAM}
Ruochen Li, Teerth Patel, Qingyun Wang, Qingyun Wang, and Xinya Du.
\newblock Mlr-copilot: Autonomous machine learning research based on large language models agents.
\newblock \emph{ArXiv}, abs/2408.14033, 2024{\natexlab{a}}.
\newblock URL \url{https://api.semanticscholar.org/CorpusID:271957477}.

\bibitem[Li et~al.(2025{\natexlab{a}})Li, Huang, Wang, Wan, Sun, and Wang]{li2025selfguideddiffusionmodelaccelerating}
Ruoyan Li, Zijie Huang, Haixin Wang, Guancheng Wan, Yizhou Sun, and Wei Wang.
\newblock Self-guided diffusion model for accelerating computational fluid dynamics, 2025{\natexlab{a}}.
\newblock URL \url{https://arxiv.org/abs/2504.04375}.

\bibitem[Li et~al.(2025{\natexlab{b}})Li, Wan, Huang, Liu, Wang, Luo, Wang, and Sun]{li2025flow}
Ruoyan Li, Guancheng Wan, Zijie Huang, Zixiao Liu, Haixin Wang, Xiao Luo, Wei Wang, and Yizhou Sun.
\newblock Flow field reconstruction with sensor placement policy learning.
\newblock In \emph{The Thirty-ninth Annual Conference on Neural Information Processing Systems}, 2025{\natexlab{b}}.
\newblock URL \url{https://openreview.net/forum?id=1esFEjMUBS}.

\bibitem[Li et~al.(2025{\natexlab{c}})Li, Marwah, Shen, Sun, Risteski, Yang, and Talwalkar]{li2025codepde}
Shanda Li, Tanya Marwah, Junhong Shen, Weiwei Sun, Andrej Risteski, Yiming Yang, and Ameet Talwalkar.
\newblock Codepde: An inference framework for llm-driven pde solver generation.
\newblock \emph{arXiv preprint arXiv:2505.08783}, 2025{\natexlab{c}}.

\bibitem[Li et~al.(2024{\natexlab{b}})Li, Song, Xiao, Lai, and Wang]{li2024harnessing}
Zhihao Li, Haoze Song, Di~Xiao, Zhilu Lai, and Wei Wang.
\newblock Harnessing scale and physics: A multi-graph neural operator framework for pdes on arbitrary geometries.
\newblock \emph{arXiv preprint arXiv:2411.15178}, 2024{\natexlab{b}}.

\bibitem[Li et~al.(2020)Li, Kovachki, Azizzadenesheli, Liu, Bhattacharya, Stuart, and Anandkumar]{li2020multipolegraphneuraloperator}
Zongyi Li, Nikola Kovachki, Kamyar Azizzadenesheli, Burigede Liu, Kaushik Bhattacharya, Andrew Stuart, and Anima Anandkumar.
\newblock Multipole graph neural operator for parametric partial differential equations, 2020.
\newblock URL \url{https://arxiv.org/abs/2006.09535}.

\bibitem[Li et~al.(2021)Li, Kovachki, Azizzadenesheli, liu, Bhattacharya, Stuart, and Anandkumar]{li2021fourier}
Zongyi Li, Nikola~Borislavov Kovachki, Kamyar Azizzadenesheli, Burigede liu, Kaushik Bhattacharya, Andrew Stuart, and Anima Anandkumar.
\newblock Fourier neural operator for parametric partial differential equations.
\newblock In \emph{International Conference on Learning Representations}, 2021.
\newblock URL \url{https://openreview.net/forum?id=c8P9NQVtmnO}.

\bibitem[Li et~al.(2023)Li, Kovachki, Choy, Li, Kossaifi, Otta, Nabian, Stadler, Hundt, Azizzadenesheli, and Anandkumar]{li2023geometryinformed}
Zongyi Li, Nikola~Borislavov Kovachki, Chris Choy, Boyi Li, Jean Kossaifi, Shourya~Prakash Otta, Mohammad~Amin Nabian, Maximilian Stadler, Christian Hundt, Kamyar Azizzadenesheli, and Anima Anandkumar.
\newblock Geometry-informed neural operator for large-scale 3d {PDE}s.
\newblock In \emph{Thirty-seventh Conference on Neural Information Processing Systems}, 2023.
\newblock URL \url{https://openreview.net/forum?id=86dXbqT5Ua}.

\bibitem[Lino et~al.(2021)Lino, Cantwell, Bharath, and Fotiadis]{lino2021simulatingcontinuummechanicsmultiscale}
Mario Lino, Chris Cantwell, Anil~A. Bharath, and Stathi Fotiadis.
\newblock Simulating continuum mechanics with multi-scale graph neural networks, 2021.
\newblock URL \url{https://arxiv.org/abs/2106.04900}.

\bibitem[List et~al.(2022)List, Chen, and Thuerey]{list2022learned}
B.~List, L.-W. Chen, and N.~Thuerey.
\newblock Learned turbulence modelling with differentiable fluid solvers: physics-based loss functions and optimisation horizons.
\newblock \emph{Journal of Fluid Mechanics}, 949:\penalty0 A25, 2022.

\bibitem[Lu et~al.(2021)Lu, Jin, Pang, Zhang, and Karniadakis]{lu2021learning}
Lu~Lu, Pengzhan Jin, Guofei Pang, Zhongqiang Zhang, and George~Em Karniadakis.
\newblock Learning nonlinear operators via {DeepONet} based on the universal approximation theorem of operators.
\newblock \emph{Nature Machine Intelligence}, 3\penalty0 (3):\penalty0 218--229, 2021.

\bibitem[Luo et~al.()Luo, Wu, Zhou, Xing, Di, Wang, and Long]{luotransolver++}
Huakun Luo, Haixu Wu, Hang Zhou, Lanxiang Xing, Yichen Di, Jianmin Wang, and Mingsheng Long.
\newblock Transolver++: An accurate neural solver for pdes on million-scale geometries.
\newblock In \emph{Forty-second International Conference on Machine Learning}.

\bibitem[McCabe et~al.(2024)McCabe, Blancard, Parker, Ohana, Cranmer, Bietti, Eickenberg, Golkar, Krawezik, Lanusse, Pettee, Tesileanu, Cho, and Ho]{mccabe2024multiple}
Michael McCabe, Bruno R{\'e}galdo-Saint Blancard, Liam~Holden Parker, Ruben Ohana, Miles Cranmer, Alberto Bietti, Michael Eickenberg, Siavash Golkar, Geraud Krawezik, Francois Lanusse, Mariel Pettee, Tiberiu Tesileanu, Kyunghyun Cho, and Shirley Ho.
\newblock Multiple physics pretraining for spatiotemporal surrogate models.
\newblock In \emph{The Thirty-eighth Annual Conference on Neural Information Processing Systems}, 2024.
\newblock URL \url{https://openreview.net/forum?id=DKSI3bULiZ}.

\bibitem[Merler et~al.(2024)Merler, Haitsiukevich, Dainese, and Marttinen]{merler-etal-2024-context}
Matteo Merler, Katsiaryna Haitsiukevich, Nicola Dainese, and Pekka Marttinen.
\newblock In-context symbolic regression: Leveraging large language models for function discovery.
\newblock In Xiyan Fu and Eve Fleisig (eds.), \emph{Proceedings of the 62nd Annual Meeting of the Association for Computational Linguistics (Volume 4: Student Research Workshop)}, pp.\  589--606, Bangkok, Thailand, August 2024. Association for Computational Linguistics.
\newblock \doi{10.18653/v1/2024.acl-srw.49}.
\newblock URL \url{https://aclanthology.org/2024.acl-srw.49}.

\bibitem[Mitchener et~al.(2025)Mitchener, Yiu, Chang, Bourdenx, Nadolski, Sulovari, Landsness, Barabasi, Narayanan, Evans, Reddy, Foiani, Kamal, Shriver, Cao, Wassie, Laurent, Melville-Green, Caldas, Bou, Roberts, Zagorac, Orr, Orr, Zwezdaryk, Ghareeb, McCoy, Gomes, Ashley, Duff, Buonassisi, Rainforth, Bateman, Skarlinski, Rodriques, Hinks, and White]{mitchener2025kosmosaiscientistautonomous}
Ludovico Mitchener, Angela Yiu, Benjamin Chang, Mathieu Bourdenx, Tyler Nadolski, Arvis Sulovari, Eric~C. Landsness, Daniel~L. Barabasi, Siddharth Narayanan, Nicky Evans, Shriya Reddy, Martha Foiani, Aizad Kamal, Leah~P. Shriver, Fang Cao, Asmamaw~T. Wassie, Jon~M. Laurent, Edwin Melville-Green, Mayk Caldas, Albert Bou, Kaleigh~F. Roberts, Sladjana Zagorac, Timothy~C. Orr, Miranda~E. Orr, Kevin~J. Zwezdaryk, Ali~E. Ghareeb, Laurie McCoy, Bruna Gomes, Euan~A. Ashley, Karen~E. Duff, Tonio Buonassisi, Tom Rainforth, Randall~J. Bateman, Michael Skarlinski, Samuel~G. Rodriques, Michaela~M. Hinks, and Andrew~D. White.
\newblock Kosmos: An ai scientist for autonomous discovery, 2025.
\newblock URL \url{https://arxiv.org/abs/2511.02824}.

\bibitem[Molinaro et~al.(2023)Molinaro, Yang, Engquist, and Mishra]{molinaro2023neuralinverseoperatorssolving}
Roberto Molinaro, Yunan Yang, Björn Engquist, and Siddhartha Mishra.
\newblock Neural inverse operators for solving pde inverse problems, 2023.
\newblock URL \url{https://arxiv.org/abs/2301.11167}.

\bibitem[Mousavi et~al.(2025)Mousavi, Wen, Lingsch, Herde, Raonić, and Mishra]{mousavi2025rigno}
Sepehr Mousavi, Shizheng Wen, Levi Lingsch, Maximilian Herde, Bogdan Raonić, and Siddhartha Mishra.
\newblock Rigno: A graph-based framework for robust and accurate operator learning for pdes on arbitrary domains.
\newblock In \emph{Advances in Neural Information Processing Systems}, volume~38, 2025.

\bibitem[Novikov et~al.(2025)Novikov, Vũ, Eisenberger, Dupont, Huang, Wagner, Shirobokov, Kozlovskii, Ruiz, Mehrabian, Kumar, See, Chaudhuri, Holland, Davies, Nowozin, Kohli, and Balog]{novikov2025alphaevolvecodingagentscientific}
Alexander Novikov, Ngân Vũ, Marvin Eisenberger, Emilien Dupont, Po-Sen Huang, Adam~Zsolt Wagner, Sergey Shirobokov, Borislav Kozlovskii, Francisco J.~R. Ruiz, Abbas Mehrabian, M.~Pawan Kumar, Abigail See, Swarat Chaudhuri, George Holland, Alex Davies, Sebastian Nowozin, Pushmeet Kohli, and Matej Balog.
\newblock Alphaevolve: A coding agent for scientific and algorithmic discovery, 2025.
\newblock URL \url{https://arxiv.org/abs/2506.13131}.

\bibitem[Ohana et~al.(2024)Ohana, McCabe, Meyer, Morel, Agocs, Beneitez, Berger, Burkhart, Dalziel, Fielding, et~al.]{ohana2024well}
Ruben Ohana, Michael McCabe, Lucas Meyer, Rudy Morel, Fruzsina Agocs, Miguel Beneitez, Marsha Berger, Blakesly Burkhart, Stuart Dalziel, Drummond Fielding, et~al.
\newblock The well: a large-scale collection of diverse physics simulations for machine learning.
\newblock \emph{Advances in Neural Information Processing Systems}, 37:\penalty0 44989--45037, 2024.

\bibitem[Paszke et~al.(2019)Paszke, Gross, Massa, Lerer, Bradbury, Chanan, Killeen, Lin, Gimelshein, Antiga, Desmaison, Köpf, Yang, DeVito, Raison, Tejani, Chilamkurthy, Steiner, Fang, Bai, and Chintala]{paszke2019pytorchimperativestylehighperformance}
Adam Paszke, Sam Gross, Francisco Massa, Adam Lerer, James Bradbury, Gregory Chanan, Trevor Killeen, Zeming Lin, Natalia Gimelshein, Luca Antiga, Alban Desmaison, Andreas Köpf, Edward Yang, Zach DeVito, Martin Raison, Alykhan Tejani, Sasank Chilamkurthy, Benoit Steiner, Lu~Fang, Junjie Bai, and Soumith Chintala.
\newblock Pytorch: An imperative style, high-performance deep learning library, 2019.
\newblock URL \url{https://arxiv.org/abs/1912.01703}.

\bibitem[Pfaff et~al.(2021)Pfaff, Fortunato, Sanchez-Gonzalez, and Battaglia]{pfaff2021learning}
Tobias Pfaff, Meire Fortunato, Alvaro Sanchez-Gonzalez, and Peter Battaglia.
\newblock Learning mesh-based simulation with graph networks.
\newblock In \emph{International Conference on Learning Representations}, 2021.
\newblock URL \url{https://openreview.net/forum?id=roNqYL0_XP}.

\bibitem[Prantl et~al.(2022)Prantl, Ummenhofer, Koltun, and Thuerey]{prantl2022guaranteed}
Lukas Prantl, Benjamin Ummenhofer, Vladlen Koltun, and Nils Thuerey.
\newblock Guaranteed conservation of momentum for learning particle-based fluid dynamics.
\newblock In Alice~H. Oh, Alekh Agarwal, Danielle Belgrave, and Kyunghyun Cho (eds.), \emph{Advances in Neural Information Processing Systems}, 2022.
\newblock URL \url{https://openreview.net/forum?id=6niwHlzh10U}.

\bibitem[Qiu et~al.(2025)Qiu, Zhang, Xu, Li, Song, Wang, and Zhang]{qiu2025aiideabench2025}
Yansheng Qiu, Haoquan Zhang, Zhaopan Xu, Ming Li, Diping Song, Zheng Wang, and Kaipeng Zhang.
\newblock Ai idea bench 2025: Ai research idea generation benchmark, 2025.
\newblock URL \url{https://arxiv.org/abs/2504.14191}.

\bibitem[Rahman et~al.(2023)Rahman, Ross, and Azizzadenesheli]{rahman2023uno}
Md~Ashiqur Rahman, Zachary~E Ross, and Kamyar Azizzadenesheli.
\newblock U-{NO}: U-shaped neural operators.
\newblock \emph{Transactions on Machine Learning Research}, 2023.
\newblock ISSN 2835-8856.
\newblock URL \url{https://openreview.net/forum?id=j3oQF9coJd}.

\bibitem[Ramachandran et~al.(2021)Ramachandran, Bhosale, Puri, Negi, Muta, Dinesh, Menon, Govind, Sanka, Sebastian, Sen, Kaushik, Kumar, Kurapati, Patil, Tavker, Pandey, Kaushik, Dutt, and Agarwal]{ramachandran2021a}
Prabhu Ramachandran, Aditya Bhosale, Kunal Puri, Pawan Negi, Abhinav Muta, A.~Dinesh, Dileep Menon, Rahul Govind, Suraj Sanka, Amal~S. Sebastian, Ananyo Sen, Rohan Kaushik, Anshuman Kumar, Vikas Kurapati, Mrinalgouda Patil, Deep Tavker, Pankaj Pandey, Chandrashekhar Kaushik, Arkopal Dutt, and Arpit Agarwal.
\newblock {{PySPH}}: {{A Python-based Framework}} for {{Smoothed Particle Hydrodynamics}}.
\newblock \emph{ACM Transactions on Mathematical Software}, 47\penalty0 (4):\penalty0 1--38, December 2021.
\newblock ISSN 0098-3500, 1557-7295.
\newblock \doi{10.1145/3460773}.

\bibitem[Rochman-Sharabi et~al.(2025)Rochman-Sharabi, Lewin, and Louppe]{rochman-sharabi2025a}
Omer Rochman-Sharabi, Sacha Lewin, and Gilles Louppe.
\newblock A neural material point method for particle-based emulation.
\newblock \emph{Transactions on Machine Learning Research}, 2025.
\newblock ISSN 2835-8856.
\newblock URL \url{https://openreview.net/forum?id=zSK81A2hxQ}.

\bibitem[Saad et~al.(2023)Saad, Gupta, Alizadeh, and Maddix]{saad2023guiding}
Nadim Saad, Gaurav Gupta, Shima Alizadeh, and Danielle~C. Maddix.
\newblock Guiding continuous operator learning through physics-based boundary constraints.
\newblock In \emph{The Eleventh International Conference on Learning Representations}, 2023.
\newblock URL \url{https://openreview.net/forum?id=gfWNItGOES6}.

\bibitem[Sanchez-Gonzalez et~al.(2020)Sanchez-Gonzalez, Godwin, Pfaff, Ying, Leskovec, and Battaglia]{sanchezgonzalez2020learningsimulatecomplexphysics}
Alvaro Sanchez-Gonzalez, Jonathan Godwin, Tobias Pfaff, Rex Ying, Jure Leskovec, and Peter~W. Battaglia.
\newblock Learning to simulate complex physics with graph networks, 2020.
\newblock URL \url{https://arxiv.org/abs/2002.09405}.

\bibitem[Sappl et~al.(2019)Sappl, Seiler, Harders, and Rauch]{sappl2019deep}
J.~Sappl, L.~Seiler, M.~Harders, and W.~Rauch.
\newblock Deep learning of preconditioners for conjugate gradient solvers in urban water related problems.
\newblock \emph{arXiv preprint}, 2019.

\bibitem[Schmidgall et~al.(2025)Schmidgall, Su, Wang, Sun, Wu, Yu, Liu, Moor, Liu, and Barsoum]{schmidgall2025agentlaboratoryusingllm}
Samuel Schmidgall, Yusheng Su, Ze~Wang, Ximeng Sun, Jialian Wu, Xiaodong Yu, Jiang Liu, Michael Moor, Zicheng Liu, and Emad Barsoum.
\newblock Agent laboratory: Using llm agents as research assistants, 2025.
\newblock URL \url{https://arxiv.org/abs/2501.04227}.

\bibitem[Shen et~al.(2024)Shen, Marwah, and Talwalkar]{shen2024ups}
Junhong Shen, Tanya Marwah, and Ameet Talwalkar.
\newblock {UPS}: Efficiently building foundation models for {PDE} solving via cross-modal adaptation.
\newblock \emph{Transactions on Machine Learning Research}, 2024.
\newblock ISSN 2835-8856.
\newblock URL \url{https://openreview.net/forum?id=0r9mhjRv1E}.

\bibitem[Skarlinski et~al.(2024)Skarlinski, Cox, Laurent, Braza, Hinks, Hammerling, Ponnapati, Rodriques, and White]{skarlinski2024languageagentsachievesuperhuman}
Michael~D. Skarlinski, Sam Cox, Jon~M. Laurent, James~D. Braza, Michaela Hinks, Michael~J. Hammerling, Manvitha Ponnapati, Samuel~G. Rodriques, and Andrew~D. White.
\newblock Language agents achieve superhuman synthesis of scientific knowledge, 2024.
\newblock URL \url{https://arxiv.org/abs/2409.13740}.

\bibitem[Song et~al.(2026)Song, Zhen, and Jiang]{song2026can}
Zeyuan Song, Xiaocong Zhen, and Zheyu Jiang.
\newblock Can large language models design effective neural operators for solving partial differential equations?
\newblock In \emph{3rd AI for Math Workshop: Toward Self-Evolving Scientific Agents}, 2026.
\newblock URL \url{https://openreview.net/forum?id=6GCIuS0oO0}.

\bibitem[Sun et~al.(2023)Sun, Yang, and Yoo]{sun2023neural}
Z.~Sun, Y.~Yang, and S.~Yoo.
\newblock A neural {PDE} solver with temporal stencil modeling.
\newblock \emph{arXiv preprint}, 2023.

\bibitem[Takamoto et~al.(2023)Takamoto, Alesiani, and Niepert]{takamoto2023learningneuralpdesolvers}
Makoto Takamoto, Francesco Alesiani, and Mathias Niepert.
\newblock Learning neural pde solvers with parameter-guided channel attention, 2023.
\newblock URL \url{https://arxiv.org/abs/2304.14118}.

\bibitem[Takamoto et~al.(2024)Takamoto, Praditia, Leiteritz, MacKinlay, Alesiani, Pflüger, and Niepert]{takamoto2024pdebenchextensivebenchmarkscientific}
Makoto Takamoto, Timothy Praditia, Raphael Leiteritz, Dan MacKinlay, Francesco Alesiani, Dirk Pflüger, and Mathias Niepert.
\newblock Pdebench: An extensive benchmark for scientific machine learning, 2024.
\newblock URL \url{https://arxiv.org/abs/2210.07182}.

\bibitem[Tang et~al.(2025)Tang, Xia, Li, and Huang]{tang2025airesearcherautonomousscientificinnovation}
Jiabin Tang, Lianghao Xia, Zhonghang Li, and Chao Huang.
\newblock Ai-researcher: Autonomous scientific innovation, 2025.
\newblock URL \url{https://arxiv.org/abs/2505.18705}.

\bibitem[Toshev et~al.(2024{\natexlab{a}})Toshev, Galletti, Fritz, Adami, and Adams]{toshev2024lagrangebench}
Artur Toshev, Gianluca Galletti, Fabian Fritz, Stefan Adami, and Nikolaus Adams.
\newblock Lagrangebench: A lagrangian fluid mechanics benchmarking suite.
\newblock \emph{Advances in Neural Information Processing Systems}, 36, 2024{\natexlab{a}}.

\bibitem[Toshev et~al.(2023)Toshev, Galletti, Brandstetter, Adami, and Adams]{toshev2023learning}
Artur~P. Toshev, Gianluca Galletti, Johannes Brandstetter, Stefan Adami, and Nikolaus~A. Adams.
\newblock Learning lagrangian fluid mechanics with e($3$)-equivariant graph neural networks, 2023.

\bibitem[Toshev et~al.(2024{\natexlab{b}})Toshev, Erbesdobler, Adams, and Brandstetter]{toshev2024neuralsphimprovedneural}
Artur~P. Toshev, Jonas~A. Erbesdobler, Nikolaus~A. Adams, and Johannes Brandstetter.
\newblock Neural sph: Improved neural modeling of lagrangian fluid dynamics, 2024{\natexlab{b}}.
\newblock URL \url{https://arxiv.org/abs/2402.06275}.

\bibitem[Tran et~al.(2023)Tran, Mathews, Xie, and Ong]{tran2023factorized}
Alasdair Tran, Alexander Mathews, Lexing Xie, and Cheng~Soon Ong.
\newblock Factorized fourier neural operators.
\newblock In \emph{The Eleventh International Conference on Learning Representations}, 2023.
\newblock URL \url{https://openreview.net/forum?id=tmIiMPl4IPa}.

\bibitem[Tripura \& Chakraborty(2023)Tripura and Chakraborty]{tripura2023wavelet}
Tapas Tripura and Souvik Chakraborty.
\newblock Wavelet neural operator for solving parametric partial differential equations in computational mechanics problems.
\newblock \emph{Computer Methods in Applied Mechanics and Engineering}, 404:\penalty0 115783, 2023.

\bibitem[Umetani \& Bickel(2018)Umetani and Bickel]{umetani2018learning}
Nobuyuki Umetani and Bernd Bickel.
\newblock Learning three-dimensional flow for interactive aerodynamic design.
\newblock \emph{ACM Trans. Graph.}, 37\penalty0 (4), July 2018.
\newblock ISSN 0730-0301.
\newblock \doi{10.1145/3197517.3201325}.
\newblock URL \url{https://doi.org/10.1145/3197517.3201325}.

\bibitem[Wang et~al.(2024)Wang, Jiaxin, Dwivedi, Hara, and Wu]{wang2024beno}
Haixin Wang, LI~Jiaxin, Anubhav Dwivedi, Kentaro Hara, and Tailin Wu.
\newblock {BENO}: Boundary-embedded neural operators for elliptic {PDE}s.
\newblock In \emph{The Twelfth International Conference on Learning Representations}, 2024.
\newblock URL \url{https://openreview.net/forum?id=ZZTkLDRmkg}.

\bibitem[Wang et~al.(2026{\natexlab{a}})Wang, Li, Xu, Sun, Han, Huang, Chang, Luo, Wang, and Sun]{wang2026fdbenchmodularfairbenchmark}
Haixin Wang, Ruoyan Li, Fred Xu, Fang Sun, Kaiqiao Han, Zijie Huang, Ching Chang, Xiao Luo, Wei Wang, and Yizhou Sun.
\newblock Fd-bench: A modular and fair benchmark for data-driven fluid simulation, 2026{\natexlab{a}}.
\newblock URL \url{https://arxiv.org/abs/2505.20349}.

\bibitem[Wang et~al.(2025)Wang, Xin, Wang, Yang, Zha, Dong, and Jiang]{wang2025mixtureofexpertsoperatortransformerlargescale}
Hong Wang, Haiyang Xin, Jie Wang, Xuanze Yang, Fei Zha, Huanshuo Dong, and Yan Jiang.
\newblock Mixture-of-experts operator transformer for large-scale pde pre-training, 2025.
\newblock URL \url{https://arxiv.org/abs/2510.25803}.

\bibitem[Wang \& Wang(2024)Wang and Wang]{wang2024LNO}
Tian Wang and Chuang Wang.
\newblock Latent neural operator for solving forward and inverse pde problems.
\newblock In \emph{Advances in Neural Information Processing Systems (NeurIPS)}, 2024.

\bibitem[Wang et~al.(2026{\natexlab{b}})Wang, Zhang, Wang, Shi, Li, Han, Tong, Deng, Taylor, Sun, Zhu, Cong, Sun, and Wang]{wang2026arlarena}
Xiaoxuan Wang, Han Zhang, Haixin Wang, Yidan Shi, Ruoyan Li, Kaiqiao Han, Chenyi Tong, Haoran Deng, Alexander~K Taylor, Renliang Sun, Yanqiao Zhu, Jason Cong, Yizhou Sun, and Wei Wang.
\newblock {ARLA}rena: A unified framework for stable agentic reinforcement learning.
\newblock In \emph{Forty-third International Conference on Machine Learning}, 2026{\natexlab{b}}.
\newblock URL \url{https://openreview.net/forum?id=90kxFi9VGP}.

\bibitem[Wen et~al.(2025)Wen, Kumbhat, Lingsch, Mousavi, Zhao, Chandrashekar, and Mishra]{wen2025goat}
Shizheng Wen, Arsh Kumbhat, Levi Lingsch, Sepehr Mousavi, Yizhou Zhao, Praveen Chandrashekar, and Siddhartha Mishra.
\newblock Geometry aware operator transformer as an efficient and accurate neural surrogate for pdes on arbitrary domains.
\newblock 2025.

\bibitem[Wu et~al.(2024)Wu, Luo, Wang, Wang, and Long]{wu2024Transolver}
Haixu Wu, Huakun Luo, Haowen Wang, Jianmin Wang, and Mingsheng Long.
\newblock Transolver: A fast transformer solver for pdes on general geometries.
\newblock In \emph{International Conference on Machine Learning}, 2024.

\bibitem[Wu et~al.(2025)Wu, Zhang, and Zhu]{wu2025automatedcodedevelopmentpde}
Haoyang Wu, Xinxin Zhang, and Lailai Zhu.
\newblock Automated code development for pde solvers using large language models, 2025.
\newblock URL \url{https://arxiv.org/abs/2509.25194}.

\bibitem[Wuwu et~al.(2025)Wuwu, Gao, Chen, Huang, Zhang, Wang, Li, Zhou, and Zhang]{wuwu2025pinnsagent}
Qingpo Wuwu, Chonghan Gao, Tianyu Chen, Yihang Huang, Yuekai Zhang, Jianing Wang, Jianxin Li, Haoyi Zhou, and Shanghang Zhang.
\newblock {PINN}sagent: Automated {PDE} surrogation with large language models.
\newblock In \emph{Forty-second International Conference on Machine Learning}, 2025.
\newblock URL \url{https://openreview.net/forum?id=RO5OGOzs6M}.

\bibitem[Yang et~al.(2025)Yang, Liu, Gao, Xie, Li, Ouyang, Poria, Cambria, and Zhou]{yang2025moosechem}
Zonglin Yang, Wanhao Liu, Ben Gao, Tong Xie, Yuqiang Li, Wanli Ouyang, Soujanya Poria, Erik Cambria, and Dongzhan Zhou.
\newblock {MOOSE}-chem: Large language models for rediscovering unseen chemistry scientific hypotheses.
\newblock In \emph{The Thirteenth International Conference on Learning Representations}, 2025.
\newblock URL \url{https://openreview.net/forum?id=X9OfMNNepI}.

\bibitem[Ye et~al.(2024)Ye, Huang, Chen, Liu, Wang, and Dong]{ye2024pdeformer}
Zhanhong Ye, Xiang Huang, Leheng Chen, Hongsheng Liu, Zidong Wang, and Bin Dong.
\newblock {PDE}former: Towards a foundation model for one-dimensional partial differential equations.
\newblock In \emph{ICLR 2024 Workshop on AI4DifferentialEquations In Science}, 2024.
\newblock URL \url{https://openreview.net/forum?id=GLDMCwdhTK}.

\bibitem[Zhan et~al.(2025)Zhan, Luo, and Khayyer]{ZHAN2025109389}
Yi~Zhan, Min Luo, and Abbas Khayyer.
\newblock Dualsphysics+: An enhanced dualsphysics with improvements in accuracy, energy conservation and resolution of the continuity equation.
\newblock \emph{Computer Physics Communications}, 306:\penalty0 109389, 2025.
\newblock ISSN 0010-4655.
\newblock \doi{https://doi.org/10.1016/j.cpc.2024.109389}.
\newblock URL \url{https://www.sciencedirect.com/science/article/pii/S0010465524003126}.

\bibitem[Zhang et~al.(2025)Zhang, Liu, Qi, Jiao, and Wu]{zhang2025m2pdecompositionalgenerativemultiphysics}
Tao Zhang, Zhenhai Liu, Feipeng Qi, Yongjun Jiao, and Tailin Wu.
\newblock M2pde: Compositional generative multiphysics and multi-component pde simulation, 2025.
\newblock URL \url{https://arxiv.org/abs/2412.04134}.

\bibitem[Zhang et~al.(2019)Zhang, Almgren, Beckner, Bell, Blaschke, Chan, Day, Friesen, Gott, Graves, Katz, Myers, Nguyen, Nonaka, Rosso, Williams, and Zingale]{Zhang2019}
Weiqun Zhang, Ann Almgren, Vince Beckner, John Bell, Johannes Blaschke, Cy~Chan, Marcus Day, Brian Friesen, Kevin Gott, Daniel Graves, Max~P. Katz, Andrew Myers, Tan Nguyen, Andrew Nonaka, Michele Rosso, Samuel Williams, and Michael Zingale.
\newblock Amrex: a framework for block-structured adaptive mesh refinement.
\newblock \emph{Journal of Open Source Software}, 4\penalty0 (37):\penalty0 1370, 2019.
\newblock \doi{10.21105/joss.01370}.
\newblock URL \url{https://doi.org/10.21105/joss.01370}.

\bibitem[Zheng et~al.(2025)Zheng, Chu, Zhang, Wu, Wang, Feng, Zou, Sun, Kovachki, Ross, Bouman, and Yue]{zheng2025inversebench}
Hongkai Zheng, Wenda Chu, Bingliang Zhang, Zihui Wu, Austin Wang, Berthy Feng, Caifeng Zou, Yu~Sun, Nikola~Borislavov Kovachki, Zachary~E Ross, Katherine Bouman, and Yisong Yue.
\newblock Inversebench: Benchmarking plug-and-play diffusion priors for inverse problems in physical sciences.
\newblock In \emph{The Thirteenth International Conference on Learning Representations}, 2025.
\newblock URL \url{https://openreview.net/forum?id=U3PBITXNG6}.

\bibitem[Zhou et~al.()Zhou, Ma, Wu, Wang, and Long]{zhouunisolver}
Hang Zhou, Yuezhou Ma, Haixu Wu, Haowen Wang, and Mingsheng Long.
\newblock Unisolver: Pde-conditional transformers towards universal neural pde solvers.
\newblock In \emph{Forty-second International Conference on Machine Learning}.

\end{thebibliography}
\bibliographystyle{iclr2027_conference}

\newpage

\appendix

\section{Broader Impact}
\label{appendix:broad}
Beyond the immediate advancements in neural PDE modeling, our work serves as a controlled, rigorous platform for investigating the capabilities of autonomous research agents in complex scientific domains. By focusing specifically on partial differential equations, the mathematical bedrock for critical applications such as weather and climate forecasting, aerospace engineering, fluid dynamics, and materials science, our benchmark grounds agentic AI evaluation in high-stakes, real-world physical challenges. We hope that by demonstrating the viability of AI-driven pipeline design, this platform will help pave the way for future autonomous research capable of accelerating scientific discovery across multiple disciplines. Finally, we do not identify any negative societal impacts associated with this work.

\section{Limitation and Future Work}
\label{appendix:limitation}
A potential criticism of our work is that the number of samples is relatively small compared to other LLM benchmarks~\citep{wang2026arlarena, jimenez2024swebench}. However, we emphasize that our benchmark is fundamentally different from traditional evaluation sets constructed by scraping vast amounts of existing text or code from the internet. Instead, our datasets are meticulously curated by domain experts who have identified these specific physical scenarios as either actively studied bottlenecks or entirely novel open problems. Because these challenges are highly specialized and recent, LLMs have no prior exposure to their solutions in their training corpora. Consequently, any successful modeling pipeline generated by an agent on these tasks represents significant, genuine progress in the field of neural PDE solvers, rather than mere data memorization.

We intend to maintain our work as a dynamic, rolling benchmark. We will progressively incorporate new datasets as novel challenges emerge in the field, and we actively encourage the broader research community to contribute their own complex datasets to our platform. In future work, we aim to collaborate with scientists across a wider spectrum of disciplines to identify and integrate even more challenging physical systems, specifically those characterized by idiosyncratic constraints that currently cause existing general-purpose solvers to fail. By continuously expanding this repository, we hope to further stress-test and advance the capabilities of autonomous AI research agents in scientific discovery.

\section{Evaluation Metrics}
\label{appendix:metrics}
All metrics have the form $\mathcal{L}(\Phi_\theta; \mathcal{D})$. They compare the prediction $\hat{y} = \Phi_\theta(x, c)$ against the ground-truth target state $y$. We write $N = |\mathcal{D}|$ for the number of evaluated samples and index them by $n$. The target and prediction share the same representation, which takes one of three forms depending on the dataset. An Eulerian grid sample is $y \in \mathbb{R}^{C \times H \times W}$, storing $C$ channels on a structured $H \times W$ grid whose nodes are fixed in space, with entries $y_{n,c,i,j}$ and predictions $\hat{y}_{n,c,i,j}$. An Eulerian mesh sample is $y \in \mathbb{R}^{M \times C}$, storing $C$ channels on $M$ fixed, unstructured mesh nodes indexed by $m$, giving $y_{n,m,c}$. A Lagrangian particle sample is $y \in \mathbb{R}^{M \times d}$, storing the $M$ particle positions that move with the flow, where $d$ is the spatial dimension; particle $i$ has position $y_{n,i} \in \mathbb{R}^d$. Particle velocities $v_{n,i}$ are obtained from the positions by finite differencing. We define these metrics in two dimensions and they generalize naturally to higher-dimensional spaces.

\textbf{Root Mean Squared Error (RMSE).} The pointwise root mean squared error measures the mean per-node discrepancy between the predicted and target states, averaged over all samples and over every channel and spatial location. Its summation domain follows the state representation. For an Eulerian grid,
\begin{equation}
\mathrm{RMSE}(\Phi_\theta; \mathcal{D}) = \sqrt{\frac{1}{NCHW}\sum_{n,c,i,j}\bigl(\hat{y}_{n,c,i,j} - y_{n,c,i,j}\bigr)^2}.
\end{equation}
For an Eulerian mesh and Lagrangian particles, the sum runs over the $M$ mesh nodes,
\begin{equation}
\mathrm{RMSE}(\Phi_\theta; \mathcal{D}) = \sqrt{\frac{1}{NMC}\sum_{n,m,c}\bigl(\hat{y}_{n,m,c} - y_{n,m,c}\bigr)^2}.
\end{equation}

\textbf{Fourier Root Mean Squared Error (FRMSE).} The Fourier RMSE compares the amplitude spectra of Eulerian-grid states, penalizing errors in how energy is distributed across spatial frequencies rather than in physical space. This is helpful because two fields can have a similar pointwise RMSE yet differ sharply in their spectral content. A model may capture the large-scale structure while smearing out fine features, or introduce spurious high-frequency noise. By comparing magnitude spectra, FRMSE exposes such discrepancies and rewards models that reproduce the correct multi-scale behavior of the flow. Let $\hat{Y} = \mathcal{F}[\hat{y}]$ and $Y = \mathcal{F}[y]$ be the discrete Fourier transforms over the spatial dimensions, with frequency indices $(k,\ell)$:
\begin{equation}
\mathrm{FRMSE}(\Phi_\theta; \mathcal{D}) = \sqrt{\frac{1}{NCHW}\sum_{n,c,k,\ell}\Bigl(\bigl|\hat{Y}_{n,c,k,\ell}\bigr| - \bigl|Y_{n,c,k,\ell}\bigr|\Bigr)^2}.
\end{equation}

\textbf{Kinetic Energy Error (KE).} With velocity components on channels $(c_u,c_v)$, the pointwise kinetic energy density is $k_{n,i,j} = \tfrac{1}{2}\bigl(y_{n,c_u,i,j}^2 + y_{n,c_v,i,j}^2\bigr)$. This metric reports the mean absolute error of that density, isolating how faithfully the model reproduces the flow's energy content, a physically meaningful quantity that pointwise state error does not directly measure. It is helpful for detecting systematic bias in the velocity magnitude, such as a model that is over- or under-energetic, which can signal unstable or overly diffusive dynamics even when the velocity fields appear qualitatively reasonable. The metric is
\begin{equation}
E_{\mathrm{KE}}(\Phi_\theta; \mathcal{D}) = \frac{1}{NHW}\sum_{n,i,j}\bigl|\hat{k}_{n,i,j} - k_{n,i,j}\bigr|.
\end{equation}

\textbf{Magnetic Energy Error (MagE).} Analogously to the kinetic energy error, with magnetic-field components on channels $(b_x,b_y)$ the pointwise magnetic energy density is $m_{n,i,j} = \tfrac{1}{2}\bigl(y_{n,b_x,i,j}^2 + y_{n,b_y,i,j}^2\bigr)$. This metric reports the mean absolute error of that density, capturing how well the model reproduces the energy stored in the magnetic field, which governs the dynamics of magnetohydrodynamic systems. It is helpful for revealing whether a model conserves or spuriously injects magnetic energy, a failure mode that pointwise field error can mask but that strongly affects the physical fidelity of the predicted evolution. The metric is
\begin{equation}
E_{\mathrm{ME}}(\Phi_\theta; \mathcal{D}) = \frac{1}{NHW}\sum_{n,i,j}\bigl|\hat{m}_{n,i,j} - m_{n,i,j}\bigr|.
\end{equation}

\textbf{Mass Error (MassE).} Specific to the advection equation, where total mass is the physically conserved quantity. The total mass of a state is its discrete spatial integral $M_{n,c} = \sum_{i,j} y_{n,c,i,j}$, and the metric is the RMSE of this quantity, reduced over all samples and channels. This is helpful because it directly probes a known conservation law: a faithful simulator should neither create nor destroy mass over time, so any deviation exposes non-physical drift that a purely pointwise metric would not single out. The metric is
\begin{equation}
\mathrm{RMSE}_{\mathrm{mass}}(\Phi_\theta; \mathcal{D}) = \sqrt{\frac{1}{NC}\sum_{n,c}\bigl(\hat{M}_{n,c} - M_{n,c}\bigr)^2}.
\end{equation}

\textbf{Boundary RMSE (BoundaryRMSE).} The RMSE restricted to the outer frame of the domain $\partial\Omega = \{(i,j): i\in\{0,H\}\ \text{or}\ j\in\{0,W\}\}$, isolating error in the enforcement of the boundary conditions. This is helpful because boundary behavior is often critical to a simulation's correctness yet contributes only a small fraction of the nodes, so its error is easily diluted in a whole-domain average. By evaluating only the boundary, the metric surfaces whether a model respects the prescribed conditions rather than trading boundary accuracy for lower interior error. The metric is
\begin{equation}
\mathrm{RMSE}_{\partial\Omega}(\Phi_\theta; \mathcal{D}) = \sqrt{\frac{1}{NC\,|\partial\Omega|}\sum_{n,c}\sum_{(i,j)\in\partial\Omega}\bigl(\hat{y}_{n,c,i,j} - y_{n,c,i,j}\bigr)^2}.
\end{equation}

\textbf{Lagrangian Kinetic Energy Error (LagKE).} Evaluated directly on particles with unit mass $m_i = 1$. The per-sample kinetic energy is
\begin{equation}
\mathrm{KE}_n = \sum_{i=1}^{N_p}\|\mathbf{v}_{n,i}\|^2,
\end{equation}
and the metric is the RMSE of this scalar across the split. This is helpful because it summarizes the total energy of the particle system in a single physically meaningful quantity, exposing whether a model produces an over- or under-energetic flow. Since the velocities $\mathbf{v}_{n,i}$ are finite-differenced from predicted positions, the metric is also sensitive to temporally inconsistent that may nonetheless have plausible instantaneous positions. The metric is
\begin{equation}
\mathrm{RMSE}_{\mathrm{KE}}(\Phi_\theta; \mathcal{D}) = \sqrt{\frac{1}{N}\sum_{n}\bigl(\hat{\mathrm{KE}}_n - \mathrm{KE}_n\bigr)^2}.
\end{equation}

\textbf{Eulerian-Aggregated Lagrangian Error (EuARMSE, EuAFRMSE).} To evaluate Lagrangian particle states with the grid metrics above, particle velocities are first interpolated onto cell-centered Eulerian grid nodes $\mathbf{x}_g$ using an SPH kernel. Writing $W(\cdot)$ for the kernel, the interpolated node velocity is 
\begin{equation}
\tilde{\mathbf{u}}(\mathbf{x}_g) = \frac{\sum_j \mathbf{v}_j\,W(\|\mathbf{x}_g - \mathbf{r}_j\|)}{\sum_j W(\|\mathbf{x}_g - \mathbf{r}_j\|)}.
\end{equation}
The grid RMSE and FRMSE are then applied to the resulting predicted and target node fields, yielding the Eulerian-aggregated RMSE (EuARMSE) and Eulerian-aggregated FRMSE (EuAFRMSE). This is helpful because it measures the practical utility of a predicted particle trajectory. Querying a physical quantity from a Lagrangian state at fixed spatial locations requires exactly this kernel aggregation, so the metric evaluates the particles as they would actually be consumed downstream.

\textbf{Lagrangian Sinkhorn Divergence (Sinkhorn).} An entropic optimal-transport metric on the particle position sets, comparing the predicted point cloud $\hat{P} = \{\hat{y}_i\}$ against the target $P = \{y_i\}$. This is helpful because it compares the two configurations as unordered distributions rather than by particle-wise correspondence, which is the natural notion of error when particles are interchangeable or when their indexing carries no physical meaning. It rewards a model that places particles in the right spatial distribution even if individual identities are permuted, capturing bulk structural agreement that a per-particle position error would penalize spuriously. With uniform weights $a = b = \tfrac{1}{N_p}\mathbf{1}$, squared-Euclidean cost $M_{ij} = \|y_i - y_j\|^2$, and entropic regularization $\varepsilon$, let $\mathrm{OT}_\varepsilon(P,Q)$ denote the Sinkhorn transport cost. The divergence debiases the raw cost with the self-transport terms:
\begin{equation}
S_\varepsilon (\Phi_\theta; \mathcal{D}) = \max\!\Bigl(\mathrm{OT}_\varepsilon(\hat{P}, P) - \tfrac{1}{2}\bigl[\mathrm{OT}_\varepsilon(\hat{P}, \hat{P}) + \mathrm{OT}_\varepsilon(P, P)\bigr],\ 0\Bigr).
\end{equation}

\textbf{Coefficient Error (Coeff).} For each sample, a scalar physical coefficient $\kappa$ (e.g., a drag or lift coefficient) is derived from the predicted state. The metric is its RMSE against the reference value across the split. This is helpful because such coefficients are the integrated, task-level quantities that practitioners ultimately care about, and reducing an entire field to a single engineering-relevant scalar tests whether the prediction is accurate where it matters for the downstream application. A model can attain low pointwise error yet still misestimate the coefficient if it misplaces the local features that dominate the integral, which this metric is designed to expose. The metric is
\begin{equation}
\mathrm{RMSE}_{\kappa}(\Phi_\theta; \mathcal{D}) = \sqrt{\frac{1}{N}\sum_{n}\bigl(\hat{\kappa}_{n} - \kappa_{n}\bigr)^2}.
\end{equation}

\section{Hardware Specifications}
We implement models in PyTorch~\citep{paszke2019pytorchimperativestylehighperformance}. All experiments can be run on servers/workstations with the following configuration:

\begin{itemize}
    \item 80 CPUs, 503G Mem, 8 x NVIDIA V100 GPUs.
    \item 48 CPUs, 220G Mem, 8 x NVIDIA TITAN XP GPUs.
    \item 96 CPUs, 1.0T Mem, 8 x NVIDIA A100 GPUs.
    \item 64 CPUs, 1.0T Mem, 8 x NVIDIA RTX A6000 GPUs.
    \item 224 CPUs, 1.5T Mem, 8 x NVIDIA L40S GPUs.
    \item 128 CPUs, 1.5T Mem, 8 x NVIDIA H200 GPUs.
    \item 128 CPUs, 480G Mem, 8 × NVIDIA RTX 4090 GPUs.
\end{itemize}

\newpage

\section{Supplementary Experiments on Transferability Across PDEs}
\label{appendix:transfer}

We present supplementary experimental results regarding transferability across PDEs in Tables~\ref{fig:coaldustexplosion}, \ref{fig:burgers}, and \ref{fig:wave}, including FRMSE and Kinetic Energy as additional evaluation metrics for Eulerian data and LagKE, EuARMSE, and EuAFRMSE for Lagrangian data. Specifically, we transfer the agent's design from \HighSpeedFlow\ to a coal dust explosion dataset, from \MixedParam\ to the 2D Burgers equation, and from \LagDynamic\ to a Lagrangian inlet wave on a slanted wall dataset. We observe that the agent's designs successfully transfer to these new environments while maintaining optimal performance.

\begin{table}[htbp]
\centering
\begin{tabular}{lccc}
\hline
 & RMSE & FRMSE & Kinetic Energy \\
\hline
GNOT        & $3.442 \times 10^{0}$  & $4.985 \times 10^{2}$ & $3.557 \times 10^{2}$ \\
Transolver  & $1.182 \times 10^{1}$  & $1.786 \times 10^{3}$ & $5.664 \times 10^{2}$ \\
Transolver++ & $1.172 \times 10^{1}$ & $1.780 \times 10^{3}$ & $5.655 \times 10^{2}$ \\
LNO         & $2.439 \times 10^{1}$  & $4.155 \times 10^{3}$ & $5.585 \times 10^{3}$ \\
AMG         & $5.930 \times 10^{0}$  & $8.538 \times 10^{2}$ & $4.740 \times 10^{2}$ \\
\textbf{Agent} & $\mathbf{1.391 \times 10^{0}}$ & $\mathbf{1.890 \times 10^{2}}$ & $\mathbf{5.134 \times 10^{1}}$ \\
\hline
\end{tabular}
\caption{Transferability of \HighSpeedFlow\ to coal dust explosion dynamics. We evaluate the agent-optimized architecture in an out-of-distribution setting on a novel coal dust explosion dataset. The agent-designed solver maintains superior accuracy across RMSE, FRMSE, and Kinetic Energy metrics compared to general-purpose baselines.}
\label{fig:coaldustexplosion}
\end{table}

\begin{table}[htbp]
\centering
\begin{tabular}{lccc}
\hline
 & RMSE & FRMSE & Kinetic Energy \\
\hline
GNOT        & $1.531 \times 10^{-2}$ & $1.954 \times 10^{0}$ & $1.426 \times 10^{-4}$ \\
Transolver  & $1.531 \times 10^{-2}$ & $1.953 \times 10^{0}$ & $1.424 \times 10^{-4}$ \\
Transolver++ & $1.531 \times 10^{-2}$ & $1.954 \times 10^{0}$ & $1.425 \times 10^{-4}$ \\
LNO         & $1.856 \times 10^{-2}$ & $2.373 \times 10^{0}$ & $1.774 \times 10^{-4}$ \\
AMG         & $1.097 \times 10^{-2}$ & $1.306 \times 10^{0}$ & $9.326 \times 10^{-5}$ \\
\textbf{Agent} & $\mathbf{7.718 \times 10^{-3}}$ & $\mathbf{9.720 \times 10^{-1}}$ & $\mathbf{9.125 \times 10^{-5}}$ \\
\hline
\end{tabular}
\caption{Transferability of \MixedParam\ to the 2D Burgers' equation. We evaluate the agent-optimized architecture in an out-of-distribution setting on the 2D Burgers' equation dataset. The agent-designed solver maintains superior accuracy across RMSE, FRMSE, and Kinetic Energy metrics compared to general-purpose baselines.}
\label{fig:burgers}
\end{table}

\begin{table}[htbp]
\centering
\begin{tabular}{lccccc}
\hline
 & RMSE & LagKE & EuARMSE & EuAFRMSE \\
\hline
GNOT        & --                     & --                     & --                     & -- \\
Transolver  & --                     & --                     & --                     & -- \\
Transolver++ & --                    & --                     & --                     & -- \\
LNO         & $5.032 \times 10^{-4}$  & $7.148 \times 10^{-3}$ & $2.779 \times 10^{-4}$ & $2.066 \times 10^{-2}$ \\
AMG         & --                     & --                     & --                     & -- \\
\textbf{Agent} & $\mathbf{1.952 \times 10^{-4}}$ & $\mathbf{2.813 \times 10^{-3}}$ & $\mathbf{1.158 \times 10^{-4}}$ & $\mathbf{7.134 \times 10^{-3}}$ \\
\hline
\end{tabular}
\caption{Transferability of \LagDynamic\ to Lagrangian wave dynamics. We evaluate the agent-optimized architecture in an out-of-distribution setting on a novel 2D wave dataset. The agent-designed solver maintains superior accuracy across RMSE, LagKE, EuARMSE, and EuAFRMSE metrics compared to general-purpose baselines.}
\label{fig:wave}
\end{table}

\newpage

\section{Supplementary Experiments on LLM Backbone  Analysis}
\label{appendix:llm}

Tables~\ref{tab:llm_highspeedflow}, \ref{tab:llm_mixedparam}, and \ref{tab:llm_lagdynamic} detail supplementary results for our LLM backbone sensitivity analysis. Beyond RMSE, we incorporate additional metrics for the Eulerian and Lagrangian datasets that measure physical consistency and practical utility.

\begin{table}[htbp]
\centering
\begin{tabular}{lccc}
\hline
Model & RMSE & FRMSE & Kinetic Energy \\
\hline
gpt-5.5 & $5.066 \times 10^{0}$ & $4.218 \times 10^{2}$ & $3.514 \times 10^{2}$ \\
gpt-4o & $8.332 \times 10^{0}$ & $7.011 \times 10^{2}$ & $7.840 \times 10^{2}$ \\
gpt-5.6-sol & $5.537 \times 10^{0}$ & $4.670 \times 10^{2}$ & $4.035 \times 10^{2}$ \\
gpt-5 & $5.771 \times 10^{0}$ & $5.147 \times 10^{2}$ & $3.308 \times 10^{2}$ \\
claude-fable-5 & $6.766 \times 10^{0}$ & $5.525 \times 10^{2}$ & $5.142 \times 10^{2}$ \\
claude-opus-4-8 & $6.687 \times 10^{0}$ & $5.473 \times 10^{2}$ & $5.676 \times 10^{2}$ \\
grok-4.5 & $5.299 \times 10^{0}$ & $4.432 \times 10^{2}$ & $5.259 \times 10^{2}$ \\
gemini-3.1-pro-preview & $9.300 \times 10^{0}$ & $7.726 \times 10^{2}$ & $7.520 \times 10^{2}$ \\
deepseek-v4-pro & $3.685 \times 10^{2}$ & $9.608 \times 10^{2}$ & $2.015 \times 10^{4}$ \\
qwen3.7-max & $6.140 \times 10^{0}$ & $5.067 \times 10^{2}$ & $5.157 \times 10^{2}$ \\
kimi-k2.7-code & $5.784 \times 10^{0}$ & $4.783 \times 10^{2}$ & $4.215 \times 10^{2}$ \\
\hline
\end{tabular}
\caption{LLM backbone sensitivity on \HighSpeedFlow. We evaluate alternative LLMs as the agent's backbone. The table reports RMSE alongside FRMSE and Kinetic Energy.}
\label{tab:llm_highspeedflow}
\end{table}

\begin{table}[htbp]
\centering
\begin{tabular}{lccc}
\hline
Model & RMSE & FRMSE & Kinetic Energy \\
\hline
gpt-5.5 & $1.345 \times 10^{0}$ & $1.078 \times 10^{2}$ & $2.784 \times 10^{-1}$ \\
gpt-4o & $4.056 \times 10^{0}$ & $3.518 \times 10^{2}$ & $2.507 \times 10^{0}$ \\
gpt-5.6-sol & $1.870 \times 10^{0}$ & $1.485 \times 10^{2}$ & $5.014 \times 10^{-1}$ \\
gpt-5 & $4.790 \times 10^{0}$ & $2.801 \times 10^{2}$ & $1.990 \times 10^{0}$ \\
claude-fable-5 & $3.435 \times 10^{0}$ & $2.691 \times 10^{2}$ & $1.693 \times 10^{0}$ \\
claude-opus-4-8 & $2.200 \times 10^{0}$ & $1.710 \times 10^{2}$ & $6.128 \times 10^{-1}$ \\
grok-4.5 & $2.002 \times 10^{0}$ & $1.594 \times 10^{2}$ & $5.295 \times 10^{-1}$ \\
gemini-3.1-pro-preview & $2.188 \times 10^{0}$ & $1.818 \times 10^{2}$ & $8.594 \times 10^{-1}$ \\
deepseek-v4-pro & $1.871 \times 10^{0}$ & $1.825 \times 10^{2}$ & $6.266 \times 10^{-1}$ \\
qwen3.7-max & $2.298 \times 10^{0}$ & $1.822 \times 10^{2}$ & $8.762 \times 10^{-1}$ \\
kimi-k2.7-code & $2.543 \times 10^{0}$ & $2.055 \times 10^{2}$ & $7.776 \times 10^{-1}$ \\
\hline
\end{tabular}
\caption{LLM backbone sensitivity on \MixedParam. We evaluate alternative LLMs as the agent's backbone. The table reports RMSE alongside FRMSE and Kinetic Energy.}
\label{tab:llm_mixedparam}
\end{table}

\begin{table}[htbp]
\centering
\resizebox{\columnwidth}{!}{%
\begin{tabular}{lcccc}
\hline
Model & RMSE & LagKE & EuARMSE & EuAFRMSE \\
\hline
gpt-5.5 & $5.530 \times 10^{-1}$ & $1.662 \times 10^{3}$ & $9.951 \times 10^{-2}$ & $2.299 \times 10^{0}$ \\
gpt-4o & $1.195 \times 10^{1}$ & $7.857 \times 10^{5}$ & $7.787 \times 10^{0}$ & $2.395 \times 10^{2}$ \\
gpt-5.6-sol & $5.647 \times 10^{-1}$ & $1.587 \times 10^{3}$ & $1.054 \times 10^{-1}$ & $2.467 \times 10^{0}$ \\
gpt-5 & $1.194 \times 10^{0}$ & $1.304 \times 10^{5}$ & $6.773 \times 10^{-1}$ & $2.140 \times 10^{1}$ \\
claude-fable-5 & $5.753 \times 10^{-1}$ & $2.149 \times 10^{3}$ & $1.073 \times 10^{-1}$ & $2.483 \times 10^{0}$ \\
claude-opus-4-8 & $5.772 \times 10^{-1}$ & $2.225 \times 10^{3}$ & $1.103 \times 10^{-1}$ & $2.550 \times 10^{0}$ \\
grok-4.5 & $5.808 \times 10^{-1}$ & $4.976 \times 10^{3}$ & $1.157 \times 10^{-1}$ & $2.768 \times 10^{0}$ \\
gemini-3.1-pro-preview & $5.848 \times 10^{-1}$ & $1.807 \times 10^{3}$ & $1.117 \times 10^{-1}$ & $2.606 \times 10^{0}$ \\
deepseek-v4-pro & $5.003 \times 10^{-1}$ & $1.252 \times 10^{3}$ & $7.332 \times 10^{-2}$ & $1.745 \times 10^{0}$ \\
qwen3.7-max & $1.195 \times 10^{1}$ & $7.856 \times 10^{5}$ & $7.787 \times 10^{0}$ & $2.395 \times 10^{2}$ \\
kimi-k2.7-code & $9.579 \times 10^{-1}$ & $1.064 \times 10^{5}$ & $5.447 \times 10^{-1}$ & $1.725 \times 10^{1}$ \\
\hline
\end{tabular}%
}
\caption{LLM backbone sensitivity on \LagDynamic. We evaluate alternative LLMs as the agent's backbone. The table reports RMSE alongside LagKE, EuARMSE, and EuAFRMSE to measure physical consistency and practical utility.}
\label{tab:llm_lagdynamic}
\end{table}

\clearpage
\newpage

\section{Supplementary Experiment on Agent Long-Horizon Scalability}
\label{appendix:long}

We present additional metrics for the agent long-horizon scalability experiments. The validation and test results at selected steps over extended iterations are detailed in Tables~\ref{tab:long_highspeedflow_valid} and \ref{tab:long_highspeedflow_test} for \HighSpeedFlow, Tables~\ref{tab:long_mixedparam_valid} and \ref{tab:long_mixedparam_test} for \MixedParam, and Tables~\ref{tab:long_lagdynamic_valid} and \ref{tab:long_lagdynamic_test} for \LagDynamic. These tables include supplementary metrics for both the Eulerian and Lagrangian datasets. We observe that the trends in these additional metrics generally align with the primary metric, RMSE. This indicates that the agent's designs consistently improve over successive iterations not only in terms of point-wise error but also in physical consistency. These findings validate the promising potential of deploying LLM-driven autonomous research agents to address domain-specific challenges.

\vspace*{\fill}

\begin{table}[htbp]
\centering
\begin{tabular}{cccc}
\hline
Iteration & RMSE & FRMSE & Kinetic Energy \\
\hline
1  & $4.654 \times 10^{0}$ & $3.976 \times 10^{2}$ & $3.373 \times 10^{2}$ \\
2  & $4.301 \times 10^{0}$ & $3.647 \times 10^{2}$ & $2.987 \times 10^{2}$ \\
3  & $4.194 \times 10^{0}$ & $3.518 \times 10^{2}$ & $2.802 \times 10^{2}$ \\
4  & $4.135 \times 10^{0}$ & $3.460 \times 10^{2}$ & $2.717 \times 10^{2}$ \\
5  & $4.112 \times 10^{0}$ & $3.421 \times 10^{2}$ & $2.678 \times 10^{2}$ \\
10 & $4.189 \times 10^{0}$ & $3.455 \times 10^{2}$ & $2.657 \times 10^{2}$ \\
15 & $4.077 \times 10^{0}$ & $3.405 \times 10^{2}$ & $2.582 \times 10^{2}$ \\
20 & $3.903 \times 10^{0}$ & $3.224 \times 10^{2}$ & $2.406 \times 10^{2}$ \\
25 & $3.977 \times 10^{0}$ & $3.298 \times 10^{2}$ & $2.421 \times 10^{2}$ \\
30 & $3.960 \times 10^{0}$ & $3.279 \times 10^{2}$ & $2.235 \times 10^{2}$ \\
\hline
\end{tabular}
\caption{Long-horizon scalability on \HighSpeedFlow\ (Validation). Supplementary metrics on the validation dataset for the agent evaluated over an extended horizon of 30 iterations. The agent directly optimizes its architecture based on validation performance and has no access to the test metrics.}
\label{tab:long_highspeedflow_valid}
\end{table}

\vspace*{\fill}

\begin{table}[htbp]
\centering
\begin{tabular}{cccc}
\hline
Iteration & RMSE & FRMSE & Kinetic Energy \\
\hline
1  & $5.683 \times 10^{0}$ & $4.725 \times 10^{2}$ & $4.360 \times 10^{2}$ \\
2  & $5.365 \times 10^{0}$ & $4.415 \times 10^{2}$ & $4.081 \times 10^{2}$ \\
3  & $5.277 \times 10^{0}$ & $4.283 \times 10^{2}$ & $3.950 \times 10^{2}$ \\
4  & $5.189 \times 10^{0}$ & $4.243 \times 10^{2}$ & $3.832 \times 10^{2}$ \\
5  & $5.136 \times 10^{0}$ & $4.189 \times 10^{2}$ & $3.829 \times 10^{2}$ \\
10 & $5.207 \times 10^{0}$ & $4.229 \times 10^{2}$ & $3.784 \times 10^{2}$ \\
15 & $5.045 \times 10^{0}$ & $4.132 \times 10^{2}$ & $3.439 \times 10^{2}$ \\
20 & $4.849 \times 10^{0}$ & $3.921 \times 10^{2}$ & $3.233 \times 10^{2}$ \\
25 & $4.998 \times 10^{0}$ & $4.050 \times 10^{2}$ & $3.310 \times 10^{2}$ \\
30 & $4.887 \times 10^{0}$ & $3.969 \times 10^{2}$ & $3.138 \times 10^{2}$ \\
\hline
\end{tabular}
\caption{Long-horizon scalability on \HighSpeedFlow\ (Test). Supplementary metrics on the test dataset for the agent evaluated over an extended horizon of 30 iterations.}
\label{tab:long_highspeedflow_test}
\end{table}

\vspace*{\fill}

\clearpage

\vspace*{\fill}

\begin{table}[htbp]
\centering
\begin{tabular}{cccc}
\hline
Iteration & RMSE & FRMSE & Kinetic Energy \\
\hline
1  & $9.274 \times 10^{-1}$ & $8.415 \times 10^{1}$ & $1.688 \times 10^{-1}$ \\
2  & $9.387 \times 10^{-1}$ & $8.286 \times 10^{1}$ & $1.716 \times 10^{-1}$ \\
3  & $9.327 \times 10^{-1}$ & $8.395 \times 10^{1}$ & $1.748 \times 10^{-1}$ \\
4  & $1.284 \times 10^{0}$  & $1.188 \times 10^{2}$ & $3.057 \times 10^{-1}$ \\
5  & $9.392 \times 10^{-1}$ & $8.450 \times 10^{1}$ & $1.780 \times 10^{-1}$ \\
10 & $9.133 \times 10^{-1}$ & $8.192 \times 10^{1}$ & $1.673 \times 10^{-1}$ \\
15 & $7.867 \times 10^{-1}$ & $7.750 \times 10^{1}$ & $1.466 \times 10^{-1}$ \\
20 & $6.570 \times 10^{-1}$ & $6.759 \times 10^{1}$ & $9.796 \times 10^{-2}$ \\
25 & $3.425 \times 10^{-1}$ & $3.087 \times 10^{1}$ & $5.273 \times 10^{-2}$ \\
30 & $2.993 \times 10^{-1}$ & $2.685 \times 10^{1}$ & $5.242 \times 10^{-2}$\\
\hline
\end{tabular}
\caption{Long-horizon scalability on \MixedParam\ (Validation). Supplementary metrics on the validation dataset for the agent evaluated over an extended horizon of 30 iterations. During this process, the agent directly optimizes its architecture based on validation performance and has no access to the test metrics.}
\label{tab:long_mixedparam_valid}
\end{table}

\vspace*{\fill}

\begin{table}[htbp]
\centering
\begin{tabular}{cccc}
\hline
Iteration & RMSE & FRMSE & Kinetic Energy \\
\hline
1  & $9.043 \times 10^{-1}$ & $8.201 \times 10^{1}$ & $1.618 \times 10^{-1}$ \\
2  & $9.171 \times 10^{-1}$ & $8.108 \times 10^{1}$ & $1.660 \times 10^{-1}$ \\
3  & $9.056 \times 10^{-1}$ & $8.203 \times 10^{1}$ & $1.671 \times 10^{-1}$ \\
4  & $1.258 \times 10^{0}$  & $1.175 \times 10^{2}$ & $2.976 \times 10^{-1}$ \\
5  & $9.101 \times 10^{-1}$ & $8.215 \times 10^{1}$ & $1.705 \times 10^{-1}$ \\
10 & $8.871 \times 10^{-1}$ & $7.983 \times 10^{1}$ & $1.608 \times 10^{-1}$ \\
15 & $7.715 \times 10^{-1}$ & $7.638 \times 10^{1}$ & $1.405 \times 10^{-1}$ \\
20 & $6.440 \times 10^{-1}$ & $6.663 \times 10^{1}$ & $9.264 \times 10^{-2}$ \\
25 & $3.237 \times 10^{-1}$ & $2.922 \times 10^{1}$ & $5.032 \times 10^{-2}$ \\
30 & $2.764 \times 10^{-1}$ & $2.476 \times 10^{1}$ & $5.018 \times 10^{-2}$ \\
\hline
\end{tabular}
\caption{Long-horizon scalability on \MixedParam\ (Test). Supplementary metrics on the test dataset for the agent evaluated over an extended horizon of 30 iterations.}
\label{tab:long_mixedparam_test}
\end{table}

\vspace*{\fill}

\clearpage

\vspace*{\fill}

\begin{table}[htbp]
\centering
\begin{tabular}{ccccc}
\hline
Iteration & RMSE & LagKE & EuARMSE & EuAFRMSE \\
\hline
1  & $5.802 \times 10^{-1}$ & $3.135 \times 10^{3}$ & $1.135 \times 10^{-1}$ & $2.698 \times 10^{0}$ \\
2  & $5.781 \times 10^{-1}$ & $1.186 \times 10^{3}$ & $1.106 \times 10^{-1}$ & $2.582 \times 10^{0}$ \\
3  & $5.779 \times 10^{-1}$ & $1.254 \times 10^{3}$ & $1.103 \times 10^{-1}$ & $2.574 \times 10^{0}$ \\
4  & $5.781 \times 10^{-1}$ & $1.773 \times 10^{3}$ & $1.112 \times 10^{-1}$ & $2.611 \times 10^{0}$ \\
5  & $5.780 \times 10^{-1}$ & $1.507 \times 10^{3}$ & $1.112 \times 10^{-1}$ & $2.609 \times 10^{0}$ \\
10 & $5.815 \times 10^{-1}$ & $4.628 \times 10^{3}$ & $1.156 \times 10^{-1}$ & $2.765 \times 10^{0}$ \\
15 & $6.144 \times 10^{-1}$ & $8.758 \times 10^{3}$ & $1.184 \times 10^{-1}$ & $2.835 \times 10^{0}$ \\
20 & $6.119 \times 10^{-1}$ & $8.826 \times 10^{3}$ & $1.183 \times 10^{-1}$ & $2.846 \times 10^{0}$ \\
25 & $5.851 \times 10^{-1}$ & $2.654 \times 10^{3}$ & $1.152 \times 10^{-1}$ & $2.705 \times 10^{0}$ \\
30 & $6.127 \times 10^{-1}$ & $4.107 \times 10^{3}$ & $1.174 \times 10^{-1}$ & $2.788 \times 10^{0}$ \\
\hline
\end{tabular}
\caption{Long-horizon scalability on \LagDynamic\ (Validation). Supplementary metrics on the validation dataset for the agent evaluated over an extended horizon of 30 iterations. During this process, the agent directly optimizes its architecture based on validation performance and has no access to the test metrics.}
\label{tab:long_lagdynamic_valid}
\end{table}

\vspace*{\fill}

\begin{table}[htbp]
\centering
\begin{tabular}{ccccc}
\hline
Iteration & RMSE & LagKE & EuARMSE & EuAFRMSE \\
\hline
1  & $5.776 \times 10^{-1}$ & $3.254 \times 10^{3}$ & $1.132 \times 10^{-1}$ & $2.698 \times 10^{0}$ \\
2  & $5.755 \times 10^{-1}$ & $1.177 \times 10^{3}$ & $1.100 \times 10^{-1}$ & $2.572 \times 10^{0}$ \\
3  & $5.753 \times 10^{-1}$ & $1.180 \times 10^{3}$ & $1.098 \times 10^{-1}$ & $2.562 \times 10^{0}$ \\
4  & $5.756 \times 10^{-1}$ & $1.877 \times 10^{3}$ & $1.108 \times 10^{-1}$ & $2.606 \times 10^{0}$ \\
5  & $5.756 \times 10^{-1}$ & $1.571 \times 10^{3}$ & $1.109 \times 10^{-1}$ & $2.605 \times 10^{0}$ \\
10 & $5.788 \times 10^{-1}$ & $4.671 \times 10^{3}$ & $1.152 \times 10^{-1}$ & $2.760 \times 10^{0}$ \\
15 & $6.124 \times 10^{-1}$ & $8.764 \times 10^{3}$ & $1.176 \times 10^{-1}$ & $2.812 \times 10^{0}$ \\
20 & $6.099 \times 10^{-1}$ & $8.834 \times 10^{3}$ & $1.174 \times 10^{-1}$ & $2.817 \times 10^{0}$ \\
25 & $5.825 \times 10^{-1}$ & $2.690 \times 10^{3}$ & $1.147 \times 10^{-1}$ & $2.694 \times 10^{0}$ \\
30 & $6.107 \times 10^{-1}$ & $4.131 \times 10^{3}$ & $1.170 \times 10^{-1}$ & $2.779 \times 10^{0}$ \\
\hline
\end{tabular}
\caption{Long-horizon scalability on \LagDynamic\ (Test). Supplementary metrics on the test dataset for the agent evaluated over an extended horizon of 30 iterations.}
\label{tab:long_lagdynamic_test}
\end{table}

\vspace*{\fill}

\clearpage
\newpage


\section{Dataset Taxonomy}
\label{appendix:taxonomy}
To characterize the challenges posed by AutoPDEBench, we organize the datasets along five categories:
\begin{itemize}
    \item \textbf{Generalization.} Each dataset mixes problem instances that differ in PDE parameters, governing equations, boundary conditions, spatial dimension (1D, 2D, or 3D), resolution, or discretization. A single model must handle all of them.
    \item \textbf{Domains \& Forcing.} The dynamics are shaped by complex domain and by external forcing.
    \item \textbf{Complex Physics.} The system consists of interacting components, develops discontinuities, or evolves randomly, properties that general-purpose solvers are not designed to capture.
    \item \textbf{Scale.} The problem is large enough, in the size of its state, the range of magnitudes it spans, or the length of its rollout, that computational cost and error accumulation become the primary obstacles.
    \item \textbf{Observational Irregularity.} The data offer only an imperfect view of the underlying system, with observations that are noisy, incomplete, irregularly sampled, or inconsistent across snapshots or between training and deployment, so the solver cannot assume complete, clean states on a fixed sampling pattern.
\end{itemize}
A dataset may fall into more than one category. Table~\ref{tab:taxonomy} summarizes the assignment of each dataset. Datasets are grouped by category so that the table forms a block-diagonal structure, and datasets spanning two categories are placed at the boundary between the corresponding blocks.

\begin{table}[htbp]
\centering
\small
\setlength{\tabcolsep}{3pt}
\renewcommand{\arraystretch}{1.15}
\begin{tabularx}{\textwidth}{@{} l *{5}{>{\centering\arraybackslash}X}}
\toprule
\textbf{Dataset} & \textbf{Generalization} & \textbf{Domains \& Forcing} & \textbf{Complex Physics} & \textbf{Scale} & \textbf{Observational Irregularity} \\
\midrule
\MixedParam      & \taxmark &          &          &          &          \\
\MixedPDE        & \taxmark &          &          &          &          \\
\MixedDomain     & \taxmark &          &          &          &          \\
\MixedRes        & \taxmark &          &          &          &          \\
\MixedDim        & \taxmark &          &          &          &          \\
\MixedBC         & \taxmark & \taxmark &          &          &          \\
\ComplexGeometry &          & \taxmark &          &          &          \\
\ExtLagOne       &          & \taxmark &          &          &          \\
\ExtLagTwo       &          & \taxmark &          &          &          \\
\LagFSI          &          & \taxmark & \taxmark &          &          \\
\MultiPhase      &          &          & \taxmark &          &          \\
\Heterogeneous   &          &          & \taxmark &          &          \\
\MultiPhysics    &          &          & \taxmark &          &          \\
\HighSpeedFlow   &          &          & \taxmark &          &          \\
\Stochastic      &          &          & \taxmark &          &          \\
\Extreme         &          &          & \taxmark & \taxmark &          \\
\HighDimensional &          &          & \taxmark & \taxmark &          \\
\ThreeD          &          &          &          & \taxmark &          \\
\LargeLag        &          &          &          & \taxmark &          \\
\LongHorizon     &          &          &          & \taxmark &          \\
\RealData        &          &          &          &          & \taxmark \\
\SimToReal       &          &          &          &          & \taxmark \\
\Partial         &          &          &          &          & \taxmark \\
\IrregularTime   &          &          &          &          & \taxmark \\
\LagDynamic      &          &          &          &          & \taxmark \\
\bottomrule
\end{tabularx}
\caption{Taxonomy of AutoPDEBench datasets.}
\label{tab:taxonomy}
\end{table}

\clearpage
\newpage

\section{Datasets}
\label{appendix:dataset}
In the sections below, we provide a comprehensive breakdown of each dataset introduced in our study. To keep the manuscript's file size manageable, all embedded figures have been intentionally compressed; however, readers can access the high-resolution, vectorized versions directly through our associated GitHub repository. Because our research spans a broad spectrum of distinct physical systems, maintaining a single, globally unified mathematical glossary across the entire paper was impractical. Therefore, mathematical symbols are strictly self-contained within their respective sections, meaning a variable used to denote one physical property in a given scenario may be redefined to represent an entirely different quantity in the next.

The datasets curated for this study are either adapted directly from peer-reviewed literature or numerically simulated utilizing well-established numerical schemes. To guarantee the highest degree of fidelity and reliability in our simulated datasets, we implemented a rigorous computational validation protocol. All underlying numerical solvers were systematically benchmarked against known analytical solutions for canonical problems and subjected to comprehensive spatial and temporal convergence studies to verify that our results are grid-independent.

\subsection{\HighSpeedFlow}
\label{appendix:highspeedflow}
\textbf{Dataset Description: } In this 2D circular blast model, a central high-pressure zone creates an outward shock wave and an inward expansion wave. The shock wave repeatedly reflects back and forth between the boundaries and the origin, losing strength with each bounce. As these reflecting shocks interact with the post-shock gas, they generate turbulent instabilities.

\textbf{Modeling Objective: } Let $x_{t_1} \in \mathbb{R}^{4 \times 128 \times 128}$ denote the state of the system at time $t_1$. The three feature channels represent the physical fields of temperature, density, and the x and y velocity fields, respectively, evaluated across a $128 \times 128$ spatial discretization grid. The initial time $t_1 \in \mathbb{R}^{1}$ and the target time $t_2 \in \mathbb{R}^{1}$ are explicitly provided as inputs to the model. Our objective is to develop a neural simulator, defined as a parameterized function $\Psi$, that maps the initial system state and the given time coordinates to the subsequent state at $t_2$. Formally, we aim to predict $\hat{x}_{t_2} = \Psi(x_{t_1}, t_1, t_2)$, where the predicted output $\hat{x}_{t_2}$ retains the same spatial and feature dimensions as the input state.

\textbf{Dataset Specific Challenges: } Existing neural solvers are designed for low-Mach-number regimes, where fluid dynamics generally exhibit smooth temporal evolution. In contrast, compressible flows approaching or exceeding the speed of sound develop abrupt, highly non-linear discontinuities, such as shock waves. Additionally, because the underlying numerical solver must adapt its resolution to capture these sudden physical changes, the time step ($\Delta t = t_2 - t_1$) between the input and output states in this dataset is inherently non-uniform.

\textbf{Dataset Generation: } We adopt the Circular Blast dataset from \citet{helwig2026twophasedeeplearningframework}. The circular blast dataset is generated using the HyBurn code, a finite volume solver that uses low-dissipation, WENO-based high-order Godunov methods with adaptive mesh refinement via the AMReX library~\citep{Zhang2019}. HyBurn implements high-order Godunov schemes and uses an explicit third-order three-stage Runge-Kutta time-stepping algorithm along with two independent CFL numbers: one for the hyperbolic and the other for the parabolic terms of the compressible Navier-Stokes equations, with the actual timestep being whichever is smaller. For the circular blast specifically, the authors initialize a circular high-pressure region over a quarter domain with symmetry boundaries, varying the initial pressure ratio from 1.99 to 50 across 99 cases. They use two Adaptive Mesh Refinement (AMR) levels triggered by predefined pressure and density ratio thresholds between adjacent cells.

\textbf{Governing Physics: } The circular blast case is governed by the two-dimensional compressible Navier--Stokes equations for a single gaseous phase, written in the conservative form used by finite-volume Godunov solvers as
\begin{equation}
    \frac{\partial \mathbf{U}}{\partial t}
    + \frac{\partial \mathbf{F}}{\partial x}
    + \frac{\partial \mathbf{G}}{\partial y}
    = \frac{\partial \mathbf{F}_v}{\partial x}
    + \frac{\partial \mathbf{G}_v}{\partial y},
\end{equation}
where $\mathbf{U}$ is the vector of conserved variables, $\mathbf{F}$ and $\mathbf{G}$ are the inviscid (hyperbolic) fluxes, and $\mathbf{F}_v$ and $\mathbf{G}_v$ are the viscous (parabolic) fluxes; this hyperbolic--parabolic split is precisely why the solver carries two independent CFL numbers. The conserved state is $\mathbf{U} = (\rho,\ \rho u,\ \rho v,\ \rho E)^{\top}$, with density $\rho$, velocity components $u$ and $v$, and total energy per unit mass $E = e + \tfrac{1}{2}(u^2 + v^2)$ for specific internal energy $e$.
The inviscid fluxes are
\begin{equation}
    \mathbf{F} =
    \begin{pmatrix}
        \rho u \\ \rho u^2 + p \\ \rho u v \\ (\rho E + p)\,u
    \end{pmatrix},
    \qquad
    \mathbf{G} =
    \begin{pmatrix}
        \rho v \\ \rho u v \\ \rho v^2 + p \\ (\rho E + p)\,v
    \end{pmatrix},
\end{equation}
and the viscous fluxes are
\begin{equation}
    \mathbf{F}_v =
    \begin{pmatrix}
        0 \\ \tau_{xx} \\ \tau_{xy} \\
        u\,\tau_{xx} + v\,\tau_{xy} - q_x
    \end{pmatrix},
    \qquad
    \mathbf{G}_v =
    \begin{pmatrix}
        0 \\ \tau_{xy} \\ \tau_{yy} \\
        u\,\tau_{xy} + v\,\tau_{yy} - q_y
    \end{pmatrix}.
\end{equation}
The system is closed with a Newtonian viscous stress tensor under Stokes'
hypothesis,
\begin{equation}
    \tau_{ij} = \mu\left(
        \frac{\partial u_i}{\partial x_j}
        + \frac{\partial u_j}{\partial x_i}
    \right)
    - \frac{2}{3}\mu\,(\nabla\cdot\mathbf{u})\,\delta_{ij},
\end{equation}
with heat flux given by Fourier's law $q_x = -k\,\partial T/\partial x$ and $q_y = -k\,\partial T/\partial y$, where $\mu$ is the dynamic viscosity and $k$ the thermal conductivity. Finally, since the gas is modeled as ideal, the equation of state, internal energy, and local sound speed are
\begin{equation}
    p = \rho R T,
    \qquad
    e = \frac{p}{(\gamma - 1)\rho},
    \qquad
    a = \sqrt{\gamma R T},
\end{equation}
where $R$ is the specific gas constant and $\gamma$ the ratio of specific heats; this sound speed $a$ feeds the maximum wave speed
$\lambda = \max(|u| + a,\ |v| + a)$ that governs the CFL timestep condition.

\textbf{Experiment Results}
The complete experimental results are presented in Table~\ref{tab:highspeedflow_full}. The dataset is represented on an Eulerian grid, and we employ RMSE, FRMSE, and Kinetic Energy as our evaluation metrics. The results demonstrate that the proposed automated research agent significantly outperforms general-purpose baseline solvers.

\begin{table}[htbp]
\centering
\begin{tabular}{lccc}
\toprule
 & RMSE & FRMSE & Kinetic Energy \\
\midrule
GNOT        & $1.826 \times 10^{1}$ & $1.529 \times 10^{3}$ & $1.538 \times 10^{3}$ \\
Transolver  & $3.232 \times 10^{1}$ & $2.372 \times 10^{3}$ & $3.067 \times 10^{3}$ \\
Transolver++ & $3.271 \times 10^{1}$ & $2.406 \times 10^{3}$ & $3.373 \times 10^{3}$ \\
LNO         & $4.482 \times 10^{1}$ & $3.282 \times 10^{3}$ & $6.092 \times 10^{3}$ \\
AMG         & $2.819 \times 10^{1}$ & $2.181 \times 10^{3}$ & $3.255 \times 10^{3}$ \\
Agent       & $\bm{5.066 \times 10^{0}}$ & $\bm{4.218 \times 10^{2}}$ & $\bm{3.514 \times 10^{2}}$ \\
\bottomrule
\end{tabular}
\caption{Performance comparison on the \HighSpeedFlow\ dataset. Best results are shown in \textbf{bold}.}
\label{tab:highspeedflow_full}
\end{table}

\begin{figure}[htbp]
    \centering
    \includegraphics[width=0.8\textwidth, page=1]{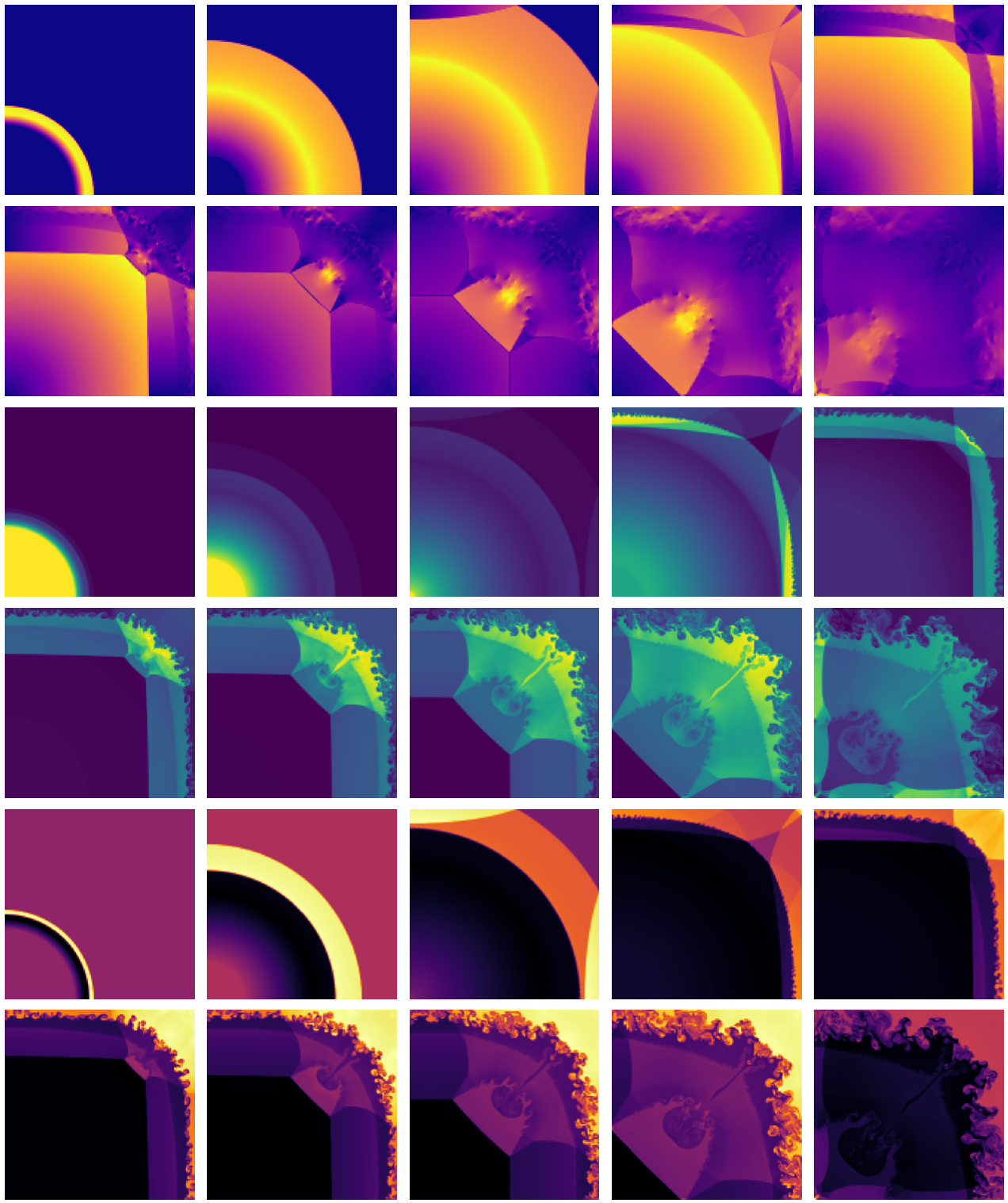}
    \caption{Visualization of the \HighSpeedFlow\ dataset. The first two rows display velocity norm, the middle two rows show density, and the bottom two rows show temperature.}
    \label{fig:highspeedflow}
\end{figure}

\textbf{Visualization: } We refer the readers to Figure~\ref{fig:highspeedflow} for visualizations of the dataset.

\newpage

\subsection{\RealData}
\label{appendix:realdata}
\textbf{Dataset Description: } This dataset captures the fluid dynamics of laminar to turbulent flow around a cylinder. While the simulation resolves the full three-dimensional flow field, we extracted only a 2D cross section for this dataset. A cylinder with a diameter 30 mm is positioned centrally within the domain, serving as a bluff body that obstructs the incoming fluid stream. As the fluid encounters this obstacle, it creates a distinct pressure gradient and a downstream wake region characterized by periodic vortex shedding.

\textbf{Modeling Objective: } The state of the system is represented by the tensor $x_t \in \mathbb{R}^{2 \times 128 \times 128}$. The two feature channels correspond specifically to the horizontal and vertical axis velocity fields, evaluated across a $128 \times 128$ spatial discretization grid. The objective is to develop a neural simulator, defined as a parameterized function $\Psi$, that maps the current system state to the state at the subsequent timestep. Formally, we aim to predict the next state $\hat{x}_{t+1} = \Psi(x_t)$.

\textbf{Dataset Specific Challenges: } The dataset's primary challenge is the inherent measurement noise typical of real-world sensors. Unlike idealized synthetic data, these fluctuations can obscure the underlying physical laws, making it difficult for models to distinguish between meaningful fluid dynamics and measurement error. Existing neural models often struggle with this data because they are typically designed for clean, high-fidelity simulations. Consequently, they may overfit to the noise or fail to maintain temporal stability.

\textbf{Dataset Generation: } We adopt the dataset from \citet{hu2026realpdebench}. The real-world Cylinder data is measured in a circulating water tunnel using Particle Image Velocimetry (PIV). The water tunnel generates an incoming flow that interacts with the cylinder, and the water is seeded with fluorescent particles that are illuminated by a continuous laser forming a thick sheet on the imaging plane, so that the fluid velocity can be inferred from the motion of these particles. The particle motions are recorded with high-speed cameras, and the resulting images are then processed with PIVLab to recover the flow-velocity field at each time step.

\textbf{Governing Physics: } In the idealized continuum limit, the wake behind the circular cylinder is a low-speed, effectively incompressible water flow and should therefore be
described by the incompressible Navier--Stokes equations,
\begin{align}
    \nabla\cdot\mathbf{u} & = 0, \\
    \frac{\partial \mathbf{u}}{\partial t}
    + (\mathbf{u}\cdot\nabla)\,\mathbf{u}
    & = -\frac{1}{\rho}\nabla p
    + \nu\,\nabla^2 \mathbf{u},
\end{align}
where $\mathbf{u}(\mathbf{x},t)$ is the velocity field, $p$ the pressure, $\rho$ the (constant) fluid density, and $\nu$ the kinematic viscosity. The first equation enforces incompressibility and the second expresses momentum conservation, with the dynamics governed by the Reynolds number $\mathrm{Re} = U D / \nu$ formed from the free-stream speed $U$ and the cylinder diameter $D$; as $\mathrm{Re}$ increases, the wake transitions from steady flow, through the periodic K\'arm\'an vortex street, toward turbulence.
 
The measured data, however, are not exact solutions of these equations. Particle Image Velocimetry does not observe $\mathbf{u}$ directly but instead estimates it from the displacement of tracer particles over a finite time interval, and it recovers only the in-plane velocity while the pressure $p$ remains unobserved. It is therefore natural to model the measured field as
\begin{equation}
    \mathbf{u}_{\text{meas}}(\mathbf{x}, t)
    = \mathbf{u}(\mathbf{x}, t) + \boldsymbol{\varepsilon}(\mathbf{x}, t),
\end{equation}
where $\mathbf{u}$ satisfies the equations above and
$\boldsymbol{\varepsilon}$ is a random measurement error. This error is approximately zero-mean but fluctuates independently from frame to frame and across interrogation windows, and it arises from several stochastic sources: sensor (camera) noise in the recorded particle images, the finite number of tracer particles sampled within each interrogation window, the loss of particles that drift out of the finite-thickness laser sheet between exposures, and sub-pixel error in locating the cross-correlation peak. In addition, the incoming flow cannot be held perfectly uniform, so the effective
inflow boundary condition itself carries a small stochastic component. Because of the random measurement noise, together with the finite spatial and temporal sampling and the spatial averaging inherent to the interrogation windows, the data satisfy the Navier--Stokes equations only approximately, which is precisely the gap between real-world and simulated data that the benchmark is designed to expose.

\textbf{Experiment Results: } Table~\ref{tab:realdata_full} details the comprehensive experimental results. Under an Eulerian grid formulation, we assess performance based on RMSE, FRMSE, and Kinetic Energy. These metrics indicate that the custom design of the automated research agent yields substantial improvements over standard, general-purpose solvers.

\begin{table}[htbp]
\centering
\begin{tabular}{lccc}
\toprule
 & RMSE & FRMSE & Kinetic Energy \\
\midrule
GNOT         & $5.638 \times 10^{-3}$ & $3.059 \times 10^{-1}$ & $5.609 \times 10^{-4}$ \\
Transolver   & $5.626 \times 10^{-3}$ & $2.916 \times 10^{-1}$ & $5.516 \times 10^{-4}$ \\
Transolver++ & $5.618 \times 10^{-3}$ & $2.906 \times 10^{-1}$ & $5.582 \times 10^{-4}$ \\
LNO          & $6.946 \times 10^{-3}$ & $4.411 \times 10^{-1}$ & $6.585 \times 10^{-4}$ \\
AMG          & $4.889 \times 10^{-3}$ & $2.944 \times 10^{-1}$ & $5.330 \times 10^{-4}$ \\
Agent        & $\bm{4.458 \times 10^{-3}}$ & $\bm{2.861 \times 10^{-1}}$ & $\bm{4.430 \times 10^{-4}}$ \\
\bottomrule
\end{tabular}
\caption{Performance comparison on the \RealData\ dataset. Best results are shown in \textbf{bold}.}
\label{tab:realdata_full}
\end{table}

\begin{figure}[htbp]
    \centering
    \includegraphics[width=0.8\textwidth, page=1]{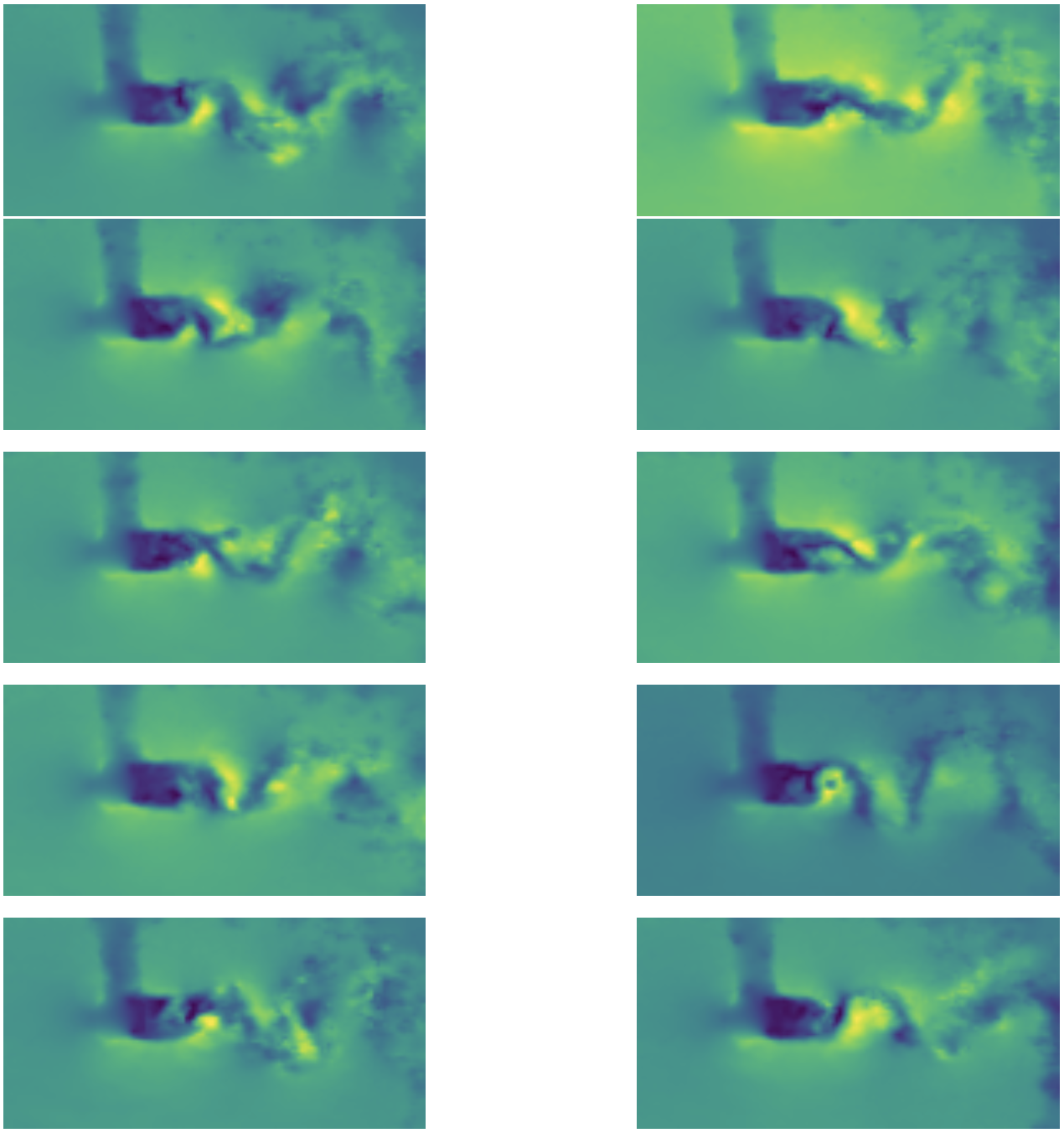}
    \caption{Visualization of the \RealData\ dataset. This dataset provides 2D cross-sectional velocity measurements of the flow around a cylinder.}
    \label{fig:realdata}
\end{figure}

\textbf{Visualization: } We refer the readers to Figure~\ref{fig:realdata} for visualizations of the dataset. We provide examples on the 2D cross-sectional velocity measurements.

\newpage

\subsection{\MixedParam}
\label{appendix:mixedparam}
\textbf{Dataset Description: } This dataset captures the parameter-dependent fluid dynamics governed by the incompressible Navier-Stokes equations. The fluid behavior spans a wide spectrum of physical regimes, transitioning from steady, laminar flows to highly unsteady, chaotic turbulence depending on the varying Reynolds number ($Re$). The training data encompasses a discrete set of simulations with Reynolds numbers sampled at intervals of 100, defined as $Re \in \{100, 200, \dots, 5000\}$. To rigorously evaluate model generalization, the validation and test sets feature a finer sampling interval of 25, defined as $Re \in \{100, 125, 150, \dots, 5000\}$.

\textbf{Modeling Objective: } The state of the fluid system is represented by the tensor $x_t \in \mathbb{R}^{1 \times 128 \times 128}$, where the single feature channel corresponds to the vorticity field evaluated across a $128 \times 128$ spatial discretization grid. The system is explicitly conditioned on the PDE parameter tensor $p \in \mathbb{R}^{1}$, representing the Reynolds number of the specific simulation. The objective is to develop a parameter-conditioned neural simulator, defined as a function $\Psi$, that maps the current system state and the governing parameter to the state at the subsequent timestep. Formally, we aim to predict the next state $\hat{x}_{t+1} = \Psi(x_t, p)$.

\textbf{Dataset Specific Challenges: } The dataset's primary challenge lies in the architectural design required to effectively embed the PDE parameter into the neural model. The model must learn to dynamically adapt its localized spatial filters based on a global scalar value. Existing neural operators often struggle with efficiently fusing this lower-dimensional parameter into the high-dimensional spatial feature space. Furthermore, because the test set contains intermediate Reynolds numbers unseen during training, the model must achieve robust zero-shot generalization to out-of-distribution parameters.

\textbf{Dataset Generation: } The dataset is generated by directly simulating the two-dimensional incompressible Navier–Stokes equations in vorticity–streamfunction form using the pseudo-spectral method. The stream function and velocity are recovered from the vorticity in Fourier space, the nonlinear advection term is evaluated in physical space and transformed back, and aliasing is controlled with the 2/3 rule. Time integration uses a semi-implicit scheme that treats the viscous term with Crank–Nicolson and the nonlinear advection plus forcing with Heun's method (a two-stage, second-order predictor–corrector), with an adaptive step size set by a CFL condition combining an advective limit and a diffusive limit and taking the smallest step across the batch.

\textbf{Governing Physics: } The dataset is governed by the two-dimensional incompressible Navier--Stokes equations on the doubly-periodic domain $\Omega = [0, 2\pi]^2$. In
velocity--pressure form these read
\begin{align}
    \nabla\cdot\mathbf{u} & = 0, \\
    \partial_t \mathbf{u} + (\mathbf{u}\cdot\nabla)\mathbf{u}
    & = -\nabla p + \frac{1}{\mathrm{Re}}\,\nabla^2\mathbf{u} + \mathbf{F},
\end{align}
where $\mathbf{u}(\mathbf{x}, t)$ is the velocity field, $p$ the pressure, $\mathbf{F}$ an external body force, and $\mathrm{Re}$ the Reynolds number. Taking the scalar curl of the momentum equation eliminates the pressure and, using incompressibility, yields the equivalent vorticity--stream-function formulation that the solver actually integrates, in which the vorticity $\omega = \nabla\times\mathbf{u} = \partial_x u_2 - \partial_y u_1$ evolves as
\begin{equation}
    \partial_t \omega + (\mathbf{u}\cdot\nabla)\omega
    = \frac{1}{\mathrm{Re}}\,\nabla^2\omega + f,
\end{equation}
where $f = \nabla\times\mathbf{F}$ is the corresponding vorticity forcing. The velocity is recovered at each instant from a stream function $\psi$ through
\begin{align}
    \mathbf{u} & = \nabla^{\perp}\psi = (\partial_y\psi,\; -\partial_x\psi), \\
    \omega & = -\nabla^2\psi,
\end{align}
which enforces $\nabla\cdot\mathbf{u} = 0$ automatically. The flow is driven by the steady Kolmogorov forcing
\begin{equation}
    f(x, y) = -4\cos(4y),
\end{equation}
a unidirectional shear forcing that is sinusoidal in the transverse coordinate $y$ with wavenumber $4$, so that increasing $\mathrm{Re}$ drives the resulting flow from a smooth laminar state toward two-dimensional turbulence. Together with periodic boundary conditions on $\Omega$ and a prescribed initial vorticity field $\omega(\cdot, 0)$, these equations fully determine the dynamics recorded in the dataset.

\textbf{Experiment Results: } Complete experimental results are detailed in Table~\ref{tab:mixedparam_full}. Performance on the Eulerian grid dataset was assessed using RMSE, FRMSE, and Kinetic Energy metrics, revealing that the automated research agent yields significantly higher accuracy than standard baseline solvers.

\begin{table}[htbp]
\centering
\begin{tabular}{lccc}
\toprule
 & RMSE & FRMSE & Kinetic Energy \\
\midrule
GNOT         & $3.417 \times 10^{0}$ & $2.744 \times 10^{2}$ & $1.845 \times 10^{0}$ \\
Transolver   & $4.359 \times 10^{0}$ & $3.514 \times 10^{2}$ & $3.593 \times 10^{0}$ \\
Transolver++ & $4.329 \times 10^{0}$ & $3.496 \times 10^{2}$ & $3.576 \times 10^{0}$ \\
LNO          & $4.346 \times 10^{0}$ & $3.553 \times 10^{2}$ & $3.605 \times 10^{0}$ \\
AMG          & $3.128 \times 10^{0}$ & $2.688 \times 10^{2}$ & $1.741 \times 10^{0}$ \\
Agent        & $\bm{1.345 \times 10^{0}}$ & $\bm{1.078 \times 10^{2}}$ & $\bm{2.780 \times 10^{-1}}$ \\
\bottomrule
\end{tabular}
\caption{Performance comparison on the \MixedParam\ dataset. Best results are shown in \textbf{bold}.}
\label{tab:mixedparam_full}
\end{table}

\begin{figure}[htbp]
    \centering
    \includegraphics[width=0.8\textwidth, page=1]{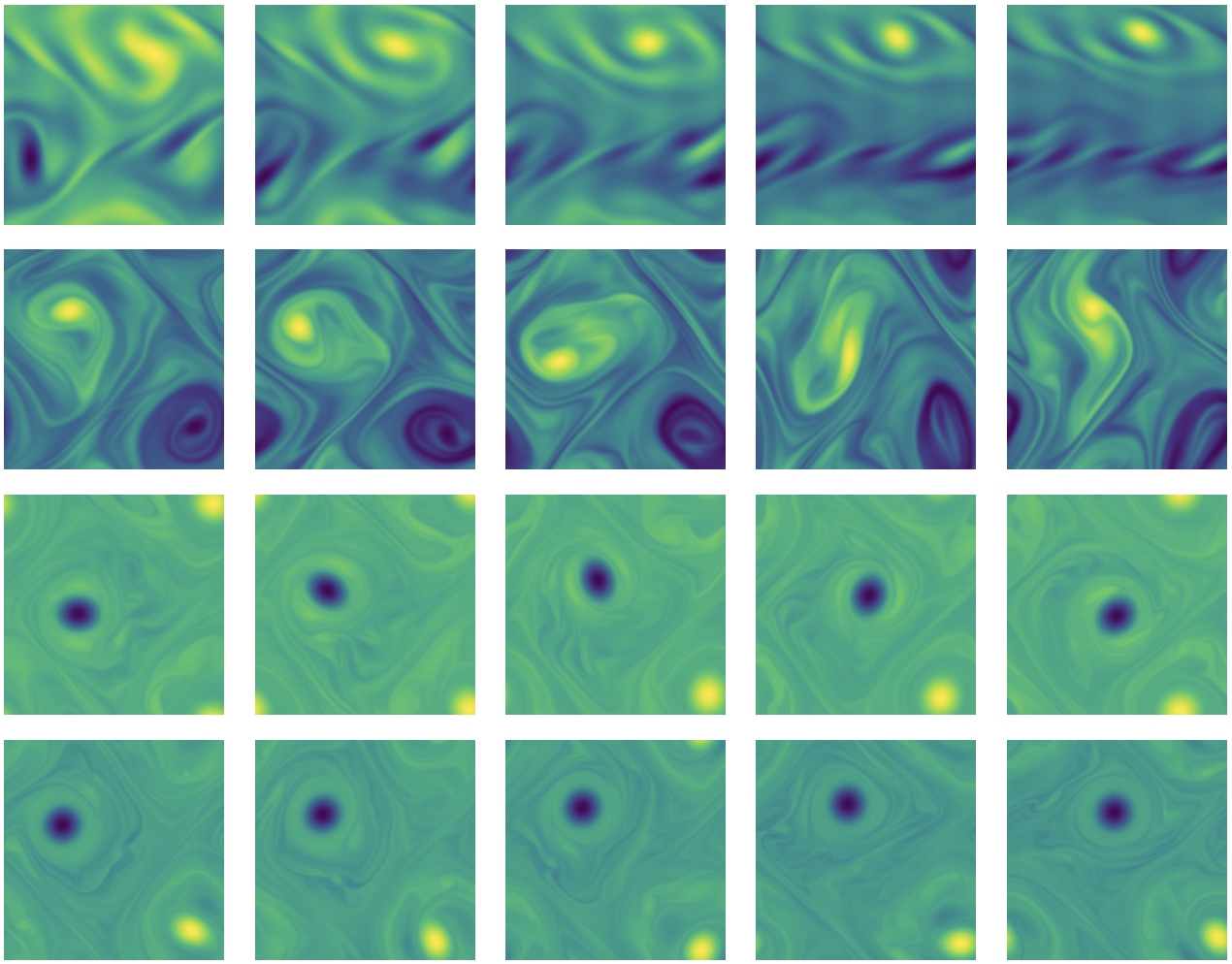}
    \caption{Visualization of the \MixedParam\ dataset. This is the Navier-Stokes Equation with mixed Reynold number. The four rows correspond to Reynold number 100, 1000, 2500, and 5000, respectively.}
    \label{fig:mixedparam}
\end{figure}

\textbf{Visualization: } We refer the readers to Figure~\ref{fig:mixedparam} for visualizations of the dataset. Each row represents a trajectory with unique Reynold number.

\newpage

\subsection{\Heterogeneous}
\label{appendix:heterogeneous}
\textbf{Dataset Description: } This dataset comprises simulations of the Rayleigh--Taylor instability, a fundamental hydrodynamic phenomenon that arises when a denser fluid is accelerated into a lighter one. Under gravity or an equivalent acceleration, small perturbations along the fluid interface grow over time, producing the characteristic morphology of the instability: rising bubbles of the lighter fluid interpenetrating with sinking spikes of the heavier fluid. A defining feature of this dataset is that the fluid densities vary across trajectories, which induces a broad spectrum of dynamical regimes. Trajectories with heavier top fluids exhibit faster sinking and develop increasingly chaotic, finely structured patterns, whereas those with lighter top fluids evolve more slowly and yield smoother, more coherent flows.

\textbf{Modeling Objective: } The goal is to learn a neural simulator $\Psi$ that advances the particle system by one timestep. At time $t$, the input is a pair of tensors $\{V, A\}$ derived from the recent history of $N$ particles connected by $M$ edges. Starting from the raw positional history $p_{t-5:t} \in \mathbb{R}^{6 \times N \times 2}$ and the time-invariant particle attributes $A \in \mathbb{R}^{N \times 1}$, we assemble the node feature matrix $V \in \mathbb{R}^{N \times 13}$ by concatenating the current position $p_t$, the five most recent velocity vectors $v_{t-4:t}$, and the particle attribute. The simulator then maps these inputs to a prediction of the position at the subsequent timestep,
\begin{equation}
    \hat{p}_{t+1} = \Psi(V, A),
\end{equation}
where $\hat{p}_{t+1} \in \mathbb{R}^{N \times 2}$ denotes the predicted per-particle position. Rolling this one-step predictor forward autoregressively then reconstructs the full trajectory.

\textbf{Dataset Specific Challenges: } The principal difficulty posed by this dataset stems from its heterogeneous physical properties, which govern markedly non-uniform dynamical behaviors that standard neural simulators struggle to resolve. Because the density contrast varies from trajectory to trajectory, a single model must internalize an entire family of attribute-conditional physical laws rather than a fixed dynamics. A conventional approach that merely concatenates these attributes into the node feature vector frequently proves insufficient: the attribute is treated as one weakly weighted input among many, and the model fails to learn how the governing dynamics should change as a function of it. Effectively conditioning the learned dynamics on the underlying physical attributes is therefore essential to capturing the full range of behaviors, from the smooth evolution of low-contrast flows to the chaotic mixing of high-contrast ones.

\textbf{Dataset Generation: } The dataset is generated by simulating the two-phase Rayleigh--Taylor instability with Smoothed Particle Hydrodynamics (SPH)~\citep{10.1093/mnras/181.3.375} using the PySPH framework~\citep{ramachandran2021a}, in which the fluid is discretized into Lagrangian particles rather than solved on a fixed grid. The two fluids occupy a rectangular domain of width $L_x = 1$ and height $L_y = 2$ under a downward gravitational acceleration $g_y = -1$, with a heavier phase of density $\rho_1$ initially resting on a lighter phase of density $\rho_2$. The interface is perturbed by a single sinusoidal mode, with particles assigned to the upper phase wherever $y > 1 - 0.15\sin(2\pi x)$, so that the instability grows from this controlled initial disturbance; the domain is enclosed by solid boundary particles built from a ghost region five particle-spacings thick. The particles are laid out on a uniform lattice at resolution $n_x = 50$ (spacing $\Delta x = L_x/n_x$) with smoothing length $h = 1.2\,\Delta x$ and per-particle mass set to reproduce each phase's reference density. The dynamics are advanced with the Transport-Velocity Formulation (TVF) scheme, a weakly-compressible SPH method that transports particles with a modified (transport) velocity to keep them evenly distributed and suppress the tensile instability; incompressibility is approximated through an artificial equation of state $p = c_0^2\rho$ with an artificial sound speed $c_0 = V_{\max}/\mathrm{Fr}$ chosen from a low Mach/Froude number $\mathrm{Fr} = 0.01$ (using $V_{\max} = \sqrt{0.5\,L_y\lvert g_y\rvert}$), and the kinematic viscosity $\nu = V_{\max} L_y/\mathrm{Re}$ is fixed by a Reynolds number $\mathrm{Re} = 420$. Time integration uses an adaptive step taken as half the minimum of a CFL (acoustic) limit $0.25\,h/(c_0 + V_{\max})$, a viscous limit $0.125\,h^2/\nu$, and a body-force limit $0.25\sqrt{h/\lvert g_y\rvert}$, integrating to a final time $t_f = 25$ and writing particle snapshots periodically.

\textbf{Governing Physics: } The dataset models the Rayleigh--Taylor instability of two immiscible,
incompressible, viscous fluids in a gravitational field, governed by the
variable-density incompressible Navier--Stokes equations on the domain
$\Omega = [0, L_x] \times [0, L_y]$ under the body force
$\mathbf{g} = (0, g_y)$,
\begin{equation}
    \nabla\cdot\mathbf{u} = 0,
    \qquad
    \frac{D\rho}{Dt} = 0,
    \qquad
    \rho\,\frac{D\mathbf{u}}{Dt}
    = -\nabla p + \nabla\cdot\!\big(2\mu\,\mathbf{S}\big) + \rho\,\mathbf{g},
\end{equation}
where $\tfrac{D}{Dt} = \partial_t + \mathbf{u}\cdot\nabla$ is the material
derivative, $\mathbf{u}$ the velocity, $p$ the pressure, $\rho$ the density,
$\mathbf{S} = \tfrac{1}{2}\big(\nabla\mathbf{u} + \nabla\mathbf{u}^{\!\top}\big)$
the strain-rate tensor, and $\mu = \rho\nu$ the dynamic viscosity for a common
kinematic viscosity $\nu$. The density is piecewise constant, taking the value
$\rho_1$ in the upper (heavy) fluid and $\rho_2$ in the lower (light) fluid
with $\rho_1 > \rho_2$, and the two phases meet across a material interface
$\Gamma(t)$ advected by the flow. The system starts from rest,
$\mathbf{u}(\cdot, 0) = \mathbf{0}$, with the interface given the single-mode
sinusoidal perturbation
\begin{equation}
    \Gamma(0) = \{\, y = \eta(x) \,\},
    \qquad
    \eta(x) = 1 - 0.15\,\sin(2\pi x),
\end{equation}
so that $\rho(\mathbf{x}, 0) = \rho_1$ for $y > \eta(x)$ and $\rho_2$ for
$y < \eta(x)$. Across $\Gamma(t)$ the velocity and the traction are continuous
(there is no surface tension),
\begin{equation}
    [\![\,\mathbf{u}\,]\!] = \mathbf{0},
    \qquad
    \big[\!\big[\,(-p\,\mathbf{I} + 2\mu\,\mathbf{S})\,\mathbf{n}\,\big]\!\big]
    = \mathbf{0},
\end{equation}
and no-slip conditions $\mathbf{u} = \mathbf{0}$ are imposed on the solid walls
$\partial\Omega$. The configuration is gravitationally unstable---a denser
fluid supported above a lighter one---so the perturbation is amplified at a
rate set by the Atwood number, while the flow regime is fixed by the Reynolds
and Froude numbers,
\begin{equation}
    A = \frac{\rho_1 - \rho_2}{\rho_1 + \rho_2},
    \qquad
    \mathrm{Re} = \frac{V_{\max} L_y}{\nu},
    \qquad
    \mathrm{Fr} = \frac{V_{\max}}{c_0},
    \qquad
    V_{\max} = \sqrt{\tfrac{1}{2} L_y\, \lvert g_y\rvert}.
\end{equation}
In the weakly-compressible formulation actually integrated, the strict
constraint $\nabla\cdot\mathbf{u} = 0$ is relaxed to the continuity equation
$\tfrac{D\rho}{Dt} = -\rho\,\nabla\cdot\mathbf{u}$ closed by the stiff
artificial equation of state $p = c_0^2\,\rho$, in which the artificial sound
speed $c_0 = V_{\max}/\mathrm{Fr}$ is chosen large enough
($\mathrm{Fr} = 0.01 \ll 1$) that density fluctuations remain of order
$\mathrm{Fr}^2$ and the incompressible limit is recovered.

\textbf{Experiment Results: } We present the complete results in Table~\ref{tab:heterogeneous_full}. Because the datasets consist of Lagrangian particles, we evaluate performance using the RMSE of particle positions, Lagrangian Kinetic Energy (LagKE), and Eulerian aggregated RMSE and FRMSE (EuARMSE, EuAFRMSE). The results demonstrate that the automated research agent yields significantly higher accuracy than standard baseline solvers.

\begin{table}[htbp]
\centering
\begin{tabular}{lcccc}
\toprule
 & RMSE & LagKE & EuARMSE & EuAFRMSE \\
\midrule
GNOT         & $4.207 \times 10^{-3}$ & $1.324 \times 10^{-1}$ & $6.836 \times 10^{-4}$ & $2.170 \times 10^{-2}$ \\
Transolver   & $4.155 \times 10^{-3}$ & $1.374 \times 10^{-1}$ & $6.655 \times 10^{-4}$ & $2.169 \times 10^{-2}$ \\
Transolver++ & $3.963 \times 10^{-3}$ & $1.622 \times 10^{-1}$ & $6.280 \times 10^{-4}$ & $2.186 \times 10^{-2}$ \\
LNO          & $3.951 \times 10^{-3}$ & $1.776 \times 10^{-1}$ & $7.030 \times 10^{-4}$ & $2.590 \times 10^{-2}$ \\
AMG          & $3.900 \times 10^{-3}$ & $1.683 \times 10^{-1}$ & $6.052 \times 10^{-4}$ & $2.155 \times 10^{-2}$ \\
Agent        & $\bm{3.311 \times 10^{-3}}$ & $\bm{8.797 \times 10^{-2}}$ & $\bm{5.280 \times 10^{-4}}$ & $\bm{1.585 \times 10^{-2}}$ \\
\bottomrule
\end{tabular}
\caption{Performance comparison on the \Heterogeneous\ dataset. Best results are shown in \textbf{bold}.}
\label{tab:heterogeneous_full}
\end{table}

\begin{figure}[htbp]
    \centering
    \includegraphics[width=0.8\textwidth, page=1]{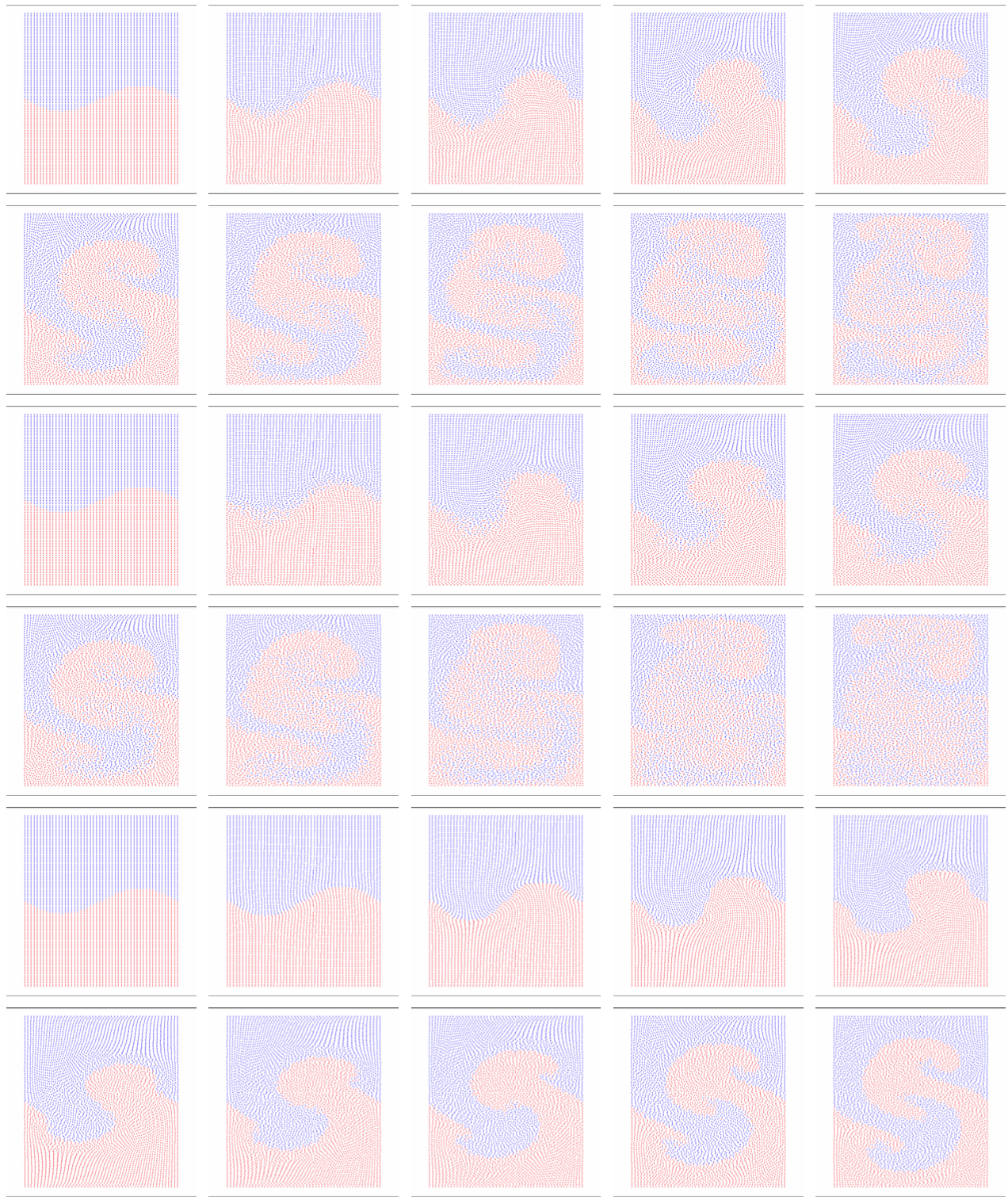}
    \caption{Visualization of the \Heterogeneous\ dataset. This is the Reyleigh Taylor Instability in Lagrangian particle representation. The density of particles differ across trajectories, resulting in distinct behaviors.}
    \label{fig:heterogeneous}
\end{figure}

\textbf{Visualization: } We refer the readers to Figure~\ref{fig:heterogeneous} for visualizations of the dataset.

\newpage

\subsection{\Partial}
\label{appendix:partial}
\textbf{Dataset Description: } This dataset captures the fluid dynamics of a two-dimensional incompressible flow governed by the Navier-Stokes equations, at Reynolds number of $Re = 1000$. This dataset models a sensor-limited scenario where the physical domain is only partially observable.

\textbf{Modeling Objective: } The underlying full physical state is represented by the tensor $x_t \in \mathbb{R}^{1 \times 128 \times 128}$, denoting the complete vorticity field. The model also receives a binary mask tensor $m \in \{0, 1\}^{1 \times 128 \times 128}$, where a value of $1$ indicates an available measurement and $0$ indicates missing data. The objective is to develop a neural simulator, $\Psi$, that maps the incomplete current observation to the complete, unmasked state at the subsequent timestep. Formally, we aim to predict the full future field $\hat{x}_{t+1} = \Psi(X_t \odot m)$.

\textbf{Dataset Specific Challenges: } The primary challenge is that partial observability transforms a deterministic physical system into a fundamentally ill-posed and non-deterministic learning problem. Because critical spatial information is hidden, multiple physically valid full-state configurations could theoretically correspond to the same masked observation. Consequently, the neural model must simultaneously solve an ambiguous spatial inverse problem to infer the unobserved regions while computing the temporal evolution of the system.

\textbf{Dataset Generation: } In the first stage, a two-dimensional incompressible Navier--Stokes flow in vorticity--stream-function form, driven by the steady Kolmogorov forcing $f(x, y) = -4\cos(4y)$ on the doubly-periodic domain $[0, 2\pi]^2$, is advanced with a pseudo-spectral solver at a fixed Reynolds number $\mathrm{Re} =
1000$; each of the $50$ trajectories is initialized from a Gaussian random field and integrated for $100$ time units so the flow reaches a statistically developed state before recording, after which the vorticity is simulated at $256 \times 256$ resolution and stored, spatially subsampled by a factor of two, on a $128 \times 128$ grid with snapshots saved every $\Delta t = 0.1$ up to $T = 20$. In the second stage the dataset is made partially observed: a binary mask $m \in \{0, 1\}^{1 \times 128 \times 128}$, with $m = 1$ where a measurement is available and $m = 0$ where data are missing, is generated at random during post-processing and applied through the element-wise product, so that only the retained entries of the field are visible while the true underlying state is preserved as ground truth.


\textbf{Governing Physics: } The governing physics is the two-dimensional incompressible Navier–Stokes equation in  vorticity–stream-function formulation similar to the one described in Section~\ref{appendix:mixedparam}

\textbf{Experiment Results: } As detailed in Table~\ref{tab:partial_full}, the complete experimental results on the Eulerian grid demonstrate the superiority of the automated research agent's architecture. Measured by RMSE, FRMSE, and Kinetic Energy, the specialized design surpasses general-purpose baselines.

\begin{table}[htbp]
\centering
\begin{tabular}{lccc}
\toprule
 & RMSE & FRMSE & Kinetic Energy \\
\midrule
GNOT         & $1.978 \times 10^{0}$ & $1.875 \times 10^{2}$ & $9.480 \times 10^{-1}$ \\
Transolver   & $3.636 \times 10^{0}$ & $4.224 \times 10^{2}$ & $4.509 \times 10^{0}$ \\
Transolver++ & $3.636 \times 10^{0}$ & $4.224 \times 10^{2}$ & $4.519 \times 10^{0}$ \\
LNO          & $3.636 \times 10^{0}$ & $4.223 \times 10^{2}$ & $4.505 \times 10^{0}$ \\
AMG          & $2.111 \times 10^{0}$ & $2.031 \times 10^{2}$ & $9.980 \times 10^{-1}$ \\
Agent        & $\bm{8.633 \times 10^{-2}}$ & $\bm{7.683 \times 10^{0}}$ & $\bm{2.100 \times 10^{-2}}$ \\
\bottomrule
\end{tabular}
\caption{Performance comparison on the \Partial\ dataset. Best results are shown in \textbf{bold}.}
\label{tab:partial_full}
\end{table}

\begin{figure}[htbp]
    \centering
    \includegraphics[width=0.8\textwidth, page=1]{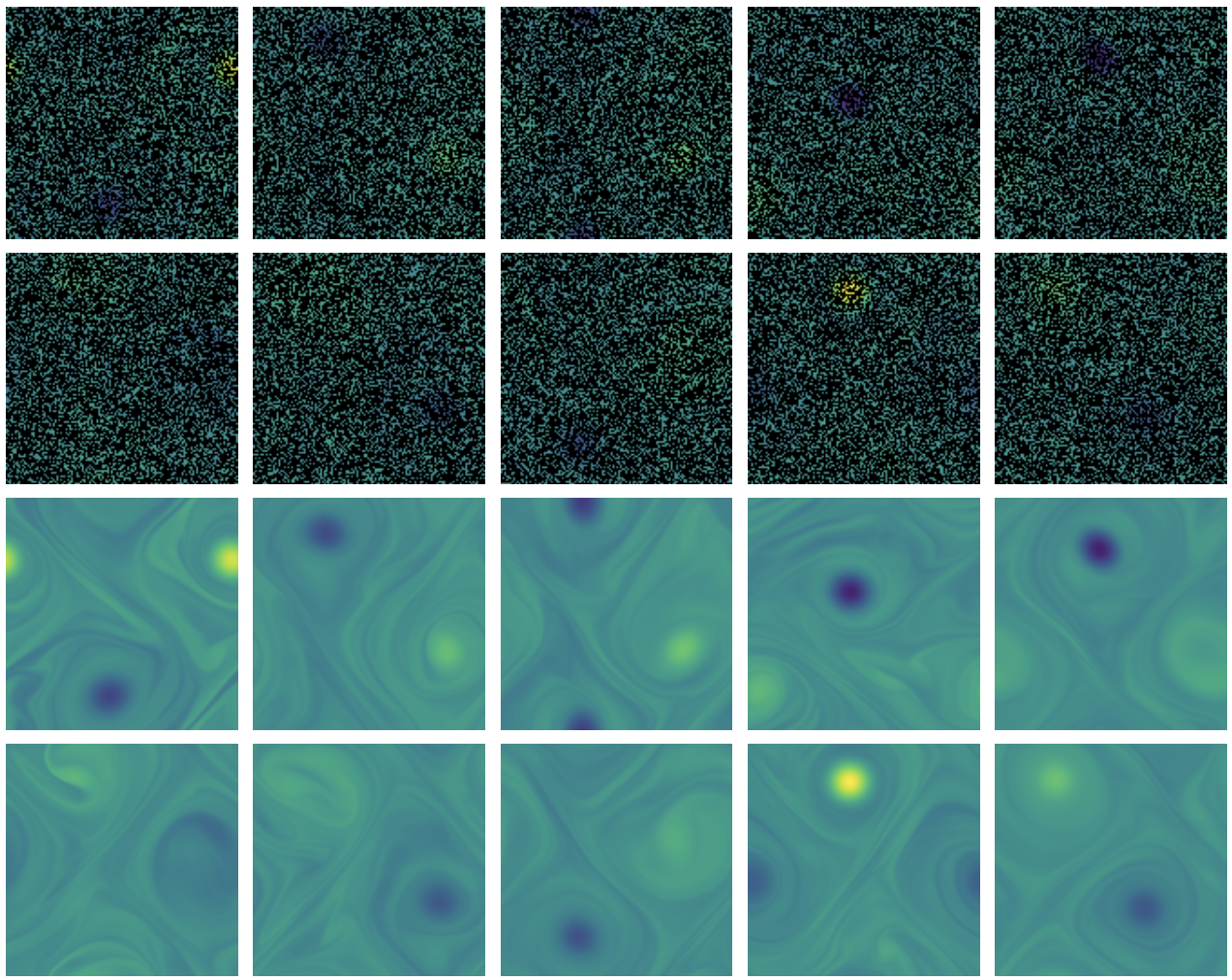}
    \caption{Visualization of the \Partial\ dataset.The first two rows show partially observed data, while the bottom two show complete data.}
    \label{fig:partial}
\end{figure}

\textbf{Visualization: } We refer the readers to Figure~\ref{fig:partial} for visualizations of the dataset.

\newpage
\clearpage

\subsection{\MixedDomain}
\label{appendix:mixeddomain}
\textbf{Dataset Description: } This dataset provides a multi-representation view of fluid flow around a cylinder. It captures the same underlying physics through three distinct lenses: a structured Eulerian grid, an unstructured Eulerian mesh, and a Lagrangian particle-based representation. The grid data consists of a $64 \times 32$ uniformly spaced domain, while the mesh data utilizes 2048 randomly scattered points to sample the flow field. For the Lagrangian perspective, the dataset tracks 16{,}000 individual particles as they advect through the domain. Because all three describe identical dynamics and differ only in how the flow field is discretized, the aim is to develop a generalized neural solver that can ingest any one of these representations and produce predictions in that same representation.

\textbf{Modeling Objective: } The objective is to develop a unified neural simulator, $\Psi$, capable of cross-domain forecasting through a single, representation-agnostic input format. Let $x_t \in \mathbb{R}^{N \times 13}$ denote the system state at time $t$, where the node count $N$ varies by representation (grid points, mesh points, or Lagrangian particles). The 13 input features per node comprise the position (2), a five-step velocity history ($2 \times 5 = 10$), and a scalar identifier indicating which representation the node belongs to (1). This scalar is encoded as $0$ for the irregular mesh, $1$ for the regular grid, and $2$ or $3$ for Lagrangian particles, where the latter two values further distinguish fluid from solid particles. For Lagrangian particles the position denotes the particle location at the current step, whereas for the grid and mesh it is the fixed node position. The goal is to learn the parameterized solver $\Psi$ such that
\begin{equation}
    \hat{y}_{t+1} = \Psi(x_t),
\end{equation}
where $\hat{y}_{t+1} \in \mathbb{R}^{N \times 2}$ is the predicted next-step velocity for every node. The model can therefore take any representation as input and forecast the velocity field in whichever representation is provided.

\textbf{Dataset Specific Challenges: } The fundamental difficulty lies in the structural divergence of the data formats, requiring a single neural solver to operate on all three representations despite their differing node counts, connectivity, and geometric structure, without a fixed grid resolution or topology to rely on. While the grid data benefits from spatial translation invariance and structured convolutions, the scattered mesh and Lagrangian particles lack a fixed global topology. Compounding this, the three representations encode the same physics, so the solver must distill a single consistent set of dynamics from all of them rather than fitting each representation independently. Crafting a unified latent space that bridges mesh-based and particle-based descriptions of the same flow, while maintaining accuracy across their disparate sampling densities, is the central hurdle for current models.

\textbf{Dataset Generation: } The dataset is constructed in two steps: a native Lagrangian dataset is generated first, and the Eulerian representations are then obtained from it by kernel interpolation. In the first step, a Smoothed Particle Hydrodynamics (SPH)~\citep{10.1093/mnras/181.3.375} simulation is run in PySPH~\citep{ramachandran2021a} for two-dimensional incompressible flow past a periodic array of cylinders: a cylinder of radius $a = 0.02$ sits at the centre of a channel of length $L = 0.12$ and height $H = 4a$ that is made periodic in the streamwise direction, so that it represents an infinite periodic array, and the flow is driven through it by a constant streamwise body force $f_x$. The fluid is discretized into roughly $16{,}000$ Lagrangian particles---together with solid particles forming the cylinder and channel walls---and advanced with the Transport-Velocity Formulation (TVF) scheme, a weakly-compressible SPH method in which incompressibility is approximated through the artificial equation of state $p = c_0^2\rho$, the kinematic viscosity $\nu = 10^{-4}$ fixes the Reynolds number, and the time step is adaptively limited by acoustic (CFL), viscous, and body-force constraints; this yields the particle trajectories that form the Lagrangian representation, recording each particle's position and velocity as it advects through the domain. In the second step, the structured $64 \times 32$ grid and the unstructured $2048$-point scattered mesh are produced from these particles by SPH kernel interpolation, which is the standard means of reconstructing field values in SPH rather than an ad-hoc resampling: any continuous field $A$ is estimated at an arbitrary location $\mathbf{x}$ by the kernel-weighted particle sum
\begin{equation}
    A(\mathbf{x}) = \sum_{j} \frac{m_j}{\rho_j}\, A_j\,
    W\!\left(\mathbf{x} - \mathbf{x}_j,\, h\right),
\end{equation}
where $m_j$, $\rho_j$, and $A_j$ are the mass, density, and field value of particle $j$, and $W$ is a smoothing kernel (a quintic spline) with smoothing length $h$. Evaluating this interpolant with $A = \mathbf{u}$ at the fixed grid nodes and mesh points gives the velocity field on each Eulerian representation directly from the same particles, so that all three representations are drawn from a single underlying simulation.

\textbf{Governing Physics: } The flow is governed by the incompressible Navier--Stokes equations, written here in Lagrangian (material) form, since the SPH discretization follows the fluid as a set of moving particles. Let a fluid particle occupy position $\mathbf{r}(t)$ and move with velocity $\mathbf{u}(t) = \dot{\mathbf{r}}(t)$; along particle paths the material derivative is $\tfrac{d}{dt} = \partial_t + \mathbf{u}\cdot\nabla$. The particle kinematics, mass conservation, and momentum balance then read
\begin{align}
    \frac{d\mathbf{r}}{dt} &= \mathbf{u}, \\[2pt]
    \frac{d\rho}{dt} &= -\rho\,\nabla\cdot\mathbf{u}, \\[2pt]
    \frac{d\mathbf{u}}{dt} &= -\frac{1}{\rho}\nabla p
    + \nu\,\nabla^2\mathbf{u} + \mathbf{f},
\end{align}
where $\rho$ is the density, $p$ the pressure, $\nu$ the kinematic viscosity, and $\mathbf{f} = (f_x, 0)$ the constant streamwise body force (per unit mass) that drives the flow through the array. In the strictly incompressible limit these are closed by the divergence-free constraint $\nabla\cdot\mathbf{u} = 0$, so that $\tfrac{d\rho}{dt} = 0$; the weakly-compressible SPH model instead retains the continuity equation above and closes the system with the stiff barotropic equation of state
\begin{equation}
    p = c_0^2\,\rho,
\end{equation}
in which the artificial sound speed $c_0$ is chosen large enough that density departures from the reference $\rho_0$ remain small and the incompressible limit is approached. The equations hold on the fluid domain $\Omega_f$---the channel $[0, L] \times [0, H]$ with the cylinder disk of radius $a$ removed---subject to no-slip conditions $\mathbf{u} = \mathbf{0}$ on the cylinder surface and the channel walls, together with streamwise periodicity of the fields, $\mathbf{u}(x + L, y, t) = \mathbf{u}(x, y, t)$ and $p(x + L, y, t) = p(x, y, t)$.

\textbf{Experiment Results: } This dataset comprises three distinct representations: an Eulerian grid, an Eulerian mesh, and Lagrangian particles. We use the averaged RMSE across all three representations as the primary evaluation metric, while also reporting performance on each individually. The complete results are presented in Table~\ref{tab:mixeddomain_full}. Notably, all models exhibit poor performance on the Lagrangian representation. This suggests that the models prioritize the Eulerian data and fail to effectively embed the Lagrangian features. In this specific context, the automated research agent performs comparably to the general baselines.

\begin{table}[htbp]
\centering
\begin{tabular}{lcccc}
\toprule
 & RMSE & RMSE (grid) & RMSE (mesh) & RMSE (particle) \\
\midrule
GNOT         & $1.784 \times 10^{-3}$ & $4.250 \times 10^{-5}$ & $4.150 \times 10^{-5}$ & $1.999 \times 10^{-3}$ \\
Transolver   & $\bm{1.739 \times 10^{-3}}$ & $\bm{3.350 \times 10^{-5}}$ & $3.560 \times 10^{-5}$ & $\bm{1.947 \times 10^{-3}}$ \\
Transolver++ & $1.775 \times 10^{-3}$ & $3.830 \times 10^{-5}$ & $\bm{3.500 \times 10^{-5}}$ & $1.988 \times 10^{-3}$ \\
LNO          & $1.785 \times 10^{-3}$ & $3.650 \times 10^{-5}$ & $3.650 \times 10^{-5}$ & $1.999 \times 10^{-3}$ \\
AMG          & $6.491 \times 10^{-2}$ & $2.583 \times 10^{-2}$ & $9.078 \times 10^{-2}$ & $6.445 \times 10^{-2}$ \\
Agent        & $1.785 \times 10^{-3}$ & $3.640 \times 10^{-5}$ & $3.640 \times 10^{-5}$ & $1.999 \times 10^{-3}$ \\
\bottomrule
\end{tabular}
\caption{Performance comparison on the \Partial\ dataset. Best results are shown in \textbf{bold}.}
\label{tab:mixeddomain_full}
\end{table}

\begin{figure}[htbp]
    \centering
    \includegraphics[width=0.9\textwidth, page=1]{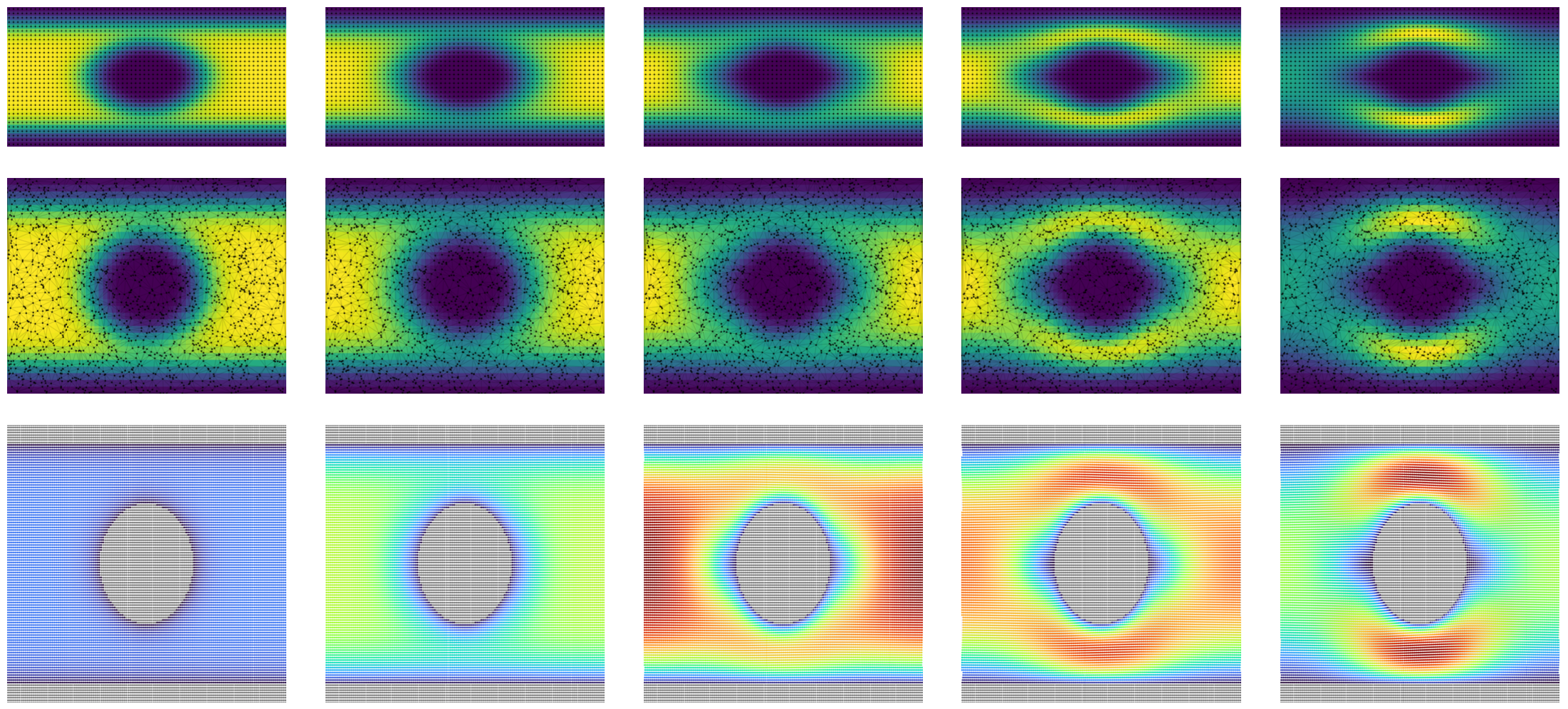}
    \caption{Visualization of the \MixedDomain\ dataset. Rows from top to bottom display grid (Eulerian), mesh (Eulerian), particle (Lagrangian) representations, respectively.}
    \label{fig:mixeddomain}
\end{figure}

\textbf{Visualization: } We refer the readers to Figure~\ref{fig:mixeddomain} for visualizations of the dataset.

\newpage
\clearpage

\subsection{\MultiPhysics}
\label{appendix:multiphysics}
\textbf{Dataset Description: } This dataset is drawn from a coupled multiphysics simulation in which three physical subsystems continuously feed back on one another, following the coupling benchmark of \citet{zhang2025m2pdecompositionalgenerativemultiphysics}. The source field is a scalar field produced by an internal generation process; it depends on the solid temperature through a feedback mechanism, on the fluid state, and on the control (boundary) parameters. The solid field is the temperature field inside a solid domain, driven by the heat source from the source field and cooled at its interface with the surrounding fluid. The fluid field is the fluid state that carries heat away from the solid, described by several channels (temperature, a velocity / momentum-like quantity, and further hydraulic variables) and driven by the heat flux leaving the solid surface. The three subsystems are solved by separate physics codes and coupled through a fixed-point (Picard) iteration: each solver is run with the current estimate of the other fields as its boundary/source condition, and the process repeats until all three fields stop changing.

\textbf{Modeling Objective: } We learn one neural surrogate per field. Each field is stored as a 5-D tensor of shape $[B, C, D, H, W]$: a batch of $B$ samples, $C$ physical channels, and a three-axis spatial grid $(D, H, W)$. Let $x \in \mathbb{R}^{C_{\mathrm{in}} \times D \times H \times W}$ denote a field's input conditioning tensor and $\hat{y} \in \mathbb{R}^{C_{\mathrm{out}} \times D \times H \times W}$ its predicted solution. For each field we develop a parameterized surrogate $\Psi_{\mathrm{field}}$ that maps the conditioning fields driving it to that field's converged solution, $\hat{y} = \Psi_{\mathrm{field}}(x)$, replacing one call to the corresponding legacy physics solver. The three fields differ in their channel roles: the source surrogate is conditioned on the boundary conditions together with the other fields and predicts the source field ($C_{\mathrm{in}} = 2 \rightarrow C_{\mathrm{out}} = 1$); the solid surrogate is conditioned on the source field and the fluid interface state and predicts the solid temperature ($C_{\mathrm{in}} = 2 \rightarrow C_{\mathrm{out}} = 1$); and the fluid surrogate is conditioned on the heat flux from the solid and predicts the 4-channel fluid state ($C_{\mathrm{in}} = 1 \rightarrow C_{\mathrm{out}} = 4$). Each surrogate is trained and validated on decoupled per-field $(\text{input}, \text{target})$ pairs, in which the target is that field's converged solution computed by the reference solver in isolation, and the reported score for a run is that field's validation accuracy on held-out decoupled data.

\textbf{Dataset Specific Challenges: } The central difficulty is that each surrogate must learn the conditional distribution of one field given the others, and these conditionals are tightly coupled: the source field is conditioned on the solid and fluid states, the solid on the source and fluid, and the fluid on the solid, so the fields are bound together through their mutual dependencies rather than being independently predictable. A surrogate is therefore useful only if it is accurate as a conditional in isolation and composes stably when its imperfect samples are fed back in as another field's conditioning input over many iterations.

\textbf{Dataset Generation: } We adopt the coupled three-field dataset from \citet{zhang2025m2pdecompositionalgenerativemultiphysics}, generated with the finite-element and finite-volume multiphysics frameworks on a two-dimensional cell geometry partitioned into adjacent solid and fluid subdomains sharing a common interface. The source field is solved on a $[64, 20]$ mesh spanning the full domain, the solid temperature on a $[64, 8]$ mesh over the solid subdomain, and the fluid state on a $[64, 12]$ mesh over the fluid subdomain; the source and solid fields are evaluated at mesh points by the finite-element method, while the fluid field is evaluated at cell centers by the finite-volume method, with interpolation used to transfer values between the differing grids. The time step is adaptively controlled by the solver, and each trajectory is stored at $16$ output steps, so the full spatio-temporal tensors have shapes $[b, 16, 64, 20]$, $[b, 16, 64, 8]$, and $[b, 16, 64, 12]$ for the three fields. Two kinds of data are produced. The decoupled data used for training and validation are generated by a pre-iteration procedure: each field is computed in isolation while the others are held at assumed conditions, sweeping the boundary condition imposed on the source field, so that each converged single-field solution is paired with the conditioning fields that drove it; the most expensive field (the fluid) is excluded from pre-iteration to reduce cost, and the remaining fields are computed in sequence. The coupled ground truth used only for the final test is generated with an operator-splitting iterative algorithm that exchanges information between the three fields at every time step and iterates to convergence.

\textbf{Governing Physics: } The dataset is governed by three coupled fields on the domain $x \in [0, L_s + L_f]$, $y \in [0, L_y]$, $t \in [0, 5]$: a source field $\phi$ obeying a diffusion--reaction balance, a solid temperature $T_s$ obeying heat conduction with a volumetric source, and a fluid state $(\vec{u}, p, T_f)$ obeying the incompressible flow and convective heat-transport equations. The source field evolves as
\begin{equation}
    \frac{1}{v}\frac{\partial \phi}{\partial t}
    = D\,\Delta \phi + \big(\nu \Sigma_f - \Sigma_a(T)\big)\phi,
    \qquad x \in [0, L_s + L_f],
\end{equation}
with $D$ a diffusion coefficient, $\nu\Sigma_f$ a generation term, and $\Sigma_a(T)$ an absorption/removal term that depends on the local temperature; a time-varying inflow condition $\phi(0, y, t) = f(y, t)$ is imposed on the left edge and homogeneous conditions on the others. This temperature dependence of $\Sigma_a(T)$ is the negative feedback that ties the source field back to the thermal state. The solid temperature is driven by the source field through a heat-generation term $A\phi_s$,
\begin{equation}
    \rho c_p \frac{\partial T_s}{\partial t}
    = \nabla \cdot \big(k_s \nabla T_s\big) + A\,\phi_s,
    \qquad x \in [0, L_s],
\end{equation}
with insulated conditions on the top and bottom edges. The fluid subdomain satisfies the incompressible Navier--Stokes equations together with convective heat transport,
\begin{equation}
    \nabla \cdot \vec{u} = 0,
    \qquad
    \rho\left(\frac{\partial \vec{u}}{\partial t} + \vec{u}\cdot\nabla\vec{u}\right)
    = -\nabla p + \mu \nabla^2 \vec{u} + \vec{f},
\end{equation}
\begin{equation}
    \rho c_p\left(\frac{\partial T_f}{\partial t} + \vec{u}\cdot\nabla T_f\right)
    = k_f \nabla^2 T_f,
    \qquad x \in [L_s, L_s + L_f],
\end{equation}
where $\rho$ is density, $\mu$ dynamic viscosity, $c_p$ specific heat, and $k_s, k_f$ the solid and fluid conductivities (both functions of temperature). The solid and fluid fields are tied together at their shared interface $x = L_s$ by continuity of temperature and of heat flux,
\begin{equation}
    T_s(L_s, y, t) = T_f(L_s, y, t),
    \qquad
    k_f \frac{\partial T_f(L_s, y, t)}{\partial x}
    = k_s \frac{\partial T_s(L_s, y, t)}{\partial x},
\end{equation}
which is precisely the strong interface coupling the surrogate must respect: the source field feeds heat into the solid, the solid exchanges heat with the fluid across the interface, and the fluid temperature feeds back into the source field through the temperature-dependent removal term $\Sigma_a(T)$.

\textbf{Experiment Results: } The complete experimental results are presented in Table~\ref{tab:multiphysics_full}.

\begin{table}[htbp]
\centering
\begin{tabular}{lccc}
\hline
 & RMSE (source) & FRMSE (source) & RMSE (solid) \\
\hline
GNOT         & $1.710 \times 10^{-2}$ & $4.983 \times 10^{-1}$ & $1.212 \times 10^{-1}$ \\
Transolver   & $1.870 \times 10^{-1}$ & $6.146 \times 10^{0}$ & $8.238 \times 10^{-2}$ \\
Transolver++ & $1.954 \times 10^{-1}$ & $6.490 \times 10^{0}$ & $7.526 \times 10^{-2}$ \\
LNO          & $3.315 \times 10^{-1}$ & $5.753 \times 10^{0}$ & $7.282 \times 10^{-1}$ \\
AMG          & $7.817 \times 10^{-3}$ & $2.438 \times 10^{-1}$ & $1.125 \times 10^{-1}$ \\
\textbf{Agent} & $\bm{6.585 \times 10^{-3}}$ & $\bm{2.061 \times 10^{-1}}$ & $\bm{3.802 \times 10^{-2}}$ \\
\hline
\multicolumn{4}{c}{} \\
\hline
 & FRMSE (solid) & RMSE (fluid) & FRMSE (fluid) \\
\hline
GNOT         & $2.446 \times 10^{0}$ & $6.879 \times 10^{-2}$ & $\bm{1.826 \times 10^{0}}$ \\
Transolver   & $1.843 \times 10^{0}$ & $1.201 \times 10^{-1}$ & $3.327 \times 10^{0}$ \\
Transolver++ & $1.697 \times 10^{0}$ & $1.330 \times 10^{-1}$ & $3.686 \times 10^{0}$ \\
LNO          & $1.648 \times 10^{1}$ & $4.316 \times 10^{-1}$ & $5.735 \times 10^{0}$ \\
AMG          & $2.447 \times 10^{0}$ & $\bm{9.876 \times 10^{-2}}$ & $2.603 \times 10^{0}$ \\
\textbf{Agent} & $\bm{7.926 \times 10^{-1}}$ & $1.096 \times 10^{-1}$ & $2.947 \times 10^{0}$ \\
\hline
\end{tabular}
\caption{Performance comparison on the \MultiPhysics\ dataset. Best results are shown in \textbf{bold}.}
\label{tab:multiphysics_full}
\end{table}

\begin{figure}[htbp]
    \centering
    \includegraphics[width=0.8\textwidth, page=1]{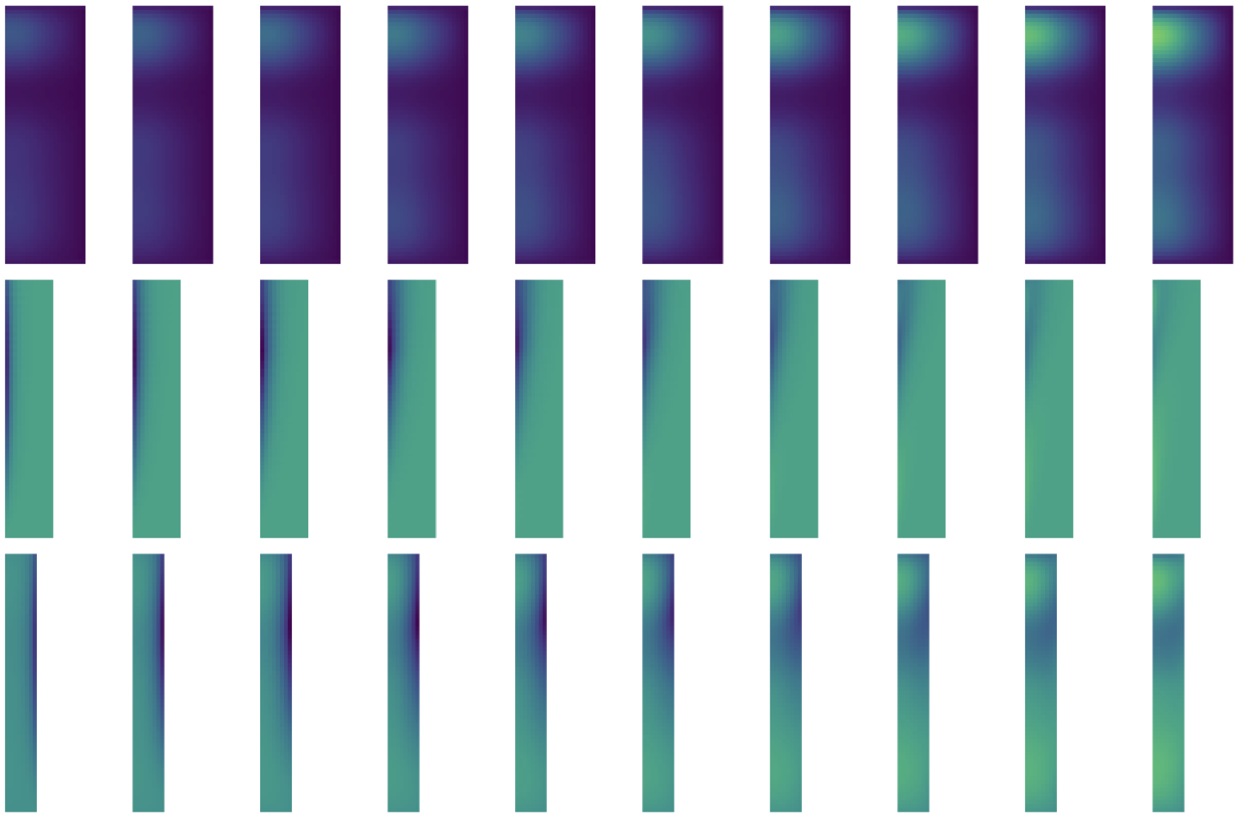}
    \caption{Visualization of the \MultiPhysics\ dataset. Rows from top to bottom display source, fluid, and solid, respectively.}
    \label{fig:multiphysics}
\end{figure}

\textbf{Visualization: } We refer the readers to Figure~\ref{fig:multiphysics} for visualizations of the dataset.

\newpage
\clearpage

\subsection{\ThreeD}
\label{appendix:threed}
\textbf{Dataset Description: } This dataset consists of three-dimensional, isothermal Magnetohydrodynamic (MHD) turbulence. It captures the complex interplay between magnetic fields and conductive fluid dynamics, essential processes for modeling the solar wind, galaxy formation, and Interstellar Medium (ISM) dynamics. The simulation represents a periodic cubic domain without self-gravity.

\textbf{Modeling Objective: } The physical state at time $t$ is represented by a high-dimensional tensor $x_t \in \mathbb{R}^{7 \times 64 \times 64 \times 64}$. This 7-channel input encapsulates the scalar density field $\rho$, the velocity vector field $\mathbf{u} = (u_x, u_y, u_z)$, and the magnetic vector field $\mathbf{B} = (B_x, B_y, B_z)$. The objective is to develop a neural simulator, $\Psi$, capable of next-step prediction, mapping the current volumetric state to the full state at the subsequent timestep: $\hat{x}_{t+1} = \Psi(x_t)$.

\textbf{Dataset Specific Challenges: } The transition from 2D to 3D PDE modeling introduces a curse of dimensionality that complicates both accuracy and computational feasibility. At a resolution of $64^3$ with 7 input channels, a single snapshot contains over 1.8 million data points, creating a massive memory footprint that often exceeds the VRAM capacity of standard hardware. This scale makes existing neural architectures prohibitively slow to train and difficult to deploy for real-time inference. Beyond the computational load, modeling the 3D spatial relationships is inherently more complex than in 2D counterparts. 

\textbf{Dataset Generation: } We directly adopt the dataset from \citet{ohana2024well}. The dataset is generated by simulating isothermal, compressible magnetohydrodynamic (MHD) turbulence without self-gravity, representative of the diffuse interstellar medium, using a second-order-accurate finite-difference code that solves the ideal MHD equations on a triply-periodic uniform Cartesian grid. The flow is initialized from a uniform density field of unit mean and a uniform magnetic field oriented along a single axis, and is driven at large scales by a random, solenoidal (divergence-free) forcing that injects energy near wavenumber $k \approx 2.5$, stirring the gas until a statistically steady turbulent cascade develops. Because the equation of state is isothermal, each trajectory is fully characterized by two dimensionless control parameters: the sonic Mach number $\mathcal{M}_s = \langle |\mathbf{v}| / c_s \rangle$, which sets the compressibility of the flow, and the Alfv\'{e}nic Mach number $\mathcal{M}_A = \langle |\mathbf{v}| / v_A \rangle$ with Alfv\'{e}n speed $v_A = |\mathbf{B}| / \sqrt{\rho}$, which sets the dynamical importance of the magnetic field. The dataset sweeps all combinations of $\mathcal{M}_s \in \{0.5, 0.7, 1.5, 2.0, 7.0\}$ and $\mathcal{M}_A \in \{0.7, 2.0\}$, spanning subsonic to strongly supersonic and sub-Alfv\'{e}nic to super-Alfv\'{e}nic regimes, with ten random initial conditions per parameter pair for a total of $100$ trajectories. Each simulation is first computed at a resolution of $256^3$ and then downsampled to $64^3$ after anti-aliasing with an ideal low-pass filter; snapshots of the density, velocity, and magnetic field are written at fixed intervals $\Delta t = 0.01$ in dimensionless units over the range $t \in [0, 1]$, yielding $100$ temporal frames per trajectory.

\textbf{Governing Physics: } The dataset models the dynamics of an electrically conducting, compressible, isothermal fluid coupled to a magnetic field, governed by the equations of ideal magnetohydrodynamics on the triply-periodic domain $\Omega = [0, L_x] \times [0, L_y] \times [0, L_z]$,
\begin{equation}
    \frac{\partial \rho}{\partial t} + \nabla \cdot (\rho \mathbf{v}) = 0,
    \qquad
    \frac{\partial (\rho \mathbf{v})}{\partial t}
    + \nabla \cdot \!\big( \rho\, \mathbf{v} \otimes \mathbf{v} - \mathbf{B} \otimes \mathbf{B} \big)
    + \nabla p = \mathbf{0},
    \qquad
    \frac{\partial \mathbf{B}}{\partial t}
    - \nabla \times ( \mathbf{v} \times \mathbf{B} ) = \mathbf{0},
\end{equation}
where $\rho$ is the density, $\mathbf{v}$ the velocity, $\mathbf{B}$ the magnetic field (absorbing a factor of $1/\sqrt{4\pi}$ so that magnetic pressure appears directly), and $p$ the gas pressure. The three relations express, respectively, the conservation of mass through the continuity equation, the conservation of momentum with the combined gas and Maxwell stresses $-\mathbf{B}\otimes\mathbf{B}$ carrying the magnetic tension and pressure, and the ideal induction equation, which advects the magnetic field with the flow and enforces the frozen-in condition of a perfectly conducting medium. The magnetic field additionally satisfies the solenoidal constraint
\begin{equation}
    \nabla \cdot \mathbf{B} = 0,
\end{equation}
which is preserved for all time given divergence-free initial data. The system is closed by an isothermal equation of state
\begin{equation}
    p = c_s^2\, \rho,
\end{equation}
with constant sound speed $c_s$, so that no separate energy equation is required and the thermal pressure is set entirely by the local density. Under this closure the flow admits no intrinsic length or time scale, and its statistics are controlled solely by the sonic and Alfv\'{e}nic Mach numbers,
\begin{equation}
    \mathcal{M}_s = \frac{\langle |\mathbf{v}| \rangle}{c_s},
    \qquad
    \mathcal{M}_A = \frac{\langle |\mathbf{v}| \rangle}{v_A},
    \qquad
    v_A = \frac{|\mathbf{B}|}{\sqrt{\rho}},
\end{equation}
together with the plasma-$\beta$ parameter $\beta = 2\,(\mathcal{M}_A / \mathcal{M}_s)^2$ measuring the ratio of thermal to magnetic pressure. The interplay of compressibility and magnetization gives rise to a rich family of behaviors, from nearly incompressible, magnetically dominated turbulence at low $\mathcal{M}_s$ and low $\mathcal{M}_A$ to highly compressible, shock-dominated flows with sharp density contrasts at high $\mathcal{M}_s$.

\textbf{Experiment Results: } The complete results are presented in Table~\ref{tab:3d_full}. Because this is Eulerian grid data, we evaluate performance using RMSE, FRMSE, and Kinetic Energy. We also incorporate a dataset-specific metric, Magnetic Energy, to compute how well the model conserves the magnetic field properties over time. Experimental results demonstrate that the agent's design achieves substantially better performance. AMG is unsuitable for this application because its graph architecture scales poorly to large 3D datasets, resulting in prohibitive training times.

\begin{table}[h]
\centering
\begin{tabular}{lcccc}
\hline
 & RMSE & FRMSE & Kinetic Energy & Magnetic Energy \\
\hline
GNOT         & $2.230 \times 10^{-1}$ & $7.444 \times 10^{1}$ & $7.545 \times 10^{-2}$ & $6.777 \times 10^{-2}$ \\
Transolver   & $2.279 \times 10^{-1}$ & $7.532 \times 10^{1}$ & $7.572 \times 10^{-2}$ & $6.807 \times 10^{-2}$ \\
Transolver++ & $2.257 \times 10^{-1}$ & $7.470 \times 10^{1}$ & $7.592 \times 10^{-2}$ & $6.801 \times 10^{-2}$ \\
LNO          & $2.238 \times 10^{-1}$ & $7.469 \times 10^{1}$ & $7.611 \times 10^{-2}$ & $6.821 \times 10^{-2}$ \\
AMG          & -- & -- & -- & -- \\
\textbf{Agent} & $\mathbf{7.182 \times 10^{-2}}$ & $\mathbf{2.577 \times 10^{1}}$ & $\mathbf{3.313 \times 10^{-2}}$ & $\mathbf{2.317 \times 10^{-2}}$ \\
\hline
\end{tabular}
\caption{Performance comparison on the \ThreeD\ dataset. Best results are shown in \textbf{bold}.}
\label{tab:3d_full}
\end{table}

\begin{figure}[htbp]
    \centering
    \includegraphics[width=0.9\textwidth, page=1]{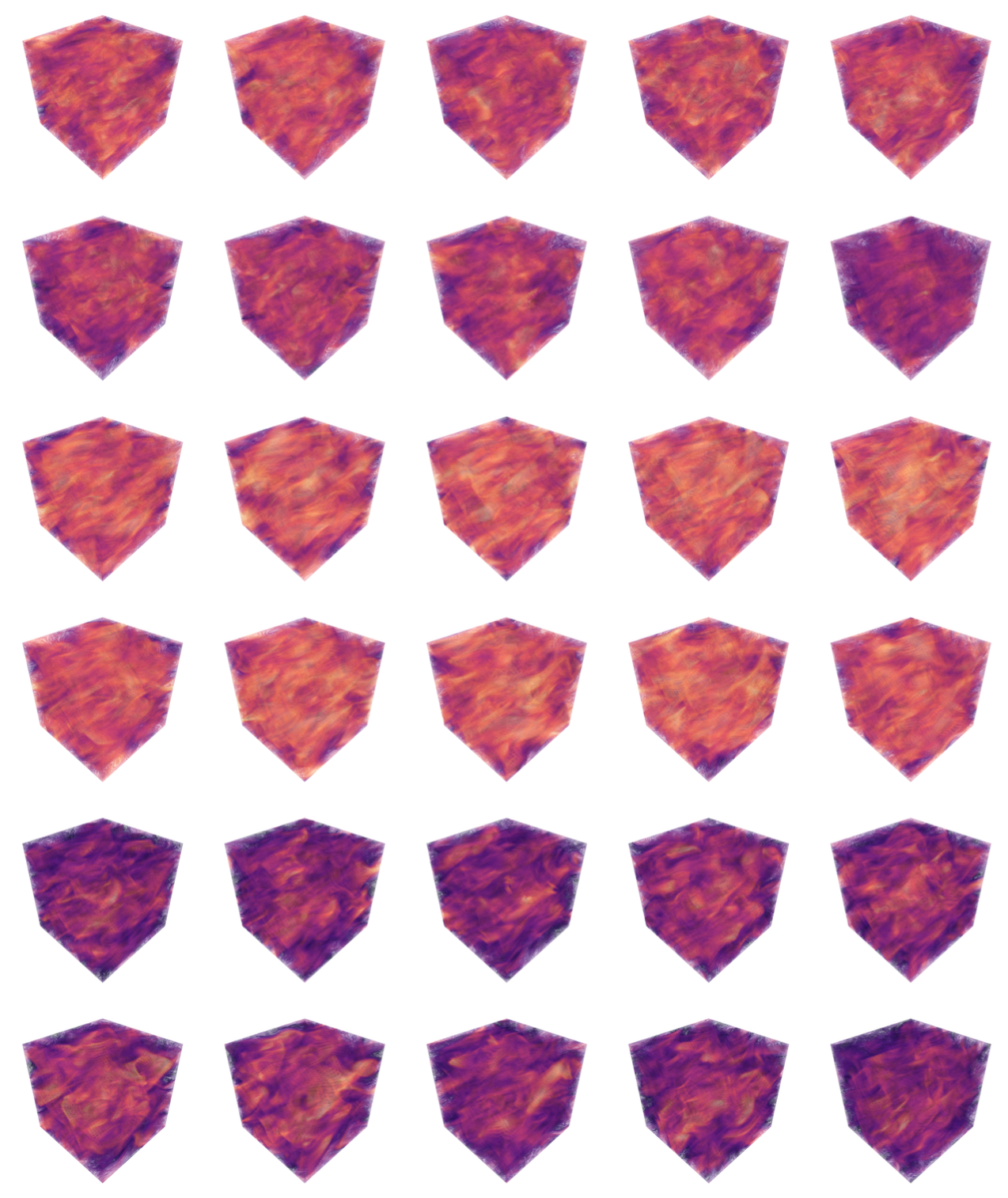}
    \caption{Visualization of the \ThreeD\ dataset. From top to bottom, the first two rows display density, the middle two the magnetic field norm, and the last two the velocity norm.}
    \label{fig:3d}
\end{figure}

\textbf{Visualization: } We refer the readers to Figure~\ref{fig:3d} for visualizations of the dataset.

\newpage

\subsection{\IrregularTime}
\label{appendix:irregulartime}
\textbf{Dataset Description: } This dataset captures two-dimensional incompressible Navier-Stokes dynamics sampled at irregular time intervals. The flow evolves on a $128 \times 128$ spatial grid under the interplay of nonlinear advection and viscous diffusion. Unlike standard benchmarks recorded on a uniform temporal grid, consecutive snapshots within each trajectory are separated by varying time gaps, reflecting practical settings where data acquisition is asynchronous or adaptively sampled.

\textbf{Modeling Objective: } The state of the fluid system is represented by the tensor $x_t \in \mathbb{R}^{1 \times 128 \times 128}$, where the single feature channel corresponds to the vorticity field evaluated across a $128 \times 128$ spatial discretization grid. The system is explicitly conditioned on a scalar time increment $dt \in \mathbb{R}^{1}$, specifying the temporal gap to the next snapshot. The objective is to develop a time-conditioned neural simulator, defined as a function $\Psi$, that maps the current system state and the elapsed time increment to the state after that interval. Formally, we aim to predict the next state $\hat{x}_{t+dt} = \Psi(x_t, dt)$, where $\hat{x}_{t+dt} \in \mathbb{R}^{1 \times 128 \times 128}$ matches the shape of the input state.

\textbf{Dataset Specific Challenges: } The dataset's primary challenge lies in the architectural design required to condition the neural model on a continuous temporal increment. Standard neural operators are trained to advance the system by a single fixed time step, implicitly baking the step size into their learned dynamics; here, the model must instead condition explicitly on $dt$ and generalize across a continuous range of temporal gaps. Furthermore, the effective difficulty of each prediction varies with $dt$: larger gaps require capturing more accumulated nonlinear evolution in a single forward pass, while smaller gaps demand near-identity behavior, and a single model must remain accurate and temporally stable across this entire spectrum.

\textbf{Dataset Generation:} The dataset is generated following the procedure outlined in Section~\ref{appendix:partial}, where recording times are randomly sampled.

\textbf{Governing Physics:} The physical setup follows the description in Section~\ref{appendix:partial}, governed by the 2D incompressible Navier-Stokes equations in vorticity form.

\textbf{Experiment Results: } Table~\ref{tab:irregulartime_full} presents the complete experimental results. We use standard metrics for Eulerian datasets to capture both pointwise error and physical consistency. The agent's design consistently achieves superior performance across all evaluated metrics.

\begin{table}[h]
\centering
\begin{tabular}{lccc}
\hline
 & RMSE & FRMSE & Kinetic Energy \\
\hline
GNOT         & $2.577 \times 10^{0}$ & $2.165 \times 10^{2}$ & $9.050 \times 10^{-1}$ \\
Transolver   & $2.623 \times 10^{0}$ & $2.092 \times 10^{2}$ & $1.376 \times 10^{0}$ \\
Transolver++ & $2.621 \times 10^{0}$ & $2.100 \times 10^{2}$ & $1.389 \times 10^{0}$ \\
LNO          & $2.622 \times 10^{0}$ & $2.124 \times 10^{2}$ & $1.431 \times 10^{0}$ \\
AMG          & $2.415 \times 10^{0}$ & $2.084 \times 10^{2}$ & $7.990 \times 10^{-1}$ \\
\textbf{Agent} & $\mathbf{6.511 \times 10^{-1}}$ & $\mathbf{6.022 \times 10^{1}}$ & $\mathbf{9.200 \times 10^{-2}}$ \\
\hline
\end{tabular}
\caption{Performance comparison on the \IrregularTime\ dataset. Best results are shown in \textbf{bold}.}
\label{tab:irregulartime_full}
\end{table}

\begin{figure}[htbp]
    \centering
    \includegraphics[width=0.9\textwidth, page=1]{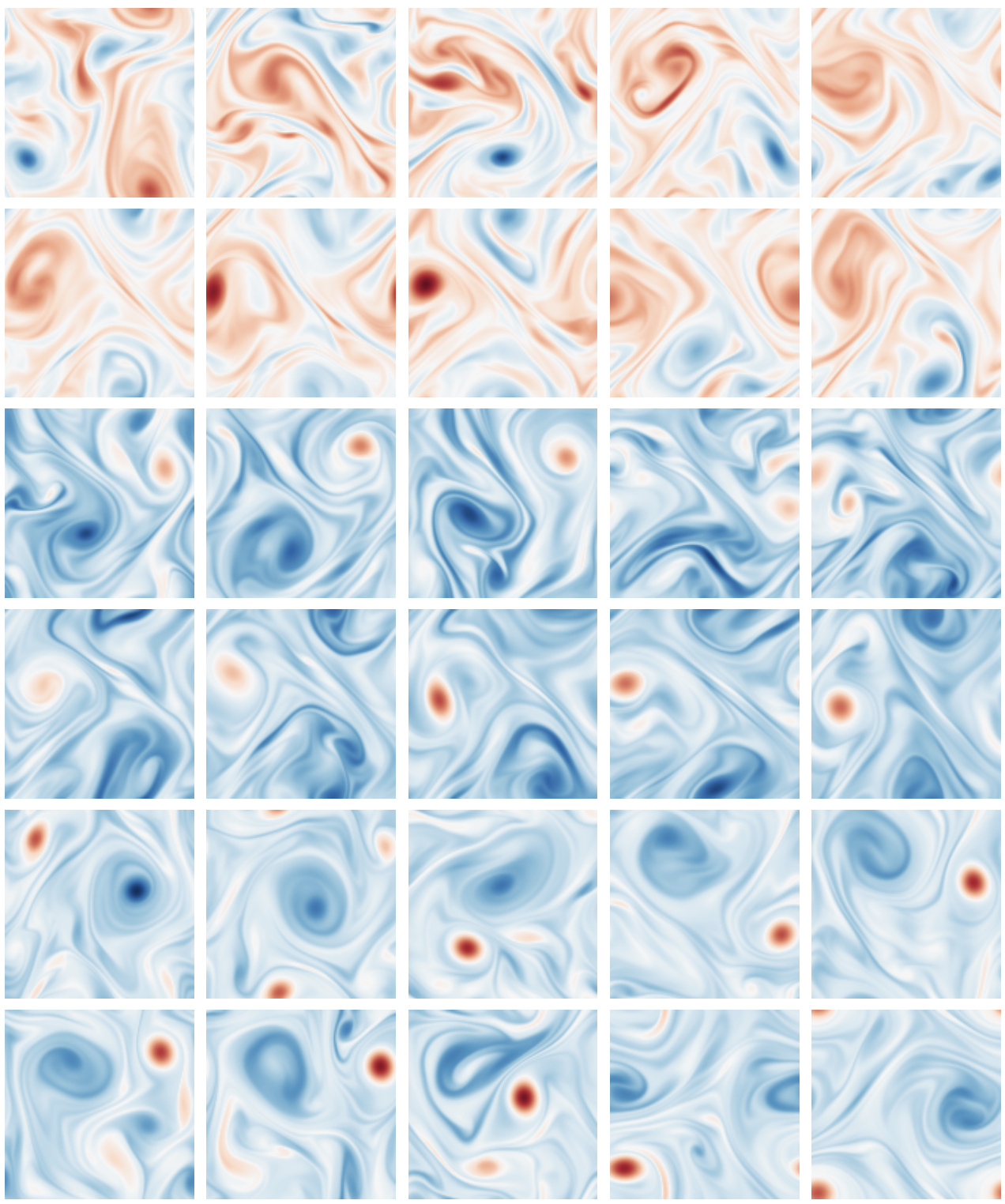}
    \caption{Visualization of the \IrregularTime\ dataset. The dataset contains solutions to the Navier-Stokes equations in vorticity form sampled at irregular time intervals.}
    \label{fig:irregulartime}
\end{figure}

\textbf{Visualization: } We refer the readers to Figure~\ref{fig:irregulartime} for visualizations of the dataset.

\newpage
\clearpage

\subsection{\LagDynamic}
\label{appendix:lagdynamic}
\textbf{Dataset Description: } This dataset contains Lagrangian particle simulations of jet engine flow in an open domain with an inlet and an outlet. The fluid is represented by a time-evolving set of particles carrying position and velocity, with new particles continuously injected at the inlet and existing particles removed once they exit through the outlet. Unlike standard Lagrangian benchmarks, where a fixed set of particles persists for the entire trajectory and the model can be supervised through a one-to-one correspondence between input and output particles, the particle set here changes identity and cardinality between consecutive time steps, so no such correspondence exists.

\textbf{Modeling Objective: } The state of the fluid system is represented by the tensor $x_t \in \mathbb{R}^{N \times 4}$, where each of the $N$ particles carries its two-dimensional position and velocity. The model is additionally conditioned on a set of query positions $Q_{t+1} \in \mathbb{R}^{M \times 2}$ at the subsequent time step. The objective is to develop a neural simulator, defined as a function $\Psi$, that encodes the input particle set into a latent representation of the flow field and decodes the fluid velocity at each query location. Formally, we aim to predict $\hat{v}_{t+1} = \Psi(x_t, Q_{t+1})$, where $\hat{v}_{t+1} \in \mathbb{R}^{M \times 2}$ is the predicted velocity at the query positions, rather than the advected state of a fixed set of tracked particles. A validity mask over the input particles supports batching across samples with differing particle counts by excluding padded entries from the encoding.

\textbf{Dataset Specific Challenges: } The dataset's primary challenge lies in the absence of particle correspondence, which removes the standard supervision signal of per-particle displacement targets. The model cannot learn a map from each input particle to its own future state, and must instead treat the particles as an unstructured, permutation-invariant sampling of an underlying velocity field. This field is queried at positions that are decoupled from the input set, requiring the model to represent the continuous flow rather than track discrete entities across time steps.

\textbf{Dataset Generation: } The dataset is generated by simulating a planar high-speed liquid jet impinging on a flat wall with weakly-compressible Smoothed Particle Hydrodynamics (SPH)~\citep{ZHAN2025109389, DOMINGUEZ2022867}, in which the fluid is discretized into Lagrangian particles advected through a fixed inlet/outlet buffer configuration rather than solved on a grid. The two-dimensional domain is centered on the origin: a thin fluid strip of width $30\,\mathrm{mm}$ and thickness $5\,\mathrm{mm}$ is placed at height $z = 0.12\,\mathrm{m}$ and driven downward through a top inlet zone that continuously injects water at a fixed velocity $v_{\mathrm{in}} = 20\,\mathrm{m/s}$, while the initial fluid block is likewise assigned a downward velocity $-v_{\mathrm{in}}$. The jet falls under gravity $g_z = -9.81\,\mathrm{m/s^2}$ and impinges on a solid bottom wall built from four boundary layers spanning the base of the domain, spreading laterally and leaving through two open outlets on the left and right sides, each realized as a buffer zone with velocity and density extrapolated from the interior ghost nodes so that fluid exits without spurious reflection. The particles are laid out at an initial inter-particle spacing $\Delta p = 2\,\mathrm{mm}$, with smoothing length $h = \Delta p\sqrt{3}$ and a reference density $\rho_0 = 1000\,\mathrm{kg/m^3}$. Incompressibility is approximated through a Tait equation of state $p = \tfrac{c_0^2\rho_0}{\gamma}\big[(\rho/\rho_0)^\gamma - 1\big]$ with polytropic exponent $\gamma = 7$ and an artificial sound speed $c_0 = 230\,\mathrm{m/s}$, chosen large enough relative to the jet velocity that density fluctuations remain small. Solid boundaries use the modified Dynamic Boundary Conditions (mDBC), in which boundary-particle densities are reconstructed by extrapolation from ghost nodes placed in the fluid to reduce the unphysical fluid--wall gap of the standard scheme; viscous stresses are modeled with a laminar viscosity $\nu = 10^{-6}\,\mathrm{m^2/s}$ combined with a Sub-Particle-Scale (SPS) turbulence closure. Time integration uses a Symplectic scheme with a Wendland kernel and an adaptive step set by a CFL condition. The simulation integrates to a final time $t_f = 12\,\mathrm{s}$, writing particle snapshots every $\Delta t_{\mathrm{out}} = 0.001\,\mathrm{s}$.

\textbf{Governing Physics: } The dataset models the flow of a single-phase, weakly-compressible, viscous liquid injected as a jet and impinging on a solid wall, governed by the Lagrangian form of the continuity and Navier--Stokes equations on the two-dimensional domain $\Omega$ under the body force $\mathbf{g} = (0, g_z)$,
\begin{equation}
    \frac{D\rho}{Dt} = -\rho\,\nabla\cdot\mathbf{u},
    \qquad
    \frac{D\mathbf{u}}{Dt}
    = -\frac{1}{\rho}\nabla p + \nu\,\nabla^2\mathbf{u} + \mathbf{g},
\end{equation}
where $\tfrac{D}{Dt} = \partial_t + \mathbf{u}\cdot\nabla$ is the material derivative following a fluid particle, $\mathbf{u}$ the velocity, $p$ the pressure, $\rho$ the density, and $\nu$ the kinematic viscosity. Rather than enforcing strict incompressibility, the weakly-compressible formulation closes the system with the barotropic Tait equation of state
\begin{equation}
    p = \frac{c_0^2\,\rho_0}{\gamma}\left[\left(\frac{\rho}{\rho_0}\right)^{\gamma} - 1\right],
    \qquad \gamma = 7,
\end{equation}
which relates pressure to density through the reference density $\rho_0$ and an artificial sound speed $c_0$. The sound speed is set well above the maximum expected flow speed so that the Mach number $\mathrm{Ma} = U/c_0$ remains small and the resulting density variations, of order $\mathrm{Ma}^2$, stay within a few percent, recovering the incompressible limit while permitting an explicit time-stepping scheme. The jet enters through a fixed-velocity inlet condition $\mathbf{u} = (0, -v_{\mathrm{in}})$ and leaves through open outlets where the velocity and density are extrapolated from the interior, while a no-penetration condition is enforced at the solid bottom wall through the modified Dynamic Boundary Conditions. The dynamics are characterized by the Reynolds number and the Froude number,
\begin{equation}
    \mathrm{Re} = \frac{v_{\mathrm{in}}\,d}{\nu},
    \qquad
    \mathrm{Fr} = \frac{v_{\mathrm{in}}}{\sqrt{\lvert g_z\rvert\, d}},
    \qquad
    \mathrm{Ma} = \frac{v_{\mathrm{in}}}{c_0},
\end{equation}
where $d$ denotes the jet thickness. With a high injection velocity and a thin jet, the flow is strongly inertia-dominated, so the impingement produces a thin spreading wall layer, lateral jetting toward the outlets, and unsteady breakup at the free surface where the fluid separates from the wall.

\textbf{Experiment Results: } Table~\ref{tab:lagdynamic_full} presents the complete results. In addition to pointwise RMSE, we introduce LagKE to evaluate physical consistency, alongside EuARMSE and EuAFRMSE to assess practical utility. We observe that while the agent achieves strong pointwise accuracy, it struggles to maintain physical consistency. Furthermore, standard transformer and graph-based models are inapplicable here because the dataset features differing numbers of input and output particles. Instead, LNO condenses the initial conditions into a fixed-size latent representation and applies the processor at this level, while the decoder allows for position-based querying.

\begin{table}[h]
\centering
\begin{tabular}{lcccc}
\hline
 & RMSE & LagKE & EuARMSE & EuAFRMSE \\
\hline
GNOT         & -- & -- & -- & -- \\
Transolver   & -- & -- & -- & -- \\
Transolver++ & -- & -- & -- & -- \\
LNO          & $6.105 \times 10^{-1}$ & $\mathbf{1.349 \times 10^{3}}$ & $1.129 \times 10^{-1}$ & $2.621 \times 10^{0}$ \\
AMG          & -- & -- & -- & -- \\
\textbf{Agent} & $\mathbf{5.530 \times 10^{-1}}$ & $1.662 \times 10^{3}$ & $\mathbf{9.951 \times 10^{-2}}$ & $\mathbf{2.299 \times 10^{0}}$ \\
\hline
\end{tabular}
\caption{Performance comparison on the \LagDynamic\ dataset. Best results are shown in \textbf{bold}.}
\label{tab:lagdynamic_full}
\end{table}

\begin{figure}[htbp]
    \centering
    \includegraphics[width=0.9\textwidth, page=1]{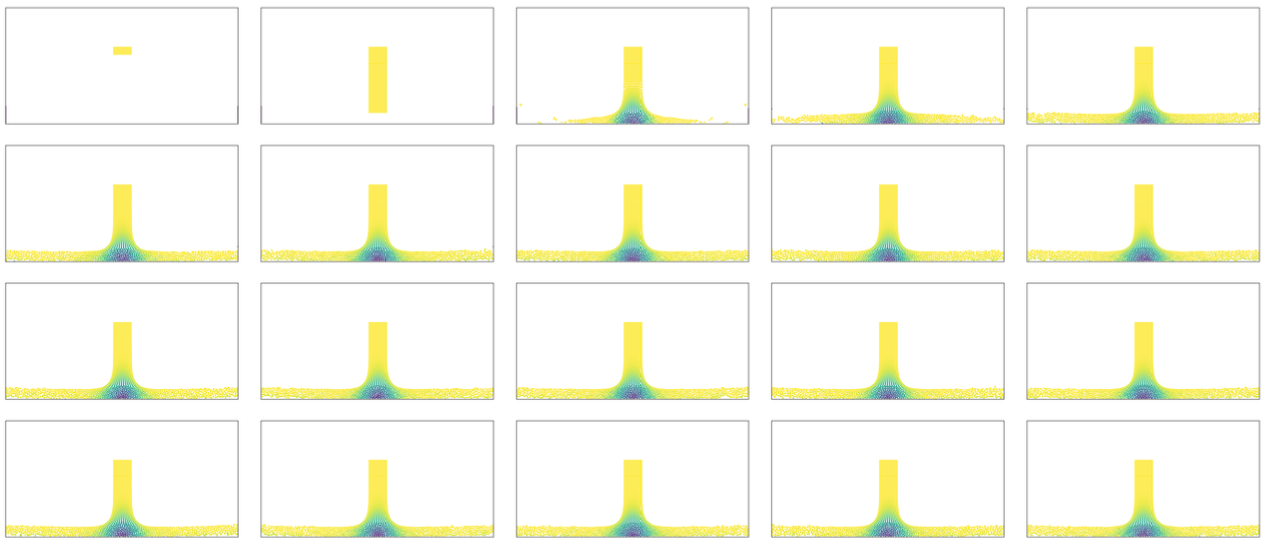}
    \caption{Visualization of the \LagDynamic\ dataset. The dataset uses Lagrangian particles to simulate high-speed exhaust from a jet engine. The airflow enters the computational domain from the top boundary, impinges on the bottom wall, and subsequently exits through the left and right boundaries.}
    \label{fig:lagdynamic}
\end{figure}

\textbf{Visualization: } We refer the readers to Figure~\ref{fig:lagdynamic} for visualizations of the dataset.

\newpage

\subsection{\LagFSI}
\label{appendix:lagfsi}
\textbf{Dataset Description: } This dataset contains two-dimensional Lagrangian particle simulations of fluid-structure interaction, in which a rigid solid sphere falls into a body of water. Both phases are discretized as particles: fluid particles follow the free-surface flow, while solid particles collectively represent the sphere and must move as a coherent rigid body. Unlike single-phase Lagrangian benchmarks, the dynamics here are governed by two-way coupling: the falling solid displaces and accelerates the surrounding fluid, while the fluid exerts buoyancy, drag, and impact forces back on the solid.

\textbf{Modeling Objective: } The state of the system is represented by the tensor $x_t \in \mathbb{R}^{N \times 13}$, where each particle's feature vector concatenates its current position in $\mathbb{R}^{2}$, a history of its $5$ most recent velocities in $\mathbb{R}^{10}$ obtained by finite-differencing consecutive positions over the preceding frames, and a particle type in $\{0, 1\}$ indicating solid or fluid. The model additionally receives a batch index vector $\bm{b} \in \{0, \dots, B-1\}^{N}$ assigning each particle to its sample, since samples with differing particle counts are concatenated by a graph data loader rather than stacked along a batch dimension, and a particle type vector $p \in \{0, 1\}^{N \times 1}$, provided as a standalone input in addition to its copy inside the feature vector. The objective is to develop a neural simulator, defined as a function $\Psi$, such that $\hat{v}_{t+1} = \Psi(x_t, \bm{b}, p)$, where $\hat{v}_{t+1} \in \mathbb{R}^{N \times 2}$ is the predicted velocity over the next time step, from which particle positions are advanced during rollout.

\textbf{Dataset Specific Challenges: } The dataset presents two intertwined challenges. First, the two phases obey qualitatively different dynamics under two-way coupling: fluid particles respond to local pressure and viscous interactions with their neighbors, while solid particles are constrained by global rigid-body kinematics yet driven by contact forces from the surrounding fluid. A single model must capture both behaviors and the momentum exchange between them, including the impulsive interaction at impact, using only the particle type label to differentiate the phases. Second, the solid has no explicitly enforced rigidity: its shape is maintained only through the predicted per-particle velocities. Small inconsistencies among solid-particle predictions accumulate over autoregressive rollout into stretching, shearing, or fragmentation of the sphere, so the model must keep the solid particles' motion mutually consistent, effectively learning the rigidity constraint, while remaining accurate for the surrounding free-surface flow.

\textbf{Dataset Generation: } The dataset is generated by simulating a rigid sphere falling into a hydrostatic water tank with weakly-compressible Smoothed Particle Hydrodynamics (SPH), in which both the fluid and the solid are discretized into Lagrangian particles rather than solved on a fixed grid. The two-dimensional domain is a rectangular tank of width $L_x = 100\,\mathrm{mm}$, filled to a variable still-water height $H$ and enclosed by solid boundary walls two particle-layers thick along the bottom and sides. A circular rigid body of radius $R$ and density $\rho_s = 500\,\mathrm{kg/m^3}$ is initialized at a prescribed center $(x_0, y_0)$ above or within the water and released from rest, falling under gravity $g_y = -9.81\,\mathrm{m/s^2}$. A defining feature of this dataset is that the sphere radius, its initial position, and the water height vary across trajectories, which induces a broad spectrum of impact and floatation regimes: spheres released higher above the surface strike with greater momentum and generate more violent splashing and cavity formation, whereas those initialized near or below the surface settle more gently toward their buoyant equilibrium. The fluid particles are laid out on a uniform lattice at spacing $\Delta x = 2\,\mathrm{mm}$ with smoothing length $h = 1.2\,\Delta x$ and are initialized at hydrostatic density through the Tait relation, while the sphere is discretized at a finer spacing to resolve its curved interface; the per-particle mass of each phase is set to reproduce its reference density. The fluid dynamics are advanced with a weakly-compressible SPH scheme, in which incompressibility is approximated through an artificial equation of state with polytropic exponent $\gamma = 7$ and an artificial sound speed $c_0 = 2\sqrt{2\lvert g_y\rvert H}$ chosen from the maximum expected flow speed, and momentum transport uses an artificial viscosity with coefficient $\alpha = 0.1$. The rigid body is integrated as a single solid whose particles move together under rigid-body motion, driven by the net force and torque accumulated from gravity and from the surrounding fluid; contact with the tank walls is handled by a penalty collision force with stiffness $k_n = 10^5$. Time integration uses a predictor--corrector (EPEC) scheme for the fluid and a Runge--Kutta rigid-body step for the solid, with a fixed step set by a CFL-type acoustic limit.

\textbf{Governing Physics: } The dataset models a two-way-coupled fluid--structure interaction problem, in which a weakly-compressible viscous fluid and a rigid solid body exchange momentum across their shared interface. The fluid is governed by the Lagrangian continuity and momentum equations,
\begin{equation}
    \frac{D\rho}{Dt} = -\rho\,\nabla\cdot\mathbf{u},
    \qquad
    \frac{D\mathbf{u}}{Dt}
    = -\frac{1}{\rho}\nabla p + \nu\,\nabla^2\mathbf{u} + \mathbf{g}
    + \mathbf{f}_{s \rightarrow f},
\end{equation}
where $\tfrac{D}{Dt} = \partial_t + \mathbf{u}\cdot\nabla$ is the material derivative, $\mathbf{u}$ the fluid velocity, $p$ the pressure, $\rho$ the density, $\nu$ the kinematic viscosity, and $\mathbf{g} = (0, g_y)$ gravity. The system is closed by the barotropic Tait equation of state $p = \tfrac{c_0^2\rho_0}{\gamma}\big[(\rho/\rho_0)^\gamma - 1\big]$ with $\gamma = 7$, so that the artificial sound speed $c_0$ keeps density fluctuations small and recovers the incompressible limit. The solid is governed by the rigid-body equations of motion, in which its particles translate and rotate as a single body according to the Newton--Euler laws,
\begin{equation}
    M\,\frac{d\mathbf{U}}{dt}
    = M\mathbf{g} + \mathbf{F}_{f \rightarrow s},
    \qquad
    \mathbf{I}\,\frac{d\boldsymbol{\omega}}{dt}
    = \mathbf{T}_{f \rightarrow s},
    \qquad
    \mathbf{u}_i = \mathbf{U} + \boldsymbol{\omega}\times(\mathbf{r}_i - \mathbf{r}_c),
\end{equation}
where $M$ is the body mass, $\mathbf{I}$ its moment of inertia, $\mathbf{U}$ and $\boldsymbol{\omega}$ the translational and angular velocities, $\mathbf{r}_c$ the center of mass, and $\mathbf{u}_i$ the velocity of constituent particle $i$ at position $\mathbf{r}_i$. The two systems are coupled through the reciprocal interfacial forces at the boundary between the phases: the fluid exerts a hydrodynamic load, comprising buoyancy, drag, and impact pressure, on the solid,
\begin{equation}
    \mathbf{F}_{f \rightarrow s} = \sum_{i \in s} \mathbf{f}_{f \rightarrow s}^{\,i},
    \qquad
    \mathbf{T}_{f \rightarrow s} = \sum_{i \in s} (\mathbf{r}_i - \mathbf{r}_c)\times \mathbf{f}_{f \rightarrow s}^{\,i},
\end{equation}
while by Newton's third law the solid exerts an equal and opposite reaction $\mathbf{f}_{s \rightarrow f} = -\mathbf{f}_{f \rightarrow s}$ back on the fluid, displacing and accelerating it as the body moves. It is this bidirectional exchange --- the solid pushing the fluid aside while the fluid buoys, retards, and deflects the solid --- that produces the characteristic dynamics of the dataset: the impact and cavity formation as the sphere enters the water, the deceleration and rebound as it interacts with the free surface, and its eventual rise toward the buoyant equilibrium set by the density ratio $\rho_s/\rho_0 = 0.5$.

\textbf{Experiment Results: } We present the complete experimental results in Table~\ref{tab:lagfsi_full}, where we not only dissect the error into fluid and solid components for finer analysis but also report metrics for physical consistency and practical utility. The RMSE reported in the main table reflects the average across both the fluid and solid domains. We observe that the agent performs poorly across all evaluated metrics.

\begin{table}[h]
\centering
\begin{tabular}{lcccc}
\hline
 & RMSE (fluid) & LagKE (fluid) & Sinkhorn (fluid) & RMSE (solid) \\
\hline
GNOT         & $\bm{9.505 \times 10^{-5}}$ & $1.043 \times 10^{-4}$ & $\bm{3.336 \times 10^{-10}}$ & $6.235 \times 10^{-5}$ \\
Transolver   & $9.731 \times 10^{-5}$ & $\bm{7.354 \times 10^{-5}}$ & $3.551 \times 10^{-10}$ & $5.417 \times 10^{-5}$ \\
Transolver++ & $1.001 \times 10^{-4}$ & $8.819 \times 10^{-5}$ & $3.617 \times 10^{-10}$ & $7.156 \times 10^{-5}$ \\
LNO          & $1.008 \times 10^{-4}$ & $8.440 \times 10^{-5}$ & $6.053 \times 10^{-10}$ & $6.830 \times 10^{-5}$ \\
AMG          & $9.554 \times 10^{-5}$ & $1.112 \times 10^{-4}$ & $4.462 \times 10^{-10}$ & $7.473 \times 10^{-5}$ \\
\textbf{Agent} & $4.024 \times 10^{-4}$ & $1.315 \times 10^{-3}$ & $1.269 \times 10^{-8}$ & $\bm{4.071 \times 10^{-5}}$ \\
\hline
\multicolumn{5}{c}{} \\
\hline
 & LagKE (solid) & Sinkhorn (solid) & EuARMSE (fluid) & EuAFRMSE (fluid) \\
\hline
GNOT         & $1.202 \times 10^{-4}$ & $3.890 \times 10^{-9}$ & $\bm{1.029 \times 10^{-4}}$ & $\bm{5.782 \times 10^{-3}}$ \\
Transolver   & $4.306 \times 10^{-5}$ & $3.045 \times 10^{-9}$ & $1.189 \times 10^{-4}$ & $6.571 \times 10^{-3}$ \\
Transolver++ & $9.993 \times 10^{-5}$ & $5.892 \times 10^{-9}$ & $1.156 \times 10^{-4}$ & $6.488 \times 10^{-3}$ \\
LNO          & $6.977 \times 10^{-5}$ & $4.417 \times 10^{-9}$ & $1.270 \times 10^{-4}$ & $7.133 \times 10^{-3}$ \\
AMG          & $1.034 \times 10^{-4}$ & $5.762 \times 10^{-9}$ & $1.051 \times 10^{-4}$ & $6.004 \times 10^{-3}$ \\
\textbf{Agent} & $\bm{4.185 \times 10^{-5}}$ & $\bm{2.208 \times 10^{-9}}$ & $5.658 \times 10^{-4}$ & $4.089 \times 10^{-2}$ \\
\hline
\end{tabular}
\caption{Performance comparison on the \LagFSI\ dataset. Best results are shown in \textbf{bold}.}
\label{tab:lagfsi_full}
\end{table}

\begin{figure}[htbp]
    \centering
    \includegraphics[width=0.9\textwidth, page=1]{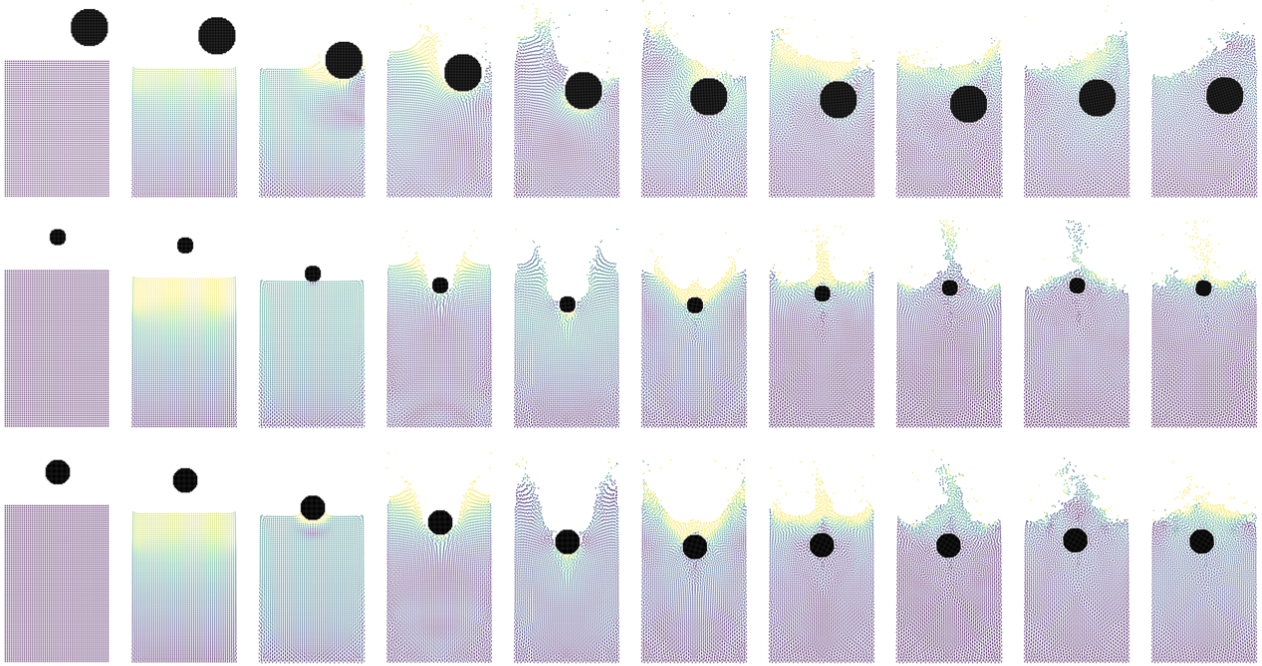}
    \caption{Visualization of the \LagFSI\ dataset. The dataset simulates a solid sphere falling into a volume of water, with both phases discretized using Lagrangian particles.}
    \label{fig:lagfsi}
\end{figure}

\textbf{Visualization: } We refer the readers to Figure~\ref{fig:lagfsi} for visualizations of the dataset.

\newpage
\clearpage

\subsection{\Extreme}
\label{appendix:extreme}
\textbf{Dataset Description: } This dataset contains three-dimensional supernova explosion simulations in a monatomic ideal gas that mocks the interstellar medium of the Milky Way Galaxy. At the beginning of each simulation, the thermal energy of a supernova is deposited at the center of the simulation box; the resulting hot ($\sim 10^{7}$~K) gas is immediately accelerated and drives a blastwave into the surrounding medium. Because the hot gas becomes supersonic, the physical quantities span roughly seven orders of magnitude.

\textbf{Modeling Objective: } Let $x_{t} \in \mathbb{R}^{6 \times 64 \times 64 \times 64}$ denote the state of the system at time $t$ on a $64 \times 64 \times 64$ spatial discretization grid. The six feature channels comprise pressure (a scalar field, 1 channel), density (a scalar field, 1 channel), temperature (a scalar field, 1 channel), and velocity (a vector field, 3 channels). Our objective is to develop a neural simulator, defined as a parameterized function $\Psi$, that maps the current system state to the subsequent state at the next timestep. Formally, we aim to predict $\hat{x}_{t+1} = \Psi(x_{t})$, where the predicted output $\hat{x}_{t+1}$ retains the same spatial and feature dimensions $(6 \times 64 \times 64 \times 64)$ as the input state.

\textbf{Dataset Specific Challenges: } The data is fully three-dimensional: each sample contains over 1.5 million values across six channels, and the blastwave propagates and interacts along all spatial axes, making both the memory footprint and the modeling of 3D couplings difficult for standard neural PDE solvers. Moreover, the physical quantities are extremely heavy-tailed and spatially localized, with values that are moderate over most of the domain but explode near the explosion center, resembling a sharply peaked Gaussian profile. For instance, density ranges from $4.7 \times 10^{-14}$ to $6.7 \times 10^{2}$ (mean $2.9 \times 10^{-1}$), pressure from $9.1 \times 10^{-17}$ to $3.2 \times 10^{5}$ (mean $3.4 \times 10^{-2}$), and temperature from $1.0 \times 10^{1}$ to $8.2 \times 10^{8}$ (mean $1.1 \times 10^{5}$), while velocity components span roughly $-2.7 \times 10^{2}$ to $2.0 \times 10^{2}$. Such multi-order-of-magnitude dynamic ranges render naive normalization ineffective and destabilize both training and rollout predictions.

\textbf{Dataset Generation: } We adopt the supernova explosion dataset from \citet{ohana2024well}. The dataset is generated with the $N$-body/Smoothed Particle Hydrodynamics (SPH). Because resolving the propagation of the supersonic blastwave demands tiny timesteps and a very large number of integration steps, the Lagrangian SPH particle data are subsequently interpolated onto a uniform Eulerian grid and temporally coarsened for storage. In this work we use the $64^{3}$ version.

\textbf{Governing Physics: } The supernova explosion is governed by the equations of self-gravitating, radiatively-cooled compressible hydrodynamics for a monatomic ideal gas, written in the Lagrangian form solved by the SPH code. The gas obeys the ideal equation of state with specific heat ratio $\gamma = 5/3$,
\begin{equation}
    P = (\gamma - 1)\,\rho\, u,
\end{equation}
where $P$ is the pressure, $\rho$ the smoothed density, and $u$ the specific internal energy. Writing $\tfrac{d}{dt}$ for the Lagrangian (material) derivative following a fluid element, the conservation of mass, momentum, and energy read
\begin{equation}
    \frac{d\rho}{dt} = -\rho\, \nabla \cdot \mathbf{v},
\end{equation}
\begin{equation}
    \frac{d^{2}\mathbf{r}}{dt^{2}}
    = -\frac{\nabla P}{\rho}
    + \mathbf{a}_{\mathrm{visc}}
    - \nabla \Phi,
\end{equation}
\begin{equation}
    \frac{du}{dt}
    = -\frac{P}{\rho}\, \nabla \cdot \mathbf{v}
    + \frac{\Gamma - \Lambda}{\rho},
\end{equation}
where $\mathbf{r}$ is the position of a fluid element and $\mathbf{v} = d\mathbf{r}/dt$ its velocity. In the momentum equation, $-\nabla P / \rho$ is the pressure-gradient acceleration, $\mathbf{a}_{\mathrm{visc}}$ is the acceleration generated by the artificial viscosity used to capture shocks, and $-\nabla \Phi$ is the acceleration due to self-gravity, with $\Phi$ the gravitational potential sourced by the gas distribution. In the energy equation, the first term is the reversible $PdV$ work done under compression or expansion, while $(\Gamma - \Lambda)/\rho$ is the net radiative source term: $\Gamma$ is the radiative heat influx per unit volume and $\Lambda$ the radiative heat outflux (cooling) per unit volume. It is this radiative sink $\Lambda$ that rapidly cools the shocked, compressed shells behind the blastwave, and the stiff, highly localized nature of $\Gamma - \Lambda$ together with the supersonic velocities is what forces the very small timesteps required to integrate the system.

\textbf{Experiment Results: } Table~\ref{tab:extreme_full} details the full experimental results, highlighting the superior performance of the proposed design. We exclude GNOT from this comparison because its training diverges due to the extreme values present in the dataset. Furthermore, AMG is omitted due to computational constraints. Its graph-based architecture scales poorly to large 3D datasets, leading to infeasible training times.

\begin{table}[htbp]
\centering
\begin{tabular}{lccc}
\hline
 & RMSE & FRMSE & Kinetic Energy \\
\hline
GNOT & - & - & - \\
Transolver & $2.289 \times 10^{5}$ & $1.090 \times 10^{8}$ & $1.736 \times 10^{1}$ \\
Transolver++ & $2.190 \times 10^{5}$ & $1.044 \times 10^{8}$ & $1.053 \times 10^{2}$ \\
LNO & $4.813 \times 10^{5}$ & $2.315 \times 10^{8}$ & $2.293 \times 10^{5}$ \\
AMG & - & - & - \\
\textbf{Agent} & $\mathbf{1.340 \times 10^{5}}$ & $\mathbf{5.065 \times 10^{7}}$ & $\mathbf{5.996 \times 10^{-1}}$ \\
\hline
\end{tabular}
\caption{Performance comparison on the \Extreme\ dataset. Best results are shown in \textbf{bold}.}
\label{tab:extreme_full}
\end{table}

\begin{figure}[htbp]
    \centering
    \includegraphics[width=0.9\textwidth, page=1]{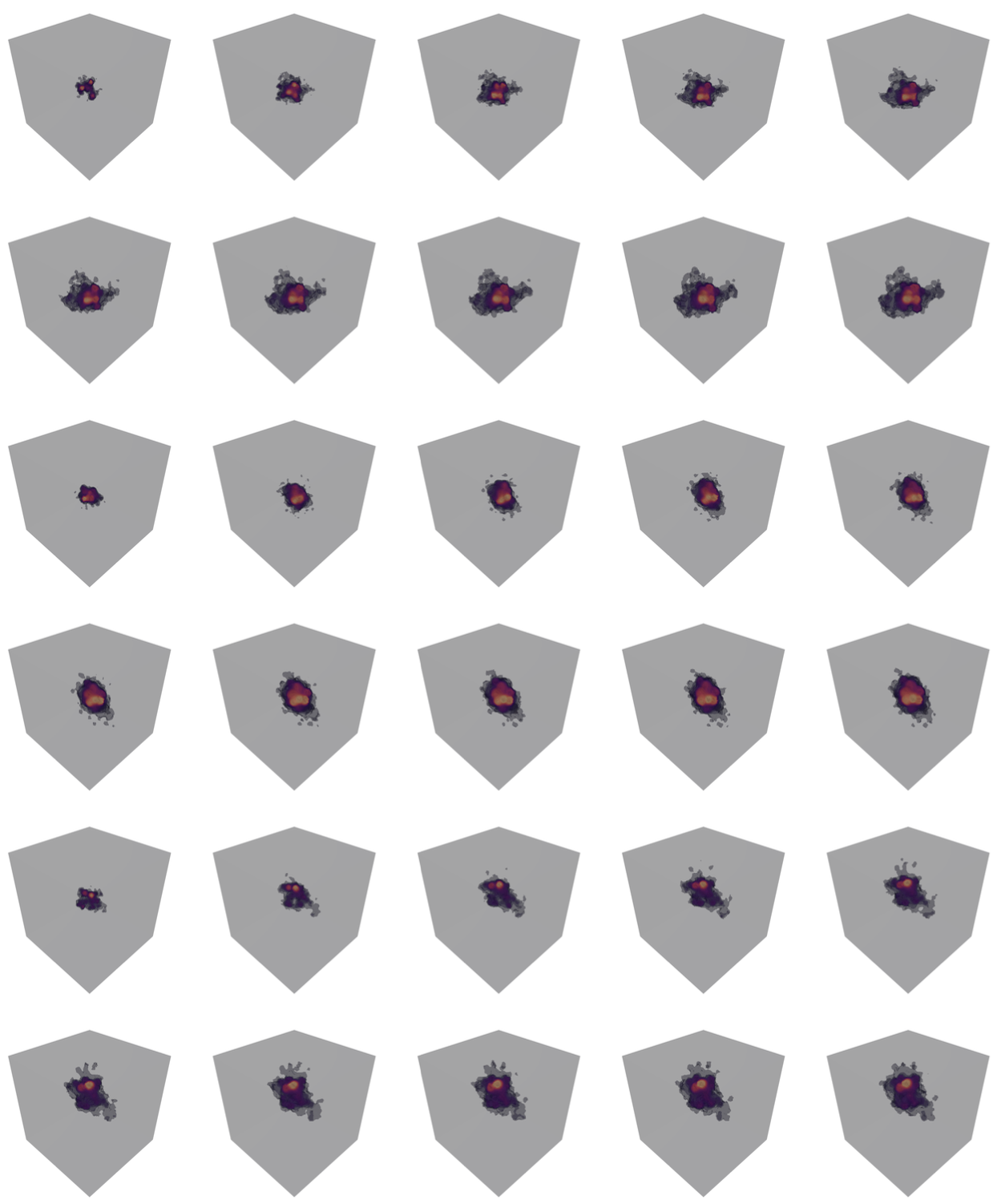}
    \caption{Visualization of the \Extreme\ dataset. It displays a supernova explosion.}
    \label{fig:extreme}
\end{figure}

\textbf{Visualization: } We refer the readers to Figure~\ref{fig:extreme} for visualizations of the dataset.

\newpage

\subsection{\MixedRes}
\label{appendix:mixedres}
\textbf{Dataset Description: } This dataset contains two-dimensional incompressible Navier--Stokes dynamics in vorticity form on periodic rectangular domains of varying size and resolution. Each trajectory is simulated at Reynolds number $500$ under a fixed sinusoidal forcing, with the domain side lengths drawn independently from $\{\pi, 2\pi, 3\pi, 4\pi\}$ and the output resolution along each axis drawn independently from $\{32, 64, 128, 256, 512\}$. All trajectories are computed on a high-resolution $1024 \times 1024$ grid and subsampled.

\textbf{Modeling Objective: } Let $x_{t} \in \mathbb{R}^{1 \times H \times W}$ denote the vorticity field at time $t$, where the spatial dimensions $H$ and $W$ vary across trajectories. The model additionally receives a positional field $p \in \mathbb{R}^{2 \times H \times W}$ containing the physical coordinates of each grid point. Our objective is to develop a neural simulator, defined as a parameterized function $\Psi$, that maps the current vorticity field and its positional coordinates to the subsequent state at the next timestep. Formally, we aim to predict $\hat{x}_{t+1} = \Psi(x_{t}, p)$, where the predicted output $\hat{x}_{t+1} \in \mathbb{R}^{1 \times H \times W}$ retains the same spatial and feature dimensions as the input state.

\textbf{Dataset Specific Challenges: } The same discrete array can represent physically different fields depending on the domain: a $128 \times 128$ grid may span a domain of length $\pi$ or $4\pi$, changing the physical wavelengths, effective grid spacing, and the amount of advection per time step. Consequently, the model cannot treat pixel indices as physical positions and must instead infer the metric of the grid from the positional input, effectively encoding $dx$ and $dy$ into its learned dynamics. Moreover, the number of grid points per sample spans a $256\times$ range, from $32 \times 32$ ($1{,}024$ points) to $512 \times 512$ ($262{,}144$ points), so the computational and memory footprint of a forward pass varies drastically across samples. The model and training pipeline must therefore remain efficient at both extremes: small samples should not be dominated by fixed overhead, while the largest samples must fit within memory.

\textbf{Dataset Generation: } The dataset generation follows Section~\ref{appendix:partial}, where we solve the 2D incompressible Navier-Stokes equations in vorticity form using a pseudo-spectral method. We randomly sample the spatial domain size and resolution from $\{\pi, 2\pi, 3\pi, 4\pi\}$ and $\{32, 64, 128, 256, 512\}$, respectively. To ensure the simulation is well-resolved, we compute all trajectories at a $1024 \times 1024$ resolution and uniformly downsample them to the target resolution.

\textbf{Governing Physics: } The governing physics follow the formulation presented in Section~\ref{appendix:partial}.

\textbf{Experiment Results: } We present the results in Table~\ref{tab:mixedres_full}. Because the data is Eulerian, we additionally report the FRMSE and Kinetic Energy. The results show that our agent consistently achieves superior performance. AMG is not applicable because our dataset includes high-resolution $512 \times 512$ data, causing its training to take 28,000 seconds per epoch, which is computationally infeasible.

\begin{table}[htbp]
\centering
\begin{tabular}{lccc}
\hline
 & RMSE & FRMSE & Kinetic Energy \\
\hline
GNOT & $3.629 \times 10^{0}$ & $8.521 \times 10^{2}$ & $1.508 \times 10^{0}$ \\
Transolver & $3.616 \times 10^{0}$ & $8.217 \times 10^{2}$ & $1.415 \times 10^{0}$ \\
Transolver++ & $3.564 \times 10^{0}$ & $7.573 \times 10^{2}$ & $1.166 \times 10^{0}$ \\
LNO & $3.481 \times 10^{0}$ & $7.548 \times 10^{2}$ & $1.125 \times 10^{0}$ \\
AMG & - & - & - \\
\textbf{Agent} & $\mathbf{3.455 \times 10^{0}}$ & $\mathbf{7.014 \times 10^{2}}$ & $\mathbf{6.830 \times 10^{-1}}$ \\
\hline
\end{tabular}
\caption{Performance comparison on the \MixedRes\ dataset. Best results are shown in \textbf{bold}.}
\label{tab:mixedres_full}
\end{table}

\begin{figure}[htbp]
    \centering
    \includegraphics[width=0.9\textwidth, page=1]{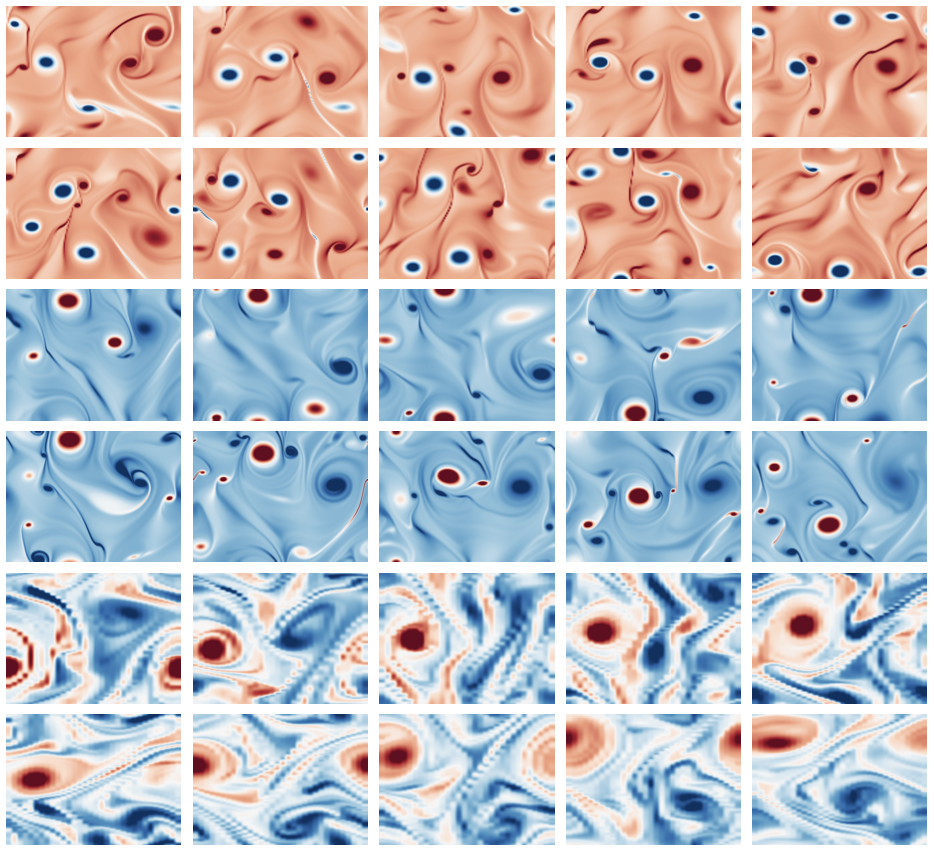}
    \caption{Visualization of the \MixedRes\ dataset. The figure displays Navier-Stokes dataset in vorticity form at different spatial sizes and resolutions.}
    \label{fig:mixedres}
\end{figure}

\textbf{Visualization: } We refer the readers to Figure~\ref{fig:mixedres} for visualizations of the dataset.

\newpage

\subsection{\ExtLagOne}
\label{appendix:extlag1}
\textbf{Dataset Description: } This dataset contains liquid sloshing dynamics in Lagrangian particle format. Fluid particles are enclosed in a sealed tank that undergoes external excitation, causing the free surface to slosh. The tank motion is prescribed as a time series of rigid-body accelerations, and the fluid responds with large free-surface deformation, wave breaking, and impacts against the tank walls. The sloshing motion is provided with the linear accelerations and the angular accelerations.

\textbf{Modeling Objective: } Let $x_{t} \in \mathbb{R}^{N \times 54}$ denote the state of the system at time $t$, where $N$ is the number of fluid particles. The $54$ features per particle comprise the last-step position ($2$), the historical velocity over $5$ frames ($5 \times 2 = 10$), and the sloshing motion over $6$ historical frames together with the next-step sloshing motion ($7 \times 6 = 42$). Our objective is to develop a neural simulator, defined as a parameterized function $\Psi$, that maps the current system state to the particle velocity at the next timestep. Formally, we aim to predict $\hat{v}_{t+1} = \Psi(x_{t})$, where the predicted output $\hat{v}_{t+1} \in \mathbb{R}^{N \times 2}$ is the particle velocity at the next time step.

\textbf{Dataset Specific Challenges: } The sloshing motion is a global, per-frame signal shared by all particles rather than a per-particle quantity. The standard approach simply concatenates it onto every particle's node features, which treats a global driving force as local information and dilutes it among high-dimensional particle states. Addressing this problem therefore requires innovative conditioning mechanisms that inject the excitation signal into the neural solver more effectively.

\textbf{Dataset Generation: } We generate the sloshing dataset using weakly-compressible Smoothed Particle Hydrodynamics (SPH) solver~\citep{ZHAN2025109389, DOMINGUEZ2022867}. The simulation is two-dimensional with an inter-particle spacing of $dp = 3\times10^{-3}\,\mathrm{m}$, yielding a smoothing length $h = 0.91924\,\sqrt{2\,dp^{2}}$, and a total of $9908$ particles ($938$ fixed boundary particles and $8970$ fluid particles). Time integration uses a Symplectic scheme together with the Wendland kernel, and the artificial-viscosity treatment is applied with coefficient $\alpha = 0.05$. The weakly-compressible pressure closure uses a polytropic exponent $\gamma = 7$ with reference density $\rho_0 = 1000\,\mathrm{kg/m^{3}}$, and the numerical speed of sound is set to $30$ times the characteristic system speed. Density evolution is stabilized with the full Fourtakas density-diffusion term (coefficient $0.1$), while particle shifting is disabled. The timestep is adaptive under a CFL coefficient of $0.2$ with $\mathrm{dtmin} = 0.05\,h/c_{s}$, and particles are rejected outside the density bounds $[700, 1300]\,\mathrm{kg/m^{3}}$. The external sloshing excitation is imposed on the fluid through a prescribed time-dependent rigid acceleration read from a driving file, applied about the tank reference centre with global gravity retained.

\textbf{Governing Physics: } The sloshing case is governed by the weakly-compressible Navier--Stokes equations for a single liquid phase, solved in the Lagrangian SPH frame. Writing $\tfrac{d}{dt}$ for the material derivative following a fluid element, the conservation of mass and momentum read
\begin{equation}
    \frac{d\rho}{dt} = -\rho\, \nabla \cdot \mathbf{v},
\end{equation}
\begin{equation}
    \frac{d\mathbf{v}}{dt}
    = -\frac{1}{\rho}\nabla P
    + \nu \nabla^{2}\mathbf{v}
    + \mathbf{g}
    + \mathbf{a}_{\mathrm{ext}}(t),
\end{equation}
where $\rho$ is the density, $\mathbf{v}$ the velocity, $P$ the pressure, $\nu$ the kinematic viscosity, and $\mathbf{g} = (0, 0, -9.81)^{\top}\,\mathrm{m/s^{2}}$ the gravitational acceleration. The final term $\mathbf{a}_{\mathrm{ext}}(t)$ is the externally imposed sloshing forcing: the fluid is driven not by moving the domain boundaries directly but by adding a prescribed, time-dependent body acceleration to every fluid element. Expressed about the tank reference centre $\mathbf{r}_{c}$, this excitation combines a translational and a rotational contribution,
\begin{equation}
    \mathbf{a}_{\mathrm{ext}}(t)
    = \mathbf{a}_{T}(t)
    + \boldsymbol{\alpha}(t) \times (\mathbf{r} - \mathbf{r}_{c})
    + \boldsymbol{\omega}(t) \times \big(\boldsymbol{\omega}(t) \times (\mathbf{r} - \mathbf{r}_{c})\big)
    + 2\,\boldsymbol{\omega}(t) \times \mathbf{v},
\end{equation}
where $\mathbf{a}_{T}(t)$ is the imposed linear acceleration of the tank, $\boldsymbol{\omega}(t)$ and $\boldsymbol{\alpha}(t) = d\boldsymbol{\omega}/dt$ are the imposed angular velocity and angular acceleration, and the successive terms represent the translational, Euler, centrifugal, and Coriolis accelerations induced by the moving reference frame. The time histories of $\mathbf{a}_{T}(t)$ and $\boldsymbol{\omega}(t)$ are supplied externally from the acceleration driving file, so that $\mathbf{a}_{\mathrm{ext}}(t)$ acts as the sole time-dependent driver that continually sloshes the liquid against the tank walls. The system is closed by the weakly-compressible equation of state relating pressure to density,
\begin{equation}
    P = B\left[\left(\frac{\rho}{\rho_{0}}\right)^{\gamma} - 1\right],
    \qquad
    B = \frac{c_{s}^{2}\,\rho_{0}}{\gamma},
\end{equation}
with reference density $\rho_{0}$, polytropic exponent $\gamma = 7$, and numerical speed of sound $c_{s}$; the pressure constant evaluates to $B = 1.0973\times10^{5}\,\mathrm{Pa}$ for this configuration.

\textbf{Experiment Results: } The comprehensive experimental results are detailed in Table~\ref{tab:exglag1_full}. Given the nature of the Lagrangian particle data, we include LagKE to evaluate physical consistency, as well as the Sinkhorn metric to quantify the optimal transport distance between the predicted and ground-truth particle distributions. Furthermore, we report Eulerian aggregated metrics (EuARMSE and EuAFRMSE) to assess the practical utility of the predictions. The agent consistently achieves strong results across all evaluated metrics, highlighting the robustness and superiority of its design.

\begin{table}[htbp]
\centering
\resizebox{\textwidth}{!}{%
\begin{tabular}{lccccc}
\hline
 & RMSE & LagKE & Sinkhorn & EuARMSE & EuAFRMSE \\
\hline
GNOT & $2.176 \times 10^{-2}$ & $1.152 \times 10^{1}$ & $3.882 \times 10^{-4}$ & $4.715 \times 10^{-3}$ & $1.431 \times 10^{-1}$ \\
Transolver & $3.526 \times 10^{-2}$ & $3.468 \times 10^{1}$ & $1.144 \times 10^{-3}$ & $6.883 \times 10^{-3}$ & $2.143 \times 10^{-1}$ \\
Transolver++ & $1.729 \times 10^{-2}$ & $6.427 \times 10^{0}$ & $2.568 \times 10^{-4}$ & $3.571 \times 10^{-3}$ & $1.071 \times 10^{-1}$ \\
LNO & $1.407 \times 10^{-2}$ & $3.814 \times 10^{0}$ & $1.868 \times 10^{-4}$ & $2.730 \times 10^{-3}$ & $8.317 \times 10^{-2}$ \\
AMG & $1.892 \times 10^{-2}$ & $5.801 \times 10^{0}$ & $3.067 \times 10^{-4}$ & $3.941 \times 10^{-3}$ & $1.152 \times 10^{-1}$ \\
\textbf{Agent} & $\mathbf{1.954 \times 10^{-3}}$ & $\mathbf{4.650 \times 10^{-2}}$ & $\mathbf{4.286 \times 10^{-6}}$ & $\mathbf{4.394 \times 10^{-4}}$ & $\mathbf{1.005 \times 10^{-2}}$ \\
\hline
\end{tabular}
}
\caption{Performance comparison on the \ExtLagOne\ dataset. Best results are shown in \textbf{bold}.}
\label{tab:exglag1_full}
\end{table}

\begin{figure}[htbp]
    \centering
    \includegraphics[width=0.9\textwidth, page=1]{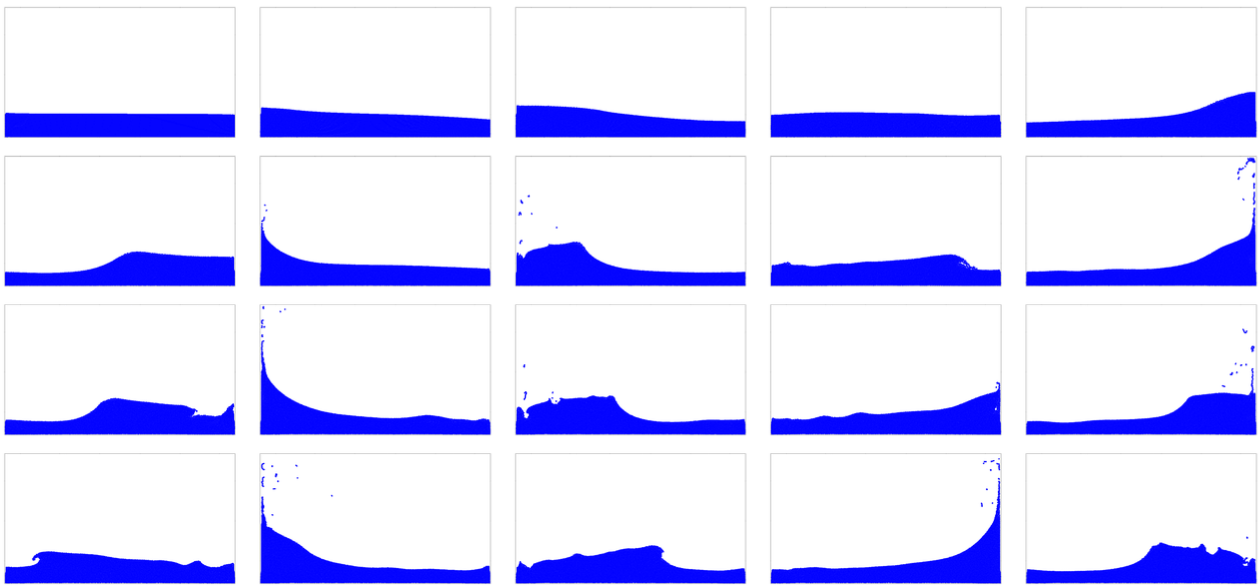}
    \caption{Visualization of the \ExtLagOne\ dataset. The dataset features Lagrangian fluid particles contained within a closed box subject to external sloshing motion.}
    \label{fig:extlag1}
\end{figure}

\textbf{Visualization: } We refer the readers to Figure~\ref{fig:extlag1} for visualizations of the dataset.

\newpage

\subsection{\ExtLagTwo}
\label{appendix:extlag2}
\textbf{Dataset Description: } This dataset contains piston-driven fluid dynamics in Lagrangian particle format. A piston at the left boundary of the tank undergoes prescribed horizontal motion, pushing the fluid and driving waves through the domain. Each particle is labeled by type ($0 =$ fluid, $1 =$ piston): fluid particles obey the free dynamics, while piston particles follow a known rigid trajectory that acts as the external forcing of the system.

\textbf{Modeling Objective: } Let $x_{t} \in \mathbb{R}^{N \times 15}$ denote the state of the system at time $t$, where $N$ is the number of particles. The $15$ features per particle comprise the last-step position ($2$), the historical velocity over $5$ frames ($5 \times 2 = 10$), the next-step velocity ($2$), and a piston indicator column ($1$). The next-step velocity slot is a known-driver feature: for piston particles it contains the true $t+1$ velocity of the prescribed motion, while for fluid particles it is zero since $t+1$ is exactly the quantity to be predicted; the indicator column lets the model distinguish a masked zero from a genuinely zero driver. Our objective is to develop a neural simulator, defined as a parameterized function $\Psi$, that maps the current system state to the particle velocity at the next timestep. Formally, we aim to predict $\hat{v}_{t+1} = \Psi(x_{t})$, where the predicted output $\hat{v}_{t+1} \in \mathbb{R}^{N \times 2}$ is the particle velocity at the next time step. The loss is evaluated on fluid particles only, since the piston motion is prescribed rather than predicted.

\textbf{Dataset Specific Challenges: } The fluid is driven entirely by the piston boundary: momentum enters the system through a thin layer of fluid particles in contact with the piston and must propagate correctly through the bulk. The model must accurately handle this external force. 

\textbf{Dataset Generation: } We generate dataset with a weakly-compressible Smoothed Particle Hydrodynamics (SPH) solver~\citep{ZHAN2025109389, DOMINGUEZ2022867}. The simulation is two-dimensional with an inter-particle spacing of $dp = 3\times10^{-2}\,\mathrm{m}$, yielding a smoothing length $h = 1.2\,\sqrt{2\,dp^{2}}$, and a total of $6865$ particles ($385$ fixed boundary particles, $201$ moving boundary particles forming the piston, and $6279$ fluid particles). Time integration uses a Symplectic scheme together with the Wendland kernel, and the artificial-viscosity treatment is applied with coefficient $\alpha = 0.01$. The weakly-compressible pressure closure uses a polytropic exponent $\gamma = 7$ with reference density $\rho_0 = 1000\,\mathrm{kg/m^{3}}$, and the numerical speed of sound is set to $20$ times the characteristic system speed. Density evolution is stabilized with the Fourtakas density-diffusion term (coefficient $0.1$), while particle shifting is disabled. The timestep is adaptive under a CFL coefficient of $0.2$ with $\mathrm{dtmin} = 0.05\,h/c_{s}$, and particles are rejected outside the density bounds $[700, 1300]\,\mathrm{kg/m^{3}}$. The waves are driven by a piston wavemaker: the moving boundary is actuated by a second-order regular-wave generation routine configured with wave height $H = 0.15\,\mathrm{m}$, wave period $T = 2\,\mathrm{s}$, and still-water depth $d = 0.66\,\mathrm{m}$, ramped in over one wave period, with the piston displacing along the $x$-direction.

\textbf{Governing Physics: } The data is governed by the weakly-compressible Navier--Stokes equations for a single liquid phase, solved in the Lagrangian SPH frame. Writing $\tfrac{d}{dt}$ for the material derivative following a fluid element, the conservation of mass and momentum read
\begin{equation}
    \frac{d\rho}{dt} = -\rho\, \nabla \cdot \mathbf{v},
\end{equation}
\begin{equation}
    \frac{d\mathbf{v}}{dt}
    = -\frac{1}{\rho}\nabla P
    + \nu \nabla^{2}\mathbf{v}
    + \mathbf{g},
\end{equation}
where $\rho$ is the density, $\mathbf{v}$ the velocity, $P$ the pressure, $\nu$ the kinematic viscosity, and $\mathbf{g} = (0, 0, -9.81)^{\top}\,\mathrm{m/s^{2}}$ the gravitational acceleration. The system is closed by the weakly-compressible equation of state relating pressure to density,
\begin{equation}
    P = B\left[\left(\frac{\rho}{\rho_{0}}\right)^{\gamma} - 1\right],
    \qquad
    B = \frac{c_{s}^{2}\,\rho_{0}}{\gamma},
\end{equation}
with reference density $\rho_{0}$, polytropic exponent $\gamma = 7$, and numerical speed of sound $c_{s}$; the pressure constant evaluates to $B = 3.3634\times10^{5}\,\mathrm{Pa}$ for this configuration.

Unlike the interior fluid, the waves are not forced through a body term but through a moving solid boundary: a subset of boundary particles forms a piston whose position $X_{p}(t)$ is prescribed in time and imposed as a Dirichlet velocity condition on the fluid. For second-order regular-wave generation, the piston displacement combines the first-order Biesel stroke with a second-order Stokes correction,
\begin{equation}
    X_{p}(t) = \frac{S}{2}\sin(\omega t + \phi)
    + \frac{S^{(2)}}{4}\sin(2\omega t + \phi),
    \qquad
    \dot{X}_{p}(t) = \mathbf{v}\cdot\mathbf{e}_{x}\big|_{\text{piston}},
\end{equation}
where $\omega = 2\pi/T$ is the angular frequency of the target wave of period $T = 2\,\mathrm{s}$ and height $H = 0.15\,\mathrm{m}$, $\phi$ is the initial phase, and $\mathbf{e}_{x}$ is the direction of piston travel. The first-order stroke $S$ follows the Biesel transfer function for a piston-type wavemaker at still-water depth $d = 0.66\,\mathrm{m}$,
\begin{equation}
    \frac{H}{S}
    = \frac{2\big(\cosh(2kd) - 1\big)}{\sinh(2kd) + 2kd},
\end{equation}
with the wavenumber $k$ fixed by the linear dispersion relation $\omega^{2} = g\,k\,\tanh(kd)$, and $S^{(2)}$ is the corresponding second-order stroke amplitude that suppresses spurious free harmonics. A ramp function multiplies $X_{p}(t)$ over the first wave period so that the excitation is applied smoothly from rest. The prescribed piston motion is thus the sole time-dependent driver that generates the propagating wave train down the flume.

\textbf{Experiment Results: } We present the complete experimental results in Table~\ref{tab:extlag2_full}. As with other Lagrangian particle datasets, we additionally report LagKE, Sinkhorn, EuARMSE, and EuAFRMSE. The results show that the agent's design achieves superior performance across all of these metrics.

\begin{table}[htbp]
\centering
\resizebox{\textwidth}{!}{%
\begin{tabular}{lccccc}
\hline
 & RMSE & LagKE & Sinkhorn & EuARMSE & EuAFRMSE \\
\hline
GNOT & $1.365 \times 10^{-3}$ & $5.047 \times 10^{-2}$ & $6.958 \times 10^{-4}$ & $2.844 \times 10^{-4}$ & $6.658 \times 10^{-3}$ \\
Transolver & $1.456 \times 10^{-3}$ & $4.405 \times 10^{-2}$ & $6.952 \times 10^{-4}$ & $2.978 \times 10^{-4}$ & $7.065 \times 10^{-3}$ \\
Transolver++ & $1.551 \times 10^{-3}$ & $7.764 \times 10^{-2}$ & $6.963 \times 10^{-4}$ & $3.396 \times 10^{-4}$ & $8.152 \times 10^{-3}$ \\
LNO & $1.640 \times 10^{-3}$ & $4.649 \times 10^{-2}$ & $6.968 \times 10^{-4}$ & $3.008 \times 10^{-4}$ & $6.991 \times 10^{-3}$ \\
AMG & $1.566 \times 10^{-3}$ & $4.806 \times 10^{-2}$ & $6.964 \times 10^{-4}$ & $2.748 \times 10^{-4}$ & $6.766 \times 10^{-3}$ \\
\textbf{Agent} & $\mathbf{5.483 \times 10^{-4}}$ & $\mathbf{1.712 \times 10^{-2}}$ & $\mathbf{6.427 \times 10^{-4}}$ & $\mathbf{9.468 \times 10^{-5}}$ & $\mathbf{2.266 \times 10^{-3}}$ \\
\hline
\end{tabular}
}
\caption{Performance comparison on the \ExtLagTwo\ dataset. Best results are shown in \textbf{bold}.}
\label{tab:extlag2_full}
\end{table}

\begin{figure}[htbp]
    \centering
    \includegraphics[width=0.9\textwidth, page=1]{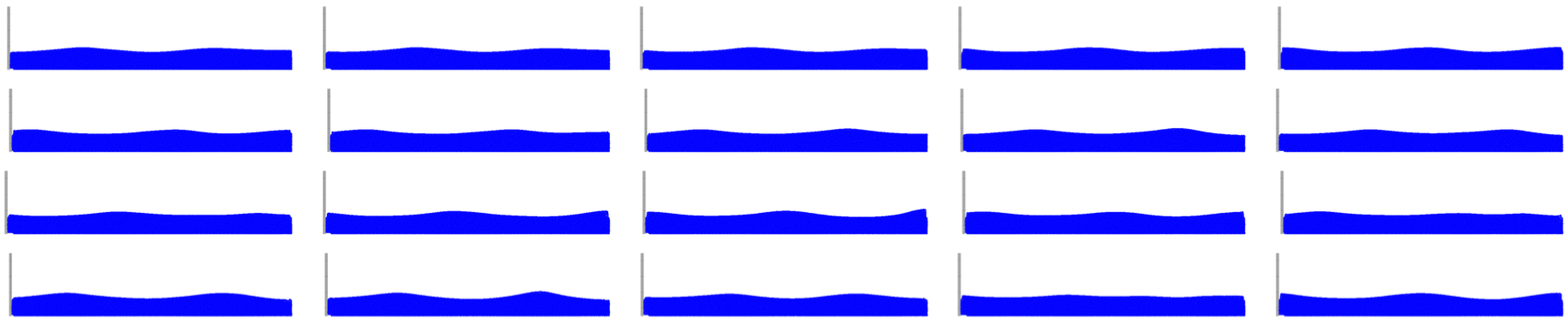}
    \caption{Visualization of the \ExtLagTwo\ dataset. The dataset features a horizontally moving piston on the right side that drives the water.}
    \label{fig:extlag2}
\end{figure}

\textbf{Visualization: } We refer the readers to Figure~\ref{fig:extlag2} for visualizations of the dataset.

\newpage

\subsection{\MixedDim}
\label{appendix:mixeddim}
\textbf{Dataset Description: } This dataset contains solutions of the Allen--Cahn equation, a reaction--diffusion PDE describing phase separation, in one, two, and three spatial dimensions. The underlying physics is identical across dimensions: a diffusion term competes with a nonlinear double-well reaction term, driving the field toward $+1/-1$ phases separated by thin diffuse interfaces. Each trajectory is discretized on a regular grid with $64$ points per spatial axis, giving $64$ nodes in 1D, $64^2 = 4096$ nodes in 2D, and $64^3 = 262144$ nodes in 3D.

\textbf{Modeling Objective: } Let $x_{t} \in \mathbb{R}^{N \times 1}$ denote the system state at time $t$, where $N$ is the number of grid nodes ($N = 64$, $64^2$, or $64^3$ depending on the dimension) and the single feature channel is the scalar phase field $u$. The data is stored flattened with shape $(\text{num\_traj}, T, N)$ and can be reshaped back to its grid form. Our objective is to learn a single parameterized neural solver $\Psi$ such that $\hat{x}_{t+1} = \Psi(x_{t})$, where the predicted output $\hat{x}_{t+1}$ retains the same shape $(N, 1)$ as the input state. Critically, one model with a single set of weights must generalize across all three dimensions, exploiting the fact that the governing equation is shared.

\textbf{Dataset Specific Challenges: } The solver must be dimension-agnostic: it cannot assume a fixed grid shape, a fixed input size, or a dimension-specific convolutional structure, and must instead learn a representation of the shared Allen--Cahn dynamics that transfers across 1D, 2D, and 3D discretizations. Compounding this, the number of nodes spans over three orders of magnitude (from $64$ to $262144$), so the model must handle both very small and very large inputs. Memory management is therefore critical: attention or global coupling mechanisms that scale quadratically in the number of nodes may be infeasible in 3D, while overly local models may underfit the 1D case.

\textbf{Dataset Generation: } We generate the Allen--Cahn dataset via Exponential Time Differencing Runge--Kutta (ETDRK) schemes. The spatial derivatives are treated spectrally: the linear diffusion operator is diagonalized in Fourier space and integrated exactly, while the nonlinear double-well reaction term is evaluated pseudo-spectrally, yielding a stiffly-accurate, high-order-in-time solver that is identical in structure across all spatial dimensions. For each of the 1D, 2D, and 3D cases we simulate on a periodic domain $[0, 1)^d$ discretized with $64$ points per axis and advance with a fixed timestep $\Delta t = 0.1$. We use a diffusivity of $\nu = 5 \times 10^{-3}$, a linear reaction coefficient $c_1 = 1.0$, and a cubic reaction coefficient $c_3 = -1.0$. Initial conditions are drawn from a random truncated Fourier series with wavenumber cutoff $5$, producing smooth random fields that subsequently coarsen into $+1/-1$ phase domains. In every dimension we roll out $30$ trajectories of $100$ frames each (excluding the initial state), giving $3000$ frames per dimension.

\textbf{Governing Physics: } The Allen--Cahn equation is a reaction--diffusion PDE modeling phase separation, in which a linear diffusion term competes with a nonlinear double-well reaction term to drive the scalar phase field $u(\mathbf{x}, t)$ toward the two stable phases $u = +1$ and $u = -1$, separated by thin diffuse interfaces. On the periodic domain $\Omega = [0, 1)^d$ it is written as
\begin{equation}
    \frac{\partial u}{\partial t}
    = \nu\,\Delta u
    + c_1\,u
    + c_3\,u^{3},
    \qquad \mathbf{x} \in \Omega,\ t > 0,
\end{equation}
where $\Delta = \sum_{i=1}^{d} \partial^2 / \partial x_i^2$ is the $d$-dimensional Laplacian and $\nu > 0$ is the diffusivity controlling the width of the interfaces. The reaction drives the field toward these two wells while the diffusion penalizes sharp gradients, and the balance between the two sets the characteristic interface thickness $\sim \sqrt{\nu}$. The governing equation is identical in form across all three spatial dimensions, differing only in the dimensionality of the Laplacian $\Delta$; this shared structure is precisely what a single dimension-agnostic solver is meant to exploit.

\textbf{Experiment Results: } We present the complete experimental results in Table~\ref{tab:mixeddim_full}. AMG is not applicable to this dataset because its graph-based architecture scales poorly to 3D data, resulting in impractical training times. Overall, the proposed agent design achieves superior performance across all dimensions.

\begin{table}[htbp]
\centering
\resizebox{\textwidth}{!}{%
\begin{tabular}{lccccc}
\hline
 & RMSE & FRMSE & RMSE (1D) & RMSE (2D) & RMSE (3D) \\
\hline
GNOT & $3.615 \times 10^{-2}$ & $3.521 \times 10^{0}$ & $6.239 \times 10^{-2}$ & $2.752 \times 10^{-2}$ & $1.855 \times 10^{-2}$ \\
Transolver & $3.381 \times 10^{-2}$ & $3.470 \times 10^{0}$ & $5.688 \times 10^{-2}$ & $2.732 \times 10^{-2}$ & $1.722 \times 10^{-2}$ \\
Transolver++ & $3.357 \times 10^{-2}$ & $3.766 \times 10^{0}$ & $5.186 \times 10^{-2}$ & $3.031 \times 10^{-2}$ & $1.853 \times 10^{-2}$ \\
LNO & $5.048 \times 10^{-2}$ & $3.937 \times 10^{0}$ & $1.014 \times 10^{-1}$ & $3.131 \times 10^{-2}$ & $1.870 \times 10^{-2}$ \\
AMG & - & - & - & - & - \\
\textbf{Agent} & $\mathbf{9.418 \times 10^{-3}}$ & $\mathbf{6.707 \times 10^{-1}}$ & $\mathbf{1.753 \times 10^{-2}}$ & $\mathbf{7.679 \times 10^{-3}}$ & $\mathbf{3.049 \times 10^{-3}}$ \\
\hline
\end{tabular}
}
\caption{Performance comparison on the \MixedDim\ dataset. Best results are shown in \textbf{bold}.}
\label{tab:mixeddim_full}
\end{table}

\begin{figure}[htbp]
    \centering
    \includegraphics[width=0.8\textwidth, page=1]{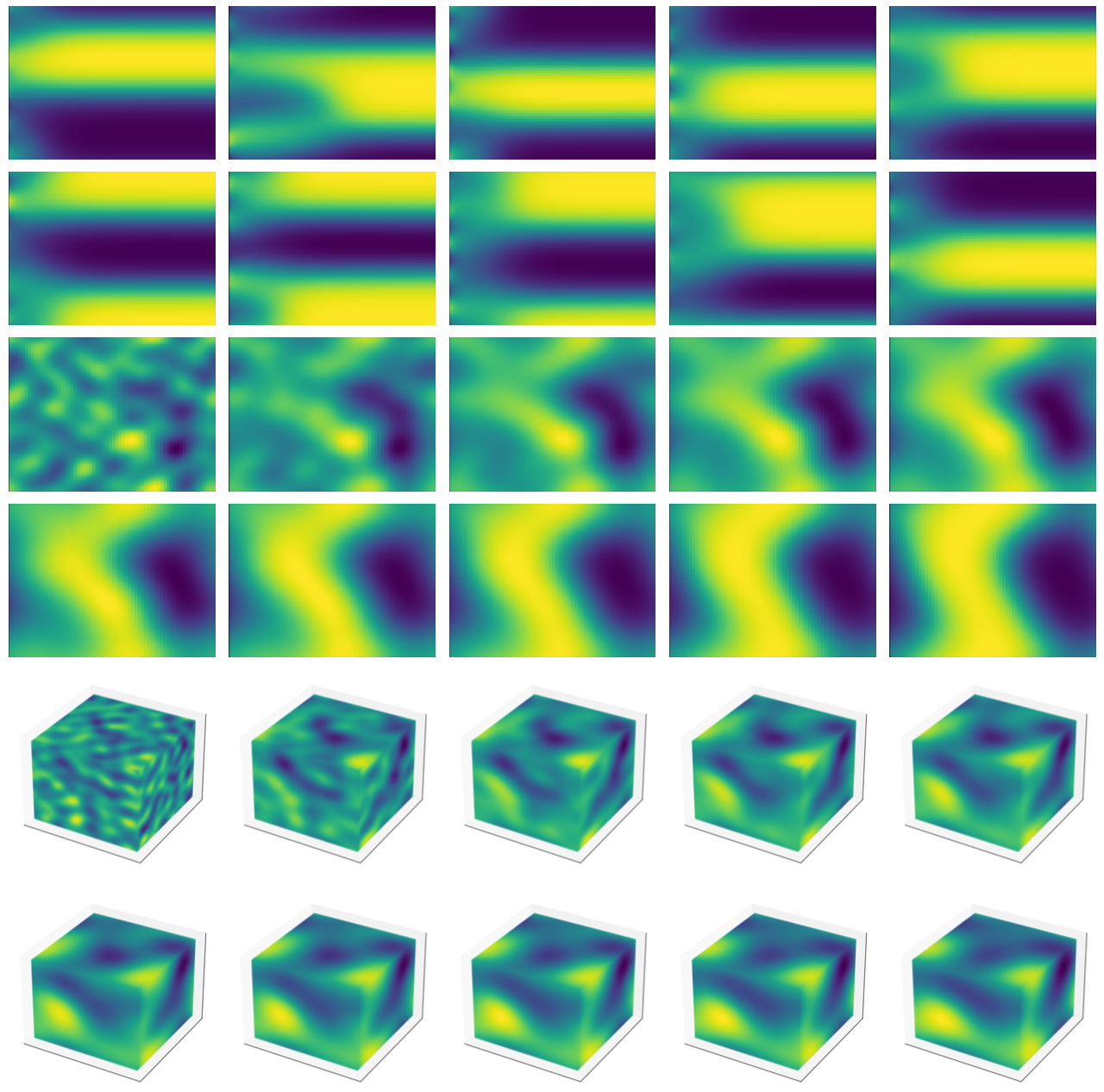}
    \caption{Visualization of the \MixedDim\ dataset. This figure displays the Allen-Cahn PDE across three spatial dimensions: 1D (first two rows), 2D (middle two rows), and 3D (final two rows). All datasets share the same governing physics and differ only in dimensionality. For visualization purposes in the 1D case, the x-axis corresponds to time and the y-axis corresponds to the spatial discretization grid.}
    \label{fig:mixeddim}
\end{figure}

\textbf{Visualization: } We refer the readers to Figure~\ref{fig:mixeddim} for visualizations of the dataset.

\newpage
\clearpage

\subsection{\MixedBC}
\label{appendix:mixedbc}
\textbf{Dataset Description: } This dataset contains two-dimensional linear advection dynamics evolving on a $128 \times 128$ spatial grid under heterogeneous boundary conditions. Each trajectory is governed by one of three boundary condition families---Dirichlet, Neumann, or periodic---with time-dependent boundary data in the non-periodic cases. Unlike standard benchmarks where a single boundary treatment is fixed across the entire dataset, trajectories here differ in both the type of boundary condition and the values imposed on each of the four domain edges, reflecting practical settings where the same interior physics must be simulated under diverse and externally driven boundary constraints.

\textbf{Modeling Objective: } Let $x_{t} \in \mathbb{R}^{1 \times 128 \times 128}$ denote the system state at time $t$. The model additionally receives a boundary type identifier $b \in \{0, 1, 2\}$, encoding Dirichlet, Neumann, or periodic conditions respectively, and a boundary value vector $v_t \in \mathbb{R}^{1 \times 4}$ specifying the boundary data on the (left, right, bottom, top) edges at time $t$: prescribed state values for Dirichlet, prescribed fluxes for Neumann, and zero-padding for periodic. Our objective is to learn a parameterized neural simulator $\Psi$ such that $\hat{x}_{t+1} = \Psi(x_{t}, b, v_t)$, where the predicted output $\hat{x}_{t+1} \in \mathbb{R}^{1 \times 128 \times 128}$ is the state at the next time step, matching the shape of the input state.

\textbf{Dataset Specific Challenges: } The three boundary condition families induce qualitatively different dynamics near the domain edges: Dirichlet conditions exactly define the state, Neumann conditions constrain its normal gradient, and periodic conditions couple opposite edges through wrap-around transport. A single model must internalize all three behaviors and switch between them based on a discrete conditioning input, rather than specializing to one edge treatment. Compounding this, the boundary value vector carries a type-dependent semantic: the same numerical entries represent state values under Dirichlet conditions, fluxes under Neumann conditions, and uninformative padding under periodic conditions. The model must therefore learn to interpret this shared input slot conditionally, using the boundary type to modulate how the boundary values influence the predicted interior dynamics.

\textbf{Dataset Generation: } We generate this dataset for the two-dimensional linear advection equation, producing $30$ trajectories of $100$ frames each for every boundary condition family on a $128 \times 128$ grid over the unit square $[0,1]^2$ with a frame interval of $\Delta t = 0.1$ (total time $T = 10$). A single divergence-free-scaled velocity field is drawn once from a Gaussian random field and reused across all trajectories and all three boundary families, so that the interior transport is held fixed and only the boundary treatment varies; its amplitude is rescaled to a target Courant number of $0.45$. For each trajectory the initial phase field is sampled independently from the same Gaussian random field prior. The non-periodic families additionally receive, on each of the four edges, an independent time-dependent boundary signal built as a low-frequency truncated Fourier series, clipped to a physical range; these per-edge signals are recorded frame-by-frame in the order (left, right, bottom, top). The Dirichlet and Neumann cases are integrated with an explicit finite-volume upwind scheme under an adaptive CFL-limited sub-stepping loop between saved frames, whereas the periodic case is integrated with a pseudo-spectral Fourier method using classical fourth-order Runge--Kutta time stepping and $2/3$-rule dealiasing. The three families are stored together with a boundary type identifier $b \in \{0,1,2\}$ (Dirichlet, Neumann, periodic) and, for the non-periodic cases, the recorded boundary value vectors; periodic trajectories carry no boundary data and are zero-padded in that slot.

\textbf{Governing Physics: } The dataset is governed by the two-dimensional linear advection (transport) equation for a scalar field $u(\mathbf{x}, t)$ carried by a prescribed, time-independent velocity field $\mathbf{a}(\mathbf{x}) = (a_x(\mathbf{x}), a_y(\mathbf{x}))$,
\begin{equation}
    \frac{\partial u}{\partial t}
    + \nabla \cdot (\mathbf{a}\, u)
    = 0,
    \qquad \mathbf{x} \in \Omega = [0,1]^2,\ t > 0,
\end{equation}
which, since the velocity field used here is constructed to be (discretely) divergence-free, is equivalent to the non-conservative form $\partial_t u + \mathbf{a} \cdot \nabla u = 0$. The equation simply transports the initial field along the characteristics $\mathrm{d}\mathbf{x}/\mathrm{d}t = \mathbf{a}(\mathbf{x})$, so that $u$ is constant along each streamline and the interior dynamics are governed entirely by the fixed advecting velocity. What distinguishes the trajectories is the closure imposed on the domain boundary $\partial \Omega$, drawn from three families:
\begin{equation}
    \underbrace{u = g(\mathbf{x}, t)}_{\text{Dirichlet}},
    \qquad
    \underbrace{\frac{\partial u}{\partial n} = h(\mathbf{x}, t)}_{\text{Neumann}},
    \qquad
    \underbrace{u(\mathbf{x} + \mathbf{L}) = u(\mathbf{x})}_{\text{periodic}},
\end{equation}
where $n$ denotes the outward normal, $\mathbf{L}$ the domain periods, and $g$ and $h$ are the prescribed time-dependent boundary data supplied edge-by-edge on the (left, right, bottom, top) walls. The Dirichlet condition pins the boundary state to $g$; the Neumann condition prescribes the normal derivative $\partial u / \partial n = h$, imposed numerically through one-sided ghost values; and the periodic condition couples opposite edges through wrap-around transport, requiring no external boundary data. Because the interior operator is identical across all trajectories, the differences observed near the domain edges are attributable solely to the boundary condition type and its associated data.

\textbf{Experiment Results: } We present the complete experimental results in Table~\ref{tab:mixedbc_full}. Because this dataset relies on an Eulerian grid, we also include the FRMSE and mass error. To isolate the effects of the boundary conditions, we report separate RMSE values for the Dirichlet, Neumann, and periodic conditions, alongside the RMSE evaluated exclusively at the boundary points. Overall, the proposed agent design achieves superior performance across all metrics.

\begin{table}[h]
\centering
\begin{tabular}{lcccc}
\hline
 & RMSE & FRMSE & Mass Error & RMSE (Dirichlet) \\
\hline
GNOT         & $1.383 \times 10^{-1}$ & $1.556 \times 10^{1}$ & $4.417 \times 10^{2}$ & $1.464 \times 10^{-1}$ \\
Transolver   & $8.500 \times 10^{-2}$ & $9.776 \times 10^{0}$ & $2.058 \times 10^{2}$ & $1.417 \times 10^{-1}$ \\
Transolver++ & $8.498 \times 10^{-2}$ & $9.771 \times 10^{0}$ & $2.025 \times 10^{2}$ & $1.418 \times 10^{-1}$ \\
LNO          & $8.511 \times 10^{-2}$ & $9.786 \times 10^{0}$ & $2.043 \times 10^{2}$ & $1.421 \times 10^{-1}$ \\
AMG          & $2.064 \times 10^{-1}$ & $2.290 \times 10^{1}$ & $8.483 \times 10^{2}$ & $1.348 \times 10^{-1}$ \\
\textbf{Agent} & $\mathbf{4.435 \times 10^{-2}}$ & $\mathbf{4.579 \times 10^{0}}$ & $\mathbf{6.910 \times 10^{1}}$ & $\mathbf{6.809 \times 10^{-2}}$ \\
\hline
\multicolumn{5}{c}{} \\
\hline
 & RMSE (Neumann) & RMSE (Periodic) & RMSE (BC) \\
\hline
GNOT         & $5.262 \times 10^{-2}$ & $1.822 \times 10^{-1}$ & $7.049 \times 10^{-1}$ \\
Transolver   & $1.550 \times 10^{-2}$ & $3.661 \times 10^{-2}$ & $4.461 \times 10^{-1}$ \\
Transolver++ & $1.505 \times 10^{-2}$ & $3.640 \times 10^{-2}$ & $4.461 \times 10^{-1}$ \\
LNO          & $1.503 \times 10^{-2}$ & $3.638 \times 10^{-2}$ & $4.467 \times 10^{-1}$ \\
AMG          & $1.057 \times 10^{-1}$ & $3.138 \times 10^{-1}$ & $9.335 \times 10^{-1}$ \\
\textbf{Agent} & $\mathbf{1.039 \times 10^{-2}}$ & $\mathbf{3.403 \times 10^{-2}}$ & $\mathbf{2.035 \times 10^{-1}}$ \\
\hline
\end{tabular}
\caption{Performance comparison on the \MixedBC\ dataset. Best results are shown in \textbf{bold}.}
\label{tab:mixedbc_full}
\end{table}

\begin{figure}[htbp]
    \centering
    \includegraphics[width=0.8\textwidth, page=1]{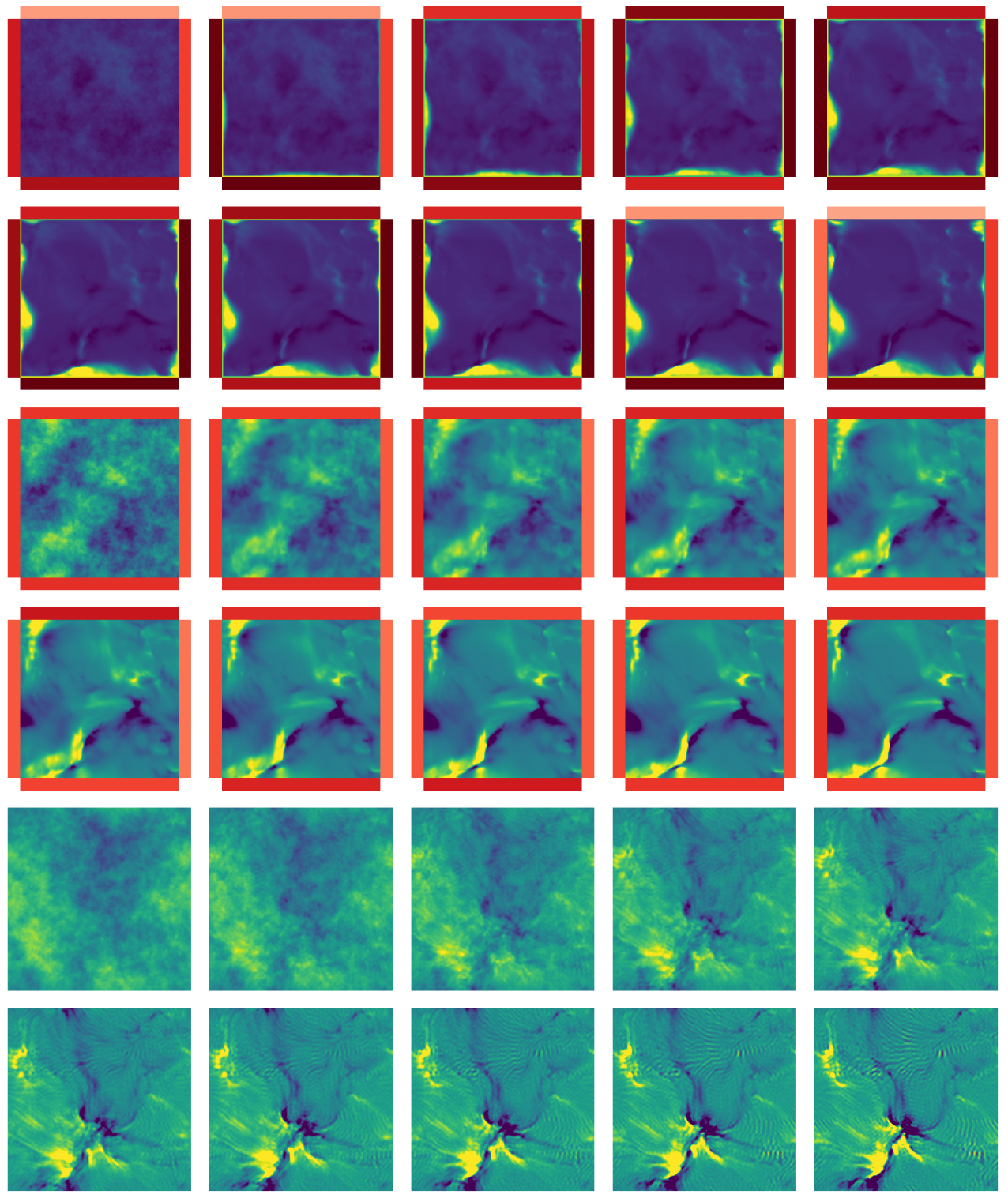}
    \caption{Visualization of the \MixedBC\ dataset. The first two rows display data for the advection equation with time-dependent Dirichlet boundary conditions. The middle and final two rows illustrate time-dependent Neumann and periodic boundary conditions, respectively. The underlying PDE is identical across all datasets. They differ solely in their boundary conditions.}
    \label{fig:mixedbc}
\end{figure}

\textbf{Visualization: } We refer the readers to Figure~\ref{fig:mixedbc} for visualizations of the dataset.

\newpage

\subsection{\LongHorizon}
\label{appendix:longhorizon}
\textbf{Dataset Description: } This dataset contains solutions of the one-dimensional Kuramoto--Sivashinsky equation, a canonical model of spatiotemporal chaos arising in flame-front propagation and thin-film instabilities. The equation combines a destabilizing second-order diffusion term, a stabilizing fourth-order hyperviscosity term, and a nonlinear advection term; their interplay produces sustained chaotic dynamics with a continuum of interacting cellular structures. The field is discretized on a uniform periodic grid of $256$ points, and each trajectory spans $5000$ time steps.

\textbf{Modeling Objective: } Let $x_{t} \in \mathbb{R}^{256 \times 1}$ denote the system state at time $t$, where $256$ is the number of spatial grid points and the single channel dimension is the scalar field $u$. Our objective is to learn a parameterized neural simulator $\Psi$ such that $\hat{x}_{t+1} = \Psi(x_{t})$, where the predicted output $\hat{x}_{t+1}$ retains the same shape $(256, 1)$ as the input state. At evaluation time, the model is applied autoregressively for the full $5000$ steps, with each prediction fed back as the next input.

\textbf{Dataset Specific Challenges: } Long autoregressive rollouts amplify any small bias or high-frequency artifact in the one-step map. Over $5000$ steps, even tiny inconsistencies in the balance between the destabilizing and hyperviscous terms can accumulate into unbounded growth or artificial damping, so temporal stability is as critical as one-step accuracy.

\textbf{Dataset Generation: } We generate the Kuramoto--Sivashinsky dataset using Exponential Time Differencing Runge--Kutta (ETDRK) schemes on periodic domains. The method treats the linear operator (the second-order and fourth-order terms) spectrally and exactly in Fourier space, while the nonlinear advection term is evaluated pseudo-spectrally, yielding a stiffly-accurate, high-order-in-time integrator well suited to the stiffness introduced by the fourth-order hyperviscosity. We simulate on a periodic domain discretized with $256$ uniformly spaced points and advance with a fixed time step $\Delta t = 0.5$. Initial conditions are drawn from a random truncated Fourier series with wavenumber cutoff $5$, producing smooth random fields that the dynamics subsequently drive into the chaotic regime. We roll out $120$ trajectories of $5000$ steps each.

\textbf{Governing Physics: } The Kuramoto--Sivashinsky equation is a fourth-order nonlinear PDE modeling spatiotemporal chaos in settings such as laminar flame-front propagation and the dynamics of thin liquid films. On the periodic domain $\Omega = [0, L)$ it is written, in the gradient-norm form integrated here, as
\begin{equation}
    \frac{\partial u}{\partial t}
    + \frac{1}{2}\left| \frac{\partial u}{\partial x} \right|^{2}
    + \frac{\partial^{2} u}{\partial x^{2}}
    + \frac{\partial^{4} u}{\partial x^{4}}
    = 0,
    \qquad x \in \Omega,\ t > 0,
\end{equation}
where $u(x, t)$ is the scalar field. The dynamics are set by the competition among three terms: the second-order term $\partial^2 u / \partial x^2$ acts as a negative diffusion that injects energy and destabilizes long-wavelength modes; the fourth-order hyperviscous term $\partial^4 u / \partial x^4$ dissipates energy at short wavelengths and stabilizes the high-wavenumber modes; and the nonlinear term $\tfrac{1}{2}|\partial_x u|^2$ transfers energy across scales and couples the growing and decaying modes. This form is equivalent, up to a spatial derivative, to the more familiar advective form $u_t + u\,u_x + u_{xx} + u_{xxxx} = 0$: differentiating the equation above in $x$ and writing $v = \partial_x u$ recovers $v_t + v\,v_x + v_{xx} + v_{xxxx} = 0$. The balance between the destabilizing second-order term and the stabilizing fourth-order term selects a band of linearly unstable wavenumbers, and once the domain is large enough to contain many such modes, the nonlinearity sustains an aperiodic, chaotic energy cascade among the resulting cellular structures.

\textbf{Experiment Results: } We present the complete experimental results in Table~\ref{tab:longhorizon_full}. As the data indicates, the agent exhibits significantly worse performance in this setting.

\begin{table}[htbp]
\centering
\begin{tabular}{lcc}
\hline
 & RMSE & FRMSE \\
\hline
GNOT         & $5.674 \times 10^{0}$ & $5.862 \times 10^{1}$ \\
Transolver   & $4.787 \times 10^{0}$ & $4.350 \times 10^{1}$ \\
Transolver++ & $4.756 \times 10^{0}$ & $4.129 \times 10^{1}$ \\
\textbf{LNO} & $\mathbf{2.807 \times 10^{0}}$ & $\mathbf{3.569 \times 10^{1}}$ \\
AMG          & $7.088 \times 10^{0}$ & $1.091 \times 10^{2}$ \\
\textbf{Agent} & $1.781 \times 10^{1}$ & $2.757 \times 10^{2}$ \\
\hline
\end{tabular}
\caption{Performance comparison on the \LongHorizon\ dataset. Best results are shown in \textbf{bold}.}
\label{tab:longhorizon_full}
\end{table}

\begin{figure}[htbp]
    \centering
    \includegraphics[width=0.8\textwidth, page=1]{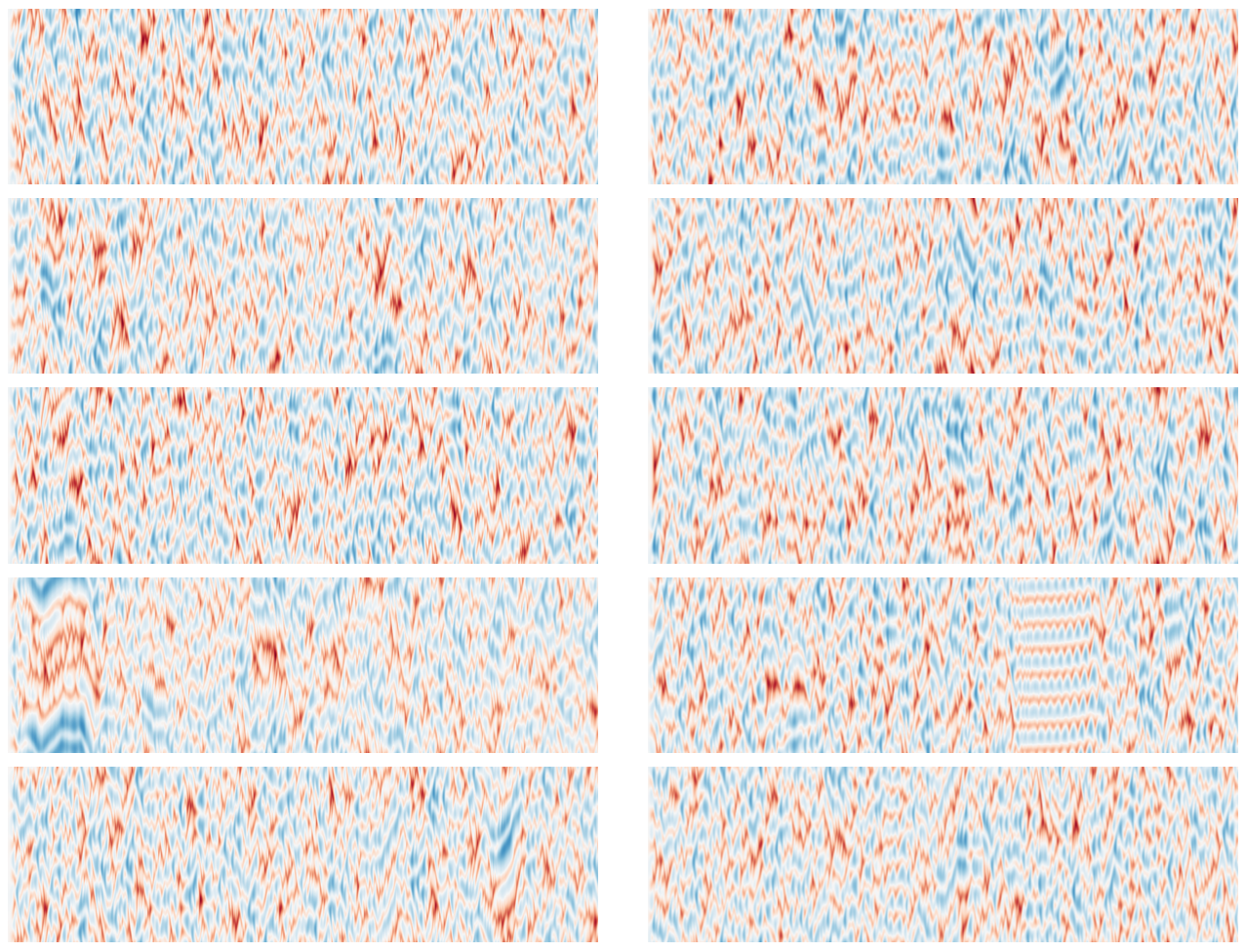}
    \caption{Visualization of the \MixedBC\ dataset. The dataset contains solutions to the 1D Kuramoto-Sivashinsky equation. For visualization purposes, the x-axis represents time and the y-axis represents the spatial discretization grid.}
    \label{fig:longhorizon}
\end{figure}

\textbf{Visualization: } We refer the readers to Figure~\ref{fig:longhorizon} for visualizations of the dataset.

\newpage

\subsection{\Stochastic}
\label{appendix:stochastic}
\textbf{Dataset Description: } This dataset contains two-dimensional stochastic $\Phi^4_2$ dynamics evolving on a $128 \times 128$ spatial grid. The state is driven by the interplay of diffusion, a double-well nonlinearity, and space-time white noise. Unlike standard benchmarks where the future state is fully determined by the current one, the evolution here depends on the particular noise path realized over the prediction window, and this driving noise is provided to the model as additional input channels.

\textbf{Modeling Objective: } Let $x_{t} \in \mathbb{R}^{21 \times 128 \times 128}$ denote the model input at time $t$, formed by concatenating the current state $u_{t} \in \mathbb{R}^{1 \times 128 \times 128}$ with the driving noise increments $W_{t}, \ldots, W_{t+19} \in \mathbb{R}^{20 \times 128 \times 128}$ over the next $20$ time steps along the channel dimension. Our objective is to learn a parameterized neural simulator $\Psi$ such that $\hat{x}_{t+20} = \Psi(x_{t})$, where the predicted output $\hat{x}_{t+20} \in \mathbb{R}^{1 \times 128 \times 128}$ is the state after the $20$-step horizon. Conditioned on the complete noise path contained in the input, the solution map is deterministic, so the task is well-posed as pathwise regression.

\textbf{Dataset Specific Challenges: } The target state depends on the entire realized noise path, not just the current state. The model must resolve how each of the $20$ stacked noise channels enters the dynamics at its respective time within the horizon and propagates through diffusion and the nonlinearity, effectively performing an implicit $20$-step stochastic integration in a single forward pass.

\textbf{Dataset Generation: } The $\Phi^4_2$ equation is discretized on a doubly-periodic $128 \times 128$ spatial grid over a square domain and advanced with an explicit finite-difference time-stepping scheme. The driving space-time white noise is synthesized on the grid from a truncated sine--cosine spectral basis: at each time slab an independent Gaussian field is drawn and band-limited at a spatial truncation scale $\epsilon$, which regularizes the otherwise distributional noise so the explicit scheme is well-defined. Because the $\Phi^4_2$ equation requires renormalization to admit a nontrivial continuum limit, we generate two versions of every trajectory: a renormalized solution, in which a mass counterterm depending on the truncation $\epsilon$ is subtracted to cancel the divergence induced by the noise, and a plain explicit solution without this correction. The initial condition is a fixed smooth sinusoidal field, $u_0(x, y) = \sin\big(2\pi(x + y)\big) + \cos\big(2\pi(x + y)\big)$, with an option to perturb it by a small random field for trajectory-to-trajectory variability.

\textbf{Governing Physics: } The dataset is governed by the two-dimensional stochastic quantization equation, the dynamic $\Phi^4_2$ model, a semi-linear stochastic PDE in which linear diffusion competes with a double-well nonlinearity while being driven by space-time white noise. On the periodic domain $\Omega = [a, b]^2$ it is written as
\begin{equation}
    \frac{\partial u}{\partial t}
    = \Delta u
    + \mu(u)
    + \sigma\, \xi(x, t),
    \qquad (x, t) \in \Omega \times (0, T],
\end{equation}
where $u(x, t)$ is the scalar field, $\Delta$ is the two-dimensional Laplacian, $\xi$ is space-time white noise (the distributional derivative of a cylindrical Wiener process), and $\sigma$ scales the noise amplitude. The reaction term is the negative gradient of a symmetric double-well potential,
\begin{equation}
    \mu(u) = 3u - u^{3}
    = -\frac{\mathrm{d} V}{\mathrm{d} u},
    \qquad
    V(u) = -\frac{3}{2} u^{2} + \frac{1}{4} u^{4},
\end{equation}
so that the deterministic part of the drift pushes the field toward the two stable phases $u = \pm\sqrt{3}$ while diffusion penalizes sharp gradients, and the white noise perpetually perturbs the field between and around these wells. In two spatial dimensions the white-noise forcing is sufficiently rough that the cubic nonlinearity is ill-defined for the resulting distributional solutions; a renormalization procedure is therefore required, replacing the nonlinearity by a Wick-ordered counterpart in which a formally infinite, truncation-dependent mass counterterm is subtracted:
\begin{equation}
    u^{3} \ \longrightarrow \ u^{3} - 3\, C_{\epsilon}\, u,
\end{equation}
where the constant $C_\epsilon$ diverges as the noise truncation scale $\epsilon \to 0$. This counterterm is precisely what distinguishes the renormalized trajectories from the plain explicit ones, and it is what renders the $\Phi^4_2$ dynamics well-posed as the discretization is refined.

\textbf{Experiment Results: } We present the experiment results in Table~\ref{tab:stochastic_full}.

\begin{table}[htbp]
\centering
\begin{tabular}{lcc}
\hline
 & RMSE & FRMSE \\
\hline
GNOT         & $3.990 \times 10^{-3}$ & $4.853 \times 10^{-1}$ \\
Transolver   & $3.689 \times 10^{-3}$ & $4.495 \times 10^{-1}$ \\
Transolver++ & $3.172 \times 10^{-3}$ & $3.818 \times 10^{-1}$ \\
LNO          & $5.447 \times 10^{-3}$ & $6.771 \times 10^{-1}$ \\
AMG          & $3.442 \times 10^{-3}$ & $4.141 \times 10^{-1}$ \\
\textbf{Agent} & $\bm{2.324 \times 10^{-3}}$ & $\bm{2.741 \times 10^{-1}}$ \\
\hline
\end{tabular}
\caption{Performance comparison on the \Stochastic\ dataset. Best results are shown in \textbf{bold}.}
\label{tab:stochastic_full}
\end{table}

\begin{figure}[htbp]
    \centering
    \includegraphics[width=0.8\textwidth, page=1]{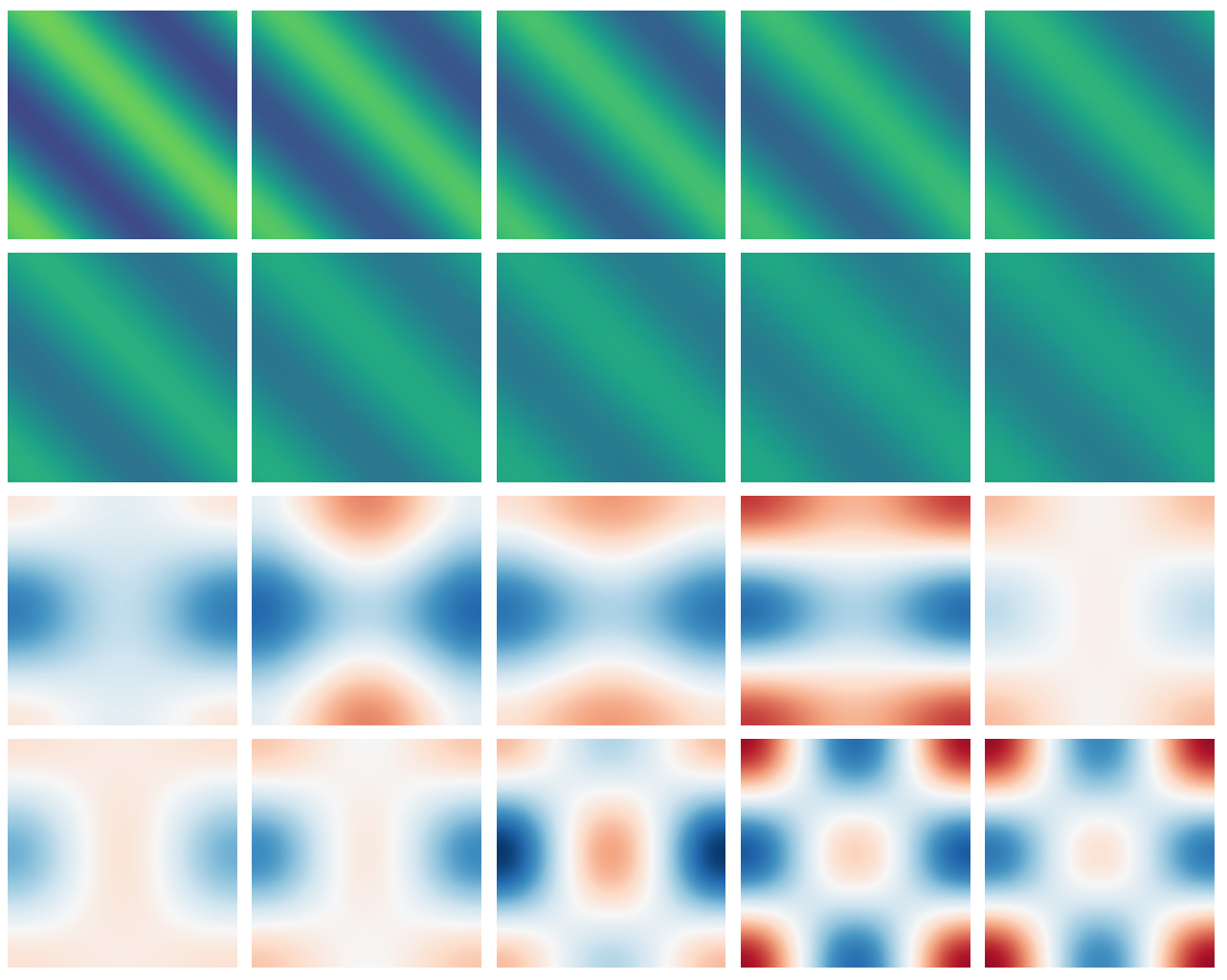}
    \caption{Visualization of the \Stochastic\ dataset. This dataset contains solutions to the dynamical $\Phi^4_2$ stochastic PDE. The first two rows display the renormalized solutions. The final two rows display the corresponding space-time white noise used to drive the system.}
    \label{fig:stochastic}
\end{figure}

\textbf{Visualization: } We refer the readers to Figure~\ref{fig:stochastic} for visualizations of the dataset.

\newpage

\subsection{\MultiPhase}
\label{appendix:multiphase}
\textbf{Dataset Description: } This dataset contains two-dimensional multi-phase Lagrangian particle simulations of liquid and gas confined in a closed box and driven by an external sloshing motion. Both phases are discretized as particles, labeled by a type indicator distinguishing liquid from gas, and per-type velocity statistics are used for normalization, reflecting the markedly different densities and motion scales of the two phases. The container undergoes a prescribed time-dependent excitation, described at each frame by linear and angular accelerations. Unlike single-phase Lagrangian benchmarks, the dynamics here arise from the interaction of two immiscible phases with a sharp, deforming interface, agitated by non-inertial forcing from the moving container.

\textbf{Modeling Objective: } Let $x_{t} \in \mathbb{R}^{N \times 55}$ denote the state of the $N$ particles at time $t$, where each particle's $55$-dimensional feature vector concatenates its current position in $\mathbb{R}^{2}$, a history of its $5$ most recent velocities in $\mathbb{R}^{10}$ obtained by finite-differencing consecutive positions over the preceding frames, a particle type in $\{0, 1\}$ indicating liquid or gas, and a sloshing feature in $\mathbb{R}^{42}$ stacking the $6$-dimensional container accelerations over a $7$-frame window covering the past $6$ frames and the upcoming frame, broadcast identically to all particles. The number of particles $N$ varies across samples, so batching is handled with a graph data loader that concatenates variable-sized particle sets. Our objective is to develop a neural simulator, defined as a parameterized function $\Psi$, that advances the system state by one time step. Formally, we aim to predict $\hat{v}_{t+1} = \Psi(x_{t})$, where $\hat{v}_{t+1} \in \mathbb{R}^{N \times 2}$ is the per-particle, per-type normalized velocity over the next time step, from which positions are advanced during rollout.

\textbf{Dataset Specific Challenges: } The two phases carry strongly contrasting physical properties: the dense liquid dominates the momentum of the system while the light gas is easily displaced, and their interaction is mediated by a sharp interface that folds, breaks, and reconnects under sloshing. A single model must capture both phase behaviors and their coupling across the interface, using only the particle type label to differentiate them, while keeping the two phases from unphysically mixing over long rollouts.

\textbf{Dataset Generation: } This dataset is generated using a weakly-compressible Smoothed Particle Hydrodynamics (SPH) solver~\citep{ZHAN2025109389, DOMINGUEZ2022867}. The simulation uses a particle spacing of $dp = 0.006$~m with a Wendland kernel and a smoothing length of $h = 0.91924\sqrt{2}\,dp$. Time integration employs a Symplectic scheme with a CFL number of $0.1$, an initial timestep of $10^{-5}$~s, and a minimum timestep of $10^{-7}$~s, running to a physical time of $60$~s with output written every $0.005$~s. Multi-phase flow is enabled with the liquid phase modeled as water ($\rho_0 = 1000$~kg/m$^3$, $\gamma = 7$, $c_0 = 20$~m/s) and the gas phase as air ($\rho_0 = 1.18$~kg/m$^3$, $\gamma = 1.4$, $c_0 = 150$~m/s), each carrying a viscosity of $0.01$. Viscosity is treated with the artificial formulation, and the Colagrossi--Landrini multiphase surface-tension coefficient is set to $0.1$. A Molteni density diffusion term (coefficient $0.1$) and particle shifting (coefficient $-2$, free-surface threshold $1.5$) are applied to suppress numerical noise. The tank walls are fixed boundaries, and the sloshing motion is driven by a time-dependent acceleration field applied to the fluid, read from an external data file about an acceleration center at $(0.45, 0, 0)$~m with global gravity disabled for the forcing term. The domain is two-dimensional, spanning approximately $[-0.55, 0.55]$~m horizontally and $[-0.118, 0.62]$~m vertically, initialized with a lower water layer beneath an air region and containing $12{,}688$ total particles.

\textbf{Governing Physics: } The sloshing case is governed by the two-phase weakly-compressible Navier--Stokes equations, in which two immiscible fluids---a heavy liquid (water) and a light gas (air)---coexist and interact across a sharp, deforming interface. Each phase $k \in \{\text{water}, \text{air}\}$ independently satisfies mass and momentum conservation, solved in Lagrangian form,
\begin{equation}
    \frac{D\rho_k}{Dt} = -\rho_k\,\nabla\cdot\mathbf{u},
    \qquad
    \frac{D\mathbf{u}}{Dt}
    = -\frac{1}{\rho_k}\nabla p_k
    + \nu_k\,\nabla^2\mathbf{u}
    + \mathbf{g},
\end{equation}
where $\rho_k$, $p_k$, and $\nu_k$ are the density, pressure, and kinematic viscosity of phase $k$, $\mathbf{u}$ is the velocity, and $\mathbf{g}$ is the body force (here the prescribed tank acceleration). The material derivative $D/Dt = \partial/\partial t + \mathbf{u}\cdot\nabla$ reflects the Lagrangian description, in which particles carry the properties of their own phase as they move; because each SPH particle is permanently tagged as either water or air, the interface is represented implicitly by the boundary between the two particle populations and requires no explicit interface tracking or reconstruction. Rather than enforcing strict incompressibility, each phase is treated as weakly compressible and closed by its own Tait equation of state,
\begin{equation}
    p_k = \frac{c_{0,k}^2\,\rho_{0,k}}{\gamma_k}
    \left[
        \left(\frac{\rho_k}{\rho_{0,k}}\right)^{\gamma_k} - 1
    \right],
\end{equation}
where $\rho_{0,k}$ is the reference density, $c_{0,k}$ the artificial speed of sound, and $\gamma_k$ the polytropic exponent \emph{specific to that phase}. The defining feature of this dataset is the stark contrast between the two phases: the water phase uses $(\rho_{0}, \gamma, c_0) = (1000, 7, 20)$ while the air phase uses $(1.18, 1.4, 150)$, an $\mathcal{O}(10^3)$ density ratio and markedly different stiffness. This jump makes the pressure and density fields discontinuous across the interface, and the momentum coupling between the dense liquid and the compliant gas is what produces the buoyancy-driven sloshing, wave breaking, air entrainment, and splashing that characterize liquid--gas free-surface flows. To keep the two populations from artificially interpenetrating and to stabilize the pressure evaluation near this high-contrast interface, a multiphase surface correction is applied, and each phase's speed of sound $c_{0,k}$ is chosen large enough that its density fluctuations stay below roughly $1\%$, approximating incompressible behavior within each fluid while keeping the interfacial dynamics explicit.

\textbf{Experiment Results: } We present the complete experimental results in Table~\ref{tab:multiphase_full}. While the main text reports the average RMSE of the liquid and gas phases, here we provide additional metrics disaggregated by phase to further analyze performance. We observe that the proposed agent architecture achieves superior results across all scenarios, demonstrating its overall effectiveness.

\begin{table}[h]
\centering
\begin{tabular}{lcccc}
\hline
 & RMSE (liquid) & LagKE (liquid) & Sinkhorn (liquid) & RMSE (gas) \\
\hline
GNOT        & $5.233 \times 10^{-4}$ & $6.509 \times 10^{-4}$ & $1.473 \times 10^{-8}$ & $1.123 \times 10^{-3}$ \\
Transolver  & $5.787 \times 10^{-4}$ & $1.189 \times 10^{-3}$ & $4.566 \times 10^{-8}$ & $1.254 \times 10^{-3}$ \\
Transolver+ & $5.471 \times 10^{-4}$ & $9.160 \times 10^{-4}$ & $2.803 \times 10^{-8}$ & $1.215 \times 10^{-3}$ \\
LNO         & $5.496 \times 10^{-4}$ & $7.371 \times 10^{-4}$ & $8.130 \times 10^{-9}$ & $1.091 \times 10^{-3}$ \\
AMG         & $5.500 \times 10^{-4}$ & $1.171 \times 10^{-3}$ & $3.026 \times 10^{-8}$ & $1.076 \times 10^{-3}$ \\
\textbf{Agent} & $\bm{4.722 \times 10^{-4}}$ & $\bm{5.109 \times 10^{-4}}$ & $\bm{3.312 \times 10^{-9}}$ & $\bm{1.044 \times 10^{-3}}$ \\
\hline
\multicolumn{5}{c}{} \\
\hline
 & LagKE (gas) & Sinkhorn (gas) & EuARMS & EuAFRMSE \\
\hline
GNOT        & $9.569 \times 10^{-3}$ & $2.184 \times 10^{-8}$ & $3.617 \times 10^{-4}$ & $6.013 \times 10^{-3}$ \\
Transolver  & $8.639 \times 10^{-3}$ & $5.797 \times 10^{-8}$ & $4.461 \times 10^{-4}$ & $7.180 \times 10^{-3}$ \\
Transolver+ & $1.051 \times 10^{-2}$ & $4.384 \times 10^{-8}$ & $4.135 \times 10^{-4}$ & $6.866 \times 10^{-3}$ \\
LNO         & $\bm{1.001 \times 10^{-2}}$ & $2.446 \times 10^{-8}$ & $3.613 \times 10^{-4}$ & $6.041 \times 10^{-3}$ \\
AMG         & $1.073 \times 10^{-2}$ & $2.058 \times 10^{-8}$ & $3.401 \times 10^{-4}$ & $5.838 \times 10^{-3}$ \\
\textbf{Agent} & $1.070 \times 10^{-2}$ & $\bm{1.711 \times 10^{-8}}$ & $\bm{3.066 \times 10^{-4}}$ & $\bm{5.264 \times 10^{-3}}$ \\
\hline
\end{tabular}
\caption{Performance comparison on the \MultiPhase\ dataset. Best results are shown in \textbf{bold}.}
\label{tab:multiphase_full}
\end{table}

\begin{figure}[htbp]
    \centering
    \includegraphics[width=0.8\textwidth, page=1]{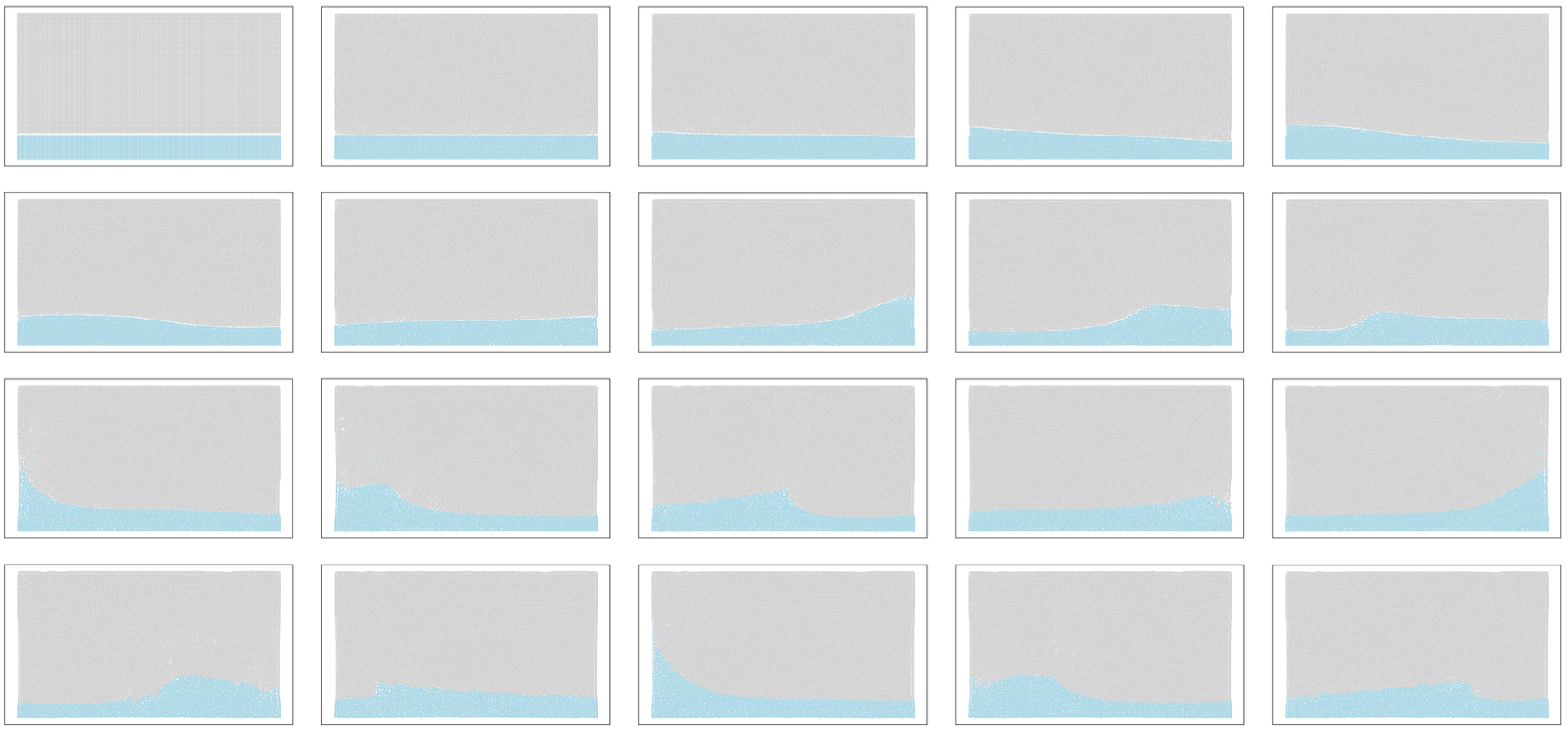}
    \caption{Visualization of the \MultiPhase\ dataset. This dataset features Lagrangian particle data modeling multiphase liquid and gas dynamics within a closed container.}
    \label{fig:multiphase}
\end{figure}

\textbf{Visualization: } We refer the readers to Figure~\ref{fig:multiphase} for visualizations of the dataset.

\newpage

\subsection{\MixedPDE}
\label{appendix:mixedpde}
\textbf{Dataset Description: } This dataset is a mixture of four two-dimensional PDE systems, each defined on a $128 \times 128$ spatial grid: advection-diffusion, Burgers' equation, Gray-Scott reaction-diffusion, and incompressible Navier-Stokes in vorticity form. The systems span qualitatively different dynamics, from linear transport and diffusive smoothing to nonlinear shock formation, pattern-forming chemical reactions, and turbulent vortex interactions. They also differ in their state dimensionality: advection-diffusion and Gray-Scott have a single physical channel, while Burgers' and Navier-Stokes have two.

\textbf{Modeling Objective: } Let $x_{t} \in \mathbb{R}^{C \times 128 \times 128}$ denote the state of the system at time $t$, where the number of channels $C$ depends on the underlying PDE ($C = 1$ for advection-diffusion and Gray-Scott, $C = 2$ for Burgers' and Navier-Stokes). The model additionally receives a PDE identifier $\mathrm{pde} \in \{0, 1, 2, 3\}$, indicating which of the four systems the trajectory belongs to ($0$: advection\_diffusion\_2d, $1$: burgers\_2d, $2$: gray\_scott\_2d, $3$: navier\_stokes\_vorticity\_2d). Our objective is to develop a single neural simulator, defined as a parameterized function $\Psi$, that advances the system state by one time step. Formally, we aim to predict $\hat{x}_{t+1} = \Psi(x_{t}, \mathrm{pde})$, where the predicted output $\hat{x}_{t+1}$ retains the same shape $(C, 128, 128)$ as the input state.

\textbf{Dataset Specific Challenges: } A single model must represent four operators with fundamentally different governing physics, characteristic time scales, and channel semantics. Conditioning on the PDE identifier requires the solver to adapt its learned dynamics rather than average across systems, which typically degrades accuracy on all of them. Compounding this, the input and output channel dimensions vary across PDEs, so the model cannot rely on a fixed-size input/output projection; the architecture must handle variable channel counts natively, and batching becomes nontrivial since samples from different PDEs have incompatible tensor shapes.

\textbf{Dataset Generation: } These datasets are generated using Exponential Time Differencing (ETD) Runge--Kutta spectral schemes. All fields are discretized on a $128 \times 128$ doubly-periodic grid and advanced with a fixed timestep, where the linear terms are integrated exactly in Fourier space and the nonlinear terms are treated with an ETDRK time-stepping scheme. For each PDE we generate $30$ trajectories of $100$ steps each, producing arrays of shape $(30, 101, C, 128, 128)$ where $C$ is the number of physical channels. Initial conditions are drawn from a random truncated Fourier series (independent per channel, with a low wavenumber cutoff and amplitude normalized to $[-1, 1]$), except for the Gray--Scott system, which is initialized with $u \approx 1$ and $v \approx 0$ perturbed by up to eight randomly placed Gaussian blobs (dipping $u$ toward $0.5$ and raising $v$ toward $0.25$) plus small symmetry-breaking noise. Per-PDE solver settings are: Advection--Diffusion (domain extent $1.0$, $\Delta t = 0.01$, velocity $(0.4, 1.0)$, diffusivity $10^{-3}$, single channel); Burgers (domain extent $1.0$, $\Delta t = 0.01$, diffusivity $10^{-2}$, two velocity channels); Navier--Stokes in vorticity form (domain extent $1.0$, $\Delta t = 0.01$, diffusivity $3\times10^{-4}$, single scalar-vorticity channel, with $10$ inner substeps per saved frame); and Gray--Scott (domain extent $1.0$, $\Delta t = 1.0$, diffusivities $2\times10^{-5}$ and $10^{-5}$, feed rate $0.04$, kill rate $0.06$, two species channels, with $50$ inner substeps per saved frame).

\textbf{Governing Physics: } The four datasets are governed by distinct semi-linear PDEs on the two-dimensional periodic domain, each written in the form $\partial_t \mathbf{u} = \mathcal{L}\mathbf{u} + \mathcal{N}(\mathbf{u})$ with a linear operator $\mathcal{L}$ integrated exactly in spectral space and a nonlinear term $\mathcal{N}$. The advection--diffusion equation transports a passive scalar $u$ at constant velocity $\mathbf{c}$ while it diffuses,
\begin{equation}
    \frac{\partial u}{\partial t}
    + \mathbf{c}\cdot\nabla u
    = \nu\,\Delta u,
\end{equation}
where $\nu$ is the diffusivity; this case is purely linear. The Burgers equation adds nonlinear self-advection to the same diffusive structure, here for a vector velocity field $\mathbf{u}$,
\begin{equation}
    \frac{\partial \mathbf{u}}{\partial t}
    + \tfrac{1}{2}\nabla\cdot(\mathbf{u}\otimes\mathbf{u})
    = \nu\,\Delta \mathbf{u},
\end{equation}
whose quadratic convection steepens gradients into sharp fronts balanced by viscous dissipation. The incompressible Navier--Stokes equations are solved in the vorticity formulation, where the scalar vorticity $\omega = \nabla \times \mathbf{u}$ is advected by the velocity field it induces,
\begin{equation}
    \frac{\partial \omega}{\partial t}
    + (\mathbf{u}\cdot\nabla)\,\omega
    = \nu\,\Delta \omega,
    \qquad
    \mathbf{u} = \nabla^{\perp}\Delta^{-1}\omega,
\end{equation}
with the velocity recovered from vorticity through a streamfunction, producing two-dimensional turbulence. Finally, the Gray--Scott model is a two-species reaction--diffusion system in which chemicals $u$ and $v$ diffuse at different rates and react nonlinearly,
\begin{equation}
    \frac{\partial u}{\partial t}
    = D_u\,\Delta u - u v^2 + F(1 - u),
    \qquad
    \frac{\partial v}{\partial t}
    = D_v\,\Delta v + u v^2 - (F + \kappa)\,v,
\end{equation}
where $D_u, D_v$ are the species diffusivities, $F$ the feed rate, and $\kappa$ the kill rate; the cubic autocatalytic coupling $u v^2$ together with differential diffusion drives the spontaneous formation of spots, stripes, and self-replicating patterns. Across all four systems the exact spectral treatment of the linear operator $\mathcal{L}$ and periodic boundaries are what make the ETDRK integrator stable and accurate at the chosen timesteps.

\textbf{Experiment Results: } We present the complete experimental results in Table~\ref{tab:mixedpde_full}, where we additionally report the individual RMSE for each of the four PDEs. The results demonstrate that the agent proposed architecture achieves strong performance across all evaluated aspects.

\begin{table}[h]
\centering
\begin{tabular}{lcccc}
\hline
 & RMSE & FRMSE & RMSE (Adv Diff) & RMSE (Burgers) \\
\hline
GNOT         & $8.792 \times 10^{-2}$ & $6.037 \times 10^{0}$ & $1.944 \times 10^{-1}$ & $2.214 \times 10^{-2}$ \\
Transolver   & $7.565 \times 10^{-2}$ & $4.685 \times 10^{0}$ & $1.745 \times 10^{-1}$ & $2.194 \times 10^{-2}$ \\
Transolver++ & $7.535 \times 10^{-2}$ & $4.352 \times 10^{0}$ & $1.736 \times 10^{-1}$ & $2.131 \times 10^{-2}$ \\
LNO          & $7.684 \times 10^{-2}$ & $4.037 \times 10^{0}$ & $1.761 \times 10^{-1}$ & $2.199 \times 10^{-2}$ \\
AMG          & $8.845 \times 10^{-2}$ & $6.221 \times 10^{0}$ & $1.922 \times 10^{-1}$ & $2.230 \times 10^{-2}$ \\
\textbf{Agent} & $\mathbf{2.701 \times 10^{-2}}$ & $\mathbf{2.264 \times 10^{0}}$ & $\mathbf{2.256 \times 10^{-3}}$ & $\mathbf{6.041 \times 10^{-3}}$ \\
\hline
\multicolumn{5}{c}{} \\
\hline
 & RMSE (Gray Scott) & RMSE (NS) \\
\hline
GNOT         & $4.709 \times 10^{-2}$ & $5.193 \times 10^{-2}$ \\
Transolver   & $4.515 \times 10^{-2}$ & $2.119 \times 10^{-2}$ \\
Transolver++ & $4.562 \times 10^{-2}$ & $2.174 \times 10^{-2}$ \\
LNO          & $5.238 \times 10^{-2}$ & $1.907 \times 10^{-2}$ \\
AMG          & $6.176 \times 10^{-2}$ & $5.106 \times 10^{-2}$ \\
\textbf{Agent} & $\mathbf{4.630 \times 10^{-2}}$ & $\mathbf{3.797 \times 10^{-3}}$ \\
\hline
\end{tabular}
\caption{Performance comparison on the \MixedPDE\ dataset. Best results are shown in \textbf{bold}.}
\label{tab:mixedpde_full}
\end{table}

\begin{figure}[htbp]
    \centering
    \includegraphics[width=0.8\textwidth, page=1]{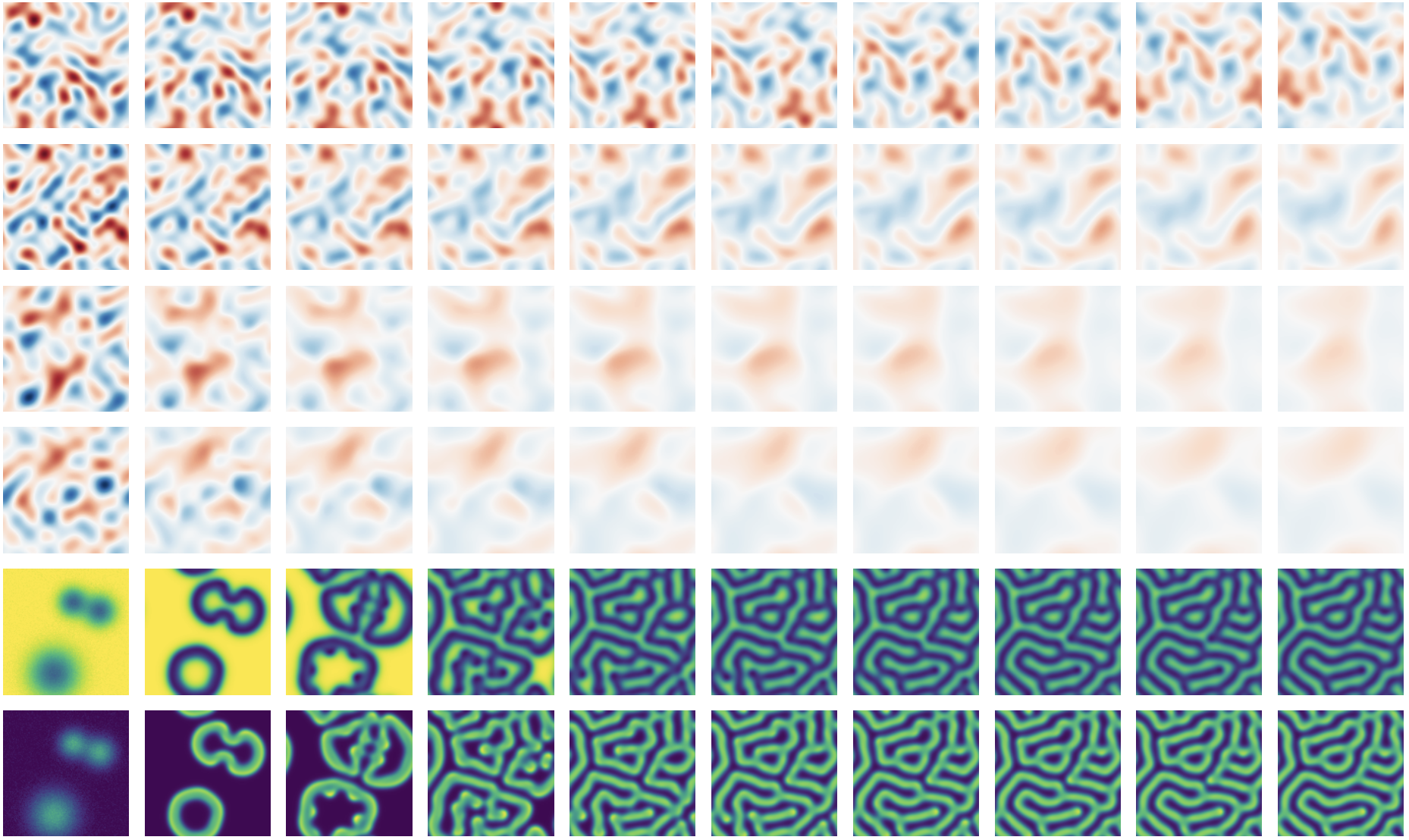}
    \caption{Visualization of the \MixedPDE\ dataset. This dataset contains a mixture of 2D PDEs. The first row corresponds to advection-diffusion, while the second row displays Navier-Stokes vorticity. The third and fourth rows represent the two channels of the Burgers' equation, and the final two rows depict the two channels of the Gray-Scott model.}
    \label{fig:mixedpde}
\end{figure}

\textbf{Visualization: } We refer the readers to Figure~\ref{fig:mixedpde} for visualizations of the dataset.

\newpage

\subsection{\LargeLag}
\label{appendix:largelag}
\textbf{Dataset Description: } This dataset contains large-scale Lagrangian simulations of flow around a cylinder. Both the fluid and the solid cylinder are discretized into small particles, with more than 2 million particles per frame. The simulation is two-dimensional, and the fluid domain is open: particles enter at the inlet and leave at the outlet.

\textbf{Modeling Objective: } Let $x_{t} \in \mathbb{R}^{N \times 13}$ denote the state of the system at time $t$, where $N$ is the number of particles in the current window. The $13$ input features per particle are the last-step position ($2$), the historical velocities over the previous $5$ steps ($2 \times 5 = 10$), and a scalar particle-type flag distinguishing fluid from solid ($1$). Our objective is to develop a neural simulator, defined as a parameterized function $\Psi$, that advances the system state by one time step. Formally, we aim to predict $\hat{x}_{t+1} = \Psi(x_{t})$, where $\hat{x}_{t+1} \in \mathbb{R}^{N \times 2}$ is the per-particle velocity at the next time step. Solid particles have zero velocity by construction and are excluded from the loss, as are fluid particles that have left the computational domain by the target frame.

\textbf{Dataset Specific Challenges: } The sheer scale of the particle set makes memory the dominant constraint. Any model must be memory efficient, since methods whose cost grows quadratically with the number of particles, such as dense global attention or dense neighborhood graphs, are infeasible at this resolution.

\textbf{Dataset Generation: } This dataset is generated using a weakly-compressible Smoothed Particle Hydrodynamics (SPH) solver~\citep{ZHAN2025109389, DOMINGUEZ2022867}, in a two-dimensional open-domain configuration with inflow/outflow boundaries. The simulation uses a particle spacing of $dp = 0.0015$~m with a Wendland kernel and a smoothing length of $h = 2\,dp$. Time integration employs a Symplectic scheme with a CFL number of $0.2$, running to a physical time of $10$~s with output written every $0.005$~s. The fluid has reference density $\rho_0 = 1000$~kg/m$^3$, $\gamma = 7$, and an artificial speed of sound of $15$~m/s. Viscosity is treated with the Laminar+SPS formulation at a kinematic viscosity of $4\times10^{-5}$~m$^2$/s. A Fourtakas density diffusion term and full particle shifting are applied to suppress numerical noise, and boundaries use the modified Dynamic Boundary Condition (mDBC) with corrected wall normals. A uniform inflow of $1.0$~m/s is imposed at the left open boundary (with entering fluid removed and re-created over four buffer layers), while the right open boundary acts as an outlet that converts fluid and extrapolates density from ghost nodes. The channel is bounded above and below by solid walls translating with the flow, and a rigid cylinder of radius $0.1$~m is placed in the stream as the bluff body. The domain spans roughly $[-0.6, 2.1]$~m streamwise and $[-1.03, 1.03]$~m cross-stream, discretized into approximately $2.31$ million total particles.

\textbf{Governing Physics: } The flow past the cylinder is governed by the weakly-compressible Navier--Stokes equations, solved in Lagrangian form where the fluid satisfies mass and momentum conservation,
\begin{equation}
    \frac{D\rho}{Dt} = -\rho\,\nabla\cdot\mathbf{u},
    \qquad
    \frac{D\mathbf{u}}{Dt}
    = -\frac{1}{\rho}\nabla p
    + \nu\,\nabla^2\mathbf{u},
\end{equation}
where $\rho$ is density, $\mathbf{u}$ the velocity, $p$ the pressure, and $\nu$ the kinematic viscosity; gravity is absent, so the dynamics are driven purely by the imposed free stream and its interaction with the obstacle. The material derivative $D/Dt = \partial/\partial t + \mathbf{u}\cdot\nabla$ reflects the Lagrangian description, in which particles carry the fluid properties as they move. Rather than enforcing strict incompressibility, the fluid is treated as weakly compressible and closed by the Tait equation of state,
\begin{equation}
    p = \frac{c_0^2\,\rho_0}{\gamma}
    \left[
        \left(\frac{\rho}{\rho_0}\right)^{\gamma} - 1
    \right],
\end{equation}
where $\rho_0$ is the reference density, $c_0$ the artificial speed of sound, and $\gamma$ the polytropic exponent; $c_0$ is chosen large enough that density fluctuations stay below roughly $1\%$, approximating incompressible behavior while keeping the pressure evaluation explicit. The defining feature of this case is the rigid circular cylinder of diameter $D$ immersed in the uniform stream $U_\infty$: the cylinder enforces a no-slip, no-penetration condition on its surface, forcing the incoming flow to separate from the curved wall.

\textbf{Experiment Results: } We present the complete experimental results in Table~\ref{tab:largelag_full}. We omit the Sinkhorn divergence and Eulerian aggregated metrics, as their computation becomes prohibitively expensive for a system of two million particles.

\begin{table}[h]
\centering
\begin{tabular}{lcc}
\hline
 & RMSE & Kinetic Energy \\
\hline
GNOT         & $1.300 \times 10^{-2}$ & $9.454 \times 10^{3}$ \\
Transolver   & $2.753 \times 10^{-2}$ & $8.507 \times 10^{4}$ \\
Transolver++ & $2.027 \times 10^{-2}$ & $4.624 \times 10^{4}$ \\
LNO          & $1.599 \times 10^{-1}$ & $7.085 \times 10^{5}$ \\
AMG          & -- & -- \\
\textbf{Agent} & $\mathbf{5.666 \times 10^{-3}}$ & $\mathbf{7.876 \times 10^{3}}$ \\
\hline
\end{tabular}
\caption{Performance comparison on the \LargeLag\ dataset. Best results are shown in \textbf{bold}.}
\label{tab:largelag_full}
\end{table}

\begin{figure}[htbp]
    \centering
    \includegraphics[width=0.8\textwidth, page=1]{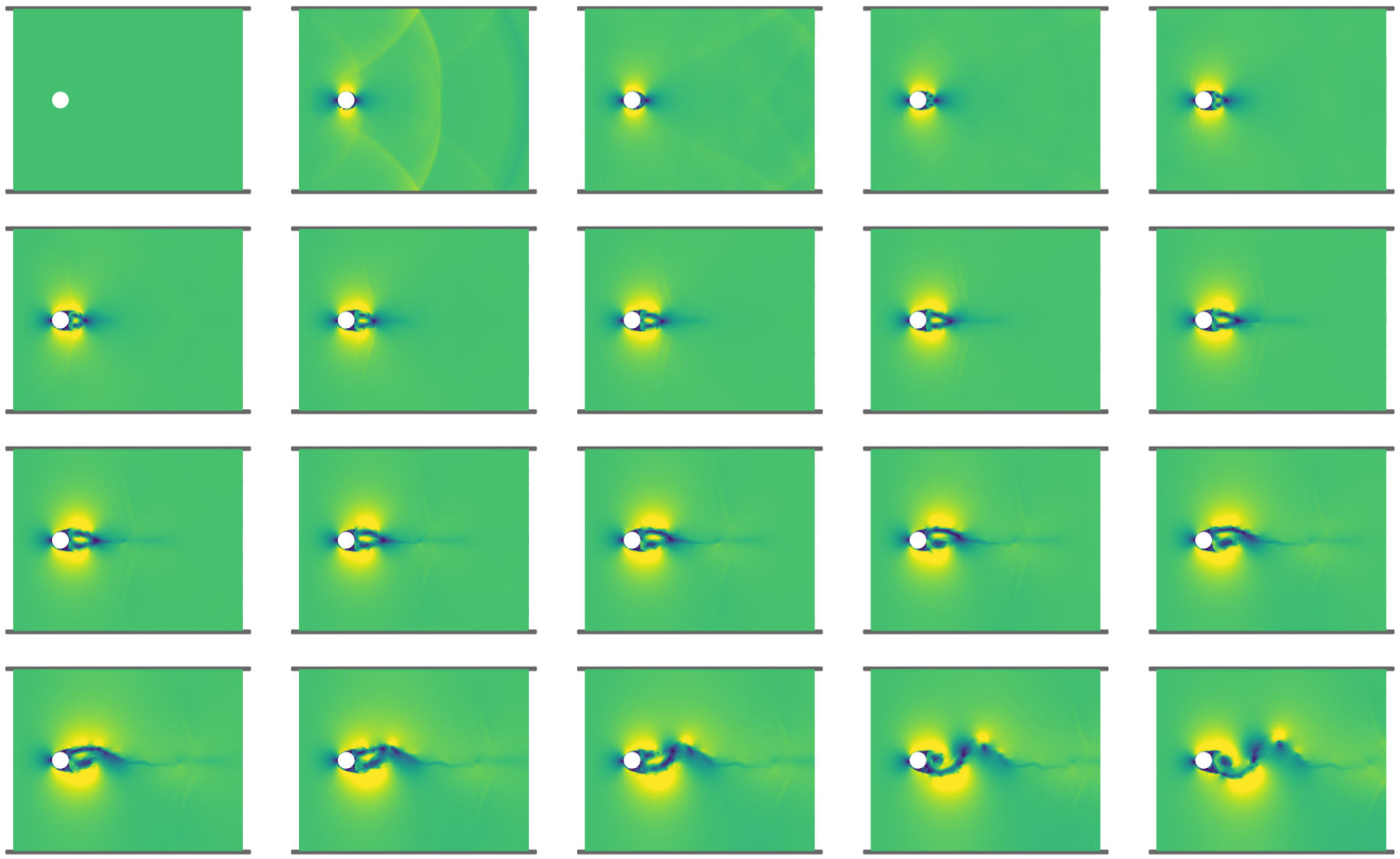}
    \caption{Visualization of the \LargeLag\ dataset. This dataset contains a high-fidelity, large-scale Lagrangian simulation of flow around a cylinder with over 2 million particles.}
    \label{fig:largelag}
\end{figure}

\textbf{Visualization: } We refer the readers to Figure~\ref{fig:largelag} for visualizations of the dataset.

\newpage
\clearpage

\subsection{\SimToReal}
\label{appendix:sim2real}
\textbf{Dataset Description: } This dataset contains fluid dynamics of vortex-induced vibration in a tandem-cylinder configuration. Two cylinders are arranged in tandem, one positioned downstream of the other, so that the wake shed by the upstream cylinder impinges on the downstream body. This coupling produces complex vortex-shedding dynamics and flow-induced structural vibration that neither cylinder would exhibit in isolation.

\textbf{Modeling Objective: } Let $x_{t} \in \mathbb{R}^{2 \times 128 \times 128}$ denote the state of the system at time $t$. The two feature channels represent the horizontal velocity and vertical velocity fields, each evaluated on a $128 \times 128$ spatial discretization grid. Our objective is to develop a neural simulator, defined as a parameterized function $\Psi$, that advances the system state by one time step. Formally, we aim to predict $\hat{x}_{t+1} = \Psi(x_{t})$, where the predicted output $\hat{x}_{t+1}$ retains the same spatial and feature dimensions $(2, 128, 128)$ as the input state.

\textbf{Dataset Specific Challenges: } This dataset follows a simulation-to-real (sim2real) setup: the model is trained on simulated data but evaluated on real-world measurements. The distribution shift between simulated and real flow fields means that a model fitting the training data well may still generalize poorly at test time. Moreover, real-world data carries measurement noise and unmodeled physical effects absent from the simulation, so a neural simulator must capture the underlying flow dynamics rather than simulation-specific artifacts in order to remain accurate and temporally stable when rolled out on real data.

\textbf{Dataset Generation: } We adopt the fluid--structure interaction (FSI) dataset from \citet{hu2026realpdebench}, using its paired numerical trajectories. The simulated data are produced with a two-dimensional incompressible finite-volume flow solver on a uniform Cartesian grid that represents the immersed cylinder through the Boundary Data Immersion Method (BDIM), a second-order immersed-boundary treatment that avoids body-fitted remeshing as the structure moves. The rigid cylinder is not held fixed but is elastically mounted: it is free to vibrate under the instantaneous fluid loading, and its structural response is advanced by a spring--mass--damper equation integrated in lockstep with the flow solver, so that the fluid and body states are updated in a two-way-coupled manner at every timestep.

\textbf{Governing Physics: } The FSI case is governed by a two-way coupling between an incompressible fluid and an elastically mounted rigid cylinder. The fluid obeys the incompressible Navier--Stokes equations,
\begin{equation}
    \nabla\cdot\mathbf{u} = 0,
    \qquad
    \frac{\partial \mathbf{u}}{\partial t}
    + (\mathbf{u}\cdot\nabla)\mathbf{u}
    = -\frac{1}{\rho}\nabla p
    + \nu\,\nabla^2\mathbf{u},
\end{equation}
with velocity $\mathbf{u}$, pressure $p$, density $\rho$, and kinematic viscosity $\nu$. The distinguishing element is the cylinder itself: unlike the stationary-cylinder wake, here the circular cylinder is a movable structure suspended on a spring and damper, and its center displacement $\mathbf{x}_s(t)$ evolves according to a forced oscillator (spring--mass--damper) equation,
\begin{equation}
    m\,\ddot{\mathbf{x}}_s
    + c\,\dot{\mathbf{x}}_s
    + k\,\mathbf{x}_s
    = \mathbf{F}(t),
\end{equation}
where $m$, $c$, and $k$ are the oscillating mass, damping, and stiffness of the mount, and $\mathbf{F}(t)$ is the hydrodynamic force the flow exerts on the cylinder, obtained by integrating the pressure and viscous stress over its wetted surface $\Gamma_{\text{cyl}}$,
\begin{equation}
    \mathbf{F}(t)
    = \oint_{\Gamma_{\text{cyl}}}
    \left[
        -p\,\mathbf{I}
        + \mu\left(\nabla\mathbf{u} + \nabla\mathbf{u}^{\top}\right)
    \right]\cdot\mathbf{n}\,\mathrm{d}S.
\end{equation}
The two systems are locked together by the no-slip, no-penetration condition on the moving cylinder wall,
\begin{equation}
    \mathbf{u} = \dot{\mathbf{x}}_s
    \quad \text{on } \Gamma_{\text{cyl}},
\end{equation}
so the cylinder both stirs the fluid and is pushed by it. The character of this coupled response is fixed by three dimensionless groups built around the cylinder: the Reynolds number $\mathrm{Re} = U D / \nu$ (with $D$ the cylinder diameter), the mass ratio $m^{*} = m / (\rho\,\tfrac{\pi}{4} D^{2})$ comparing structural to displaced-fluid inertia, and the damping ratio $\zeta = c / (2\sqrt{k m})$.

\textbf{Experiment Results: } We present the complete experimental results in Table~\ref{tab:sim2real_full}. The findings indicate that the agent proposed architecture exhibits poor performance in this setting.

\begin{table}[h]
\centering
\begin{tabular}{lccc}
\hline
 & RMSE & FRMSE & Kinetic Energy \\
\hline
GNOT         & $1.987 \times 10^{-3}$ & $1.739 \times 10^{-1}$ & $1.474 \times 10^{-4}$ \\
Transolver   & $\bm{1.783 \times 10^{-3}}$ & $\bm{1.535 \times 10^{-1}}$ & $1.432 \times 10^{-4}$ \\
Transolver++ & $1.809 \times 10^{-3}$ & $1.568 \times 10^{-1}$ & $\bm{1.430 \times 10^{-4}}$ \\
LNO          & $8.306 \times 10^{-3}$ & $9.829 \times 10^{-1}$ & $1.035 \times 10^{-3}$ \\
AMG          & $2.705 \times 10^{-3}$ & $2.424 \times 10^{-1}$ & $2.257 \times 10^{-4}$ \\
\textbf{Agent} & $6.734 \times 10^{-3}$ & $6.740 \times 10^{-1}$ & $5.157 \times 10^{-4}$ \\
\hline
\end{tabular}
\caption{Performance comparison on the \SimToReal\ dataset. Best results are shown in \textbf{bold}.}
\label{tab:sim2real_full}
\end{table}

\begin{figure}[htbp]
    \centering
    \includegraphics[width=0.8\textwidth, page=1]{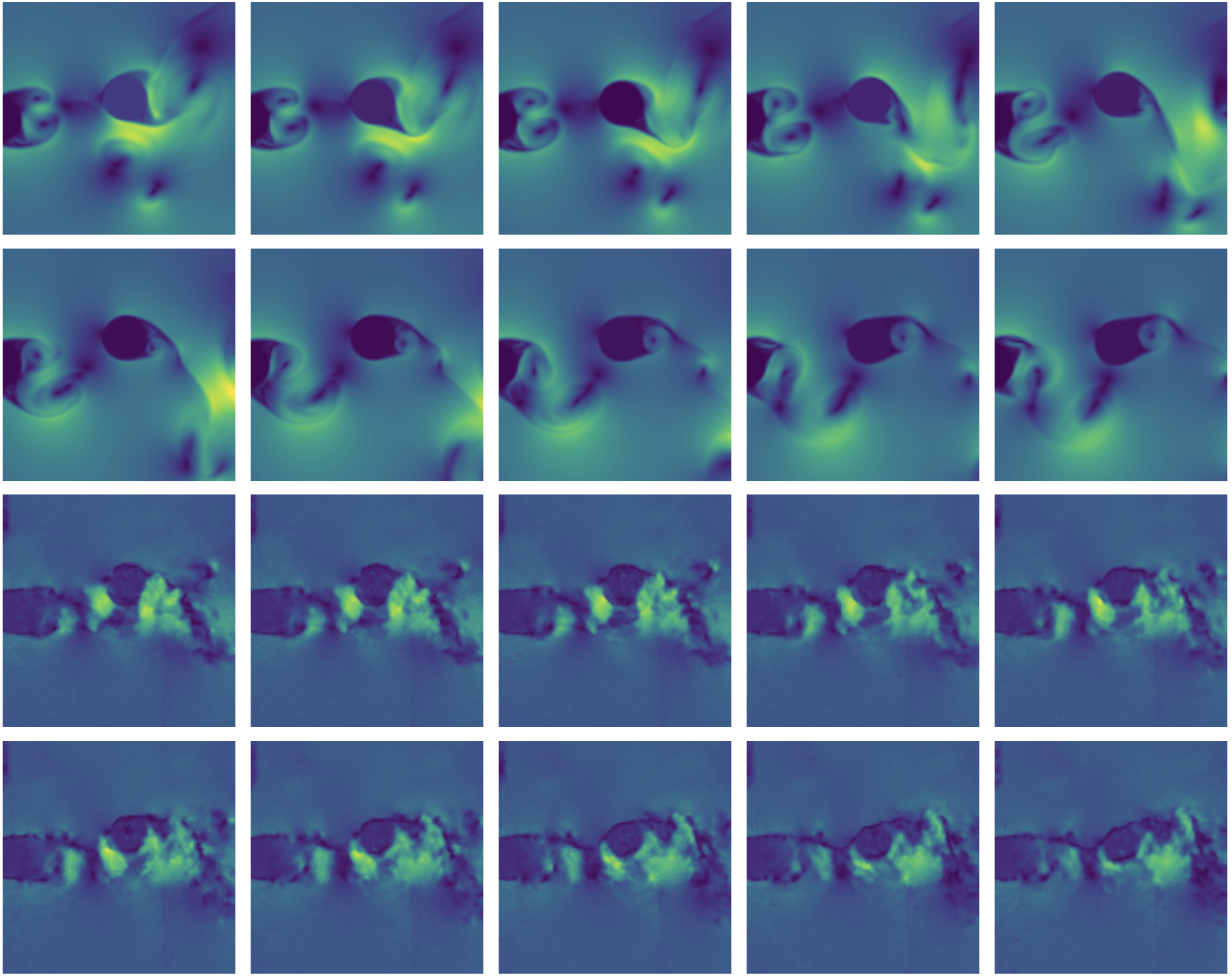}
    \caption{Visualization of the \SimToReal\ dataset. The top two rows present numerically simulated data for vortex-induced vibration in a tandem-cylinder arrangement, while the bottom two rows display real-world experimental data.}
    \label{fig:sim2real}
\end{figure}

\textbf{Visualization: } We refer the readers to Figure~\ref{fig:sim2real} for visualizations of the dataset.

\newpage
\clearpage

\subsection{\HighDimensional}
\label{appendix:highdimensional}
\textbf{Dataset Description: } This dataset contains gyrokinetic simulations of plasma micro-turbulence in a magnetized fusion plasma. The system evolves the perturbed gyrocenter distribution function on a five-dimensional phase-space grid comprising two velocity-space coordinates, the parallel velocity $v_\parallel$ and the magnetic moment $\mu$, and three spatial coordinates: the position $s$ along the magnetic field line and the two perpendicular directions represented spectrally by wavenumbers $k_x$ and $k_y$. As the turbulence develops, fine-scale structures in the distribution function couple across velocity space, the field-aligned direction, and the perpendicular spectral modes. Each trajectory records 266 time steps of the turbulent dynamics.

\textbf{Modeling Objective: } Let $x_{t} \in \mathbb{R}^{2 \times 32 \times 8 \times 16 \times 85 \times 32}$ denote the state of the system at time $t$. The dimensions are $(2, n_{v_\parallel}, n_\mu, n_s, n_{k_x}, n_{k_y})$, where the leading channel dimension of size $2$ holds the real and imaginary parts of the spectral distribution function. Our objective is to develop a neural simulator, defined as a parameterized function $\Psi$, that advances the system state by one time step. Formally, we aim to predict $\hat{x}_{t+1} = \Psi(x_{t})$, where the predicted output $\hat{x}_{t+1}$ retains the same shape $(2, 32, 8, 16, 85, 32)$ as the input state.

\textbf{Dataset Specific Challenges: } The state lives on a five-dimensional phase-space grid rather than the 2D or 3D spatial grids for which most neural PDE surrogates are designed, with a single state containing roughly 22.3 million degrees of freedom. The five axes are physically heterogeneous: two are velocity-space coordinates, one is a field-aligned spatial coordinate, and two are spectral (Fourier) coordinates that carry no translational structure in the usual convolutional sense. Consequently, standard architectural assumptions such as 2D convolutions, isotropic attention over pixels, and spatial FFT layers do not transfer directly and must be rethought for mixed position/velocity/spectral axes. Compounding this, the data is too large for in-memory processing, since each state occupies roughly 85 MB in float32, so training demands small batch sizes and memory-efficient architectures.

\textbf{Dataset Generation: } The trajectories are produced with GKW, a nonlinear gyrokinetic flux-tube  code that solves the perturbed 5D gyrokinetic equation. To keep the cost of generating a large dataset manageable, the simulations use the adiabatic-electron approximation, so only ion-temperature-gradient-driven turbulence is evolved and the electron distribution need not be solved; each nonlinear run is carried out on a 5D grid of resolution over the two velocity-space coordinates (parallel velocity $v_\parallel$ and magnetic moment $\mu$) and three spatial coordinates (the field-line coordinate $s$, the radial direction $x$, and the binormal direction $y$), with the field stored spectrally in $(k_x, k_y)$ and each snapshot carrying two channels for the real and imaginary parts of the ballooning transform. Four operating parameters are varied to span turbulent regimes---the ion temperature gradient $R/L_T \in [1, 12]$, the density gradient $R/L_n \in [1, 7]$, the safety factor $q \in [1, 9]$, and the magnetic shear $\hat{s} \in [0.5, 5]$. Each simulation is integrated for $31{,}920$ steps, averaged every $40$ steps and subsampled every third step to yield $266$ snapshots.

\textbf{Governing Physics: } The dataset is governed by the nonlinear gyrokinetic equation, the reduced kinetic description of magnetized-plasma turbulence. Starting from the 6D Vlasov equation for the distribution function $f(\mathbf{r}, \mathbf{v}, t)$ and transforming to guiding-center coordinates $(\mathbf{R}, v_\parallel, \mu, \theta)$, one exploits the scale separation between the Larmor radius $\rho$ and the background gradient length $L$ ($\rho \ll L$) and the low-frequency ordering (frequencies well below the cyclotron frequency) to gyroaverage over the gyrophase $\theta$. This removes one velocity dimension and leaves a 5D distribution function $f = f(k_x, k_y, s, v_\parallel, \mu)$ evolving over time. Its perturbed evolution, for each species, takes the split form
\begin{equation}
    \frac{\partial f}{\partial t}
    + 
        (v_\parallel \mathbf{b} + \mathbf{v}_D)\cdot\nabla f
        - \frac{\mu B}{m}\,\frac{\mathbf{B}\cdot\nabla B}{B^2}\,
        \frac{\partial f}{\partial v_\parallel}
    + 
        \mathbf{v}_\chi \cdot \nabla f
    = S,
\end{equation}
where $\mathbf{b} = \mathbf{B}/B$ is the unit vector along the magnetic field, $\mathbf{v}_D$ is the magnetic drift arising from gradients and curvature in $\mathbf{B}$, $S$ collects sources and collisions, and $\mathbf{v}_\chi = \tfrac{c}{B}\,\mathbf{b}\times\nabla\chi$ is the drift induced by the gyroaveraged generalized potential $\chi = \langle \phi - \tfrac{v_\parallel}{c}A_\parallel \rangle$. The nonlinear term $\mathbf{v}_\chi\cdot\nabla f$ is the $\mathbf{E}\times\mathbf{B}$ advection of the distribution by its own self-consistent potential: it is the central driver of turbulence and the most expensive term to evaluate, and it is precisely the physics that reduced (quasilinear) models omit. The system is closed self-consistently, with the electrostatic potential recovered from a velocity-space integral of $f$ through the gyroaveraging Bessel factor $J_0$,
\begin{equation}
    \phi = A \int J_0\, f \, \mathrm{d}v_\parallel\, \mathrm{d}\mu,
\end{equation}
so that the field feeding $\mathbf{v}_\chi$ is itself a moment of the state it advects. In practice the distribution is split into an equilibrium and a perturbation, $f = f_0 + \delta f$, with $\delta f$ separating into a passive adiabatic response $-\tfrac{Z\phi}{T}f_0$ and a non-adiabatic kinetic part that carries the irreversible, transport-producing dynamics; the adiabatic-electron approximation used here treats the electron response as purely adiabatic, leaving only the ion kinetic equation to be solved. The physically decisive quantity is the radial heat flux transported toward the reactor wall,
\begin{equation}
    Q = \int_C \int_v 2\,\phi\, f \, \mathrm{d}v_\parallel\, \mathrm{d}\mu\, \mathrm{d}x\, \mathrm{d}y\, \mathrm{d}s .
\end{equation}
A simulation first passes through a linear phase in which unstable $k_y$ modes grow, then saturates into a statistically steady turbulent state regulated by zonal flows---the $k_y = 0$ component of $\phi$---which shear apart eddies and set the final transport level.

\textbf{Experiment Results: } The complete experimental results are detailed in Table~\ref{tab:highdimensional_full}. For this analysis, RMSE serves as the only evaluation metric. Furthermore, due to the high dimensionality of the 5D data, there are no general-purpose baselines available for comparison.

\begin{table}[h]
\centering
\begin{tabular}{lc}
\hline
 & RMSE \\
\hline
GNOT         & -- \\
Transolver   & -- \\
Transolver++ & -- \\
LNO          & -- \\
AMG          & -- \\
\textbf{Agent} & $\mathbf{3.412 \times 10^{-4}}$ \\
\hline
\end{tabular}
\caption{Performance comparison on the \HighDimensional\ dataset. Best results are shown in \textbf{bold}.}
\label{tab:highdimensional_full}
\end{table}

\begin{figure}[htbp]
    \centering
    \includegraphics[width=0.8\textwidth, page=1]{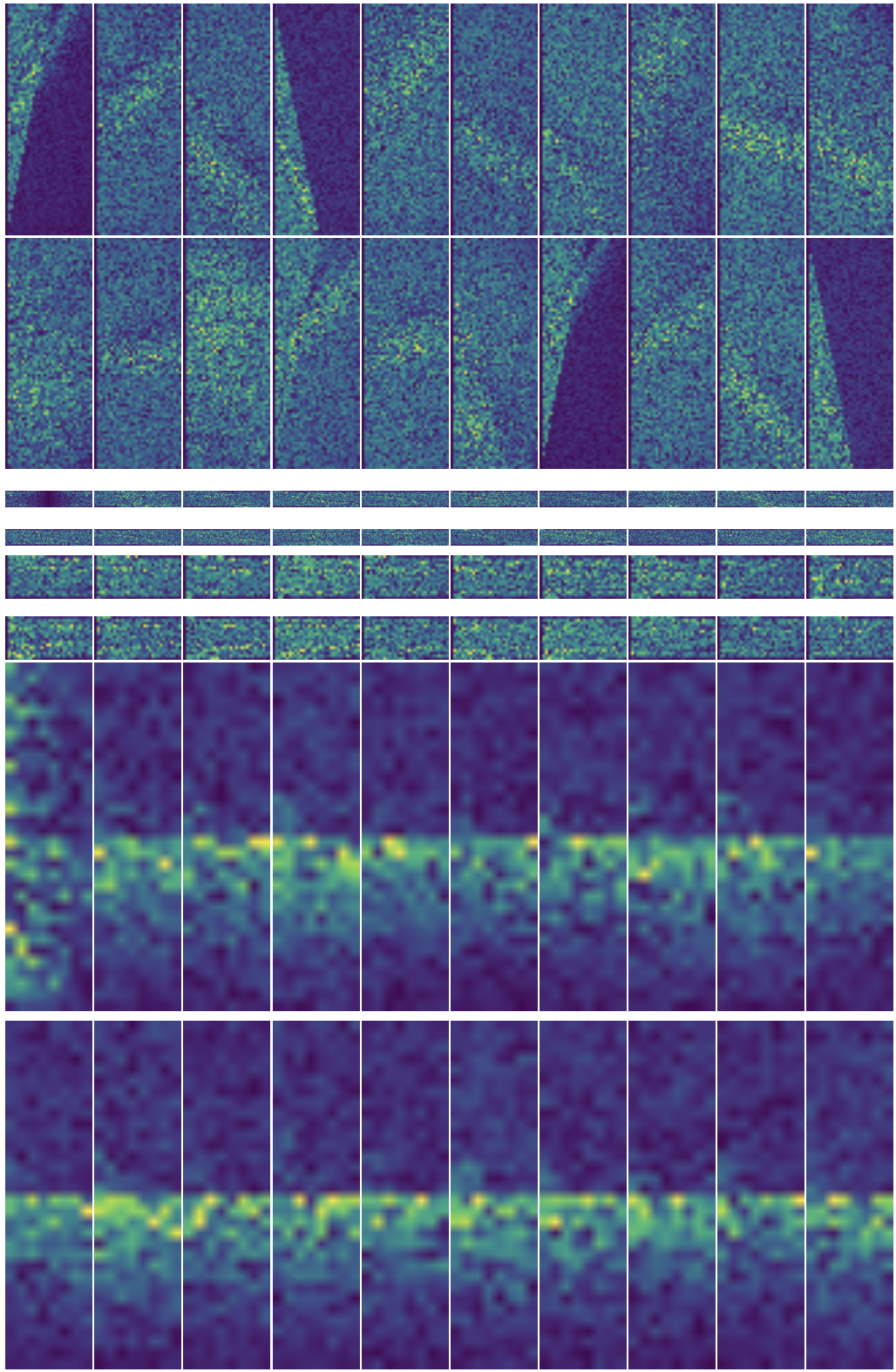}
    \caption{Visualization of the \HighDimensional\ dataset. This dataset contains high-dimensional gyrokinetic PDE data, represented as a complex-valued 5D array of the form \texttt{(2, nvpar, nmu, ns, nkx, nky)}. The first dimension denotes the real and imaginary components. The variables \texttt{nvpar} and \texttt{nmu} represent the two velocity-space coordinates (parallel velocity, $v_{\parallel}$, and magnetic moment, $\mu$), while \texttt{ns}, \texttt{nkx}, and \texttt{nky} represent the three spatial coordinates: the position $s$ along the magnetic field line and the two perpendicular directions represented spectrally by wavenumbers. In the figure, the first two rows display the \texttt{kx\_ky} cross-section, the next two show the \texttt{s\_kx} cross-section, followed by two rows for \texttt{s\_ky}, and the final two rows displaying the \texttt{vpar\_mu} cross-section.}
    \label{fig:highdimensional}
\end{figure}

\textbf{Visualization: } We refer the readers to Figure~\ref{fig:highdimensional} for visualizations of the dataset.

\newpage

\subsection{\ComplexGeometry}
\label{appendix:complexgeometry}
\textbf{Dataset Description: } This dataset contains steady-state aerodynamic simulations of turbulent airflow around 3D car geometries, derived from the ShapeNet ``car'' category. Each sample is an automobile shape immersed in an incompressible fluid stream. As air flows over and around the vehicle, it produces a pressure distribution across the car surface and a surrounding velocity field in the volume, including a downstream wake. The data is provided as an unstructured point cloud: exterior (volume) points sampled around the car and surface points sampled on the car body, with connectivity given as a graph.

\textbf{Modeling Objective: } Let $x \in \mathbb{R}^{N \times 7}$ denote the input features for a single car, where $N$ is the number of points (volume and surface). The seven input channels are the 3D point position $(x, y, z)$, the signed distance to the car surface ($\mathrm{sdf}$), and the 3D surface normal $(n_x, n_y, n_z)$. Our objective is to develop a neural simulator, defined as a parameterized function $\Psi$, that predicts the flow fields $\hat{y} = \Psi(x)$, where $\hat{y} \in \mathbb{R}^{N \times 4}$ has four output channels per point: the 3D fluid velocity $(v_x, v_y, v_z)$ and the pressure $(p)$.

\textbf{Dataset Specific Challenges: } The domain is a large, irregular 3D point cloud rather than a regular grid, with points spanning both the thin car surface and the surrounding volume at very different densities. Models must therefore handle this unstructured, multi-scale geometry while respecting the sharp distinction between on-surface and off-surface behavior.

\textbf{Dataset Generation: } We adopt the ShapeNet-Car dataset of \citet{umetani2018learning}, whose car geometries are drawn from the ``car'' category of ShapeNet~\citep{chang2015shapenet}. The flow around each car is computed with OpenFOAM, using its steady-state incompressible solver based on the SIMPLE algorithm (a segregated pressure--velocity coupling that iterates the Reynolds-averaged equations to a converged steady state) with a finite-volume discretization on an unstructured mesh. Each car is placed in a virtual wind tunnel with a uniform inflow of $20$~m/s (i.e. the $72$~km/h driving condition), a no-slip condition imposed on the car surface, and Reynolds-averaged turbulence closure to represent the effect of unresolved turbulent fluctuations. Overall, $889$ samples with distinct car shapes are simulated; the domain is discretized into an unstructured mesh with $32{,}186$ mesh points, and each simulation records the air flow field (velocity and pressure) in the volume surrounding the car together with the pressure distribution over the car surface.

\textbf{Governing Physics: } The ShapeNet-Car case is governed by the steady, incompressible Reynolds-Averaged Navier--Stokes (RANS) equations for the airflow around the vehicle. Decomposing the velocity into a mean and a fluctuating part, $\mathbf{u} = \bar{\mathbf{u}} + \mathbf{u}'$, and time-averaging the incompressible Navier--Stokes equations gives
\begin{equation}
    \nabla\cdot\bar{\mathbf{u}} = 0,
    \qquad
    (\bar{\mathbf{u}}\cdot\nabla)\,\bar{\mathbf{u}}
    = -\frac{1}{\rho}\nabla \bar{p}
    + \nu\,\nabla^2 \bar{\mathbf{u}}
    + \nabla\cdot\boldsymbol{\tau}^{R},
\end{equation}
where $\bar{\mathbf{u}}$ and $\bar{p}$ are the mean velocity and pressure, $\rho$ the air density, and $\nu$ the kinematic viscosity. The time-averaging leaves an extra term, the Reynolds stress $\boldsymbol{\tau}^{R} = -\overline{\mathbf{u}'\otimes\mathbf{u}'}$, which represents the mean momentum transport by turbulent fluctuations and must be modeled to close the system. Under the Boussinesq eddy-viscosity hypothesis it is related to the mean strain rate through a turbulent viscosity $\nu_t$,
\begin{equation}
    -\overline{u_i' u_j'}
    = \nu_t\left(
        \frac{\partial \bar{u}_i}{\partial x_j}
        + \frac{\partial \bar{u}_j}{\partial x_i}
    \right)
    - \frac{2}{3}k\,\delta_{ij},
\end{equation}
where $k$ is the turbulent kinetic energy and $\nu_t$ is supplied by a RANS turbulence model. The defining feature of this benchmark is the car itself as a bluff body immersed in the uniform stream $U_\infty$: the no-slip condition on the vehicle surface forces the flow to develop thin boundary layers, separate over the rear and other curved regions, and leave a turbulent wake, and it is this flow--geometry interaction that sets the pressure field on the body. The quantity of engineering interest is the aerodynamic drag, obtained by integrating the surface pressure and viscous (skin-friction) stress over the car surface $S$ and projecting onto the streamwise direction $\mathbf{e}_x$,
\begin{equation}
    F_d = \oint_{S}
    \left[
        -\bar{p}\,\mathbf{n}
        + \boldsymbol{\tau}_w
    \right]\cdot\mathbf{e}_x \, \mathrm{d}S,
    \qquad
    C_d = \frac{F_d}{\tfrac{1}{2}\rho\,U_\infty^{2}\,A},
\end{equation}
where $\boldsymbol{\tau}_w$ is the wall-shear stress, $A$ a reference frontal area, and $C_d$ the drag coefficient. Because $C_d$ depends on the full separated, turbulent flow established by the car's shape, accurately predicting it requires capturing how geometry variations reshape the surface pressure and wake---the core physical challenge the dataset poses.

\textbf{Experiment Results: } We present the complete experimental results in Table~\ref{tab:complexgeometry_full}. In addition to the volumetric RMSE reported in the main text, we provide the surface RMSE and the RMSE of the drag coefficient. The results indicate that the agent proposed architecture exhibits sub-optimal performance in this setting.

\begin{table}[h]
\centering
\begin{tabular}{lccc}
\hline
 & RMSE & RMSE (surface) & RMSE (drag) \\
\hline
GNOT         & $\bm{7.638 \times 10^{-2}}$ & $\bm{1.741 \times 10^{-1}}$ & $6.132 \times 10^{-3}$ \\
Transolver   & $8.295 \times 10^{-2}$ & $1.785 \times 10^{-1}$ & $\bm{5.690 \times 10^{-3}}$ \\
Transolver++ & $9.702 \times 10^{-2}$ & $2.044 \times 10^{-1}$ & $9.061 \times 10^{-3}$ \\
LNO          & $1.032 \times 10^{-1}$ & $2.058 \times 10^{-1}$ & $9.151 \times 10^{-3}$ \\
AMG          & $1.448 \times 10^{-1}$ & $2.728 \times 10^{-1}$ & $1.855 \times 10^{-2}$ \\
\textbf{Agent} & $1.366 \times 10^{-1}$ & $2.729 \times 10^{-1}$ & $1.855 \times 10^{-2}$ \\
\hline
\end{tabular}
\caption{Performance comparison on the \ComplexGeometry\ dataset. Best results are shown in \textbf{bold}.}
\label{tab:complexgeometry_full}
\end{table}

\begin{figure}[htbp]
    \centering
    \includegraphics[width=0.8\textwidth, page=1]{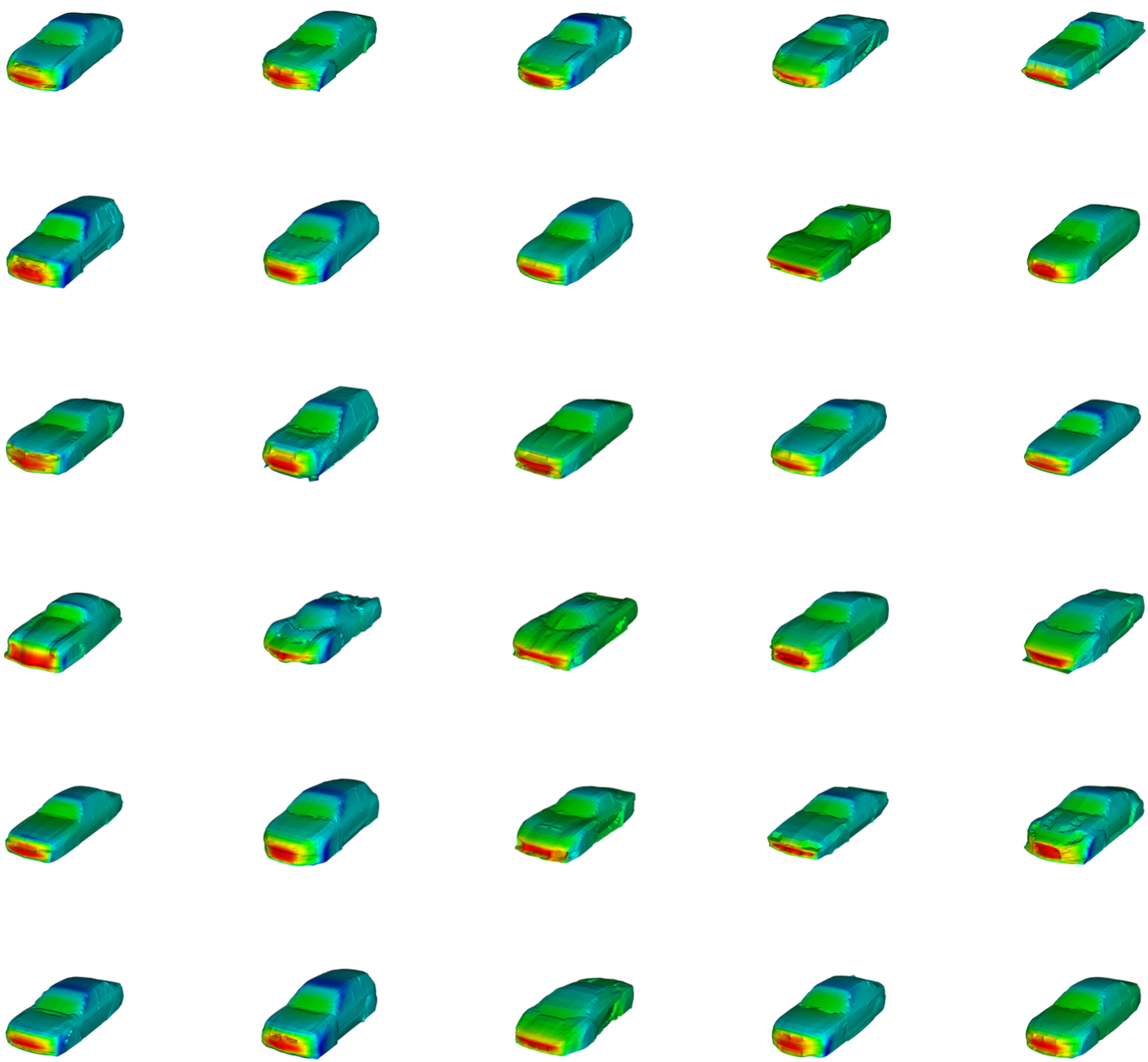}
    \caption{Visualization of the \ComplexGeometry\ dataset. This dataset comprises steady-state aerodynamic simulations of 3D car models. he visualizations show surface pressure, and the full data also includes volumetric velocity.}
    \label{fig:complexgeometry}
\end{figure}

\textbf{Visualization: } We refer the readers to Figure~\ref{fig:complexgeometry} for visualizations of the dataset.

\newpage

\end{document}